%% file: ms.tex
\documentclass[a4paper,11pt,phdthesis,oneandhalfspace,twoside,nocoursecode]{cssethesis}
\input{settings}

\begin{document}   
\renewcommand{\onlyinsubfile}[1]{} 

\pagenumbering{roman}                   

\thesistitle{Beyond Social Media Analytics: Understanding Human Behaviour and Deep Emotion using Self Structuring Incremental Machine Learning}
\thesisauthor{Tharindu Rukshan Bandaragoda}
\thesisauthorlastname{Bandaragoda}
\thesisdepartment{Research Centre for Data Analytics and Cognition}     
\thesisuniversity{La Trobe University\\Victoria, Australia}
\thesisauthoremail{T.Bandaragoda\@@latrobe.edu.au} 
\thesisdate{ 25} 
\thesismonth{February} 
\thesisyear{2020} 
\thesisauthorpreviousdegrees{MPhil (Monash University, 2015)\\ BSc. Eng. (Honours) (University of Moratuwa, 2011)}
  
\thesissupervisor{Prof. Damminda Alahakoon}
\thesissupervisoremail{D.Alahakoon\@@latrobe.edu.au} 
\thesisassocsupervisor{Dr Daswin de Silva} 
\thesisassocsupervisoremail{D.Desilva\@@latrobe.edu.au} 
\thesisdedication{To Amma and Thaththa}

\frontmatter					

\thesistitlepage
\thesiscopyrightpage

\thesisdedicationpage
    
\begin{thesisabstract}
\subfile{Abstract}
\end{thesisabstract} 

\thesisdeclarationpage

\begin{thesisacknowledgments}

\subfile{Acknowledgements}

\end{thesisacknowledgments}                            

\tableofcontents    
\listoftables
\listoffigures

\subfile{listpublications}
\newpage        


 

\mainmatter

\subfile{Chapter1}

\subfile{Chapter2}

\subfile{Chapter3}

\subfile{Chapter4}

\subfile{Chapter5}

\subfile{Chapter6}

\subfile{Chapter7}


\bibliographystyle{dcu}
\bibliography{library}{}

\end{document}

%% file: settings.tex
\usepackage{hyperref}
\usepackage{natbib}
\usepackage{bibentry}
\usepackage{pdfpages}
\usepackage[ruled,vlined]{algorithm2e}
\usepackage{titlesec}
\usepackage{csquotes}
\usepackage{setspace} 
\usepackage{amssymb}
\usepackage{amsmath}
\usepackage{epstopdf}
\usepackage{enumerate}
\usepackage{caption}
\usepackage{multirow}
\usepackage[group-separator={,},group-four-digits=true]{siunitx}
\usepackage{subfig}
\usepackage{subfiles}
\usepackage[titletoc]{appendix}
\usepackage{xr}
\usepackage{epigraph}
\usepackage{xspace}
\usepackage{fixltx2e}
\usepackage{tabulary}
\usepackage{textcomp}
\usepackage{xcolor}
\usepackage{color,soul}
\newcommand{\ikaslext}{IKASL\textit{text}\xspace}
\newcommand*{\rom}[1]{\expandafter\@slowromancap\romannumeral #1@}

\bibpunct{(}{)}{;}{a}{,}{,} 
\renewcommand{\cite}{\citep} 
\newcommand{\onlyinsubfile}[1]{#1}

%% file: Abstract.tex
Recent advances in mobile and social media technologies has led to a rapid increase in online social interactions among individuals from all walks of life. For instance, by 2019, 48\% of the world population were active on at least one online social media platform. Social media platforms archive all aspects of online social interactions. The data representing social interactions (or social data) can be used to analyse social behaviours and underlying causalities. Unlike the data collected conventionally from censuses or controlled experiments, this data is neither retrospectively collected nor controlled, and accumulated in large volumes leading to more accurate insights.

Existing techniques for gaining insights from social data are mainly adaptations of standard machine learning and natural language processing techniques. The performance of such techniques on social data is suboptimal as social data is highly unstructured, due to brevity and out of vocabulary terms, while also dynamic and bursty. Specialised techniques are required to handle issues related to the unstructured and dynamic nature of these data streams. This thesis aims to develop such techniques that can be used to gain in-depth insights from social data. 

In the first phase of the thesis, a conceptual framework has been developed considering social data as representing the surface layer of a hierarchy of human social behaviours, needs and cognition. This framework is subsequently employed to transform social data into representations that preserve social behaviours and their causalities. In the second phase existing machine learning and natural language processing techniques have been extended to overcome the challenges of social data. Two platforms were built to capture insights from fast-paced and slow-paced social data. For fast-paced, a self structuring and incremental learning technique was developed to automatically capture salient topics, and corresponding dynamics over time. An event detection technique was developed to automatically monitor those identified topic pathways for significant fluctuations in
social behaviours using multiple indicators such as volume and sentiment. The capabilities of this platform are demonstrated using two large datasets with over 1 million tweets each. The separated topic pathways were representative of the key topics/discussions of each entity and coherent against topic coherence measures. The identified events were validated against contemporary events reported in news.

Secondly for the slow-paced social data, a suite of new machine learning and natural language
processing techniques were developed to automatically capture self-disclosed information
of the individuals such as demographics, emotions and timeline of personal events of significance. This platform was trialled on a large text corpus of over four million posts collected from online support groups. This was further extended to transform prostate cancer related online support
group discussions into a multidimensional representation and investigated the self-disclosed
quality of life of patients (and partners) against time, demographics and clinical factors.
The capabilities of this extended platform have been demonstrated using a text corpus
collected from 10 prostate cancer online support groups comprising of 609,960 prostate
cancer discussions and 22,233 patients.

%% file: Acknowledgements.tex
	First and foremost, I would like to express my sincere gratitude to my primary supervisor Prof. Damminda Alahakoon for his continues support and guidance throughout my PhD journey. Thank you for all the inspiration, motivation, enthusiasm and immense wisdom shared with me. I could not have imagined having a better supervisor and a mentor to support my PhD journey.
	  
	\noindent Secondly, I would like to thank my co-supervisor Dr Daswin De Silva for his continues support guidance and feedback. Thank you for always encouraging me to write and publish; and keeping my research journey on track until the very end.  Also thank you for all the collaboration opportunities provided.
	 
	\noindent This work was supported by an Australian Government Research Training Program Scholarship, a La Trobe University Full Fee Research Scholarship and a La Trobe University Postgraduate Scholarship. Also, I would like to thank Data to
	Decisions Cooperative Research Centre (D2D CRC) for providing the top-up scholarships. 
	
	I would like to thank Dr Buddhi Jayathilaka for introducing me to this research group and providing guidance and support during the initial phase of my PhD. Moreover, I would like to make this an opportunity to thank Dr Weranja Ranasinghe,  A/Prof. Nathan Lawrentschuk and Prof. Damien Bolton for research collaboration and providing clinical expertise to explain the analytical findings of my research.  Also, I would like to thank Dr Evgeny Osipov and Luleå University of Technology, Sweden for research collaboration and supporting my research visit. 
	
	\noindent A big thank to Latrobe Business school and its support staff for all kinds of support.
	
	\noindent I would like to thank all my colleagues at the Center for Data Analytics and Cognition for all the support extended and the cherish memories. 
	
	\noindent I would like to express my sincer gratitude to my parents who always encouraged me to take this PhD journey. Thank you amma and thaththa for all the support and encouragement throughout. None of this would be possible without you. 
	
	\noindent I was so fortunate to have the unconditional love and support of my wife Chammi throughout my PhD journey. Thank you all your motivation, encouragements and tolerance. Also, I would like to thank my little daughter, Amelie, for her cuddles and smiles that kept me going during the latter part of my PhD.  
	
	\noindent Finally, I would like to thank everyone who supported me during my PhD journey.

%% file: listpublications.tex
	
\chapter*{List of Publications}
\subsection*{Journals}

\begin{enumerate}
  	\item T. Bandaragoda, D. De Silva, and D. Alahakoon. Automatic event detection in microblogs using incremental machine learning. \textit{Journal of the Association for Information Science and Technology}, 68(10):2394--2411, 2017
  	
  	\item T. Bandaragoda, D. De Silva, D. Alahakoon, W. Ranasinghe, and D. Bolton. Text mining for personalized knowledge extraction from online support groups. \textit{Journal of the Association for Information Science and Technology}, 69(12):1446--1459, 2018
  	
    \item T. Bandaragoda, W. Ranasinghe, A. Adikari, D. de Silva, N. Lawrentschuk, D. Alahakoon, R. Persad, and D. Bolton. The patient-reported information multidimensional exploration (PRIME) framework for investigating emotions and other factors of prostate cancer patients with low intermediate risk based on online cancer support group discussions. \textit{Annals of surgical oncology}, 25(6):1737--1745, 2018
    
    \item W. Ranasinghe, D. de Silva, T. Bandaragoda, A. Adikari, D. Alahakoon, R. Persad, N. Lawrentschuk, and D. Bolton. Robotic-assisted vs. open radical prostatectomy: A machine learning framework for intelligent analysis of patient-reported outcomes from online cancer support groups. \textit{Urologic Oncology: Seminars and Original Investigations}, 36(12):529.e1--529.e9, 2018
    
    \item D. De Silva, W. Ranasinghe, T. Bandaragoda, A. Adikari, N. Mills, L. Iddamalgoda, D. Alahakoon, N. Lawrentschuk, R. Persad, E. Osipov, et al. Machine learning to support social media empowered patients in cancer care and cancer treatment decisions. \textit{PLoS One}, 13(10):e0205855, 2018
    
    \item W. Ranasinghe, T. Bandaragoda, D. De Silva, and D. Alahakoon. A novel framework for automated, intelligent extraction and analysis of online support group discussions for cancer related outcomes. \textit{BJU International}, 120:59--61, 2017
    	
    \item D. Nallaperuma, R. Nawaratne, T. Bandaragoda, A. Adikari, S. Nguyen, T. Kempitiya, D. De Silva, D. Alahakoon, and D. Pothuhera. Online incremental machine learning platform for big data-driven smart traffic management. \textit{IEEE Transactions on Intelligent Transportation Systems}, 20(12):4679--4690, 2019
    
    \item M. Sherwood, M. Lordanic, T. Bandaragoda, E. Sherry, and D. Alahakoon. A new league, new coverage? comparing tweets and media coverage from the first season of AFLW. Media International Australia, 172(1):114--130, 2019
    
    \item T. Bandaragoda, A. Adikari,R. Nawaratne, D. Nallaperuma, A. K. Luhach, T. Kempitiya, S. Nguyen, D. Alahakoon, D. De Silva. Artificial intelligence based commuter behaviour profiling framework using Internet of things for real-time decision-making. \textit{Neural Computing and Applications}, 1-15, 2020.
  \end{enumerate}

\subsection*{Conference proceedings}
\begin{enumerate}
	\item T. Bandaragoda, D. De Silva, D. Kleyko, E. Osipov, U. Wiklund, and D. Alahakoon. Trajectory clustering of road traffic in urban environments using incremental machine learning in combination with hyperdimensional computing. In 2019 IEEE Intelligent Transportation Systems Conference (ITSC), pages 1664--1670. IEEE, 2019
	
	\item R. Nawaratne, T. Bandaragoda, A. Adikari, D. Alahakoon, D. De Silva, and X. Yu. Incremental knowledge acquisition and self-learning for autonomous video surveillance. In IECON 2017-43rd Annual Conference of the IEEE Industrial Electronics Society, pages 4790--4795. IEEE, 2017
\end{enumerate}

%% file: Chapter1.tex
\onlyinsubfile{
\tableofcontents 	
}
	
\chapter[]{Introduction}\label{chap:1}

\epigraph{{\textit{Man is by nature a social animal; an individual who is unsocial naturally and not accidentally is either beneath our notice or more than human. Society is something that precedes the individual. Anyone who either cannot lead the common life or is so self-sufficient as not to need to, and therefore does not partake of society, is either a beast or a god.\\}}{\hfill Aristotle, Politics (\textasciitilde 384 BC)}}	

This chapter initiates the thesis by postulating an expansive context for its research questions, objectives and contribution. From the early origins of human societies to contemporary habituation of digital environments, the manifestation of social behaviours and interactions has eventuated in diverse schools of thought and distinct domains of knowledge. The persistence of social data in digital environments has significantly transformed this scholarly pursuit. This thesis is an endorsement, empowerment and advancement of that pursuit. Let us begin. 

\section{Background}
Humans are social animals. Social interactions continue to be paramount for modern humans, just as it was the determinant of survival of our first ancestors against natural selection, the Hominin hunter-gatherer groups (1.8-1.3 million years ago)~\cite{charles1859origin}. Over the course of human civilisation, the level of sophistication of social interactions, as well as the number of social interactions among humans have advanced from simple survival to complex social needs. 
In modern humans, social interactions are controlled by cognitive processes in the brain, which perceive the social environment (social actions of others) based on sensory information and existing knowledge. Evolution has led to biological adaptations in brain, sensory and motor neurons to optimise our social interactions in the natural world~\cite{Kock,Kock2005}. For instance, human ears are more sensitive to human voices than any other acoustic stimuli~\cite{nass2005wired} and the human brain has a high volume ratio for the neocortex (relative to the entire brain) in order to handle the computational demand of  social interactions~\cite{Dunbar1998}. Moreover, a study on cognitive skills among chimpanzees, orangutans, and two year-old children uncovered that although there is no significant difference in cognitive skills for physical activities, human children possess superior cognitive skills for social activities~\cite{herrmann2007humans}.
\\\\
Social behaviour is a goal directed action to achieve a perceived goal~\cite{ajzen1985intentions}. Social behaviours are formally defined as abstractions of different types of human social interactions that are executed to achieve similar perceived goals~\cite{ajzen1985intentions}. Some examples of social behaviours are interpersonal communication, self disclosure, cooperation and social comparison. For instance, expressing emotions  and discussion of political views are distinct social interactions. However, both interactions disclose different types of information i.e., emotions and political views about oneself to others in society. Therefore, such interactions can be abstracted the behaviour of self-disclosure. Developing an intricate understanding of social behaviours is crucial for all domains of knowledge and academic disciplines. This expands across humanities, social sciences and biological sciences more directly, and indirectly into physical sciences, formal sciences and applied sciences.   
\\\\
Conventional studies of human social behaviours rely on \textit{surface data}: limited information on a large number of individuals or \textit{deep data}: extensive information on few individuals~\cite{manovich2011trending}. \textit{Surface data} is the population level data collected using censuses which collects data about individuals, households and businesses. Such datasets are well structured and often analysed using statistical methods.~\textit{Deep data} is collected from a small group of individuals using controlled and natural experiments~\cite{neuman2013social}. Such datasets are detailed and unstructured, and analysed using qualitative methods. Both these data collection approaches have merits and long served as the key approaches to study social behaviours. However, the population level \textit{surface data} is often too shallow with few attributes about each individual, thus leads to over-generalised insights that are only valid at the population level. Also, such studies are costly and time consuming, hence, not frequent (e.g., US census is every 10 years).\textit{Deep data} from social experiments are limited to few individuals and mostly obtained using controlled environments. As pointed-out by~\citet{Meshi2015}, the major issues of \textit{deep data} are lack of generalisability to a larger population, lack external validity due to differences in controlled experiment environment and natural environment, and various biases in data collection such as \textit{recall bias}.  
\\\\
Given these inherent limitations in conventional methods of studying human social behaviours, it is timely that recent technological advancements have collectively led to the invention and development of new digital environments for enriched social interactions, that also encapsulate the full spectrum of social behaviours. It is opportune that each interaction and behaviour is digitally represented, captured and archived to facilitate  much more meticulous approaches to this field of study.

\begin{figure}[!htb]
	\centering
	\includegraphics[clip=true, width=1.0\linewidth]{{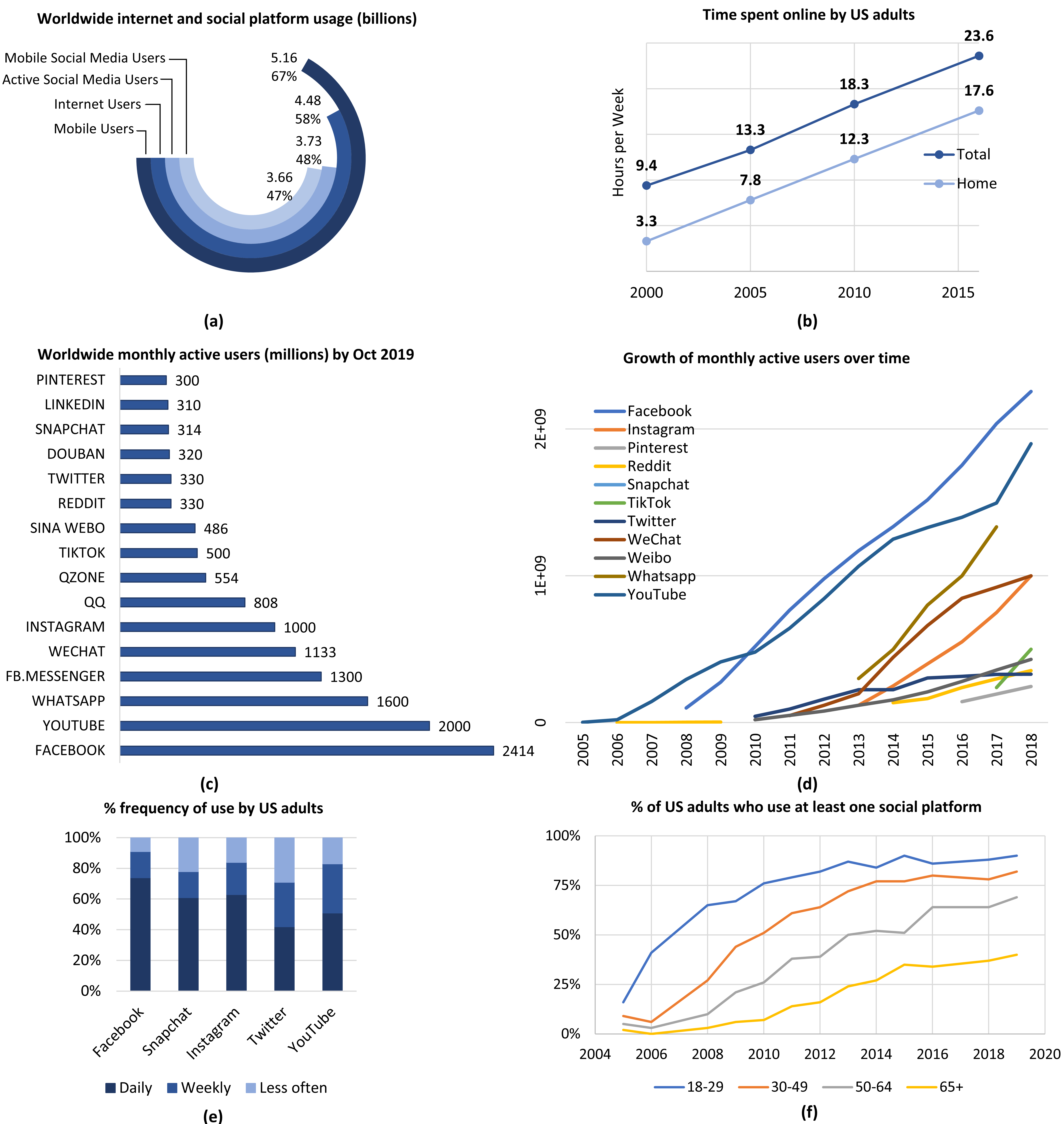}}
	\caption[Recent trends in internet and social media platform usage]{Recent trends in internet and social media platform usage: \\(a) World internet and social media platform usage by fourth quarter of 2019 (source:~\cite{kemp2019global}), \\(b) Time spent online by US adults from 2000 to 2016 (source:~\cite{cole2017surveying}), \\(c) Worldwide monthly active users (MAU) in top online social media platforms by fourth quarter of 2019 (source:~\cite{kemp2019global}), \\(d) Growth of worldwide monthly active users (MAU) in top online social media platforms from 2005 to 2018 (source:~\cite{kemp2019global}), \\(e) Percentage frequency use of different online social media platforms by US adults in 2019 (source:~\cite{SocialFactSheetPew2019}, and \\(f) Percentages of US adults by age group who use at least one social media platform from 2005 to 2019 (source:~\cite{SocialFactSheetPew2019}))
	}
	\label{fig:social_platform_stats}
\end{figure}  

\section{Social media}

These (established and emerging) digital environments are widely referred to as social media or social media platforms. 
Technological developments in affordability, availability and ease of use have reduced the digital divide, leading to an exponential increase in social media platforms for social interactions. As shown in Figure~\ref{fig:social_platform_stats}.a in 2019, worldwide mobile usage was 67\% and internet usage was 58\% which highlights worldwide adaptation of mobile and internet technologies~\cite{kemp2019global}. 
A recent survey of US adults~\cite{cole2017surveying} shows that the average weekly internet usage has risen from 9.4 hours in 2000 to 23.6 hours in 2016 (Figure~\ref{fig:social_platform_stats}.b). This increase was mainly due to domestic internet use for leisure and social activities, which has risen five-fold over the past 16 years~\cite{cole2017surveying}. 
Although the functionality of modern online social media platforms are diverse, and often specialised to facilitate the demands of different socio-demographic segments, they are in general web and mobile based platforms that facilitate users to create, share and co-create multimodal (text, image, audio and video) user generated content as well as engage in online social discussions with other members on the platform~\cite{Kaplan2010,Kietzmann2011}. 

\subsubsection{A vibrant history}
Usenet (an online discussion system started in 1979) and Bulletin Board Systems (a dial-up internet based computer resource sharing system introduced in the late 1970s) are the likely precursors of modern day online social media platforms. The first recognisable online social media platform (SixDegrees.com) launched in 1997~\cite{Boyd2008} which allows the users to create profiles, friend lists and browse the profiles of their friends. As people started to flock into the internet in the early 2000s, there has been a rapid increase in the launch of diverse online social media platforms. As shown in Figure~\ref{fig:social_platform_stats}.a in 2019, out of the 7.73 billion of world population, 3.73 billion individuals are active users of at least one social media platform, which accounts for 48\% of the world population. 
Figure~\ref{fig:social_platform_stats}.c depicts the worldwide Monthly Active User (MAU) counts in October 2019 of the top social media platforms. It shows that Facebook has 2.414 billion MAU which means 31\% of the world population are active Facebook users. Similarly, video sharing platform YouTube has 2.0 billion MAU (25.9\% of world population). Most popular messenger applications are WhatsApp (1.6 billion MAU), FB.Messenger (1.3 billion MAU), and WeChat (1.1 billion MAU). Popular microblogging platforms are QQ (808 million MAU), Sina-Weibo (486 million MAU) and Twitter (330 million MAU). This massive scale of active users highlights the popularity of online social media platforms and the increasing need for online social interactions, across the world. Moreover, Figure~\ref{fig:social_platform_stats}.d shows that all those online social media platforms have been launched in last 15 years and managed to get a massive number of active users within a very short period of time. For instance, TikTok a video sharing platform launched in 2017 has reached 500 million MAU by end of 2019. Furthermore, as shown in Figure~\ref{fig:social_platform_stats}.e, most of those active users engage with the social media platforms on daily basis. For instance in US, percentage of daily active users are close to 75\% for Facebook, over 60\% for Instagram, over 50\% for YouTube and over 40\% for Twitter. This phenomenon highlights that most of the active users use social media platforms frequently to participate in various online social activities.   

\subsubsection{A diversity of social behaviours}
The social interactions on these online social media platforms are highly diverse representing the variety of interests of the users. In addition to those common online social media platforms, social forums (or fora), which are also known as  online support groups, self-help groups are a further innovation in this space. These are organised as discussion threads, where each thread starts with a question, comment or experience and other users of the forum respond to it. Such forums are mainly devoted for specialised areas such as health (WebMed, Patient.info), recreation (TripAdvisor, Yelp, Zomato), academia (Academia.edu), data science (Kaggle), and photography (EyeEm). Moreover, similar specialised social groups are often formed within mainstream online platforms such as Facebook groups,  sub-reddits (in Reddit), LinkedIn groups to social interact with individuals with specific criteria (e.g, occupation, hobby, medical condition).   
\\\\
As online social media platforms continue to expand and develop, individuals and societies have become increasingly intrigued by the unprecedented opportunities provided to  interact, share and collaborate globally. Even older individuals are increasingly using online social media platforms. For example, according to an annual survey in US by Pew Research~\cite{SocialFactSheetPew2019} (Figure~\ref{fig:social_platform_stats}.f) use of online social media platforms are increasing rapidly among older age groups, and as of 2019 40\% 65+ adults (was 2\% in 2005 a 20 fold increase) and 69\% among age 50-64 (was 5\% in 2005, a 14 fold increase) use any form of social media platforms. In contrast, the increase was less for 18-29 and 30-49 age groups. It was a 5.6 fold increase (16\% in 2005 to 90\% in 2019) for 18-29 age group, and a 9 fold increase (9\% in 2005 to 82\% in 2019) for 30-49 age group.    
\\\\
Face-to-face interactions in the physical world that served as the main outlet of social behaviours during most of human history is gradually giving way to increased online interactions that occur on online social media platforms.  This trend was apparent in  a recent study~\cite{wallsten2013we} based on data from the American Time Use Survey (ATUS), which shows that the time spent socialising online has increased at the expense of the time spent on other activities such as face to face interactions, sleep, and travel.  
Initially, this trend of increased online social interactions was seen as a negative impact on the social life of individuals which lead to social isolation by decreasing physical social interactions~\cite{kraut1998internet,nie2002internet}. However, later research points that although offline socialisation activities are degraded, online socialisation increased which leads to similar or even better overall social outcomes~\cite{valenzuela2009there,ellison2007benefits,wellman2001does}.  In fact, online social interactions overcomes some barriers of physical social interactions such as (i) \textit{geographic proximity} where the individuals have to be in close proximity, (ii) \textit{synchronicity} where interactions have to timely,  and (iii) \textit{accessibility} where some individuals and groups are not accessible to others~\cite{McFarland2015}. 
\\\\
However, online social media are \textit{lean mediums} in contrast to the \textit{rich medium} of face-to-face communication which allows the transmission of multiple communication cues~\cite{daft1987message}. In particular, non-verbal communication cues are not transferable through  most of the online social media platforms, excepts for those that allow video based interactions.  This claim is disputed by other studies which argue that without non-verbal communication cues, humans are capable of adapting to utilise the available cues in \textit{learner mediums} more effectively with prolonged use~\cite{Walther1992,Kock,Tidwell2002a}. For instance, in text-based mediums emoticons are used to expresses specific emotions and elongated terms (e.g., \lq soooo\rq) are used to emphasise the meaning of the base term. 
\\\\
Although a variety of social interactions (that represent a number of social behaviours) take place on online social media platforms, the baseline profile of salient social interactions is generally focused on the characteristics of each social media platform. Such characteristics of the social media platform are sometimes \textit{by design} to intentionally encourage certain type of interactions, or may have emerged as a consequence of frequently having certain type of user interactions over others. For instance, mainstream social media platforms such as Twitter are fast-paced social media platforms, that are used for rapid dissemination of current and trending information through the social network using functions such as \textit{re-tweeting} and \textit{sharing}~\cite{Kwak2010,Lotan2011,lovejoy2012engaging}, which may even spread faster than seismic waves during an earthquake~\cite{Sakaki2010}.  Due to this focus of information dissemination rather than having continuing discussions, the ties between dyads are relatively weaker (unless known by other means), and often lacks homophily, reciprocity and interpersonal self-disclosure. Also, the emotions expressed in \textit{tweets} are often shallow and intense, clearly leaning towards  being positive or negative. In contrast, forums/online support groups such as online health support groups are slow-paced with relatively smaller groups where the participants has high degree of homophily (e.g., the patients of a specific medical condition). These social media platforms are used to seek and provide, informational and emotional support~\cite{preece1999empathic}. Thus, users tend to publish longer posts providing high degree of self-disclosure and express complex emotions.  

\subsubsection{Digital representations of social data}
Social data can be broadly defined as any form of information that represents and characterises social interactions. Every social interaction releases information into the environment in which it  occurs, which can be both explicit or implicit as well as intentional or unintentional to the participants of the interaction. For instance, in a face-to-face conversation in the physical environment, some of the information that characterised the interaction are  the theme of the conversation, duration, location, emotions/opinions expressed, facial expressions of the  participants, tone variation of the conversation, and the appearance of participants (e.g., what they wear). In addition, there can be pre- and post-interaction information that characterises what leads to the interaction and what follows.  
\\\\
Social data from social interactions in the physical world (mainly face-to-face communications) is seldom recorded. The interactions in the physical environment are generated by different forms of energy transformations which eventually dissipate into the environment without leaving any traces. A cognitive representation of social interaction exists in the memory of the participants, which gradually fades with time and often leads to inaccurate accounts~\cite{schacter1998cognitive}. Also, the recollection of such interactions are subjected to various forms of recall biases yielding a distorted recollection~\cite{kahneman1999objective}.  
\\\\
In contrast, almost all aspects of online social interactions are recorded. For instance, in a social conversation on Facebook or Twitter, its content (text, emojis, shared images or videos), duration, geolocation of participants and other actions (e.g., shares, likes, re-tweets) are recorded in the digitised form. In addition, pre- and post- online interactions of the individuals are recorded and each participant has a self-authored profile which often contains their socio-demographic information. All these information are stored in massive databases of online social media platforms, and often backed up as multiple copies. Theoretically, those digitised records of the online social interactions live indefinitely on the Internet. For instance, there are currently over 500 billion \textit{tweets} archived in Twitter.

\section{Motivation}

Building upon this digest of social interactions, social behaviours, digital environments, and the social data accumulating on social media platforms, it is now pertinent to delineate the motivation of this thesis. 
\\\\
Conventional methods of studying human social behaviours are impacted by the challenges of collecting relevant data, in sufficient volumes. A distinctive opportunity lies within the massive amount of social data accumulated in online social media platforms. As noted earlier, social data are the archived records of online social interactions, thus, provides an unparalleled  aperture into understanding various human social behaviours as well as underlying causalities. This data is contributed by a large number of individuals often in the scale of millions, thus they are in par or even larger than the population level datasets collected in censuses. Also, prolonged social interactions of individuals in online social media platforms generate sufficient information to conduct an in-depth analysis of their social behaviour. 
\\\\
Existing computational approaches of using these large volumes of social data are mainly adaptations of standard machine learning and natural language processing techniques. Table~\ref{table:application_socialdata} shows some example uses of social data in different applications alongside the techniques employed and social media platforms used for data collection. These applications mainly use sentiment analysis techniques~\cite{liu2012survey} to evaluate  public opinion, topic extraction techniques~\cite{aggarwal2012mining} to identify salient discussion topics and entity extraction techniques~\cite{ritter2011named} to identify related entities (e.g., people, brands, locations) involved. 
\\\\
These adaptations of standard machine learning and natural language processing techniques demonstrate two high-level fundamental limitations when utilised for the exploration and understanding of social behaviours. Firstly, the constrained focus on basic insights of social data, when there are deep insights on human social behaviour that are encapsulated but largely untapped due to the challenging nature of pre-processing, analysis and synthesis. For instance, the shallow focus on surface-level meaning, frequency, and basic sentiment, and lack of consideration for deeper insights such as emotions, psycholinguistics, and personal traits, that provides a better representation of the underlying social behaviours in social data. 
Secondly, sub-optimal computational performance ~\cite{Eisenstein2013} due to the highly unstructured nature of social data~\cite{Aggarwal2012b}. The dynamic and bursty nature of social data~\cite{Barabasi2005}, requires techniques that are time sensitive and can learn incrementally from a dynamic data stream, in contrast to conventional machine learning techniques which are mostly designed for static datasets.
\\\\

\begin{table}[!htb]
\caption{Applications of social data based on recent literature.}\label{table:application_socialdata}
\begin{tabulary}{\linewidth}{|>{\raggedright}p{2cm}|>{\raggedright\arraybackslash}p{12cm}|}\hline  
Application	& Technique (social media platform) \\ \hline 

Natural disaster detection& \citet{Sakaki2010}- event detection (Twitter)\newline ~\citet{middleton2014real}- event detection, clustering (Twitter)\newline ~\citet{abel2012twitcident}- event detection, entity extraction (Twitter)  \\ \hline 

Planned protest prediction &  ~\citet{Ramakrishnan2014}- event detection, entity extraction (multiple sources) \newline ~\citet{becker2012identifying}- event detection, entity extraction (multiple sources) \\ \hline

First story detection & ~\citet{osborne2012bieber}- event detection, entity extraction (Twitter, Wikipedia) \newline ~\citet{lau-etal-2012-line}- topic extraction (Twitter) \\ \hline

Traffic incident detection & ~\citet{pan2013crowd}- anomaly detection, graph analysis (Weibo) \newline ~\citet{d2015real}- event detection, classification (Twitter) \\ \hline

Sales forecasting & ~\citet{Rui2013}- sentiment analysis, regression (Twitter) \newline ~\citet{mishne2006predicting}- sentiment analysis (Blogs)\\ \hline

Stock market prediction& ~\citet{si2013exploiting}- sentiment analysis, topic extraction (Twitter) \newline ~\citet{das2007yahoo}- sentiment analysis, classification (stock message boards)\\ \hline

Public opinion prediction & ~\citet{Tumasjan2010}- sentiment analysis (Twitter) \newline ~\citet{bollen2011modeling}- sentiment analysis (Twitter) \\ \hline

Public health insights & ~\citet{Huber2017}- thematic analysis (online support groups) \newline ~\citet{paul2011you}- topic extraction (Twitter) \\ \hline
\end{tabulary} 
\end{table}

On motivation, this thesis is firstly inspired by the unprecedented opportunity to comprehensively study human social behaviours represented in social data accumulated on online digital social media platforms. Secondly, it is stimulated by the limitations of current computational and artificial intelligence approaches in addressing the scientific, semantic and technical challenges of online social data. And finally, the ambition of common good and contribution to human society through the transformation of novel artificial intelligence approaches into technology platforms that deliver actionable insights for societal advancement.


\section{Research objectives}
This thesis develops an understanding of social behaviours through the study of online social data using novel computational and artificial intelligence approaches that model, transform, analyse and generate insights on social behaviours and underlying causalities. The research objectives are:

The first objective is to explore existing theories on human social behaviour, social needs and cognition; and develop a conceptual framework to understand social behaviours through representative online social data. This conceptual framework considers social data as representing the surface layer of a  hierarchy of human social behaviour, needs and cognition. 

The second objective is to advance new self-structuring artificial intelligence approaches to address the challenges and limitations of existing machine learning and natural language processing techniques in the study of social behaviours based online social data. This exploration will focus on self-structuring and incremental learning techniques to represent, adapt to and evolve with text based social data streams, and automatically monitor those structures for changes in social behaviours and causes.

The third objective is to transform the aforementioned approaches into technology platforms that capture insights from text based social data of different online social media streams and use such insights to gain descriptive understandings and predictive insights for decision-making.   

\section{Research questions}
Based on the above research objectives, the following research questions will be addressed in this thesis.
\begin{enumerate}
	
	\item \textit{What are the limitations of existing artificial intelligence algorithms and natural language processing techniques in the study of social interactions and social behaviours using representative online social data in digital environments?}
	
	
	\item \textit{How can theories of social behaviour from social sciences contribute towards a conceptual model of enhanced understanding of social interactions in digital environments, as well as the representative online social data?}  
	

	\item \textit{How can new incremental machine learning algorithms, founded on the principles of self-structuring artificial intelligence, address the challenges of using social data to understand social behaviours?}
	

	\item \textit{How can the research contributions that address research questions 2. and 3. be formulated into a technology platform that delivers actionable insights for societal advancement?} 
	
	 
\end{enumerate}
\section{Research contributions}
Based on the above research objectives and research questions, this thesis yields the following contributions as the outcomes:
\begin{enumerate}
\item A comprehensive investigation on current state-of-the-art machine learning and natural language processing techniques employed to generate insights from social data followed by an exhaustive analysis on their limitations due to distinct challenges present in social data.  

\item A conceptual framework based on existing social theories to explicate the complex relationships between social behaviours and online social data. 

\item Development of two novel self-structuring artificial intelligence algorithms for generating insights from online social data. (1) A new unsupervised, self-structuring and incremental learning technique to structure a dynamic and diverse text based social data stream to automatically capture salient topics, and their dynamics over time. 
(2) An automated, intelligent event detector to monitor the identified topic pathways for significant fluctuations in social  behaviours using multiple indicators such as volume and sentiment to detect social events of interest. 

\item Demonstration of the two novel self-structuring artificial intelligence algorithms on two large microblogging data streams.  

\item Transformation of the two new algorithms, along with an ensemble of related machine learning and natural language processing techniques, into a technology platform. This platform was further adapted for slow-paced online social data, to  automatically capture and incrementally learn self-disclosed information. 

\item Extending the above mentioned platform into an application domain with high social impact, patient-centred healthcare. The  platform was demonstrated to better facilitate the diverse information needs of its stakeholder groups: consumers, researchers and health professionals. 

\item Further materialising the patient-centred healthcare technology platform for prostate cancer related online support groups, with real-life outcomes that have advanced clinical knowledge of patient needs and expectations. 
This was demonstrated using a text corpus collected from 10 prostate cancer online support groups comprising of 609,960 prostate cancer discussions and 22,233 participants. 
\end{enumerate}

\section{Thesis outline}
\begin{figure}[!h]
	\centering
	\includegraphics[clip=true, width=0.6\linewidth]{{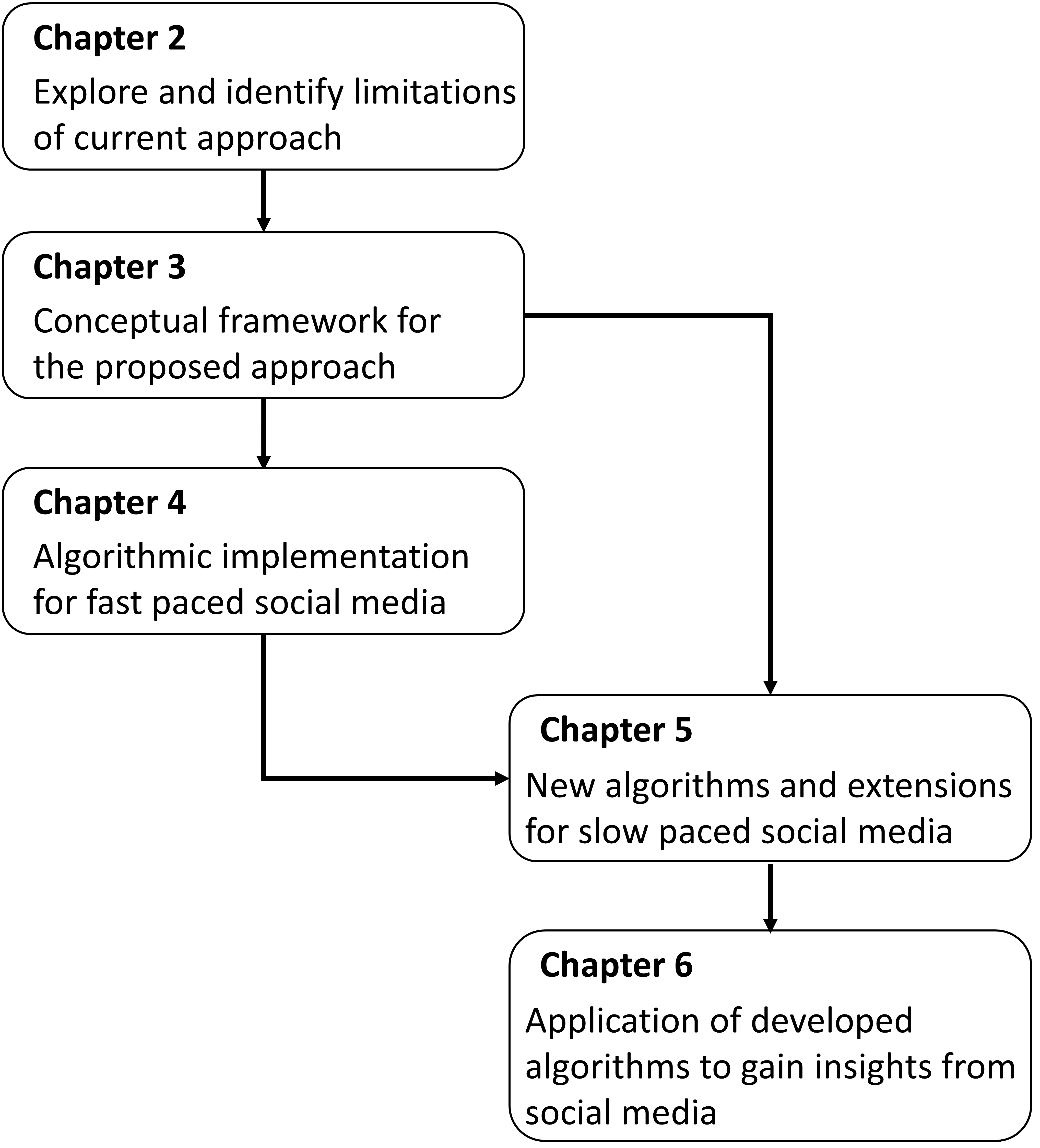}}
	\caption{Chapter structure of the thesis}
	\label{fig:thesis_structure}
\end{figure}  

Figure~\ref{fig:thesis_structure} shows how this thesis is organised.

Chapter two discusses the key literature on machine learning and natural language processing techniques that are used to capture insights from textual sources which includes text clustering, topic modelling, event detection, sentiment and emotion capturing, self-structuring and incremental learning. Subsequently, it presents the major challenges such techniques would encounter when applied to social data. 

Chapter three presents the proposed conceptual framework for understanding the underlying mechanisms that generate social data using the existing physiological and social theories on cognition, social needs and social behaviour as the foundation. This conceptual framework is then extended to a platform that can be employed with artificial intelligence techniques to represent, self-structure and learn insights from social data. 

Chapter four presents two novel techniques (i) a self structuring and incremental learning  technique that is capable of automatically structuring a text based social data stream into topic pathways over time, and (ii) an automatic event detection technique that automatically monitors the above topic pathways  for behavioural changes using multiple change detectors. These techniques were trialled using two large Twitter datasets and demonstrated  their capabilities to identify salient topic pathways, track their evolution over time, detect  new topic pathways, and  detect significant events based on sentiment and volume based event indicators.    

Chapter five presents the development of information structuring platform for online support group platforms to better facilitate the diverse information needs of its stakeholder groups: consumers, researchers and health professionals. It presents the details of a set machine learning and natural language processing techniques to capture self disclosed information such as demographics, emotions, psycholinguistics. The capabilities of this platform are demonstrated using a large text corpus extracted from two large online health support groups. 

Chapter six presents the application of the techniques developed in chapters four and five to gain insights from prostate cancer related online support group discussions. In addition it discusses the development of a specialised set of technique to capture treatment type, side effects, treatment decision making behaviour and numeric pathological information specific to prostate cancer from the self disclosed free-text. It then showcases the capabilities of the captured multidimensional representation (prostate cancer related information, topics/topic pathways, demographics, emotions) to analyse the self disclosed quality of life against the other factors.

Chapter seven concludes the thesis by providing a summary of work presented in the preceding chapters, how the research questions formulated above were addressed in the body of work and ideas for future research based on this thesis.

\onlyinsubfile{	
\bibliographystyle{dcu}
\bibliography{library}{}
}

%% file: Chapter2.tex
\onlyinsubfile{
\tableofcontents 	
\setcounter{chapter}{1} 
}

\chapter[]{Literature review and reflections}\label{chap:2}

\epigraph{{\textit{If I have seen further it is only by standing on the shoulders of giants.\\}}{\hfill Sir Isaac Newton (1675)}}

This chapter reviews existing machine learning and natural language processing techniques that are relevant to harness insights from social data. Also, it discusses the limitations and gaps of such techniques to effectively capture insights from social data. 

This chapter is organised as follows: Section~\ref{sec:manual_assessment} reports on manual assessment of social data and its limitations; Section~\ref{sec:topic_extraction} reviews the techniques for topic extraction and text clustering; Section~\ref{sec:event_detection} reviews event detection from text data focussing on both specified and unspecified event detection techniques; Section~\ref{sec:emotion_extraction_from_text} reviews emotions theories, models and sentiment/emotion extraction techniques from text; Section~\ref{sec:self_structuring} discusses the self-structuring techniques covering biological inspiration and its computational adaptations; Section~\ref{sec:incremental_learning} reviews techniques developed for unsupervised incremental learning;
and Section~\ref{sec:social_data_analysis_challenges} discusses the challenges and limitations of existing techniques in relation to gaining insights from social data. 

\section{Manual assessment of social data}\label{sec:manual_assessment}
Social research using social data is a relatively emerging field of study that was developed in conjunction with the rapid growth of social media platforms and their use. During the initial stages, social data was analysed manually by the researchers. Such attempts include (i) manually coding social messages based on the topic discussed in the content, (ii) manually assessing of the sentiment polarity and strength, and (iii) manual inference of socio-demographic information of the authors of social data. 

The key advantage of manual assessment of social data is that it does not require challenging technical skills. Also, humans are skilled at decoding the complex social expressions such as sarcasm which are often present in social data. However, personal judgement on social data varies across assessors as such judgements are impacted by the personal experience of each assessor. Such variations specifically occur in the assessment of sentiment or emotions expressed in social data. These variations were often minimised by using multiple assessors to assess the same social data and producing an aggregated assessment. Also, statistical agreement measures such as kappa coefficient ($\kappa$) are measured to evaluates the inter assessors agreement. 

However, the key issue with the manual assessment of social data is the associated human cost that limits its applicability to small social datasets often in the scale of hundreds of data points. Use of crowd sourced manual assessment techniques such as Amazon Mechanical Turk~\cite{buhrmester2011amazon} enables relatively inexpensive manual assessment possible. However, even such approaches can only assess social datasets in the scale of thousands. One of the key utility of the manual assessment is for the generation of reference assessment results which are often known as \textit{gold standard records} that are used as the reference social data points for training and evaluation of the machine learning based models to automatically learn the underlying patterns of inferences made by human assessors.

%

\subfile{Chapter2-ml-techniques}

\subfile{Chapter2-sentiment-emotion}

\subfile{Chapter2-so-il}

\section{The formidable challenges to social data analysis}\label{sec:social_data_analysis_challenges}
Previous sections discuss the existing machine learning and natural language processing techniques that are used to gain insights from social data generated in online social media platforms. As discussed within the above sections gaining insights from social data is highly challenging. Such challenges come from several avenues: (i)  scale of data generation, (ii) time sensitivity, (iv) diversity and (ii) the unstructured nature of data.  

\subsection{Scale of data}
As discussed in the previous chapter, over the past decade, online social media platforms have been increasingly embraced by hundreds of millions of people worldwide. Moreover, users are more frequently active in major social media platforms mainly due to the availability of mobile devices.  

These large number of active users in prominent social media platforms generate large volumes of new social data continuously resulting in a high velocity data stream. The analysis of data streams of this scale requires specialised computational approaches that are optimised to handle large volumes of data. 

\subsection{Unlabelled data}
As discussed in previous sections, supervised machine learning techniques were mainly employed in classification tasks such as sentiment, topic and event classification. Such tasks were carried out using two types of techniques: (i) dictionary based approach where a term dictionary is engineered to represent the concepts of each class to be classified and (ii) model based approach, where a labelled training dataset is employed to learn a predictive model using supervised machine learning techniques. The first approach requires a term dictionary or a lexicon and the second approach requires a labelled dataset with relevant classes. 

Since social data is not structured with any taxonomy or explicitly labelled, it does not carry any labels that are needed for such supervised or semi-supervised learning tasks. There are a handful of human labelled social data samples, however, such samples are disproportionate and not representative enough when considering the scale and variation in social data.

There are two key approaches to handle this issue. The first is to use unsupervised learning techniques which do not require any form of supervision. Unsupervised techniques leverage underlying patterns in data to separate them into groups based on similarity. Current use of unsupervised techniques in text (mainly topic extraction) is covered in Section~\ref{sec:text_clustering}.  

Another approach to overcome this issue is to use folksonomies or user generated tags as a proxy for labels. For example, (i) 20 Newsgroups~\cite{joachims1996probabilistic} which is used for topic classification is taken from a news discussion forum where the news category is used as the label, (ii) Movie Review Dataset~\cite{pang2005seeing, pang2008opinion} used for sentiment classification is a set of user created movie reviews on the movie review website Rotten Tomatoes~\footnote{www.rottentomatoes.com} in which the labels were obtained based on the user provided star rating for the movie.

\subsection{Time sensitivity}
Social data can be considered as a near real-time data stream which gets updated by freshly generated content from its users. For instance, Facebook users frequently publish status updates about how they feel and Twitter users frequently tweet their opinion about their current issues of interest. 

Moreover, as pointed out by~\citet{Aggarwal2012b} these social data streams are not uniformly distributed, but bursty in nature. These bursts are mostly due to current events of interest that have disrupted~\cite{Sakaki2010a,Zhou2014a} or captured the attention~\cite{Becker2011} of a significant swath of human population. Hence, it is apparent that social data is tightly coupled with the time of its publication and that time sensitivity needs to be considered for any analytic task.

Most of the above discussed conventional machine learning algorithms require the dataset to be a priori which is not possible in social data streams as new patterns appear in the data streams as it progresses over time. Therefore, machine learning algorithms need to be extended to incrementally learn from the data stream as new patterns appear over time.   

\subsection{Diversity}
Diversity in social data is an important feature that is often overlooked in other studies. Social data is diverse due to multitude of reasons. It contains traces of multiple social behaviours. Also, even the same behaviour could be actioned in different depths and intensities. Moreover, at the highest granularity, contrasting differences among different individuals may appear due to individuality resulting from differences in socio-demographics. 

All these reasons induce diversity to social data. For instance, social data consists of diverse linguistic patterns, a large number of discussion topics and differences in expression of emotion.~\citet{Eisenstein2014} found that there are distinct linguistic patterns exist in Twitter among groups of similar geographic proximity as  well as similar socio-demographics. This diversity in social data act as noise on patterns that exist in social data. Hence, the machine learning techniques have to first reduce the noise due to diversity by separating diverse social data into coherent groups which can then be used to extract insights.  

\subsection{Unstructured nature of data}
Social data mainly consists of unstructured text posted by users in online social media platforms. This text corpora of social data are substantially different from the professional discourse (found in news articles and other formal documents) in many aspects such as brevity, lack of syntactic structure and use of out-of-vocabulary terms. 

\subsubsection{Brevity} 
Short length of social data partially due to restrictions imposed by some social media platforms (e.g., tweet is limited to 140 characters). Even in platforms without such restrictions, users prefer brevity since it allows them to express more efficiently. This brevity is mostly achieved by using shortened forms of terms (e.g., \textit{u} for \textit{you}), and relaxation of grammaticality~\cite{Baldwin2013}. 
This brief nature of social data result in highly sparse feature representations that is challenging for conventional natural language processing techniques.

\subsubsection{Lack of syntactic structure}
Discourse in documents often follows certain syntactic structure which is driven by a set of language rules that governs the formation of clauses, phrases and sentences. Such rules were often established through repeated documentation over time. However, the discourse in social data loosely follows such syntactic structures. This is mainly because although a written language social data is the outcome of social conversations where individuals prefer to make it more relaxed from syntactic structures.

The above discussed differences in social data to any other formal discourse are highly challenging most of the conventional natural language processing techniques that rely on such language rules to capture different constructs from the discourse such as noun, noun phrases, co-references etc. Since those elementary constructs are then feed into other techniques to capture complex insights such as sentiment, these errors due to lack of syntactic structures often impacts most of the conventional natural language processing techniques. 

\subsubsection{Out of vocabulary terms} 
Users tend to coin new terms during social conversations in online social media platforms. Such terms are mostly developed as social tagging or folksonomy highlighting important aspects i.e., topics or emotions in the conversation text. Hashtags used in Twitter is a good example of such terms, where users create \textit{hashtags} to represent certain events or topics (e.g., \textit{\#WinterOlympics}). Another type of new terms is constructed by repeating certain characters of standard terms (e.g., \textit{cooolll}) which are mostly used to emphasise the expressed emotion~\cite{Brody2011}. Use of such out-of-vocabulary terms is a challenge as such terms are often not included in thesauruses that are often used by the natural language processing techniques to derive certain properties of each word (e.g., sentiment dictionaries). Filtering out such terms not an option as they are tightly coupled to the intended meaning of the post.        

From the above examples, it is apparent that as is application of natural language techniques which are designed for standard discourse on social data would yield sub-optimal outcomes~\cite{Baldwin2013}. Such techniques have to be significantly extended to capture different above discussed aspects of social data.

\section{Chapter Summary}
This chapter has conducted a comprehensive literature review on current state-of-the-art machine learning and natural language processing techniques employed to generate insights from social data. It focused around three types of insights often captured from social data which are topics, events and emotions. Topic capturing includes topic modelling to model topic distributions and text clustering to capture topically coherent clusters. Event detection is twofold where specified event detection is used to capture events with known prior information and unspecified event detection is used to capture events with unknown or partially known information. The emotion detection approaches use emotion models based on physiological and cognitive theories of emotion.  

As highlighted in the review most of the techniques employed are adaptations and extensions of machine learning and natural language processing algorithms developed for conventional text datasets such as news articles. However, as discussed in Section~\ref{sec:social_data_analysis_challenges} social data comes with formidable challenges that are less prevalent in conventional text datasets due to its scale of data generation, time sensitivity, unstructured nature and diversity. Hence this approach leads to suboptimal outcomes.

The next chapter proposes an alternative approach to this conventional approach of social media analytics. This proposed novel approach goes beyond considering social data as just another data source by considering social data as traces of online human social interactions. It enables social data to be represented by the underlying drives of human social interactions which can then be transformed to generate more meaningful insights. 

\onlyinsubfile{	
\bibliographystyle{dcu}
\bibliography{library}{}
}

%% file: Chapter2-ml-techniques.tex
\section{Topic extraction}\label{sec:topic_extraction}
Topic extraction (or mining) is one of the key areas in natural language processing which attempts to organise a large set of documents (e.g., news articles or tweets) into semantically meaningful groups which are often denoted as \textit{topics} or \textit{themes}. Each identified topic is represented as a set of keywords that semantically represents a particular topic. 

Topic extraction is mainly handled as an unsupervised task as in most of the scenarios underlying topics in a document corpus is not a priori knowledge. Two key unsupervised approaches for the topic extraction are (i) text clustering and (ii) topic modelling. 

\subsection{Text clustering}~\label{sec:text_clustering}
Clustering is the classic data mining task of finding groups of similar records in the data based on a similarity measure. Similarly, text clustering intends to separate the text corpus into groups with similar meaning~\cite{Aggarwal2012}. It is assumed that a set of documents clustered together contains a  common topic/theme/subject which is represented by the salient terms of that cluster. Text clustering techniques have been applied to organise documents~\cite{cutting1992scatter, kohonen2000self}, organise web search results~\cite{zamir1997fast}, and generate an abstractive summary of a given corpus~\cite{schutze1997projections}.

Text clustering consists of two key phases~\cite{larsen1999fast}. First the documents are converted into representative feature vectors and then clustering techniques are employed to cluster those feature vectors into groups based on an appropriate distance function. These two phases are required because the clustering techniques are designed for datasets with real-valued features. Hence, the text corpus has to be transformed into representative real-valued feature vectors.

Document feature vectors are often obtained based on the bag-of-words (BOW) concept~\cite{jurafsky2000speech} which simplifies a document as a bag of words, relaxing word order based relationships. Based on BOW, documents are transformed into feature vectors using vector space model (VSM)~\cite{salton1975vector} where each term in the corpus vocabulary is a feature and the feature value is the term weight. Feature values of a document could be simply set to binary indicating presence or absence of a respective term in the document. However, term weighting techniques based on statistical specificity of a term are employed to apply higher feature values to more specific terms which have higher information value. A popular such weighting scheme is \textit{term frequency inverse document frequency} tf-idf~\cite{salton1988term,sparck1972statistical} which gives high feature value to terms that exist more on a given document (proportional to term frequency) and less across the corpus (inverse document frequency). The size of the vocabulary often grows with the corpus size resulting in very high-dimensional feature vectors. Therefore, document frequency based pruning techniques are used to remove less informative terms which are either common terms among most of the documents or rare terms that exist in a small set of documents~\cite{salton1975vector}. 

The techniques employed in text clustering are either extended or adapted from the clustering techniques used for quantitative data such as k-means~\cite{lloyd1982least}, self-organising-maps (SOM)~\cite{Kohonen1998}, and hierarchical clustering~\cite{ward1963hierarchical}. Scatter/Gather~\cite{cutting1992scatter} one of the well-known text clustering algorithm which uses hierarchical clustering to determine the cluster centres and further refines them using k-means clustering technique.~\citep{larsen1999fast} proposed a variant of k-means which uses a damping technique during the centre adjustment that improves the cluster quality.~\citep{steinbach2000comparison} proposed a \textit{bisecting} k-means variant that iteratively clusters the largest cluster (initially the entire dataset) into two clusters using k-means until \textit{k} clusters are generated.~\citep{kohonen2000self} employs self-organising-maps to cluster a large collection of patent abstracts which uses randomly projected document vectors to reduce dimensions.

\subsection{Topic modelling}
Topic modelling is another approach widely employed to extract salient topics from a document corpus. It employs a probabilistic generative model to represent documents in a corpus. Topic models estimate probabilities for each document being generated by the identified topics, where the most salient topics of a document can be chosen based on a threshold. It means a document could belong to multiple topic clusters. This is in contrast to the previously discussed clustering approaches which exclusively partition a set of documents into topic clusters resulting in a single topic for a document. Having multiple topics is more intuitive for larger documents which may contain multiple topics.    

Topic models assume that the corpus is generated by $N$ number of \textit{latent} topics, and each document $d_i$ has a probability $P(t_j|d_i)$ for belonging to each topic $t_j$. Similarly, each term $w_k$ in the corpus vocabulary $V$ being part of a topic $t_j$ is given by the probability $P(w_k|t_j)$. Both topic-document $P(t_j|d_i)$ and term-topic $P(w_k|t_j)$ cannot be directly estimated as topics are \textit{latent} and the only relationship that can be estimated is the term-document probability $P(w_k|d_i)$ for a term $w_k$ and document $d_i$ which is given by the term frequency of $w_k$ in $d_i$ divided by number of terms in $d_i$. Therefore, the topic-document and term-topic probabilities are estimated based on its relationship with term-document probability given bellow:

$$P(w_k|d_i) = \sum_{j=1}^{N} P(w_k|t_j) \times P(t_j|d_i)$$

Probabilistic Latent Semantic Analysis (PLSA)~\cite{hofmann1999probabilistic,hofmann2001unsupervised} is one of the initial generative topic models, which used expectation maximization (EM) algorithm to estimate the probabilities  of the topic-document and term-topic relationships. Initially, the probabilities are randomly initialised and then iteratively conduct \textit{expectation} (E) and maximization (M) steps. E-step estimates the posterior probabilities of latent variables given the current values of them and term-document probabilities. Then M-step estimate new values for the probabilities based on the probabilities of latent variables estimated in E-step. These two steps alternatively carried out until the likelihood converges to a local maximum. The key parameter to PLSA is the number of topics $N$ that the generative model is optimised. The key limitation of PLSA is its tendency to overfit when estimating large number of parameters.  

The most popular variant of probabilistic topic models is the Latent Dirichlet Allocation (LDA)~\cite{Blei2012} algorithm. LDA closely follows PLSA but, is a complete generative model which models topic-document and term-topic probabilities using Dirichlet distribution as priors using hyperparameters $\alpha$ and $\beta$ respectively. $\alpha$ controls the mixture of topics in a document and higher values lead to higher topic-document probabilities i.e. more topics are assigned to a document. $\beta$ controls the term-topic probabilities and high values lead to higher term-topic probabilities i.e., more terms are assigned to a topic. In contrast to PLSA, LDA has less number of parameters, thus less susceptible to overfitting  problem. Moreover, LDA is capable of estimating the topic distribution of a new document that was not present during learning.      

\subsection{Topic extraction from social data}~\label{sec:topic_extraction_social_data}
The two key challenge for topic extraction techniques on social data is the high sparsity and time sensitivity. Text data is often sparse, however, social data is overly sparse mainly due to brevity and diversity. Social media posts are often left brief as it allows authors to construct them efficiently and with less concern about the flow of it. On the other hand, social data is constructed by many individuals who use diverse writing styles and discuss diverse topics resulting a large vocabulary across a corpus of social data. Both these factors contribute to sparse document feature vectors (from \textit{vector space model}) containing only a handful of activated features. This increased sparsity makes topic extraction techniques susceptible to noise, degrading their performance. 

The topics discussed in social data changes over time as the interests of the author changes. Existing topics drop in popularity while new unseen topics emerge over time. Therefore, the topic mixture changes dynamically over time. However, both text clustering and topic modelling techniques assume a static mixture of topics thus have to be extended to facilitate changes in topic mixture.

Sparsity issue is often addressed using various aggregation strategies to increase the document size used for topic extraction.~\citet{Rosen-Zvi2004,rosen2010learning} proposed the Author-Topic model. It is a generative model similar to LDA, except that, instead of topic-document relationship modelled in LDA, topic-author relationship is modelled in Author-Topic model. Hence, it estimates the topic-author association probabilities yielding a set of topics discussed by each author. This approach addresses the issue of sparsity as it learns a topic mixture for an aggregated corpus of social data of each author, thus has been utilised for topic mining in social data~\cite{tang2008arnetminer,Hong2010a}. However,~\citet{Xu2011} pointed out that this strategy would only work for authors who are consistently discussing about a set of topics, and also it does not account for changes of authors interest over time.

~\citet{yan2013biterm} present \textit{biterm topic model} which builds a generative model to represent topic to word-pair (biterm) relationship. The assumption here is that if two words co-exists frequently it is likely that those two words are part of the same topic. This model is claimed to more robust to brevity in social data as it does not model the topic-document relationship which tends to suffer more from sparsity issues.   

~\citet{Mehrotra2013} attempted different aggregation strategies such as aggregate tweets based on hashtags, authors and time-window (aggregate hourly tweets). They empirically showed that hashtag based aggregating is the most effective for LDA based topic extraction. However, the granularity of hashtag based aggregation varies widely based on the usage of each hashtag. For example, \#obama is used to discuss a wide spectrum of topics, so aggregating tweets of that hashtag might not yield meaningful topics. Another approach to the sparsity problem is to use relevant external resources to extend social data.~\citet{Banerjee2007} and~\citet{hu2009exploiting} incorporate text in related Wikipedia articles with the short text to generate joint document feature vectors to reduce sparsity before applying text clustering techniques. Similarly,~\citet{petrovic2012using} employed Wordnet~\cite{Miller1995} ontology to enrich short text with paraphrases. These approaches are significantly time and resource extensive as they have to query a big knowledge graph for each text document, making them expensive to use in the scales of social data.

Topic modelling techniques are also extended to incorporate time sensitivity of text data.~\citet{wang2006topics} present \textit{topic over time} (ToT) which is an extension to the LDA topic modelling algorithms to consider time in the generative model. The time-topic relationship added to the generative model of LDA with document-topic and term-topic relationships. The time-topic relationship is modelled using a beta distribution as the prior in contrast to document-topic and term-topic relationships modelled using Dirichlet priors. However, in ToT time is employed only to get a better fit to a fixed topic mixture and the topic trends such as topic evolution and new topic are not captured. Moreover, it is not trained in incremental fashion to use in a text stream. 

~\citet{alsumait2008line} and~\citet{lau2012line} present \textit{online} versions of LDA (OLDA) which are incremental versions of LDA, to facilitate topic mining from a text stream. These online variants take time-sliced input of a text stream and incrementally update a LDA topic model at each time-slice. They use the LDA model generated from text corpus at $t-1$ time-slice as the prior to learn the LDA model at $t$. This process generates an evolutionary matrix over each topic to capture the evolution of that topic over time. Based on this evolutionary matrix, the topics that are significantly different at $t$ in contrast to $t-1$ are marked as an outlier or new topics.  
  

\section{Event detection}\label{sec:event_detection}
An \textit{event} can be broadly defined as an occurrence that can be bounded by space and time~\cite{allan1998line}. Events could happen anytime and anywhere which could impact a single person, a handful of individuals, or a large number of individuals (e.g., natural disaster, election). Events alter the behaviour of the individuals impacted.    

An event in a social media platform is due to either a real world event (e.g., natural disaster, election) discussed in the social media platform or it can be originated from the social media platform itself without a link to a real world event (e.g., a controversial tweet from an influential author). An event often results a change of online social behaviour of the impacted individuals for the duration of the event (e.g., change of discussion topics, changes of emotions). Such changes due to an event often happen in near-realtime as a result of frequent use and availability of mobile devices. For instance, during natural disasters such as  earthquakes and floods, related discussions instantly start to flow in social media platforms with firsthand individuals accounts, opinions, emotions and arguments~\cite{Java2007,Zhao2009}. 

Identifying these events and also analysing the related online social behaviour yield valuable information for individuals, corporations, and law enforcement agencies. Corporations increasingly monitor social media platforms to understand customer opinions, concerns and sentiment on their product portfolio; and to provide insights for business decisions~\cite{Jansen2009,Pak2010}. Social media platforms can be further monitored to identify breaking news on events reported in realtime by firsthand individual accounts~\cite{Phuvipadawat2010, Sankaranarayanan2009}. Also, social media platforms are monitored for events that may lead to public unrest~\cite{Ramakrishnan2014}.

Event detection techniques can be broadly separated into \textit{unspecified} and \textit{specified} events~\cite{atefeh2015survey} on the event type. Unspecified event detection is employed when prior information about the event is not available, and specified event detection is employed when details about the event looking for is partially known.

\subsection{Unspecified event detection}\label{sec:unspecified_event_detection}
Unspecified events are events with no prior information available. Such events can range from extreme events such as natural disasters or terrorist act, unplanned political events such as a controversial statement from a key politician, and unplanned corporate events such as product malfunction or service outage. Since no prior information is available unspecified events can only be detected by monitoring the relevant social data streams for significant temporal changes. Therefore, unspecified events are mostly detected using unsupervised techniques.  

Detecting and tracking unspecified events in a stream of news stories has been comprehensively studied during the DARPA funded Topic Detection and Tracking (TDT) research program~\cite{allan1998topic}. Both~\citet{allan1998line} and~\citet{yang1998study} employed single pass text clustering techniques to cluster a time-ordered news article stream where each cluster is denoted as an event. Each news article is checked for its membership in an existing cluster and if it significantly different such articles forms a new cluster i.e., new event.     

However, event detection in social data streams is more challenging than the detection tasks in traditional media. As discussed in Section~\ref{sec:topic_extraction_social_data}, in contrast to the professionally composed news articles, social data is unstructured, short and written with diverse styles. In addition, often social data streams are several magnitudes higher in scales compared to news steams. 

Events with a significant public interest results increased number of related discussions by the individuals which lead to appear as sudden bursts~\cite{Kleinberg2003,yang1998study} in social data streams. Those discussion starts as bursts and then diminished over time with a heavy tail as public loses the interest on that. The discussions in social data streams are driven by events, thus can be considered as a mixture of event related discussions happening at different time windows. Detecting such bursts of activity is often the key to detecting events from a social data stream.

Initial burst detection techniques look for bursty keywords or tag words (hashtags in Twitter)~\cite{atefeh2015survey} with the hypothesis that such burst keywords are indicative of an event related to the topic inferred by the meaning of those keywords.~\citet{Mathioudakis2010} developed a queuing technique to detect \textit{bursty} terms from tweets and subsequently clustered the identified such terms into topics/events.~\citet{fung2005parameter} modelled the individual word distributions over time as binomial distributions and identified the bursty terms. Those terms with sufficiently overlapping active time windows are grouped together and identified as events.~\citet{he2007analyzing} applied Discrete Fourier Transform (DFT) and searched for bursty terms in the frequency domain which appears as a spike in the frequency spectrum. However, the time window of the event cannot be retrieved from the frequency domain.~\citet{Weng2011a} developed \textit{Event Detection with Clustering of Wavelet-based Signals} (EDCoW) which applied the Wavelet Transform to word distributions, filters out the trivial terms based on autocorrelation, and then employed a graph clustering technique to group the event related terms into events. ~\citet{Sankaranarayanan2009} trained a Naive Bayes classifier to classify tweets as news or junk, which was trained on term vectors of a hand-labelled tweet dataset as junk or news. The identified \textit{news} tweets are then online clustered to identify the salient topics among the news tweets.

Although change of volume based signals are widely employed for event detection, there are other changes that could happen to a social data stream due to an event. An event can significantly alter the opinion or emotions expressed in a discussion~\cite{tan2014interpreting,Thelwall2011}. For example, a social media post by a politician on a certain issue could lead to a backslash with heavy criticism resulting significant increase in negative opinion and expression of emotions such as \textit{anger} or \textit{sarcasm}. Detecting such opinion or emotions related signals, supplement the volume based events detections, as well as provides more detailed insights about the genre of the event.

~\citet{Paltoglou2015} developed a sentiment based event detection technique that monitored the temporal variation of sentiment polarity of tweets with particular keywords (e.g., hashtags) and identified instances with significant changes of sentiment as events that are associated with that keyword. The events captured from this technique was shown to be comparable to the events captured from a frequency based approach. One key limitation of this technique is the assumption that there is a one-to-one mapping between keywords and topics, whereas in social data streams there can be multiple keywords associated with topics and vice-versa. 

\subsection{Specified event detection}\label{sec:specified-event}
Specified event detection is the detection of known events and its related information such as location, time, participants, outcomes etc.~\cite{atefeh2015survey}. In contrast to unspecified events, the key terms that denote or associated with the event(s) are \textit{a priori} information and can be employed to detect them. Hence, in contrast to the unsupervised change detection techniques used for unspecified events, specified events are mainly detected using supervised techniques which uses either an engineered feature dictionary or labelled datasets to train a classifier. Also, unlike the change detection techniques which requires a substantial change in social data stream from large scale events, specified event detection techniques can be employed to detect events at different scales, from large events such as riots, earthquakes to individual events such as report of an illness from a personal social data stream.

Specified event detection on different types of text sources has been studied in different application domains. For instance, in financial domain, financial documents and business news are mined for finance related events such as catastrophic financial events~\cite{cecchini2010making}, mergers and acquisitions~\cite{lau2012web}, and competitor relations~\cite{ma2011mining}. Similarly, social data streams are mined for mentions of relevant entities (e.g., organisation names, brands, movie titles) and their associated mood (or sentiment) which can be used to determine market volatility such as stock price movements~\cite{mao2012correlating,bollen2011twitter} and movie revenues~\cite{Rui2013,Asur2010a}. 
In the health domain, electronic health records (EHR) such as admission notes, treatment plans, patient summaries, radiology reports, pathology reports and discharge notes are mined using natural language processing techniques to capture different clinical events of patients such as diagnosis, treatments, symptoms and side effects~\cite{Murdoch2013,meystre2008extracting}. These events are used with machine learning techniques to gain further insights. For example, association rules such as symptom-disease, drug-disease~\cite{chen2008automated}, drug-side effects (adverse drug events)~\cite{wang2009active} and disease co-occurrence~\cite{Jensen2012} are automatically extracted from EHR text. These associations are further employed to build machine learning models to predict outcomes such as cancer survival~\cite{miotto2016deep}, and hospital readmissions~\cite{rajkomar2018scalable}. These insights can be employed enhance the capabilities of clinical decision support systems which are used to assist clinicians by providing relevant and reliable information as well as automatically monitor, alert and even predict clinical conditions of patients~\cite{Jensen2012,demner2009can}. 

In law enforcement domain social data streams are often monitored to capture events that could impact social stability (e.g., protests, riots) since social media are often used to organise and propagate information about such events~\cite{Lotan2011}.~\cite{popescu2010detecting} developed a supervised model to detect controversial events based on the expressed public opinion in Twitter. Similarly,~\citet{Sakaki2010a} developed a model to detect natural disaster related event such as earthquakes. These techniques capture the events retrospectively while it is more beneficial to capture any planned events before it occurs. To capture future planned events~\citet{Ramakrishnan2014} developed EMBERS platform that mines online free-text data (social data streams, websites, news and Wikipedia) to capture events that could lead to civil-unrest in 10 Latin American countries.

Irrespective of the problem domain, the key steps of specified event extraction can be abstracted to (i) identify the relevant documents/web pages/posts that contain event related information, and (ii) identify relevant entities and their relations to the event. 

\subsubsection{Identifying documents that contain event information}
The techniques that mainly employed to extract event related documents are (i) thesaurus based techniques and (ii) classification model based techniques. 

Thesaurus based techniques assumes that presence of certain terms in documents indicate that such documents contain information about the events of interest. Therefore, such thesaurus is engineered with event related terms and use it to identify the documents which contain such terms. For instance, EMBERS planned protest model~\cite{Ramakrishnan2014} employs a multi-lingual lexicon of planned protest related terms (e.g., \textit{plan to strike} in English or \textit{preparaci\textasciiacute on huelga} in Spanish)  to capture the documents (e.g., social media posts or web pages) that discuss planned protests. These thesauruses are developed based on the domain knowledge and often based on knowledge bases or glossaries that exists in each domain. For example, in the health domain, the specified event detection techniques often employ the health related knowledge bases such as UMLS Metathesaurus~\cite{Bodenreider2004} which contains multiple health thesauruses including SNOMED CT which contains normalised health concepts, synonyms and descriptions;  and  RxNORM~\cite{Nelson2011} which is a drug thesaurus all drugs approved in USA.  

The classification model based techniques use a labelled document set where positives discuss the event of interest and negatives do not, to train a machine learning classification model. The documents are first feature transformed, in order to convert text into real-values features that are needed for machine learning techniques. Generally employed feature transformation techniques include one-hot-encoding which transform presence/absence of a term into a binary feature and \textit{term frequency inverse document frequency} tf-idf~\cite{salton1988term,sparck1972statistical} techniques discussed in Section~\ref{sec:text_clustering}. However, recent advances in semantic text-representations lead to the development of word-embedding techniques~\cite{mikolov_word2vec} that are often used as a feature transformation in deep neural network based techniques. The features transformed positive and negative samples are then used to train a classification model which learns to differentiate positives from the negatives. This trained model is then used to identify the event related documents from an unlabelled document collection or stream. For instance,~\citet{Sakaki2010a} trained an SVM (Support Vector Machine)~\cite{cortes1995support} based classifier using feature transformed tweets where positives are tweets that refer to an actual earthquake. This technique is mainly employed when reliable domain knowledge bases do not exist for the events of interest. 

\subsubsection{Identifying relevant entities of an event}

Once an event is identified the next step is to capture more information about the event. This task is generally achieved by capturing the relevant named entities that are related to the identified event. 

Named entities refer to terms or phrases that consistently stands for the same referent. This definition is derived based on \textit{rigid designator}~\cite{kripke1972naming}: a designator is rigid when it denotes the same thing in all possible worlds. The relevant named entities defer across application domains. For instance, in a public event, the named entities are often location, time and people/organisations involved, whereas in the biomedical/health domain the named entities could be health related entities such as drugs, illnesses, symptoms, treatments, genes etc.

Recognition of named entities is achieved through detection of entity terms (words or phrases) and subsequently classifying them into meaningful categories. The recognition part is often handled using text chunking, regular expressions and noun phrase extraction techniques~\cite{nadeau2007survey} to extract potential entity phrases. 

The categorisation of the identified entities is either achieved based on  relevant ontologies or using supervised learning techniques. The ontologies contain term dictionaries for high-level categories and the identified entities present in those dictionaries are assigned to the relevant categories. Those ontologies are often engineered for the requirements of different application domains. WordNet~\cite{Miller1995} and YAGO~\cite{mahdisoltani2013yago3} are popular examples of such ontologies available for general purposes which contain a hierarchy of high-level categories for each entity in the ontology. Similarly, there are specialised domain specific ontologies i.e., UMLS Metathesaurus~\cite{Bodenreider2004} for health domain. These ontologies are employed by the state-of-the-art health text-processing tools such as Apache cTAKES~\cite{Savova2010} and MetaMap~\cite{Aronson2010}.

The supervised learning techniques attempt to learn the semantic patterns around the entities using a labelled corpus and incorporate those patterns to identify the entities from an unlabelled corpus. The early attempts of supervised learning techniques are mainly based on a set of hand-crafted features to captures named entities of different categories. For example, rules like \enquote{Mr. XXX} can be used to identify person names~\cite{nadeau2007survey} in text. More recent attempts are based on machine learning techniques such as Conditional Random Fields (CRF)~\cite{lafferty2001,finkel2005incorporating} to learn generalised semantic patterns of the entities from the training corpus. One of the popular training corpus for supervised NER is MUC-6~\cite{grishman1996message} which categorised the named entities into \textit{person}, \textit{organisation}, \textit{location}, \textit{time/date} and \textit{money}. These categories are formulated for information extraction tasks form news articles. This categorisation has been followed by most of the current state-of-the-art NER tools in Standford NLP~\cite{finkel2005incorporating} and Open NLP~\cite{baldridge2005opennlp}.

%% file: Chapter2-sentiment-emotion.tex
\section{Emotion extraction from text}\label{sec:emotion_extraction_from_text}
Emotions are an important aspect of human social life and social behaviour. In psychology, emotions are defined as a complex state of mind that influence thought process and behaviour. Emotions play an integral part in human social interactions, especially in the interpersonal communication process. In face-to-face communication, humans express emotions verbally (emotional words and tone) as well as using non-verbal cues such as facial expressions and hand gestures. In written communication expression of emotions is limited to the use of emotional terms (words, phrases and emoticons). 

Understanding the emotions conveyed is an integral part of  understanding the communicated message, during the communication process. The emotions conveyed in a message supplements its discussion topic by adding opinions/feelings of the communicator towards that discussion topic. Such expression of emotion is ubiquitous in social data during social conversations where the individuals freely express their opinions and feelings on different discussion topics. Capturing these emotions encapsulated in social data provides invaluable insights about the conversations happens in social media platforms, which is important to understand the opinions/feelings of individuals towards different topics.

In contrast to topic extraction and event detection, techniques on emotion extraction from text are still in the early stages. This is mainly because emotions extraction from text only became relevant and useful recently with the rise of social media platforms, while topic extraction and event detection techniques were started to develop much earlier to capture topics/events from news articles. 

The rest of this section is organised as follows: firstly, emotion theories and relevant emotion models were discussed which are being used as the basis for the emotion extraction techniques. Subsequently, sentiment extraction techniques  (two dimensional emotions extraction) were discussed and finally, multi-granular emotion extraction techniques were discussed.

\subsection{Theories of emotion}\label{sec:theories_emotion}

The key theories of emotion can be categorised into physiological and cognitive. Physiological theories suggest that emotions are due to physical changes in the body. A prominent physiological theory is James-Lange theory of emotions~\cite{cannon1927james} which was formulated by two 19th-century scholars, William James and Carl Lange. This theory states that emotions are nothing more than conscious feelings about bodily changes. This idea that emotions occur solely due to physiological reactions to events is been criticised by others. Another similar theory is Cannon-Bard theory~\cite{cannon1927james} which states that after an event physiological reactions and emotions occurs at the same time. Two-factor theory~\cite{schachter1962cognitive} a more recent theory of similar genre which states that emotions are generated by a two factor process. First, the physiological event occurs and then the individuals refer to his past experience or the immediate environment to find emotional cues to provide an emotion label to that event. A key commonality in those theories is that the stimulus for emotions is physical changes.

\begin{figure}[!htb]
	\centering
	\includegraphics[clip=true, width=0.8\linewidth]{{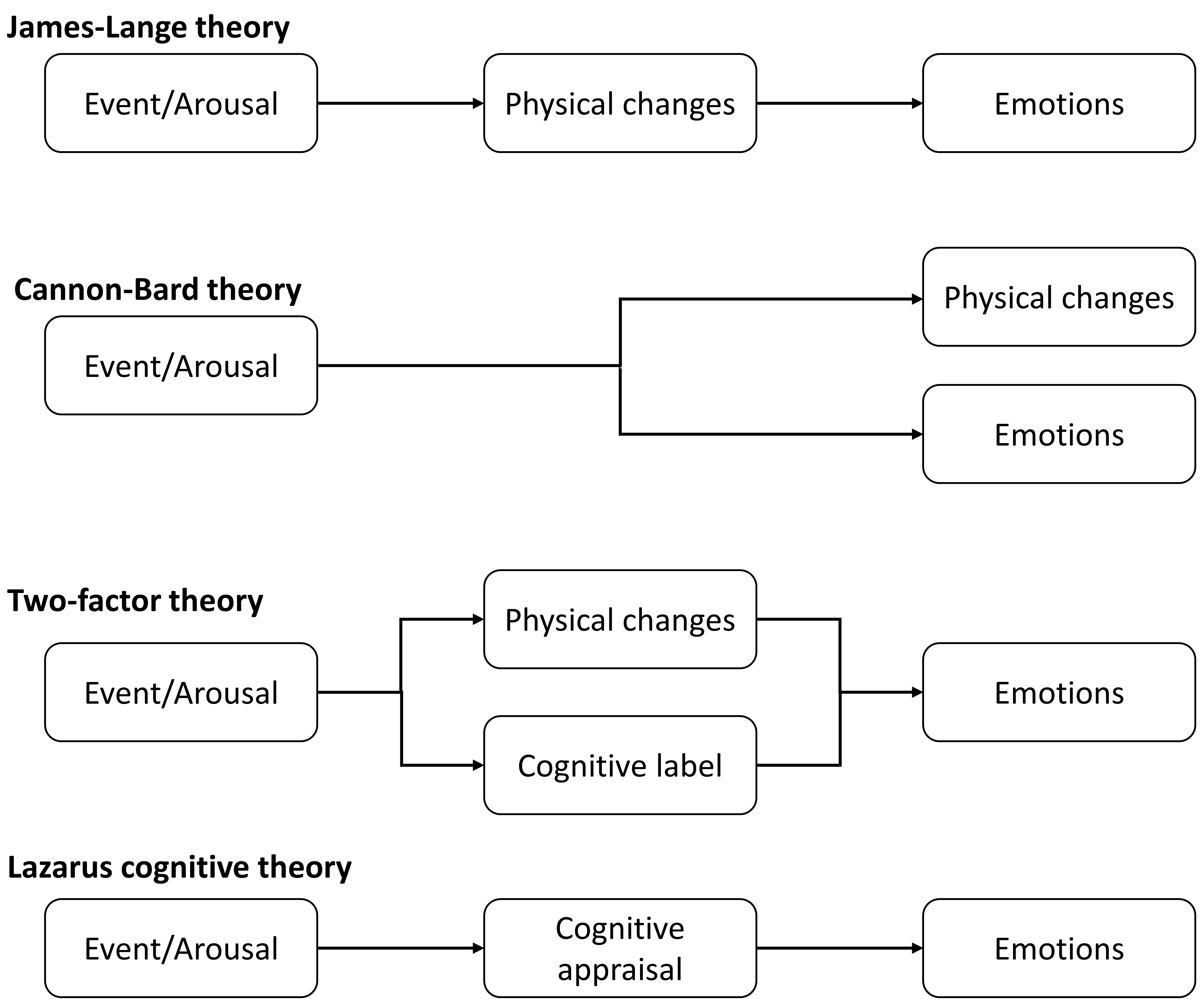}}
	\caption{The process that generates emotions, according to several prominent emotion theories.}
	\label{fig:emotion_theories}
\end{figure}

Cognitive theories, on the other hand, argue that emotions occur due to conscious cognitive activities such as thoughts, judgements, and evaluations. One of the key theories in this line of thought is Lazarus cognitive theory~\cite{lazarus1991emotion} proposed by Richard Lazarus. This theory states that emotions are determined by the cognitive appraisal of an event or stimuli in the environment. It argues that the quality and intensity of the emotions are controlled by the cognitive process. Therefore, cognitive appraisal mediates the effect of the event in the environment. Since the cognitive appraisal process is personalised (based on their past experience), same event often yield different emotional responses from different individuals. Figure~\ref{fig:emotion_theories} presents how emotions are generated following an event in the environment based on different theories of emotion discussed above.

The term \textit{emotion} is used alongside and interchangeably with terms \textit{sentiment}, \textit{affect}, and \textit{feeling}.~\citet{shouse2005feeling} points out that \textit{affect} is the most general and abstract of the above terms. Affect represents the intensity of experience as either positive or negative. It is often not conscious and not explainable using language~\cite{munezero2014they}. In contrast, feelings are conscious and labelled sensations that are checked against the past experience. Hence, feelings are personal to each individual as the labelling of sensations is based on the past experience of the individual. Moreover,~\citet{shouse2005feeling} states that emotions are the display or projection of feelings.~\citet{friedenberg2011cognitive} state that emotions are short episodes that involves both brain and body to respond to an event of interest. The sentiment is defined as the positive or negative mental predisposition in an individual towards an object (e.g., another person, event). Sentiment is often formed when an individual constantly perceive or think about an object, which leads to building up a dispositional idea about the object.   

\subsection{Modelling emotions}\label{sec:emotion_models}
Characterising or assessing emotions is a challenging task. This is because, firstly, as discussed in above theories emotions originated by a cognitive process, which makes emotions harder to measure as an entity. Secondly, emotions are personalised to individuals based on their personal experiences, thus the same event/experience may often yield different emotions on different individuals. These challenges lead to the development of emotional models to conceptualise human emotions. In contrast to emotional theories which describes origin of emotions, emotion models provide a conceptual model for characterising emotions. Such conceptual models are often employed in machine learning applications as the basis for capturing insights from emotion expressions. 
 
The development of emotion models has taken two distinct approaches:
\begin{enumerate}
	\item Dimensional models, which model emotions as a continuous $n$ dimensional space.
	\item Discrete models, which model all emotions into a number of basic categories.
\end{enumerate}

\noindent Dimensional models argue that emotions are overlapping and interrelated affective states generated by a common neurophysiological system~\cite{posner2005circumplex}. Therefore, all human emotions can be represented in a conceptual continuous $n$ dimensional space (often two or three dimensional). Each of these dimensions is modelled to represents a different aspect of the underlying neurophysiological system. The key dimensional models are (i) positive activation negative activation (PANA) model~\cite{watson1985toward}, (ii) Vector model~\cite{bradley1992remembering} and (iii)  Circumplex model~\cite{russell1980circumplex}. 

\begin{figure}[!htb]
	\centering
	\includegraphics[clip=true, width=0.6\linewidth]{{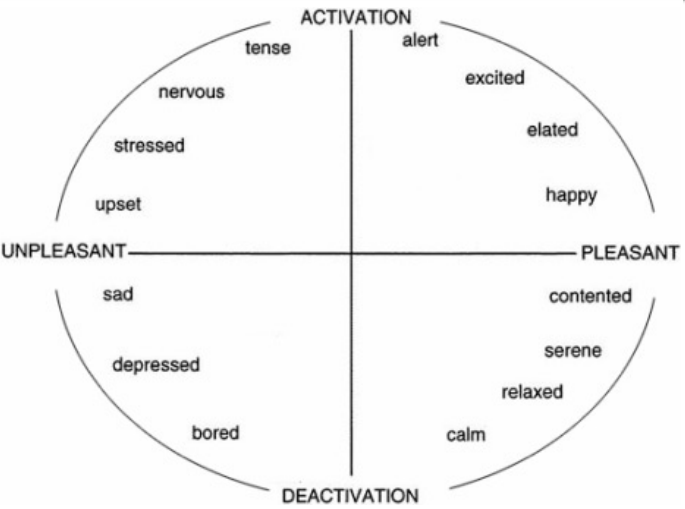}}
	\caption{A graphical representation of the Circumplex model consisting of valence and arousal dimensions. Multiple emotions are placed in the axis system based on their valance and arousal. Image adapted from~\citet{posner2005circumplex}.}
	\label{fig:circumplex_model}
\end{figure}

PANA model~\cite{watson1985toward} represents emotions in two dimensions where the horizontal dimension is the strength of positive affect and the vertical dimension is the strength of negative affect. Among those models, the most popular dimensional model for emotions is the Circumplex model~\cite{russell1980circumplex}. It argues that all emotions originate from two neurophysiological systems, \textit{valence} (pleasant to unpleasant continuum) and \textit{arousal} (active to passive continuum). According to Circumplex model, each emotion can be represented as a linear combination of valance and arousal. For instance, \textit{excited} is a pleasant and active emotion while \textit{relaxed} is a pleasant and passive emotion. Figure~\ref{fig:circumplex_model} shows a graphical representation of the Circumplex model and how different emotions can be represented along the valance and arousal dimensions of the model.

In contrast to dimensional models, discrete emotion models argue that human emotional expression consists of a set of distinct basic emotions. These basics emotions are common to all humans and even recognisable across different cultures. The hypothesis of distinct basic emotions stems from~\citet{darwin1872expression} that different emotions are distinct neuropsychological phenomena shaped by the natural selection process to provide an organised physiological and cognitive responses to challenges and opportunities in the environment~\cite{Plutchik1980}. In other words, emotions are formulated as part of the evolution process and thus, it is argued that there should be a set of basic emotions common to all humans.  

This hypothesis of a universal set of basic emotions was initially experimented by Darwin~\cite{darwin1872expression} in which he showed an array of photographs with different facial expressions to people and asked them to recognise them. This experiment had been advanced by~\cite{ekman1979facial} where they showed photographs of Europeans with different facial expressions to two isolated tribal communities (have never met Europeans) in Borneo and New Guinea. Those people (including children) from the tribal communities were able to understand the emotions expressed in the facial expressions of the photographs with significant accuracy. Also, photographs with facial expression taken from the tribes were recognised by US college students with a similar level of success. These experiments lead to the assertion that there exists a basic set of discrete emotions present in all humans.
~\citet{ekman1992argument} further argues that basic emotions are distinguishable as they have distinct universal signals such as facial expressions common across all humans and also often common across other primates as well.~\citet{ekman1992argument} identified that there are six such distinct basic emotions which are: anger, disgust, fear, happiness, sadness, and surprise. 

\begin{figure}[!htb]
	\centering
	\includegraphics[clip=true, width=0.8\linewidth]{{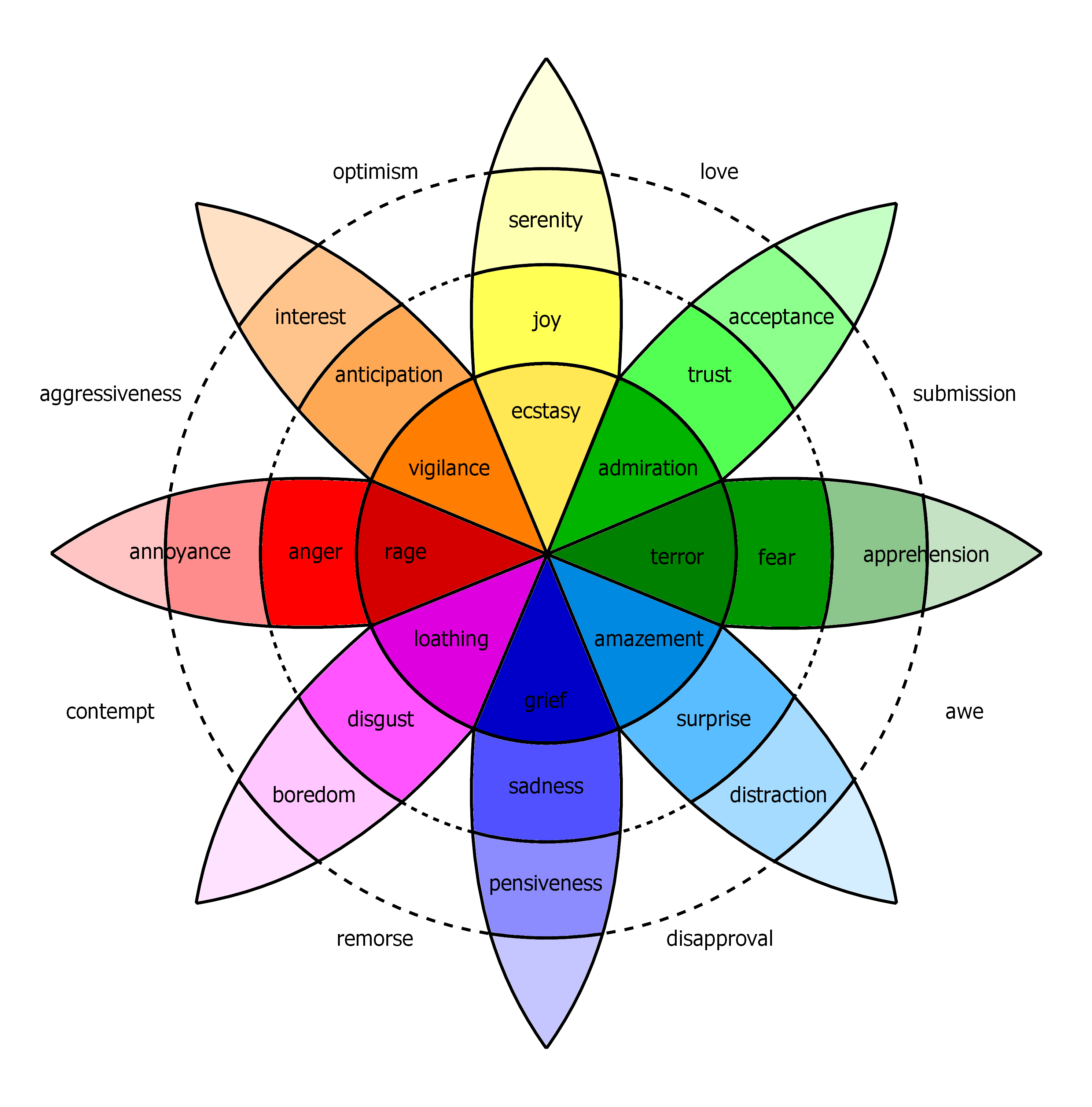}}
	\caption{Plutchiks wheel of emotions adopted from~\cite{plutchik2001nature}}
	\label{fig:plutchik-wheel}
\end{figure}

Similar findings on discrete basic emotions were published in other contemporary studies~\cite{izard1993four, lazarus1991emotion, tomkins1984affect, Plutchik1980}. Out of these studies, emotion wheel proposed by~\cite{Plutchik1980} is of particular interest. Plutchik presented a psychoevolutionary theory of basic emotions which argues that basic emotions evolved with the evolution thus, applies to all animals. The basic emotions provide support to all animals against the challenges from the environment. The theory further postulates that emotions can be conceptualised into pairs of polar opposites, emotions have a varying degree of similarity to others, and emotions have different levels of intensity in arousal. Based on these postulates,~\citet{Plutchik1980} presented a set of eight basic emotions anger, fear, sadness, disgust, surprise, anticipation, trust, and joy. These emotions exist as four bipolar pairs such that anger vs fear, joy vs sadness, trust vs disgust and surprise vs anticipation. Also, as shown in Figure~\ref{fig:plutchik-wheel} those emotions were placed in a circumplex model based on the degree of similarity among emotions creating a wheel of emotions known as \textit{Plutchiks emotion wheel}.     

The next sections discuss how those emotions models are utilised to detect sentiment and emotions from text. 

\subsection{Sentiment detection from text}\label{sec:sentiment_from_text}
As discussed in Section~\ref{sec:theories_emotion} sentiment is often defined as affect (or emotion) towards an object or topic. Sentiment detection (or opinion mining) is the computerised analysis of discourse to capture emotions/opinions expressed towards an object (or topic) by the author. Most of the sentiment extraction techniques provide a positive or negative intensity score for a given text based on the emotional expression in that text. 

Sentiment detection techniques are often considered as the computational implementation of the Circumplex model~\cite{russell1980circumplex} discussed in the previous section, which models emotions in a two dimensional space as linear combinations of valance and arousal. 

Most of the early sentiment detection techniques were limited to detecting valance (positive/negative) dimension of the text~\cite{pang2008opinion}. Those techniques consider the task as a binary classification task and classify a given text as positive or negative based on the sentiment expressed in that text. The key issue with these techniques is the lack of arousal or strength of the sentiment expressed. For instance, both~\enquote{food is good} and~\enquote{food is exceptional} would be just rated as positive overlooking that the latter is relatively more positive.  
  
Recent sentiment detection techniques attempt to capture both valence (positive/negative) and arousal (strength of the sentiment) and provide a polarised sentiment score~\cite{Mohammad2016b} where the polarity represents valance and the number represents the amount of arousal. These approaches consider the problem as either a regression problem or a multi-class classification problem (sentiment in a Likert-type scale). 

Sentiment detection techniques can be broadly categorised into two approaches~\cite{medhat2014sentiment}: (i) rule based classifiers based on sentiment lexicons and (ii) machine learning models learned using a text corpus with human annotated sentiment.      

Rule based classifiers depend on a sentiment lexicon, which is a dictionary of lexical features (e.g., words, phrases) that has a semantic orientation as either positive or negative. Some lexicons only provide a list of positive and negative terms. Two prominent such lexicons are General Inquirer~\cite{stone1966general} and LIWC~\cite{pennebaker2001linguistic}. General Inquirer~\cite{stone1966general} is the oldest sentiment lexicon that is still in use, which has been manually constructed by social scientists, political scientists, and psychologists to capture different aspects of text messages. It is a thesaurus of more than 11,000 terms in 183 categories, in which the terms categorised as positive and negative can be used as a sentiment lexicon. LIWC~\cite{pennebaker2001linguistic} which stands for \textit{Linguistic Inquiry and Word Count} is a thesaurus constructed by sociologists and linguists. It contains more than 4,500 terms organized into 76 categories, in which categories \textit{positive emotion} and \textit{negative emotion} can be used as a sentiment lexicon. 

There are also sentiment lexicons with both polarity and valence scores for each term. SentiWordNet~\cite{Esuli2006} is sentiment lexicon which is created by annotating over 115,000 synsets in WordNet~\cite{Miller1995} lexical database. Each synset is annotated with three numeric scores indicating its \textit{positivity}, \textit{negativity}, and \textit{objectivity} where the sum of the three scores equals to 1.00 for each synset. Affective Norms for English Words (ANEW)~\cite{bradley1999affective} is another lexicon with valance scores. It has 1,034 English terms which were given numeric scores ranging from 1-9. These scores indicate both valence and arousal where $<5$ is negative, $=5$ is neutral and $>5$ is positive. Valance score is given by $\Vert score - 5.0 \Vert$, where a score of $1.0$ provides the highest negative valance and score of $9.0$ gives the highest positive valence.

A simple rule based classifier would search a given text for the presence of sentiment related terms based on the employed sentiment lexicon and assign the polarity based on the most prominent sentiment polarity present in the text. Similarly, sentiment valence can be determined by aggregating the valence scores of the terms present in the given text. Advances versions of the rules based sentiment classifiers are equipped with rules to capture linguistic patterns when used with sentiment related terms, would alter valence and/or arousal of that term given in the lexicon. The state-of-the-art rule based techniques such as SentiStrength~\cite{Thelwall2012} and VADER~\cite{Hutto2014} uses a set of engineered rules with a thesaurus of terms relevant to each rule to alter the designated sentiment score of sentiment related terms. 

For instance, VADER~\cite{Hutto2014} uses a sentiment lexicon constructed by combining lexicons General Inquirer~\cite{stone1966general}, LIWC~\cite{pennebaker2001linguistic} and ANEW~\cite{bradley1999affective} as well as social media related emotion artefacts such as emoticons and slang (e.g., LOL). Valence scores for this lexicon were determined by crowdsourcing. Moreover, a set of four rules were used to alter the sentiment score of a term as follows:
\begin{enumerate}
	\item Degree modifiers: those terms when used in conjunction with a sentiment related term alter (increase or decrease) the valance. For instance, the term \textit{really} increases the valence of \textit{good} when used together (e.g., good vs really good). Similarly, \textit{somewhat} decreases the valence (e.g., good vs somewhat good). A dictionary of degree modifiers was employed or this rule.
	\item Negation: negations terms like \textit{not}, \textit{hardly} reverse the polarity (arousal) of a sentiment term (e.g., not good). A negation term dictionary is used to capture such occurrences. 
	\item Capitalisation: increase the valence of a term (e.g., good vs GOOD)
	\item Exclamation mark: increases the valence (e.g., good vs good!)
\end{enumerate} 

Similarly, sentiment lexicon in SentiStrength~\cite{thelwall2010sentiment,Thelwall2012} is constructed based on General Inquirer~\cite{stone1966general}, LIWC~\cite{pennebaker2001linguistic}; as well as a set of slang, emoticons and idioms commonly used in social data. The initial valence scores for each term is determined by a pool of raters. Moreover, since SentiStrength is mainly optimised for brevity in social data, the valence scores were further refined for that domain using a supervised machine learning technique which was based on a corpus of sentiment labelled MySpace posts. Similar to the previous technique, SentiStrength uses rules such as negation, degree modifiers, punctuations, capitalisation and repeated letters (good vs goooood) to alter the sentiment scores.

The rule based classifier approach involves layers of engineering efforts which includes the construction of sentiment lexicons, assigning valence scores to each sentiment term and setting up linguistic rules to alter the sentiment score of sentiment terms. This process result a set of sentiment scores for a given text based on its content, which is often aggregated to produce the final sentiment score of that text. 

In contrast, the techniques that train a machine learning model mainly dependent on a sufficiently large text corpus with sentiment labels/scores. Using sentiment label/scores as the target, machine learning models were trained on the features extracted from the text corpus. The features vectors often constructed based on words and n-grams in the document using vector-space model. In addition, the syntactic structure of the document is often captured by using part of the speech tags and punctuations as features. Moreover, the classifiers designed for social data often capture features that represent hashtags, user mentions, and emoticons related features.

In an early work,~\citet{Pang2002} employed several machine learning techniques to train a sentiment classifier on a labelled movie review dataset using unigram, bigram and POS features. Labels were automatically generated as positive, neutral and negative based on the star rating provides with a movie review. It was found that Support Vector Machine (SVM) technique performs better than other techniques. Other contemporary works employed similar machine learning techniques such as Naive Bayes~\cite{melville2009sentiment} and Random Forest~\cite{da2014tweet}. Some approaches used feature selection techniques such as PCA and information gain~\cite{riloff2006feature} to reduce the sparsity by selecting important features.

~\citet{Socher2013} build the Stanford Sentiment Treebank, a corpus of movie reviews (a subset of~\cite{Pang2002}) with fully labelled parse trees. Such parse trees can be used to capture the context of the sentiment language in contrast to assessing the sentiment of individual words or phrases.~\citet{Socher2013} further present a recurrent neural network (RNN) based sentiment classifier learned from the above corpus which is learned to represent sentiment arousal and valence. It was shown that RNN is capable of capturing the context specifics of sentiment. 

Lexicon based approaches are more generalisable to different application domains as they are based on sentiment lexicons constructed using emotional terms that are generally employed in any domain. However, such approaches may not sensitive to the sentiment expressions that are specific to a particular domain. For instance, some sentiment lexicons do not include sentiment terms that are widely used in social data (e.g., LOL), but not part of the standard English vocabulary. In contrast, machine learning models learn specific sentiment expressions encapsulated in their training corpus. Such techniques also often capture more complex sentiment expressions that may span across multiple words and semantic patterns that are used to alter the sentiment expression. Hence, machine learning models which learned from a sufficiently large corpus often yield better performance on a similar corpus in comparison to lexicon based methods. However, they are less generalisable to capturing sentiment from text corpora in a different domain, as the patterns learned by the machine learning model may not be generally applicable.  

There are attempts on hybrid approaches which employs both sentiment lexicons and machine learning models to build better sentiment detection techniques. Lexicon based techniques employ machine learning models to optimise for a given application domain. For instance, SentiStrength~\citet{Thelwall2012} optimise the valence scorers of its sentiment lexicon to social data using a machine learning model trained on MySpace text corpus. Similarly, SocialSent~\citet{Hamilton2016} employed to build a domain specific sentiment lexicon from a word-embedding learned from a corpus of a given application domain. This technique first builds a word embedding from a large corpus of text from the relevant application domain. Subsequently, it starts with a generic seed lexicon of positive and negative terms and uses a random walk technique to expand the lexicon by looking for similar terms in the word embedding. This approach result a sentiment lexicon that better represent the sentiment expressed in the given application domain. Machine learning model based techniques also employed sentiment lexicons to improve its generalisability. For instance,~\citet{chikersal2015sentu} incorporated some features based on sentiment lexicons in their machine learning model. These lexicon based features consist of the number of negative and number of positive terms of each sentiment lexicon present in the given text. Such inclusions make the model more generalisable as it considers sentiment polarity assessment from different lexicons.  

\subsection{Emotion detection from text}
Emotion detection from text has been a relatively new avenue of research in contrast to sentiment detection, mainly due to higher complexity and lack of resources such as emotion labelled corpora and emotion lexicons. In fact, sentiment detection is a simplified version of emotion detection which assess emotion only in one or two dimensions, while emotion detection generally denotes assessing emotions with increased granularity (more than two dimensions). Because of this close resemblance, from the technical aspect, emotion detection techniques follow the same approaches taken by the sentiment detection techniques.   

Current state-of-the-art emotion detection techniques are extensions to sentiment detection techniques. For instance, in rules based emotion detection techniques were developed extending rules based sentiment detection techniques by simply replacing sentiment lexicon with an emotions lexicon. Similarly, machine learning models designed for sentiment detection were trained on an emotion labelled text corpora to detect emotion.

Emotion lexicons were constructed in different granularities which often adherer to different emotional models discussed in Section~\ref{sec:emotion_models}. Most of these emotion lexicons are based on~\citet{ekman1992argument} emotion model of six basic emotions or~\citet{Plutchik1980} emotion model of eight basic emotions. NRC Emotion Lexicon~\cite{mohammad2013crowdsourcing} is one of prominent emotion lexicon which has emotion terms related to eight Plutchik emotions joy, sadness, fear, anger, anticipation, trust, surprise, and disgust. It has close to 14,000 emotion terms generated using crowd sourcing techniques.

Similarly, emotion labelled dataset were created mainly based on the emotion models~\citet{ekman1992argument} or~\citet{Plutchik1980}.~\citet{alm2005emotions} annotated sentences from children's stories with eight emotions (~\citet{ekman1992argument} six basic emotions and positively and negatively surprised). Similarly,~\citet{strapparava2007semeval} annotated news headlines based on ~\citet{ekman1992argument} six basic emotions.~\citet{brooks2013statistical} annotated 27,344 chat messages based on 13 emotions (eight basic emotions and five secondary emotions from~\citet{Plutchik1980}). Unlike sentiment annotation, these emotion annotations are multi-label where each text can be labelled as containing multiple emotions. These annotated corpora are used to learn machine learning models (by the original authors and others) to detect emotions as multi-class classification problems.

%% file: Chapter2-so-il.tex



\section{Self structuring techniques}\label{sec:self_structuring}
Self structuring or self organising techniques automatically learn a structure from a given dataset and then maps each data-point to the most similar structural element. This structure can be used to develop a coherent grouping of the dataset. Such techniques are essential to learn the underlying structure of social data. 

\subsection{Biological inspiration}
Self structuring techniques were mainly inspired by the self-structuring capabilities of cells (neurones) in human cortex. Those cells apart from passing information to upper layers, self-structure horizontally as well. This self-structuring happens through short-range excitatory interactions between the neighbouring cells and inhibitory interactions between distant cells~\cite{ratliff1965mach}. These inhibitory and excitatory actions lead to competition and correlative learning which is the basis for self-structuring. Topographically ordered maps are observed in many parts of the cortex~\cite{kertesz1983localization} including visual cortex~\cite{van1985functional}, auditory cortex~\cite{reale1980tonotopic} and somatotopic cortex~\cite{kaas1979multiple}. In these topographically ordered maps, different sensory input receptions are mapped to different parts of the cortex. Although the primary structure of the cortex is determined before birth, these subsequent topographical orderings are due to self-structuring based on the sensory receptions.

This self-structuring learning mechanism in the cortex is studied and theorised by many research works. One of the most pioneering and prominent work is Hebbian theory~\cite{hebb2005organization} which states that~\enquote{When an axon of cell A is near enough to excite a cell B and repeatedly or persistently takes part in firing it, some growth process or metabolic changes take place in one or both cells such that A's efficiency as one of the cells firing B, is increased}. This theory explains that synaptic strength among neurone cells strengthen overtime if activation of one cell repeatedly and persistently leads to the activation of another cell. This can be mathematically stated as the synaptic weight between input cell and output cell is proportional to the correlation between input and associated output. Similar prominent theories are proposed elsewhere. Marr's theory of the cerebellar cortex~\cite{marr1991theory} states that cerebellum is an associative memory which maps the state of the body learned from cutaneous and proprioceptive receptors into motor commands. Malsburg's theory on self-organisation in visual cortex~\cite{von1973self} states that retinotopic organization (a mapping from retina to visual cortex) is learned through self-organisation process where neighbouring cells in visual cortex mapped to similar input cells in retina. 

\subsection{Self organising maps}
The biological process of self-structuring in the brain has inspired the design of several unsupervised computational techniques such as  Willshaw-Malsburg Neural Network Model~\cite{willshaw1976patterned} and Kohonen's Self Organising Map (SOM)~\cite{Kohonen1998, kohonen1997exploration}. However, simplicity, computational efficiency and scalability of the Self Organising Map is unparalleled to any other such techniques which have lead SOM to be the only successful computational technique inspired by the biological self-organising process.      

\begin{figure}[!htb]
	\centering
	\includegraphics[clip=true, width=0.8\linewidth]{{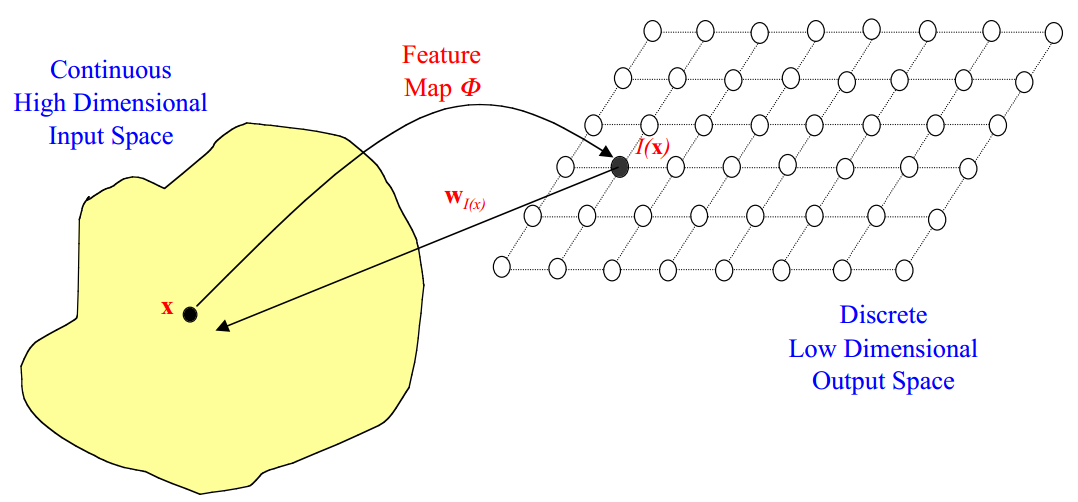}}
	\caption{Self Organising Map(SOM) creates a low dimensional structure from a high dimensional input space while preserving the topological relations exist in high dimensional input space. Source:~\cite{haykin1994neural}}
	\label{fig:som_mapping}
\end{figure}

Kohonen's Self Organising Map (SOM)~\cite{Kohonen1998, kohonen1997exploration} follows the Hebbian learning rule which emulates the self-structuring process in brain, but is simplified to reduced the computational complexity by replacing pre- and post-synaptic layers with the computational efficient two (or one) dimensional map structure. As illustrated in Figure~\ref{fig:som_mapping}, this map self-learns a structure from a high dimensional input space to a low dimensional map space while preserving the topological relations exist in the high dimensional input space. 

The learned SOM structure consists of nodes where each is denoted by a vector which has the same dimensionality as the input space. As shown in Figure~\ref{fig:som_mapping}, once learned from the input data each node $I(x)$ in SOM structure maps to a space $x$ in the input space and represent that space in the SOM map. Since SOM is learned while preserving the topological relations in the input space neighbouring nodes in the learned structure maps to adjacent spaces in input space.  

SOM structure is initialised as a fixed 2-dimensional grid of nodes where each node is randomly initialised with a $d$ dimensional  vector (where $d$ is the dimensionality of the input space). Once initialised SOM uses a combination of competition, cooperation, and adaptation phases to self learn its topology preserving low-dimensional structure from the input space. 

The competitive phase is used to find the \textit{winning node} or the \textit{best matching unit} for each input. This \textit{winning node} $N_w$ for each input is selected using a similarity function which assesses all nodes against a given input and selects the closest node to the given input:  
$$sim(X,N_w)\: \geq \: sim(X,N),\: \forall N \in \{ N \}$$
\noindent where $\{ N \}$ includes all nodes in the SOM structure. Euclidean distance is often used as this similarity function and select the node with the lowest Euclidean distance to the input. This competitive process closely resembles the process in the brain where input neuron excites an output neurone.

Next is the cooperative phase in SOM which emulates the  lateral interactions formulated among the neighbourhood of the neurones where a fired neurone excites its adjacent neurones more than the distance ones. In SOM, the topographic neighbourhood of a node is determined using a neighbourhood function $\mathcal{N}(d_{i,j},t)$ that decays with the lateral distance $d_{i,j}$ between winning node $i$ and neighbouring node $j$. Also, to reach the convergence, the neighbourhood function often decays with the number of iterations $t$, so that  winning nodes excite a wider neighbourhood during the initial stages, and a smaller neighbourhood in later stages. A Gaussian function is often used as the neighbourhood function $\mathcal{N}$ as it peaks at the centre and decays exponentially with distance from centre. 

Finally, the adaptive phase updates the weights of winning node and its neighbourhood based on the input vector. The weight vectors of winning node and its neighbourhood gets updated to reduce the distance to the given input. Let $w_1,\ldots,w_d$ be the weight vector (in a $d$ dimensional space) of the SOM node $j$  and $x_1,\ldots,x_d$ be the input vector, weight updates in each dimension $k$ in iteration $t$ is determined as follows:

$$ w_{j,k}(t) = w_{j,k}(t-1) + \alpha(t) \times \mathcal{N}(d_{i,j},t) \times (x_k- w_{j,k}(t-1))$$

\noindent where $\alpha(t)$ is the learning rate which determines the amount of learning captured into the winning node at each iteration. $\mathcal{N}(d_{i,j},t)$ is the neighbourhood function where winning node is $i$ and $\mathcal{N}(d_{i,i},t) = 1$. In SOM, learning rate is a decaying function over the number of iterations ($t$), leading to substantial adjustments during the initial iterations and minor adjustments in later iterations.  This weight adapting process emulates the Hebbian theory~\cite{hebb2005organization} by adapting the weights of the winning neurone to reduce the distance between input and winning neurone so that the efficiency of exciting that neurone to same (or similar) input increases.

Observations from the input space are iteratively used in the SOM learning process. During initial iterations, with higher learning rate and bigger neighbourhood, rapid self-structuring of the SOM grid takes place. Over the iterations learning rate drops and neighbourhood size diminishes leading to a smooth convergence of the SOM grid to the learned topological structure. The number of iterations required for convergence depends on the complexity of the underlying structure of the input space. Convergence in SOM can be assessed by monitoring the quantization error, which is the sum of absolute distances between input and the winning node $\sum^{d}\vert x_i- w_i\vert$. Quantization error rapidly drops during initial iterations and converges to a minimum value as SOM converges. The learning can be stopped when quantization does not show significant improvements over iterations.    

Self organising map has been used in many application domains in both academia and industry over the last few decades~\cite{oja2003bibliography, kohonen2013essentials}. It has been frequently used as a tool for exploratory analysis of large datasets as it projects high dimensional input data into two dimensional grid which can be easily visualised to understand the underlying patterns exist in a dataset. In addition, it has been used in applications such as dimensionality reduction, clustering, classification, anomaly detection, and information organisation/retrieval.   
   
\subsection{Dynamically expanding self organising maps}
A key limitation in self organising map is its fixed grid size which needs to be pre-determined. SOM grid size determines the number of nodes trained which is indicative of the learning capacity of SOM. Using a smaller grid on a large complex dataset would result in over-generalisation of the learned structure where some nodes may represent more than one distinct patterns. On the other hand using an unnecessarily large grid is expensive in terms of run time and memory usage. 

The key solution to overcome this issue is to use an expandable grid which dynamically expands during the training phase if more nodes are required to represent the underlying structure of the input dataset. 
There are several extended versions of self organising map developed to start with a small number of nodes and dynamically expand during training based on error metrics that indicates more nodes are required. 

Growing Grid~\cite{fritzke1995growinggrid} is a dynamically expanding version that starts with 4 nodes. In every iteration, it updates error of the winning node based on the distance to the input. Expansion of the grid happens after every $k \times m \times \lambda$ iterations ($k$ and $m$ are the dimensions of the grid), where it expands the grid based on the node $q$ with the highest accumulated error. A new column or row is added to the grid between $q$ and its furthermost connected node. This addition of an entire column or row may lead to adding too many nodes if trained on a large dataset (a large number of iterations). 

Growing Neural Gas (GNG)~\cite{fritzke1995neuralgas} creates a dynamically growing graph (instead of a grid). It also keeps track of the accumulated error and after $\lambda$ iterations select the node with the highest error for further growth. Similar to Growing Grid, a new node is added in between node with the highest error and its furthest neighbour. 

Growing Self Organising Map (GSOM)~\cite{Alahakoon2000} is another dynamically growing version of SOM. Similar to the above methods it keeps track of the error accumulated in winning nodes, but unlike above methods it triggers growth when the largest accumulated error exceeds a threshold. Growth happens from the node with the largest error and nodes added to all its remaining edges in the grid. Unlike Growing Grid, GSOM does not maintain a rectangle shape of the structure and nodes are only added to the impacted node. 

The self structuring techniques were mainly designed for conventional datasets where it expects a real-valued dense feature matrix. Extended versions are required to handle social data which are often unstructured, sparse and high-dimensional.  

\section{Unsupervised incremental learning techniques}\label{sec:incremental_learning}
Incremental learning is the paradigm of learning where existing learned knowledge incrementally gets extended and updated as new data comes in. Incremental learning techniques are the machine learning techniques that are capable of learning knowledge incrementally from new data. Such techniques are mostly applied to scenarios where the entire dataset is not known priorly but available over time. A naive approach for such scenarios is to discard previously learned knowledge/model and re-learn from scratch using both old and new data. However, such an approach is highly resource consuming in terms of both processing (to process the entire dataset) and memory (model has to keep old data for retraining). Incremental learning techniques overcome these challenges by only learning from new data and updating the existing model to reflect knowledge in new data.

Incremental learning is a must for social data, as it is a constantly evolving data stream. There will always be new unseen patterns appear on social data streams. Also, previously appeared patterns may appear later, so learning techniques have to keep the previously learned knowledge intact without discarding. 

\citet{Polikar2001} have specified four key characteristics of an incremental learning technique as follows:
\begin{enumerate}
	\item It should learn additional information from new data.
	\item It should not require access to the past data that it has already processed.
	\item It should not suffer from catastrophic forgetting, thus should preserve the previously acquired knowledge 
	\item  It should be able to accommodate new classes that may be introduced with new data
\end{enumerate}   
\noindent As mentioned in the first characteristic incremental learning techniques should capture any additional previously unseen information from the new data. Second characteristic delineates that  learning from new data should not require access to previously processed data. This characteristic enables such techniques to be scalable to handle big data streams where retaining old data is not feasible. Although some techniques may keep samples of previous data that represent each class or group seen in the data. The third characteristic is related to a phenomenon called catastrophic forgetting~\cite{french1999catastrophic} where models discard previously learned knowledge completely when new knowledge is presented. This ability to learn new knowledge while preserving the existing knowledge is known as the \textit{stability-plasticity dilemma}~\cite{Carpenter1988}. It is a dilemma because the incremental learning techniques have to sufficiently stable to be resilient to the noise in new data while flexible enough to learn new knowledge encapsulated in new data~\cite{gama2014survey}. If it learns noise from new data, that could lead to dropping relevant previous knowledge causing catastrophic forgetting. On the other hand, if the technique is too conservative it won't learn important new knowledge from new data. The fourth characteristic is to accommodate previously unseen classes that may come with new data. This is primarily for supervised learning techniques. However, in the context of unsupervised learning, it refers to accommodating previously unseen patterns in data. 

There are numerous supervised techniques that support \textit{supervised} incremental learning, however, unsupervised incremental learning techniques were relatively less. Unsupervised incremental learning techniques maintain a dynamically evolving cluster structure reflecting the dynamics of the data stream, where new clusters may appear and others may disappear. Such incremental clustering techniques were mainly adaptations from the standard clustering techniques. 

Leader~\cite{spath1980cluster} is one of the earliest partitioned based incremental clustering technique. It employs a user defined threshold to partition data into clusters. Every new input is assigned to the closest cluster if its distance to the cluster centroid is less than the given threshold, otherwise a new cluster centroid (leader) is added based on that input. This approach is simple but unstable as it depends on a user defined threshold for separating clusters. Another technique is the single pass k-means~\cite{farnstrom2000scalability} which is an incremental learning adaptation of k-means. It keeps a fixed sized buffer and when the buffer is full from incoming data, runs k-means to identify $k$ cluster. After running k-means, only cluster centroids are retained in the buffer, and when it is full again runs a weighted k-means using new data and existing centroids. This technique incrementally update clusters as new data arrives, however, fixed $k$ value limits capturing of new emerging clusters. 

~\citet{aggarwal2003framework} introduced CluStream, a stream clustering framework which separates incremental clustering into two phases online micro-clustering and off-line macro-clustering. Micro clusters are groups of data points which keeps statistical information of data and time stamps. A new data point absorbed into an existing micro-cluster if its distance to the centroid falls within that micro-cluster boundary. Only a fixed number of micro-clusters are maintained, and when it exceeds, older clusters were merged into others. In off-line macro-clustering, k-means runs over the set of micro-cluster centroids formulating the final clusters. 

~\citet{Furao2006} introduced self-organizing incremental neural network (SOINN) which is an incremental clustering technique based on self-organising-maps~\cite{kohonen2000self}. SOINN uses a two layer network where the first layer represents density distribution of inputs and second layer separates clusters by detecting low dense areas in the first layer. For each input presented, first and second winner is identified from the first layer, and if the distance to those exceeds a threshold, then the new input is added as a new node (between class insertion). Otherwise, standard SOM weight updates occur updating first winner and its neighbours (within class insertion). 

Another recent technique is Incremental Knowledge Acquisition and Self-Learning (IKASL) \cite{DeSilva2010a,DeSilva2010} which self learns a layered structure across time generalising the knowledge embodied in data. Each layer learns from a buffered batch of data using GSOM self-structuring technique~\cite{Alahakoon2000}. IKASL preserves the acquired knowledge in a generalised form therefore, it does not require access to the past data that it has already processed. The generalised version of the acquired knowledge from each layer (n) is used as the basis for the knowledge acquisition from the subsequent layer (n+1), thus it avoids catastrophic forgetting of the past knowledge. Moreover, while using the past acquired knowledge as the base, it incrementally acquires new knowledge that is embodied in the upcoming data.

Similar to self structuring techniques discussed in the previous section, unsupervised incremental learning techniques were mainly designed for conventional datasets where it expects a real-valued dense feature matrix. Extended versions are required to handle social data which are often unstructured, sparse and high-dimensional. Also, existing techniques do expect all features to be previously known, which is not the case in social data as new features will appear and existing features may disappear over time.

%% file: Chapter3.tex
\onlyinsubfile{	
\tableofcontents 
\setcounter{chapter}{2} 
}

\chapter[]{The Intellection of a Conceptual Framework}\label{chap:3}

\epigraph{{\textit{Nothing in life is to be feared, it is only to be understood. Now is the time to understand more, so that we may fear less.\\}}{\hfill Marie Curie}}

The previous chapter presented current work in machine learning and natural language processing techniques that are being used to process, analyse and learn insights from social data. As shown, the majority of such approaches for social media analytics were adaptations and extensions of machine learning and natural language processing methods developed for conventional text datasets. Although such approaches generate useful results, they tend to overlook the human behaviours, emotions and thought process which are embedded in social data. Therefore a wealth of rich information which can be used to infer potential causality of events and behaviours lie unused. This thesis proposes a novel approach that goes beyond conventional social media analytics. It considers the data generated in social media platforms a.k.a social data as the traces of human social interactions online and thus can be better represent based on theories on social behaviours, which can then be used by machine learning and natural language processing techniques to generate more meaningful insights. This chapter presents the theoretical foundations for this novel approach which has been materialised in subsequent chapters.  
     
This chapter begins with a discussion on the importance of social data  in advancing an understanding of social  behaviours (Section~\ref{sec:why_study_social_data}). In Section~\ref{sec:conceptual_framework_social_data}, behavioural theories from the social sciences are used to propose a new multi-layered conceptual framework that addresses this need. Section~\ref{sec:proposed_platform} concludes the chapter delineating  the materialisation of this conceptual framework based on the paradigms of self-structuring artificial intelligence, which also lays the foundation for the rest of the thesis.
\\

\section{On the importance of social data}\label{sec:why_study_social_data}

Social data is generated by humans through their social actions on online social media platforms in cyberspace. In fact social data can be considered as  archived traces of human social actions/interactions in cyberspace. For instance, \textit{tweets} are the outcome of social conversations and the expression of opinion that happens in the Twitter online social media platform. 

Social data and online social media platforms are relatively new paradigms which have only been in existence for the past two decades. However, online social actions that generate social data are in fact adoptions from human social actions in the physical world. For example, computer mediated communication techniques used in online social conversations are adaptations from the fact-to-face conversations in the physical world. Social actions were vital for the survival of human hunter gathers since the early ages of human civilisations. However, social actions and exchange of social information increased significantly with the use of language as it enables the individuals to convey complex expressions and have lengthy conversations.  

Online social media platforms are also digital adaptations from the social media platforms in the physical worlds. The first form of a social media platform could simply be a \textit{campfire} where the hunter-gatherer groups gathered around at night. The relaxing, warm and safe environment around a campfire leads to social chit-chat in-contrast to work related discussions during day-time~\cite{wiessner2014embers}, leading to increased social interactions among the group. Similar social media platforms were present in coffee-houses (mainly in Europe) and tea-houses (mainly in Asia) which serves as centres of social interactions~\cite{standage2013writing} where people chat with other visitors while sipping tea or coffee for hours. These interactions although seems casual, were important to formulate ties with individuals for economic or political benefit.    

These social actions and social media platforms in the physical world have been studied for decades and theorised by scientists across many disciplines that have developed an in-depth understanding of social actions and their underlying causalities. Such causalities have been abstracted into multiple layers of the human decision making process, where abstract social actions are represented as a handful of social behaviours driven by social needs perceived through human cognition. This understanding has developed incrementally over many decades and collectively by many theories which provide explanations to the social actions and their causalities based upon cognition, social needs and social behaviours. Considering online social behaviours are adaptations of social behaviours in the physical world and the causalities that drive such behaviours are the same, this research proposes to incorporate those theories to enhance the computational techniques (e.g., machine learning techniques, natural language processing techniques) that captures insights from social data in online social media platforms. For instance, sharing of different types of personal information (e.g., demographics, emotions) can be better understood in terms of the social behaviour called \textit{self disclosure}, while providing advice in online support groups represent the social behaviour \textit{altruism}. Such incorporation would enable the computational techniques to effectively capture patterns from social data as well as meaningfully aggregate such patterns based on the underlying causalities.     

Moreover, capturing insights into human behaviours and their causalities from social data, would enable social studies on human behaviour to use traces of human behaviours encapsulated in massive volumes of social data accumulated in online social media platforms. 
Conventionally, such studies on human behavioural causalities are conducted using controlled social experiments and natural experiments (observational studies)~\cite{neuman2013social}. Controlled experiments are designed to study the cause and effect of a certain condition on individuals against a control group while in natural experiments the behaviour of a group of individuals are observed in their natural environment. In both types of studies the outcomes are acquired from the individuals either by recording their behaviour or using interviews/survey instruments. 


~\citet{Meshi2015} point out several key advantages of the use of social data to understand human behavioural causalities, in contrast to the use of conventional social experiments.
\begin{enumerate}
	\item \textbf{External validity:} It is the generalisability of the casual inference~\cite{drost2011validity} from the \textit{experimental setting} to the \textit{natural setting}. External validity is one of the key challenges of controlled social experiments, mainly because human behaviour is inherently complex and irregular. It is difficult to exclusively isolate cause and effect of a certain behaviour in a controlled experiment. Therefore, the casual inference obtained from experimental settings may differ in a natural setting. In contrast, archived social data is the retrospective accounts of social actions which occurred in a \textit{natural setting}.Thus, causal inferences obtained using social data does not have the bias of any experimental settings.
	\item \textbf{Less recall bias:} Is the error caused by the human recollection of past experiences. Self experiences are stored in the \textit{episodic memory} as emotional and contextual details of the experience. However, \textit{episodic memory} is gradually forgotten over the passage of time.~\citet{Kahneman2012, kahneman1999objective} argue that recollection of an experience is biased towards the  most intense aspects of that experience. Therefore, retrospective report of an experience is different than being reported shortly afterwards~\cite{robinson2002belief}. 
	This recall bias affects the conventional studies since the data collection often happens periodically and the participants were asked to recollect their accounts of experiences that happen sometime back. In contrast, social data is often recorded near real-time, mainly due to the ease of access to online social media platforms using mobile devices. Therefore, social data is more immune to recall bias.
	\item \textbf{Large sample size:} Conventional social studies were often conducted using a handful of individuals, mainly due to the associated cost. Since human behaviour is so diverse across individuals, the casual inference learned from a small sample could be over-fitting to that selected cohort. On the other hand, social data of millions of individuals are already being archived in online social media platforms, who are from diverse socio-economic backgrounds. Hence, social data enables social studies using large and diverse cohorts.                     
\end{enumerate}

\begin{figure}[!htb]
\centering
\includegraphics[clip=true, width=1.0\linewidth]{{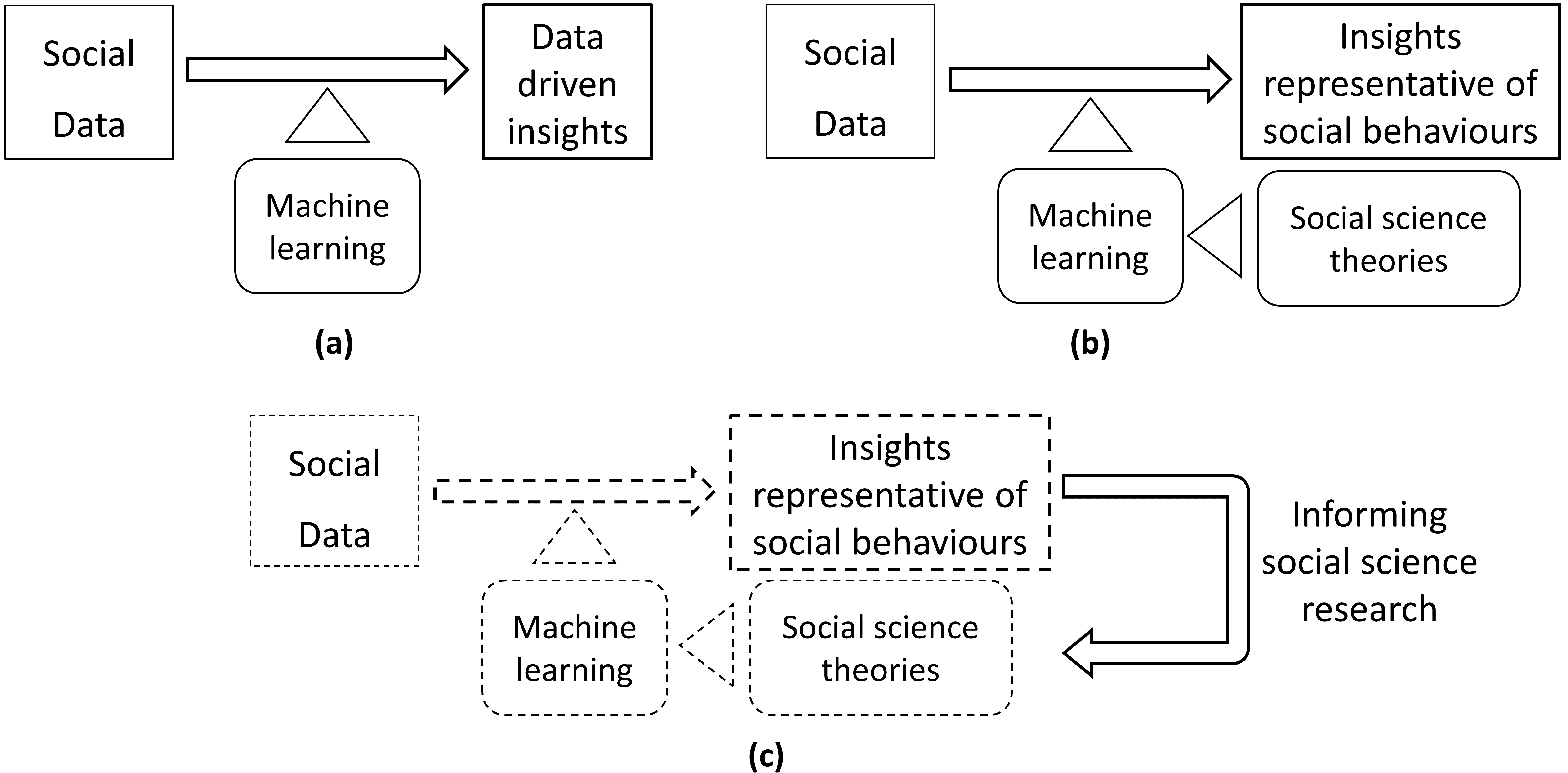}}
\caption{The approaches of transforming social data: a. the conventional approach of using machine learning techniques on social data which yields data-driven insights, b. combining machine learning techniques with social theories to transform social data into insights representative of underlying social behaviours, c. using insights representative of underlying social behaviours to better understand human social behaviours.}
\label{fig:Chapter3_Figure_SD_SB}
\end{figure}

In a nutshell, social data can be considered as traces of online social interactions which enable the use of theories on social behaviours to better understand social data.  This understanding can be leveraged to develop alternative approaches that better transform social data into insights representative of underlying social behaviours. These two approaches to transform social data is depicted in Figure~\ref{fig:Chapter3_Figure_SD_SB}, which shows a. the conventional approach of applying machine learning techniques on social data to produce data-driven insights, and b. the alternative approach of combining machine learning techniques with social theories to transform social data into insights representative of underlying social behaviours. Furthermore, Figure~\ref{fig:Chapter3_Figure_SD_SB}.c shows that the insights representative of underlying social behaviours learned from the proposed approach can be leveraged by social sciences to better understand human social behaviours in general and especially in online social media platforms. However, this feedback loop to improve the understanding of human social behaviours is beyond the scope of this thesis which is focused on building technical capabilities for harnessing insights from social data.

\begin{figure}[!htb]
	\centering
	\includegraphics[clip=true, width=0.8\linewidth]{{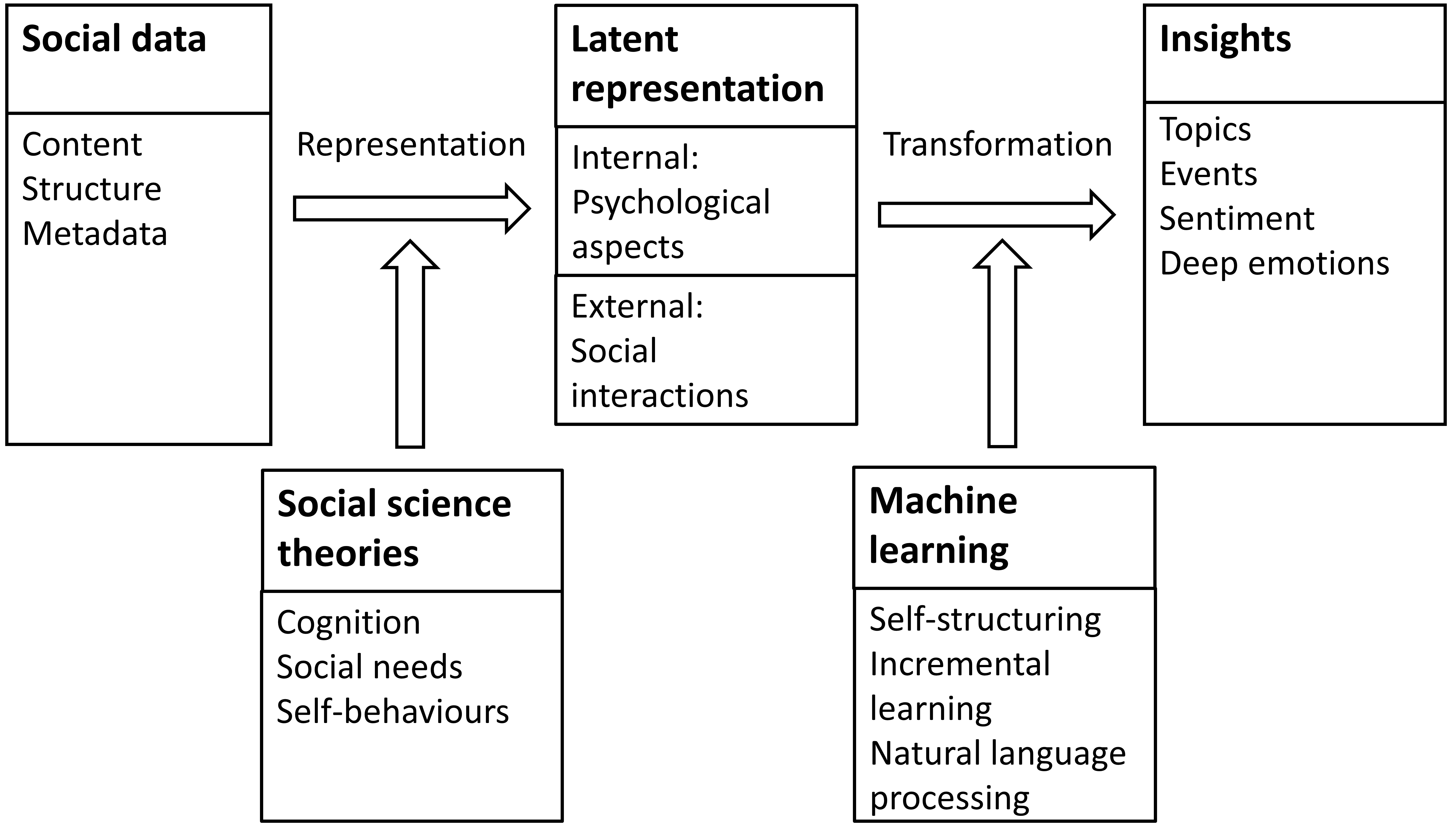}}
	\caption{Incorporating social science theories and machine learning techniques to generate insights from social data.}
	\label{fig:Chapter3_Figure_SD_SB_B_EXP}
\end{figure}

The new approach specified in Figure~\ref{fig:Chapter3_Figure_SD_SB}.b is further elaborated in Figure~\ref{fig:Chapter3_Figure_SD_SB_B_EXP}. Social science theories can be used to build latent representations from social data. Such representations can be based on either internal or psychological aspects such as emotions as well as external aspects such as topics of conversation or behaviours or events. Subsequently, machine learning and natural language processing techniques can be used to transform those latent representations into insights.

The next section presents a hierarchical conceptual model that enables to represent social data as a layered latent representation which is derived from prominent social science theories.

\section{The proposed conceptual framework}~\label{sec:conceptual_framework_social_data}

\begin{figure}[!htb]
	\centering
	\includegraphics[clip=true, width=1.0\linewidth]{{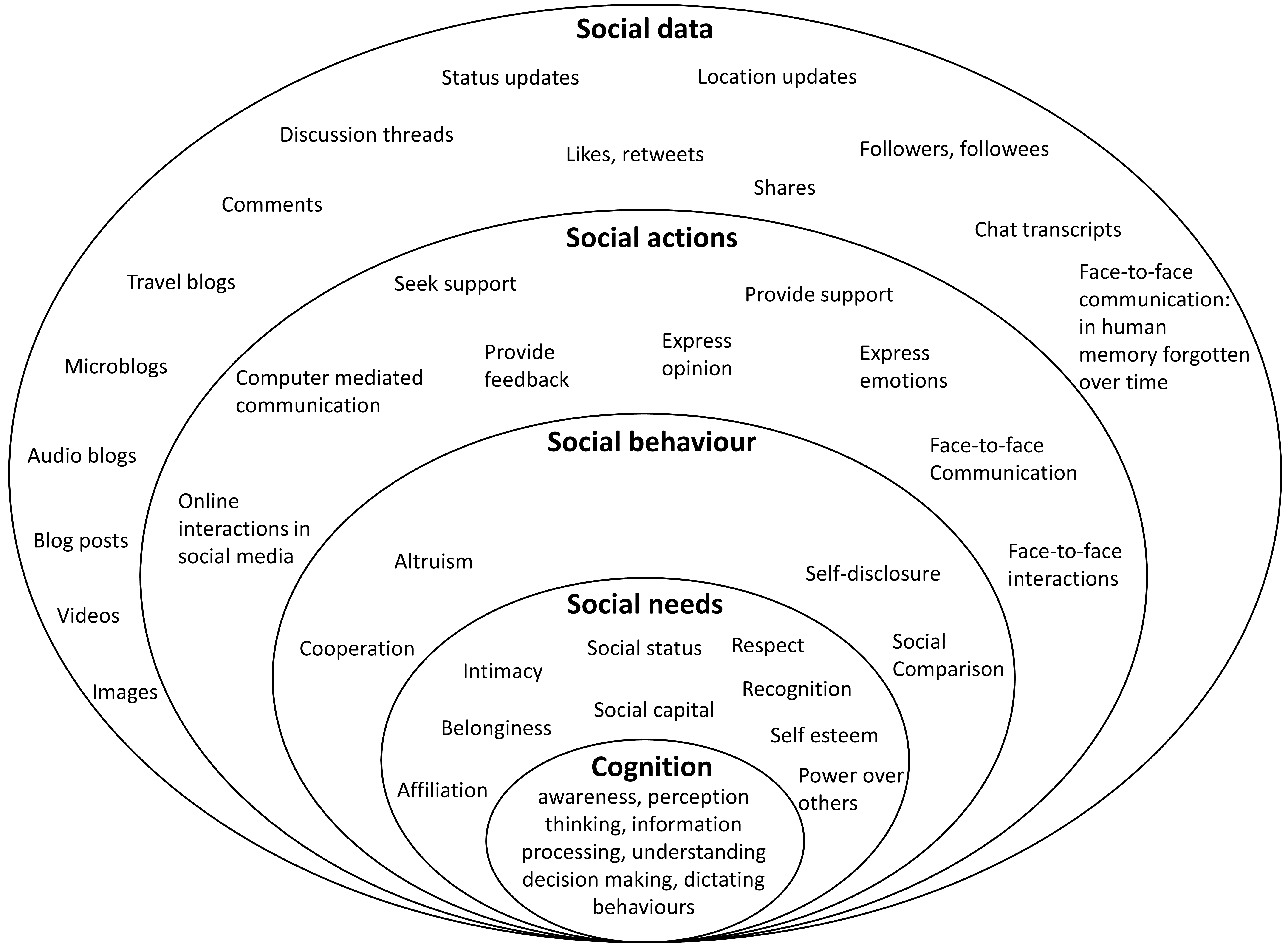}}
	\caption{The conceptual framework for understanding social data using social theories of human social behaviours, social needs and cognition as the foundation.}
	\label{fig:conceptual_model}
\end{figure}

Figure~\ref{fig:conceptual_model} presents the proposed conceptual framework that can develop a latent representation of social data. The following subsections describe the layers of the conceptual framework and supporting theories from the social sciences.

\subsection{Cognition}~\label{sec:cognition}
Human cognition is the mental process of acquiring or perceiving knowledge from thought, experience and senses. Cognition is thought to have physically based in the neural networks of the human brain within its over 100 billion neuron cells and their synaptic interconnections. In fact it is argued that  cognition represents the collective functionality of all neural networks. 

The cognitive process of acquiring or perceiving knowledge identifies behaviours that lead to positive or desired outcomes. This cognitive process dictates the execution of such desired behaviours by stimulating the motor neurons. These characteristics of cognitive process lead to the cognitive theory of human needs~\cite{rosenfeld1992human}, which states that, from a cognitive perspective, a need is an abstraction of the desired outcome obtained using a set of similar behaviours. Therefore, needs originate during the cognitive process of perceiving knowledge.  

   
\subsection{Social needs}~\label{sec:social_needs}
Based on this understanding of how cognition formulates human needs, this section discusses prominent human needs that involves social aspects i.e., the needs that have an emphasis on social interactions with other individuals or groups. 

\subsubsection{Evolutionary and biological perspective}
It is argued in Darwinian evolution theory~\cite{charles1859origin} that social needs developed as a survival trait in human hunter-gather groups. Since, humans do not possess exceptional features such as teeth, strength or speed compared to other hominoid relatives such as gorillas or chimpanzees, their survival against the selective pressure of natural selection is probably the behavioural adaptation of belonging to a group. Although humans are now an ecologically dominant and no longer under threat of a predatory species, effective social interactions remains a key factor to survive or rather thrive among others in the modern complex society.
     
A similar idea is proposed by the social brain hypothesis, where~\citet{Dunbar1998} states that human intelligence initially evolves as a behaviour adaptation to survive as groups. The larger brains in humans is a result of increased computational demands of living in large social groups. In fact it is found that the optimum group size of primate social groups correlates with the volume ratio of the neocortex (volume of neocortex relative to the brain)~\cite{Dunbar1998,Dunbar1993, Dunbar2007}.  

The attachment theory~\cite{bretherton1992origins, bowlby1969attachment} also describes that social needs of a human originate from the filial bonds developed between an infant and his close caregivers. It describes that an infant needs to develop a close and continuing relationship with at least one primary caregiver i.e., mother, in order for safety as well as cognitive, social and emotional development of the infant. 
        
\subsubsection{Psychological perspective}

One of the earliest mentions of social needs appears in \textit{pyramid of needs} presented in Maslow\rq s theory of motivation~\cite{Maslow1943,Maslow1962} which presents a hierarchy of human needs. The hierarchical formulation represents that humans look for needs in a particular layer only when the needs in respective lower layers are satisfied. In this pyramid of needs, socially relevant needs are placed third and fourth after physiological and safety needs. The third layer captures the human needs to feel a sense of belongingness among other individuals or social groups. Such relationships can range from close intimate relationships to  work-related companionship. The fourth layer is self-esteem which is a sense of being respected and valued by others. It includes acquiring social status, fame or recognition from the groups that the individual is a member.

~\citet{alderfer1969empirical} formulated the Existence, Relatedness and Growth theory (ERG) by simplifying the Maslow\rq s pyramid of needs. In ERD, relatedness is the desire to establish and positively maintain interpersonal relationships with other individuals. Such relationships could serve the need for belongingness as well as the need for self-esteem, thus relatedness is similar to third and fourth layers of \textit{pyramid of needs}. However, in contrast ERD is not hierarchical and expect all three needs to exist simultaneously.


The Three Needs theory~\cite{mcclelland1967achieving, mcclelland1987human}, describes three types of social needs: the need for achievement, affiliation and power. The \textit{achievement} need requires work on doable tasks and receive frequent feedback from others on the accomplishments. The \textit{affiliation} is the need to establish and maintain social relationships; and feeling of belonging to social groups. The \textit{power} need is the need to have power over others, social status and recognition    

These different psychological theories of needs highlight the key social need as being belonging to social groups and maintain social relationships with other individuals, and within those relationships there exists need to receive feedback and recognition of achievement as well as the need to influence others.  

In summary, social needs are a set of human needs that often need to be nurtured by others in society. Some needs such as love and belongingness often provided by those who are intimate to the individual (often family and friends of that individual). In contrast, social needs such as achievement, affiliation and power are related to receiving a form of recognition from society. Such needs are often satisfied by being part of different social groups. 
  
\subsection{Social capital}
Social capital looks at social needs in the economic perspective, as a capital that each individual should possess to survive and thrive in modern society. Social capital describes the resources embedded in social groups as a form of capital i.e. an asset that can be economically useful and can be invested for future economic gain. Unlike other forms of \textit{capitals}, that can be measurable, social capital is more conceptual and less measurable. 

Over the past decade, many sociologists have provided slightly contrasting definitions on social capital and its utility~\cite{woolcock2000social, lin2002social, Adler2002a, putnam2001bowling, nahapiet2000social}.~\citet{Adler2002a} define social capital as the \textit{goodwill}  (trust, reciprocity and sympathy) extended to an individual due to his social relations with individuals or groups, that will be available as information, influence, and solidarity.

~\citet{nahapiet2000social} describe three types of social capital based on the form of existence as \textit{structural, cognitive}, and \textit{relational}. The structural dimension is based on the structure of the social groups which includes roles, rules and procedures of the social group that determines the approachability to the resources in the group. The relational dimension is based quality and nature of the relationships which  determines the trust, obligations, expectations and sanctions~\cite{putnam2001bowling}. The cognitive dimension is based on the shared reality~\cite{bourdieu2011forms} which determines the shared values, beliefs and attitudes.   

Another dimension of social capital is often defined based on the type of the relationship, which is initially defined as \textit{bonding} or \textit{internal} and \textit{bridging} or \textit{external} social capital~\cite{putnam2009better,putnam2001bowling, Adler2002a}, base on the work \textit{strong and weak ties} by~\citet{Granovetter1973}. Bonding social capital arises from the relationships between people with similar characteristics i.e., homophily. Such individuals often have strong relationships with each other and peruse collective goals, thereby bonding capital facilitate support in terms of material, emotional and safety. In contrast bridging social capital arises from relationships between different social groups or individuals with different social identities. Bridging capital often operates on trust and reciprocity, and it provides access to important resources that are not available by bonding capital~\cite{Granovetter1973}. 

Later~\citet{Szreter2004,woolcock2001place} introduced \textit{linking} social capital which is argued to be a special form of \textit{bridging} social capital but the relationships and interactions occur across power or authority gradients of the society. The \textit{linking} social capital mainly refers to the relationships with who have the decision making power over the individual, thus such capital provides access to services, jobs and resources~\cite{Szreter2004}.

Social capital needs maintenance, where the individuals have to periodically renew the social links with others to make such links effective~\cite{Adler2002a}. Moreover, leaving social groups reduces the social capital associated with that group (e.g., leaving an organisation)~\cite{Glaeser2002}.

Initially, researchers speculated that the use of online social media platforms will reduce the time available for gaining and maintaining social capital. However, as new evidence emerge, now it is argued that use of online social media platforms not only provide an avenue to maintain social capital gained offline, it enables users to gain new social capital which were inaccessible offline due to barriers such as geographic proximity~\cite{ellison2007benefits}. It is believed that online social media platforms mainly contributes to forming and maintaining \textit{weak ties} with friends, colleagues and others with similar interests. Hence, most of the online social capital gains are in the form of bridging social capital~\cite{resnick2001beyond}. However, some of the \textit{weak ties} formed online may become stronger overtime leading to bonding social capital. Also, social media platforms help to maintain bonding capital when the individuals are geographically distanced~\cite{burke2011social}.~\citet{ellison2007benefits} also provide evidence that online social media platforms enables users with low self-esteem to achieve bonding social capital and improve their  psychological well-being.

The Table~\ref{table:summary_social_needs} summaries the key theories discussed in relation to social needs.

\begin{table}[!htb]
\caption{A summary of the key theories related to social needs discussed in this section.}
\label{table:summary_social_needs}
\centering
\begin{tabulary}{\linewidth}{|p{4cm}|p{10cm}|}\hline 
Darwinian evolution theory 
\newline \cite{charles1859origin} 
& Social needs developed to support belonging to a group as a survival trait against the selective pressure of natural selection. \\\hline
Social brain 
\newline \cite{Dunbar1998} 
\newline \cite{Dunbar2007} 
& Human intelligence evolves as a behavioural adaptation to survive as groups and larger brains is a result of increased computational demands of living in large social groups. \\\hline
Attachment theory 
\newline \cite{bretherton1992origins} 
\newline \cite{bowlby1969attachment} 
& Human social needs originate from the filial bonds developed between an infant and close caregivers. \\\hline
Theory of motivation (pyramid of needs) 
\newline \cite{Maslow1943}
\newline \cite{Maslow1962}
& Humans needs are hierarchical, reach higher order needs only when low order needs are satisfied. \\\hline
Three Needs theory
\newline \cite{mcclelland1967achieving} 
\newline \cite{mcclelland1987human}
& Humans have three types of social needs: need for achievement, affiliation and power.
\\\hline
Social capital
\newline \cite{woolcock2000social}
\newline \cite{Adler2002a}
\newline \cite{putnam2001bowling}
& Social capital describes resources embedded in social groups as a form of asset that can be economically useful and can be invested for future economic gain.
\\\hline
\end{tabulary}
\end{table}

\subfile{Chapter3-behaviour}

\subsection{Social actions}~\label{sec:social_actions}
Social action can be defined as an execution of a particular social behaviour detailed in the previous section. Social action for a particular behaviour is a specialised execution of that behaviour considering different functional aspects and environmental aspects. Functional aspects are determined by the cognition of the individual based on the perceived costs and benefits of that action. Functional aspects are often intensity, depth and duration, which determines the desired scale of the execution. For example, in self-disclosure, the depth of self-disclosure which determine the level of the self to expose, which can be a shallow disclosure of political views or a much deeper disclosure of emotional expression, and the intensity of the self-disclosure determines how much information to disclose. 

Environmental aspects are the aspects that are external to the individual and the individual has no or minimal control over. Environmental aspects are mainly relevant to the medium of execution. Social actions need to be executed considering the characteristics of the medium of execution, where it needs be optimised to employ the favourable characteristics of a given medium to exercise the social behaviour effectively to yield the desired outcome. For instance, in a rich communication medium such as F2F communication, social actions can leverage both verbal and non verbal to exercise the social behaviour, while in a leaner medium such as Twitter, social actions have to use different strategies to achieve similar outcomes. 

The next section describes social data, which are the archived accounts of social actions that are accumulated in online social media platforms.

\subsection{Social data}~\label{sec:social_data}
Social data is the archived accounts of social actions. Before the era of digitisation, accounts of social actions only exist in the memory of the participants and forgotten over time, unless being recorded on paper or dramatised as folk stories, which only happens to a handful of important social actions. 

The development of digitisation and digital storage techniques have enabled the archival of social actions in the physical world in digitised format. Initially, digitisation techniques were expensive, devices were bulky, and required technical expertise. Thus, such techniques and devices were mostly possessed by professionals and only used to record a few significant social actions. However, over the last few decades, (i) digital storage techniques have become cheaper and accessible, and (ii) digitisation devices has become cheaper, compact and easier to use. For instance, cloud storage has become cheaper and globally accessible through internet services. Also, smart phones packed with multiple digitisation techniques (e.g., text, voice recording, camera) has become affordable to the general public. The widespread availability of these techniques has enabled the general public to record and archive the social actions in the digitised form. 

\subsubsection{Accumulation of social data}
\begin{figure}[!htb]
	\centering
	\includegraphics[clip=true, width=0.7\linewidth]{{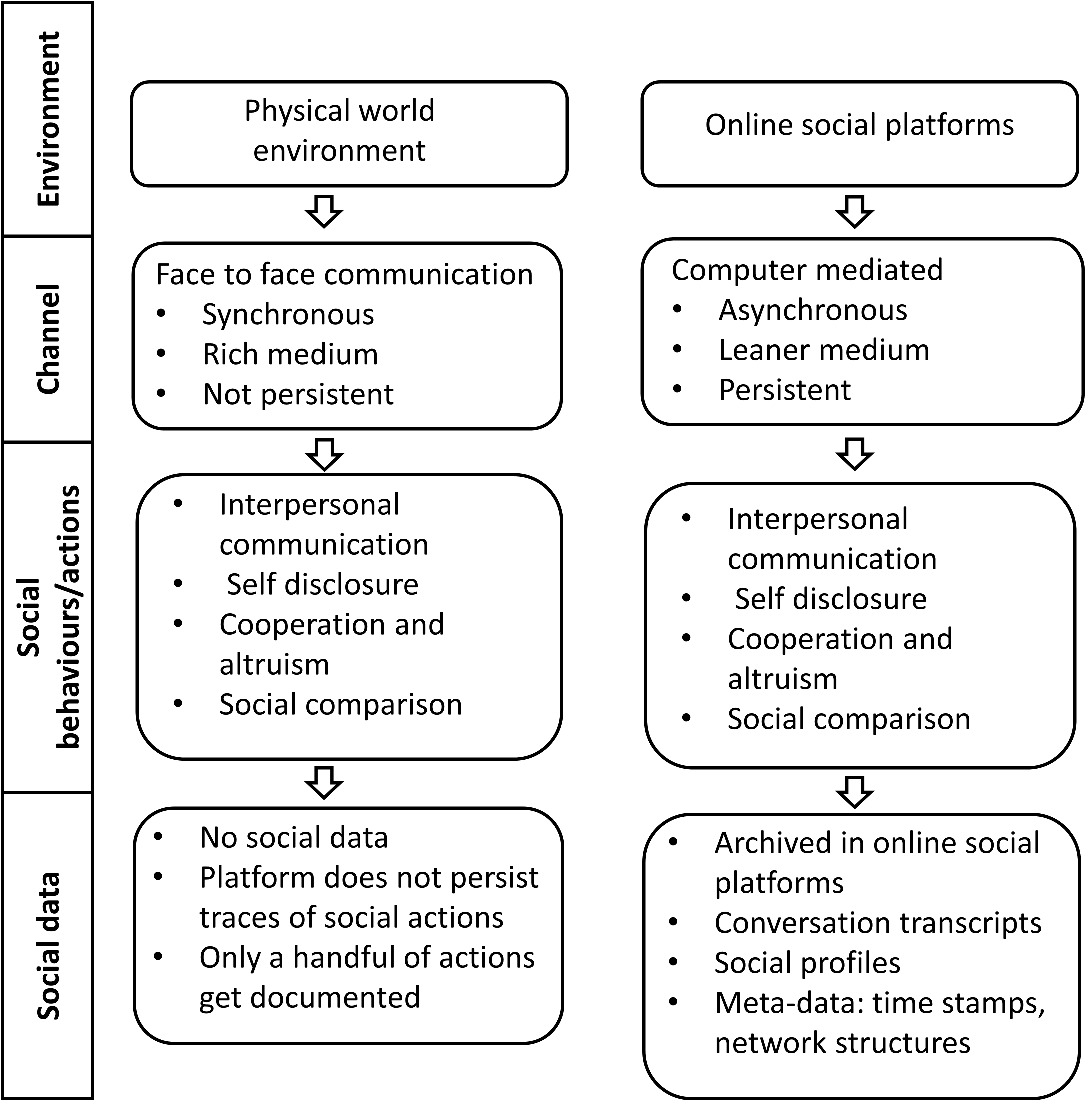}}
	\caption{Accumulation of the traces of social behaviours in online social media platforms as social data.}
	\label{fig:social_data}
\end{figure}

Although the digitisation has enabled the archival of social actions digitally, such data are stored and distributed privately. Moreover, in most of the cases those are snapshots of selected moments of social actions which occur in the physical world. 

The next paradigm shift comes with the rise of online social media platforms over the last decade. As discussed in previous sections, people start interacting with others on those online social media platforms and participate in social behaviours that are founded upon the social behaviours observed in the physical world. As shown in Figure~\ref{fig:social_data} the CMC based channel in online social media platforms persist all interactions that happen through the platform, which include conversation transcripts, social profiles and metadata such as timestamps and social network structures. Those archived data can be considered as the traces of the corresponding social behaviours or social actions that took place on the online social media platforms.

\subfile{Chapter32}

\section{Chapter Summary}
This chapter developed a layered conceptual model for interpreting social data generated in online social media platforms and a self-structuring artificial intelligent framework to capture insights from social data.  

The conceptual model is developed after an extensive analysis on existing theories of cognition, social needs and social behaviour. It depicts that social data is generated by social interactions happening in online social media platforms, and such interactions reflect different social behaviours driven by different social needs. It provides a deeper meaning to social data as archived traces of online social behaviours rather than just a corpus of unstructured text. In addition, four key social behaviours were conceptualised based on existing literature and a comprehensive comparison is provided comparing and contrasting exhibition of each behaviour in the physical world and online social media platforms.  

The proposed self-structuring artificial intelligent framework is designed to overcome the challenges of social data highlighted in Chapter~\ref{chap:2}. It consists of four key elements (i) representing social data into useful latent representations based on above conceptual model , (ii) unsupervised self-structuring to automatically structure such representations into semantically coherent groups, (iii) incremental learning to learn temporal changes, and (iv) natural language processing to extract semantic information. The next chapter presents an algorithmic development of this self-structuring artificial intelligent framework to capture insights from fast-paced social media platforms. 
   
\onlyinsubfile{	
\bibliographystyle{dcu}
\bibliography{library}{}
}

%% file: Chapter3-behaviour.tex
\subsection{Social behaviour}~\label{sec:social_behaviour}
Behaviour is defined as an intentional act to achieve a perceived goal~\cite{ajzen1985intentions}. This goal oriented nature of the behaviour arises from having perceived needs which individuals plan to satisfy through their own behaviour. The theory of planned behaviour~\cite{ajzen1985intentions,ajzen1991theory} states that execution of a behaviour is guided by the individual perceived attitude towards that behaviour, perceived subjective norms of that behaviour and finally the perceived ability to control its execution. 

Constructing on top of the general theory of behaviour, social behaviour can be defined as an intentional act that involves interactions between one or more individuals to achieve the perceived social needs of those who involved. Therefore, social behaviours are intentional acts of achieving social needs discussed in the previous section. Among social behaviours communication or interpersonal communication is the foundational social behaviour which is essential for any form of human social interaction. Other prominent social behaviours include self-disclosure, cooperation and social comparison. All these social behaviours are observed in the natural world as well as in online social media platforms. However, differences in the online social environments have led each social behaviour to have several adaptations from the natural environment to online social environments. The following subsections discuss the above mentioned social behaviours in terms of theoretical perspective, social needs they attempt to achieve and the different adaptations of them for online social media platforms.

\subsubsection{Interpersonal communication}\label{sec:communication}
Interpersonal communication or simply communication is the act of transmitting information among individuals or groups. Any form of social interaction between humans involves passing information from an individual to others (intentionally or unintentionally). Therefore, interpersonal communication can be considered as a foundational social behaviour that itself is a social behaviour as well as required for all other social behaviours. The key dimensions of communication are sender, channel, message, and the receiver~\cite{Shannon1948,schram1954process}. Sender encodes the desired information in a message and sends it to the receiver using an appropriate communication channel. The receiver decodes the message to receive the information. The encoding and decoding schemes have to be the same for the receiver to get the intended message, while differences in encoding and decoding may lead to unintended interpretations of the message received~\cite{schramm1954communication}. 

Although communication happens among all living organisms, this discussion is exclusive to the behaviour of \textit{interpersonal communication} which occurs between individuals or groups of humans. Conventionally, interpersonal communication happens face-to-face (FtF) where individuals visually see each other and exchange information using a combination of both verbal (using natural language) and non-verbal (e.g., body language, facial expression) communication. Verbal communication is often less ambiguous as language rules and the meaning of the words/phrases are well defined and common knowledge to all communicators. In contrast, the meaning of non-verbal communication often deviates based on different attributes (e.g., past experience) of the receiver.         

Interpersonal communication as a social behaviour can be explained using the \textit{uncertainty reduction theory}~\cite{Berger1975,berger1982language}. It states that during social interactions, individuals have a need to reduce the uncertainty about others, thus use interpersonal communication as a mean of acquiring more information about them.~\citet{berger1982language} further state that uncertainties exist as (i) \textit{cognitive} which are the uncertainties on values, ideas and attitudes of each other; and (ii) \textit{behavioural} which are the uncertainties on behaviours of each other in a given situation. 
Both verbal and non-verbal communication with each other leads to the exchange of information that leads to the reduction of such uncertainties and improvement of trust. Such reduction of uncertainties and improvement of trust is essential to initiate and  establish social relationships with the society. Effective and frequent communication with closed ones (family and friends) is essential to maintain intimate relationships to achieve the social needs love and belongingness (bonding social capital). Similarly, communication with different social groups and broader society is the key to establish affiliations thereby improve bridging social capital. Also, interpersonal communication is used as a way of influencing others which yields influencing power. Therefore, interpersonal communication is a fundamental behaviour to achieve most of the social needs.      

\subsubsection{Computer mediated communication}\label{sec:cmc}
Computer mediated communication (CMC) is the process of two or more individuals communicate using a computer platform. It is the communication process used in all online social media platforms. In early days it is primarily text based, however, recent CMC platforms allow a range of audio and video features to enhance the communication experience. Interaction in CMC can be either synchronous (e.g., Internet Relay Chat, Skype, Facebook Messenger) as well as asynchronous (e.g., email). Modern online social media platforms support both approaches. For instance, in Facebook platform, posts and status updates are mostly used for asynchronous communication while Facebook Messenger can be used for synchronous communication.

Most of the earlier research looked down upon CMC as an inferior communication channel in contrast to face-to-face communication (F2F). One of the prominent theory on that line of thought is the Media Richness Theory by~\citet{Daft1986} and~\citet{trevino1987media}, which defines the richness of information as~\enquote{the ability of information to change understanding within a time interval}. The theory delineates that the \textit{richness} of a communication medium is its capability to effectively transmit information to alter the understanding of an individual or a group~\cite{daft1987message, Daft1986, trevino1987media}. The theory further states that richness of medium is a function of the following capabilities of a medium: (i) transmit multiple communication cues, (ii) provide rapid responses, (iii) have a personal focus and (iv) support natural/conversational language for communication. These four capabilities of a rich medium is formulated considering F2F communication as the baseline standard and any other medium is evaluated based on how close it can resemble F2F communication. 

One of the key criticism on media richness theory is that it focuses on the communication medium objectively and haven't accounted for different context/topic of the communication or the differences in the human actors involved in the communication process~\cite{Kock2005,Kahai2003}. In order to accommodate those features~\citet{Carlson1999} proposed Channel Expansion Theory, which states that individuals perceptions on the \textit{richness} of a particular communication channel dependent on his relevant experience of using that channel and familiarity with the topic of discussion. Hence, frequent and prolonged use of a particular CMC channel in a particular context makes it \textit{richer} for the associated individuals.~\citet{Tidwell2002a} report that such improvements of the perception of richness of a CMC channel may result similar of greater richness than F2F communication.

Another key theory that discusses about the CMC medium is the Media Naturalness Theory which is proposed by~\citet{Kock,Kock2005} as a psychobiological model. It argues that humans (hominoids) have conducted F2F communication inside social groups as an act of survival. Hence, the evolutionary pressures over a long time period would have developed adaptations in both brain (e.g., sensory and motor neurons) and body (e.g., facial muscles, visual and auditory organs) to make F2F communication cognitively efficient, less ambiguous, and result in better physiological arousal. For example, human ears are sensitive to human voices than any other acoustic stimuli~\cite{nass2005wired}.


~\citet{Kock,Kock2005} further argues that due to the adaptive capability of the human brain, with frequent and prolonged use of a leaner CMC medium, human brain is capable of adapting to use such medium naturally with lesser cognitive effort, less ambiguity and increased physiological arousal. This phenomenon aligns with Channel Expansion Theory~\cite{Tidwell2002a} which states similar outcomes with frequent and prolonged use of a CMC medium.

Several such channel expansion attempts have been clearly visible across the recent history of social media platforms such as emoticons and hashtags. Emoticons in short for \textit{emotion icon} are a pictorial representation of facial expressions using a combination of characters. The most famous emoticons- smiley face :-) and frowning face :-( are introduced to online communication in 1982 as a suggestion to mark statements as jokes or serious statements. The use of emoticons has since spread across all text based communications channels such as emails, SMS and all social media platforms. A large set of emoticons are currently in use and also extended to multiple cultural variants such as Japanese \textit{kaomoji}. Also, emojis are pictorial versions of emoticons that often provides a more lively feeling of facial expression. The main use of emotions/emojis is to reduced the ambiguity of discourse resulted by not having facial expressions and improve the social presence in text based CMC~\cite{Walther1992,derks2007emoticons}. Therefore, emotions/emojis expand CMC channel by providing a means to transmit facial expressions.

Another prominent channel expansion attempt is social tagging or folksonomy which is the practice of adding user generated tags to content published online. Such content published in social media platforms are mostly unstructured text and published in a less organised manner without adhering to any form of a taxonomy. Hence it is a daunting task to find content or discussions that are relevant to a given topic of interest. It also limits the formation of online social groups with similar interests. Social tagging is one of the solutions adopted by the online community to mitigate this limitation. Social tags start with \# symbol and often comprises of a word or a concatenated phrase that the user wants to highlight. Use of social tags originated within the IRC (Internet Relay Chat) community to signify different chat rooms. However, its adaptation into Twitter~\footnote{https://www.cmu.edu/homepage/computing/2014/summer/originstory.shtml} in 2007 has lead to its widespread use across most of the prominent social media platforms.  As pointed out in~\cite{zappavigna2015searchable}, social tags serves different purposes such as: (i) highlighting the semantic domain of the post (e.g., \#bigdata), (ii) linking the post to an existing social discussion (e.g., \#auspol), and (iii) highlighting the affect of the post (\#sad).

These examples highlight that although CMC is a leaner medium compared to F2F, humans have invented different channel expansion strategies to improve its richness and provide a better social presence. 

\subsubsection{Self-disclosure}\label{sec:self_disclosure}
Self-disclosure is another social behaviour, which is the act of disclosing self to an individual or a group~\cite{jourard1958some}. Such disclosure can be intentional or unintentional, voluntary or by request, and verbal or non-verbal. The disclosure of information includes any type of information about the self such as socio-demographic information, emotions, opinions and needs~\cite{cozby1973self}. Self-disclosure a key social behaviour that reportedly consumes around 30-40\% of everyday human conversations~\cite{dunbar1997human}. No other species including other primates engage such high level of self-disclosure, which suggests that humans may have intrinsic motivation to self-disclose personal information. In fact, recent studies on brain reveal that self-disclosure lights up mesolimbic dopamine system that relates to reward and pleasure indicating that self-disclosure is a rewarding experience for an individual~\cite{tamir2012disclosing}.

Self disclosure is often studied on long terms strategic relationships such as romantic relationships~\cite{sprecher2004self,derlaga1987self,greene2006self} and therapeutic (therapist-patient)~\cite{derlaga1987self, knox1997qualitative, hill2001self}. In addition, there are studies on spontaneous self disclosure to a complete stranger which is commonly known as the \lq stranger on the train\rq~phenomena~\cite{Rubin1975}.  

\begin{figure}[!htb]
	\centering
	\includegraphics[clip=true, width=0.7\linewidth]{{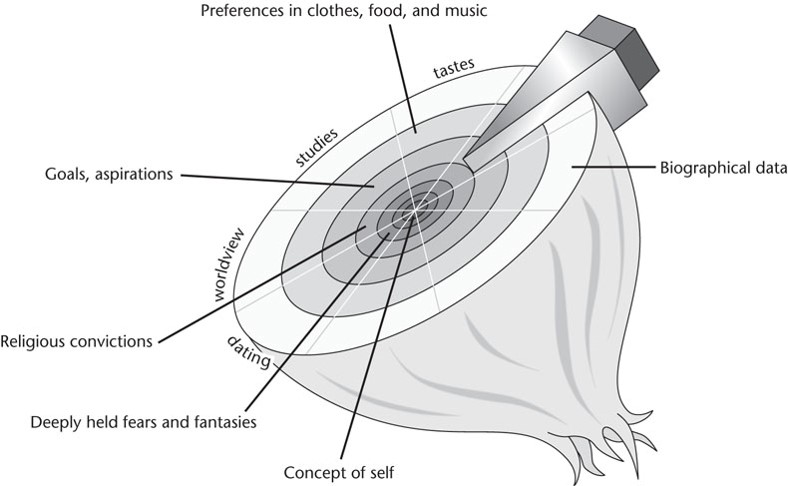}}
	\caption{The onion like nature of personality~\cite{altman1973social}, where outer layers contain more visible information and inner layers contain more intimate information. \textit{image source: Wikipedia}}
	\label{fig:SPT_onion}
\end{figure}

There are several theories that describe self-disclosure process and its objectives. Among them, social penetration theory (SPT) proposed by~\citet{altman1973social} is the most widely regarded. The social penetration theory (SPT)~\cite{altman1973social} was developed to explain how interpersonal communication develops from shallow to intimate through reciprocal exchange of self-disclosed information. SPT highlights two key dimensions of self-disclosure- \textit{breadth} and \textit{depth}. Breadth is the variety of topic in self-disclosure which could include topics related to profession, political affiliation etc. Depth is how deep an individual would reveal on a certain topic, where some disclosures are more of surface information while others disclose deeper information about the individual.
SPT describes this multi-dimensional and multi-layered nature of personality using the \lq onion metaphor\rq (Figure~\ref{fig:SPT_onion}), where each layer of the onion represents different layers of personality. The outermost layer contains more visible and public information such as socio-demographics, while inner layers progressively contain more intimate information such as values, attitudes, emotions and goals. SPT states that self-disclosure process is similar to peeling the onion layer-by-layer, where it initiates by disclosing more visible information and then progressively disclose more intimate information as the interpersonal relationship grows.         

Self-disclosure serves multiple social needs discussed in Section~\ref{sec:social_needs} which includes affiliation, intimacy, belongingness and recognition. Self-disclosure is a key behaviour required in formulating and maintaining any form of affiliation. Different topics are self-disclosed in different affiliations. For example, work related information such as skills and qualifications could be disclosed in professional affiliations, while sports team or player could be disclosed for affiliation related to fan clubs. The depth of self-disclosure is often shallow for affiliations. However, with reciprocity and over time self-disclosure goes deeper in some affiliations, making them more intimate relationships (become close friends or family) which often serves social needs of intimacy and belongingness. Also, self-disclosing information about achievements is essential to get recognition from individuals and social groups. 

Self-disclosure is a prominently observed social behaviour in social media platforms as well. It is reported that sometimes self-disclosure amounts to around 80\% of the content in online social media platforms~\cite{naaman2010really}. This is known as the \textit{online disinhibition effect}~\cite{suler2004online} where individuals self-disclose more frequently and intensely in online communication in contrast to face-to-face communication.~\citet{suler2004online} identified few factors that lead to this online disinhibition effect, which are  dissociative anonymity, invisibility, asynchronicity, empathy deficit, dissociative imagination, and minimization of authority. Anonymity and invisibility are similar to the \textit{stranger in the train} phenomenon where the individuals feel safe to self-disclose more in online social media platforms when their identity is not or loosely linked into the actual person. Similarly, asynchronous nature of the online social media platforms let the individuals self-disclose without worrying about an immediate response of others.
	
~\citet{suler2004online} further claims that the online disinhibition effect can be both benign and toxic. Benign online disinhibition is when the individuals use online social media platforms to express their feelings, issues, concerns etc. which they are reluctant to disclose in face-to-face conversations, potentially due to either embarrassing nature of the disclosure or introvert personality. It is found that such disclosures improve the mental well-being by reducing depression and anxiety among such individuals. Also,~\citet{davis2012friendship} found that some individuals (especially with introvert personality) can perform social interactions better in online social media platforms which enable them to achieve their belongingness and affiliation social needs as well as to improve their social capital.

Toxic online disinhibition happens when individuals use online social media platforms disclose hostility, hatred and threats targeted towards an individual or even a particular socio-ethnic group. Dissociative anonymity and invisibility in online social media platforms enable individuals to ignore the social norms that they often conform to in face-to-face conversations and disclose such toxic thoughts without any restrain in online social media platforms~\cite{lapidot2012effects}.

Although self-disclosure is more frequent in online social media platforms, depth of self-disclosures varies across different platforms depending on a multitude of factors such as visibility (whether posts are public or only visible to members), strength of ties between dyads, homophily, and reciprocity.~\citet{bazarova2014self} have investigated self-disclosure in wall-posts and peer-to-peer messages in Facebook and shown that public wall-posts contain more peripheral information and less intimate information while peer-to-peer messages contain more intimate information and less peripheral information. This highlights that although individuals self-disclose more in social-platforms, they mostly control the depth of the self-disclosure for different audiences or different platforms.

\subsubsection{Cooperation and altruism}   
Cooperation is the behaviour of working as a group to attain a goal that benefits everyone involved. Altruism is defined as an act which only benefits the recipient while incurring some cost to the actor~\cite{hamilton1964genetical}. Both cooperation and altruism exist in the animal world in tasks such as hunting and bringing up offspring, however, their scale and scope in humans society are unmatched to any other species. Humans have the capacity to cooperate as large groups of genetically unrelated individuals~\cite{boyd2009culture} and the capability to cooperate in complex tasks. 

In social brain hypothesis~\cite{Dunbar2007} it is argued that the increased cooperation and altruism in humans is an evolutionary trait for survival as complex social groups. Another argument is that cultural and social norms~\cite{fehr2004social}, supporting prosocial emotion and the establishment of the legal system to punish those who violate such norms~\cite{boyd2003evolution,boyd2010coordinated} have contributed to increased cooperation and altruism. 

Cooperation and altruism are often practised as a way to develop affiliations with social groups. It enables social interactions with the group members which often improve and establish interpersonal relationships with the other group members, thus improves the bridging social capital of the individual. Such behaviours also lead to achieving recognition from the affiliated social groups. Also, cooperation and altruism with social groups enable the individual to take a leading role and influence others.

Behaviours related to cooperation and altruism are often observed in online social media platforms as well. Individuals form online social groups based on different interests and perform cooperative tasks. In contrast to social groups in physical world, online social groups are not bound by any geographic boundaries. Hence, online social media platforms enable individuals to perform cooperative tasks beyond geographical boundaries.

Large social media platforms such as Facebook and Twitter often becomes the birthplace of social movements~\cite{golder2014digital} which originates as people discussing or sharing their views on certain socio-political issues. Others with similar views join those discussions. Some of those discussions transition into social movements either facilitating or demanding a corrective action for the issues. Tunisian and Egyptian revolutions that has overthrown the then governments are prominent examples for such social movements. Those social movements prominently used Twitter for organising protests/demonstrations (e.g., logistic coordination) and spreading news and updates to both local and global audience in near real-time~\cite{Lotan2011}.

Another avenue for cooperative and altruistic behaviours are the online platforms for knowledge sharing  where individuals from across the globe, contributes voluntarily by publishing user generated content. Such content is mostly based on their personal experience or professional experience. One of the well-known examples for such projects is Wikipedia which is a massive online encyclopaedia contributed by thousands of volunteers with different areas of expertise~\cite{golder2014digital}. In Wikipedia its community is delegated with multiple roles covering content generation, editing, validation and administration which drivers content generation as well as the integrity of the content.

Similarly, in other virtual communities such as self-help groups, cooperation and altruism displayed when individuals seek support by providing informational support and emotional support based on their personal experiences~\cite{Ziebland2012}. Unlike in Wikipedia most of the other virtual communities does not have set roles. However, each piece of information is either supported or opposed by other participants leading to community validated response.

Cooperative acts in online social media platforms primarily provide a sense of affiliation to the members of that online community. Especially when such contributions are commended by other members of the community. Moreover, online knowledge sharing communities have different approaches to provide social recognition to the contributors. For example, some platforms maintain contributor ratings either granted automatically based on the number of articles/posts or provided by the readers based on the usefulness of the contributions. In addition highest contributors receive privileged administration and moderation roles to assess the suitability of user content and user behaviour in the community. They have the power to remove inappropriate content as well as suspend users for inappropriate behaviour. Such roles provides a sense of power and influence over others in that online community.

\subsubsection{Social comparison}\label{sec:social_comparison}
Social comparison is another key behaviour which is an act of self-evaluation of an individual\lq s attributes such as opinions, abilities, appearance, wealth, performance etc. against those of other individuals~\cite{suls2002social}. Such comparisons enable the individual to understand self-worth as well as reduce any uncertainties about the compared attributes.~\cite{festinger1954theory} proposed the initial social comparison theory which denotes social comparison as a behaviour that happens against individuals similar to oneself. People use social comparison for self-enhancement and to gain self-esteem.

Social comparisons can be  broadly categorised as upward (compare against someone better) and downward (compare against someone worse)~\cite{collins1996better}.  In upward social comparisons, the comparison happens against individuals who are doing better in the compared aspects, such comparisons yield motivation and inspiration for self-improvement. However, it may lead to poor self-evaluation, and negativity as well. In contrast, downward comparisons happen against individuals who are doing worse in compared aspects. Such comparisons improve their self-worth by comparing to others who are worse off~\cite{taylor1989social, wills1981downward}.

Traditionally, social comparison happens among family, friends and co-workers whom an individual often encounter. In contrast, there are significantly more opportunities for social comparison in online social media platforms. This is because most of those platforms allow and encourage users to publish a user profile where user can self-created their own persona. For instance, in Facebook, there are millions of user profiles which can be freely accessible to anyone which contains personal information as well as collections of photos/videos that provides a sneak peek to the lifestyle of each individual. Similarly, professional social media platforms such as LinkedIn contains comprehensive work related profiles. Therefore, social comparison is significantly high on online social media platforms~\cite{vogel2014social} where individuals compare their \textit{offline} self to \textit{online} self of others perceived based on the information presented on online social media platforms. However, most of the profiles in online social media platforms are built using selected content to attractively present the ideal self of the individuals. For example, people only publish their best looking photos and highlights of their life; and avoid publishing any negative looking content~\cite{walther2007selective, vogel2014social}. Because of this selective self-presentation in online social media platforms social comparison is not \textit{like for like}, but rather a disadvantage for the \textit{offline} self. This phenomenon leads to excessive upward comparisons where more realistic \textit{offline} self gets compared to meticulously presented \textit{online} self of others; and often results negative self-evaluations, diminished self-worth and even depressive symptoms~\cite{vogel2014social, nesi2015using}.~\citet{chou2012they} report that the impact of such comparison is higher when compared to an \textit{online} self of people never or less frequently met in person, and lower for people with frequent encounter.

\begin{table}[!htb]
\caption{A summary of the key theories related to social behaviour discussed in this section.}
\label{table:summary_social_behaviour}
\centering
\begin{tabulary}{\linewidth}{|p{4cm}|p{10cm}|}\hline 
Theory of planned behaviour 
\newline \cite{ajzen1985intentions}
\newline \cite{ajzen1991theory} 
& A behaviour is an intentional act to achieve goal(s) perceived based on the needs of individuals. \\\hline
Uncertainty reduction theory
\newline \cite{Berger1975}
\newline \cite{berger1982language} 
& During social interactions, individuals have a need to reduce the uncertainty about others, thus use interpersonal communication as a mean of acquiring more information.
\\\hline
Media richness theory
\newline \cite{Daft1986}
\newline \cite{daft1987message} 
&  Richness of a communication medium is its capability to effectively transmit information to alter the understanding of an individual or a group. A rich medium can transmit multiple communication cues, provide rapid responses, have a personal focus and support natural language for communication. 
\\\hline
Channel expansion theory
\newline \cite{Carlson1999}
\newline \cite{Tidwell2002a}
&  Perception on the richness of a communication channel dependent on the experience of using that channel. Frequent and prolonged use of a channel makes it richer for the associated individuals. 
\\\hline
Social penetration theory
\newline \cite{altman1973social}
&  Self-disclosure is like peeling an onion layer-by-layer, where it initiates by disclosing more visible information and then progressively disclose deeper information as relationship grows.
\\\hline
Social comparison theory
\newline \cite{festinger1954theory}
&  Individuals compare themselves to others as a form of self-evaluation to reduce uncertainty about the opinions and abilities of themselves.
\\\hline

\end{tabulary}
\end{table}  

Table~\ref{table:summary_social_behaviour} provides a summary of the key theories discussed in this section. In summary, the above discussed social behaviours are designed to achieve different social needs. Interpersonal communication is the foundational behaviour that is required to enable any form of social interaction with others while the other discussed social behaviours utilise communication. Social behaviours are observed in both the physical world and the online environment (online social media platforms) however, some differences exist due to the differences in the environments. Table~\ref{table:compare_social_behaviour} summarises such differences of social behaviours in physical and online social environments.

\begin{table}[!htb]
\caption{Comparison of social behaviours in physical world and online social media platforms.}
\label{table:compare_social_behaviour}
\centering
\begin{tabulary}{\linewidth}{|p{2cm}|L|L|}\hline 
		Social behaviour & Physical world & Online social media platforms \\\hline
	Interpersonal communication
		 & Face to face social conversations \newline
		   Verbal and non-verbal cues\newline
		   Facial expressions, eye contact\newline
		   Participants in close proximity\newline
		   Immediate response required  
		& Computer mediated social conversations\newline
		  Mostly text based, no access to non-verbal cues\newline
		  No access to facial expressions or eye contact\newline
		  Not bounded by geographic proximity\newline
		No immediate response required\\\hline
	Self disclosure
	   & Often disclose to the participants in close proximity\newline
		 More control over the recipients of the disclosure
	  & Disclose to all the members or a selected group in social media platform\newline
	    Often self-disclose in a higher breadth and depth due to online disinhibition effect \\\hline
  Cooperation and altruism
      & Groups are bounded by geographic proximity \newline
	    Group have to work synchronously 
	 & Groups are often formed based on interest, not bounded by geographic proximity\newline
	  Groups can work synchronously and asynchronously\newline
	  Often form relatively large groups\\\hline
  Social comparison
     & Compare different attributes of oneself to others
	   Compare against friends, peers, and colleagues 
	& Compare different attributes of oneself to the online presented self of others \newline
	  Access to personal information of large number of individuals in social media platforms to do comparisons\\\hline
\end{tabulary}
\end{table}    
  
As shown in the conceptual model in Figure~\ref{fig:conceptual_model}, the above discussed social behaviours are mainly abstractions of human actions originated to achieve the perceived social needs, Those abstracted behaviours are executed in a multitude of ways as \textit{social actions} depending on different functional and environmental aspects which will be discussed in the following section.

%% file: Chapter32.tex
\section{An embodiment in self-structuring artificial intelligence }~\label{sec:proposed_platform}
As discussed in Chapter~\ref{chap:2} there have been attempts of qualitative methods as well as conventional machine learning approaches to transform social data into patterns that provide data-driven insights into social behaviours. However, such attempts yield suboptimal results due to a multitude of challenges present in social data that are discussed in Section~\ref{sec:social_data_analysis_challenges}. 
This thesis proposed a novel approach, the multi-layered conceptual model in Section~\ref{sec:conceptual_framework_social_data}. It is now pertinent to delineate the embodiment of this model, based on the paradigm of self-structuring artificial intelligence.



\begin{figure}[!htb]
\centering
\includegraphics[clip=true, width=1.0\linewidth]{{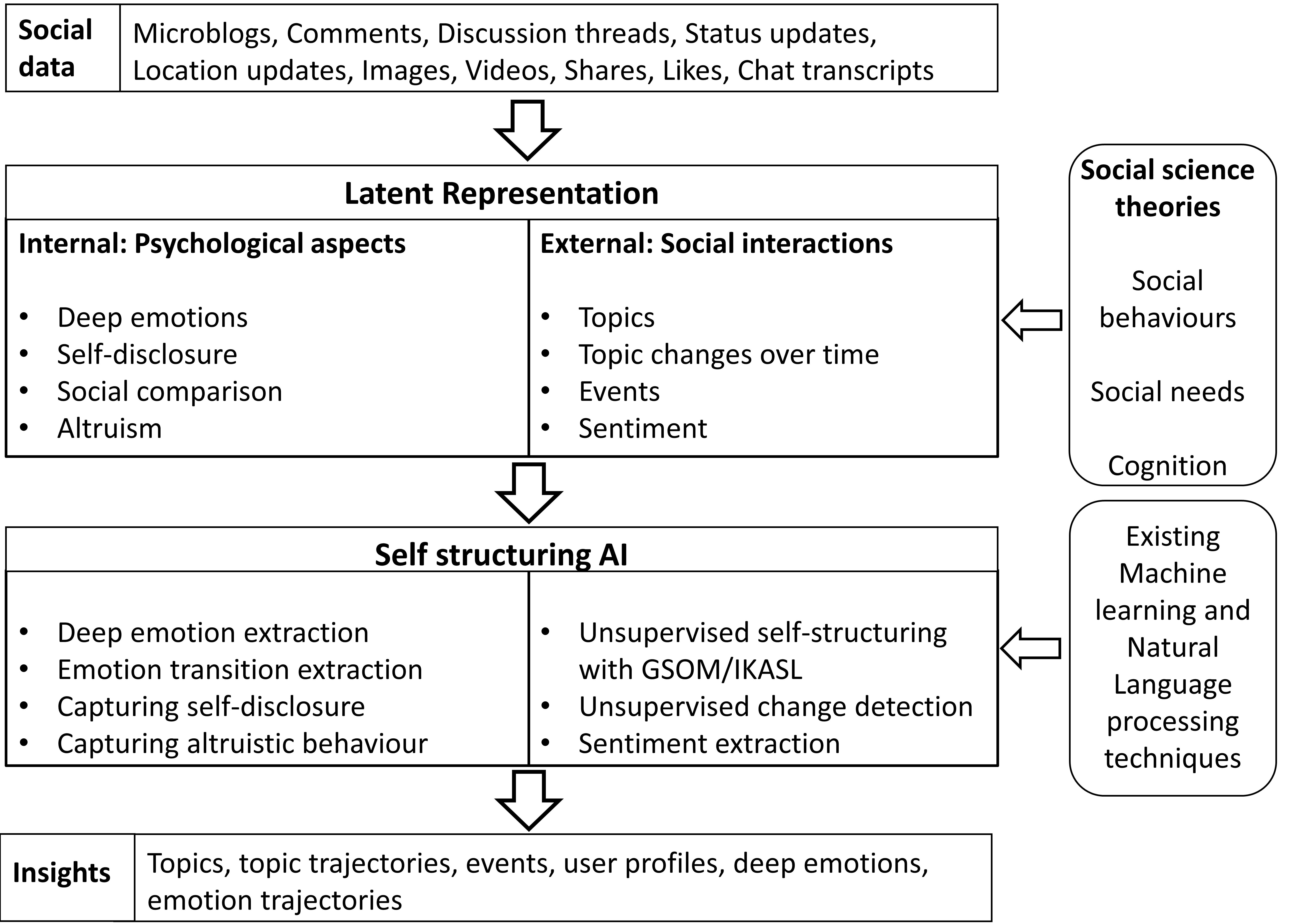}}
\caption{The proposed self-structuring artificial intelligence framework for generating insights from social data in online social media platforms.}
\label{fig:proposed_platform}
\end{figure}

Figure~\ref{fig:proposed_platform} shows the key elements of the proposed framework which consists of (i) representing social data based on existing social sciences theories into useful latent representations such as emotions, topics and events,
(ii) unsupervised self-structuring to automatically structure representations of social data into semantically coherent groups, (iii) incremental learning from social data streams to learn temporal changes, and (iv) natural language processing to extract semantic information about individuals and groups. 

The unsupervised self-structuring machine learning techniques are employed to automatically generate representative structures from the feature transformed social data, based on the underlying similarities of the social behaviours present. Such structuring of social data highlights the common patterns of social behaviours among different social groups. These unsupervised algorithms are extended from the conventional machine learning algorithms to suit the challenges of the social data such as high velocity, diversity and unstructured nature. 

Incremental learning techniques helps to capture temporal patterns which occur in social data streams at different granularities. For instance, at the individual level, changes are related to internal individual behavioural changes over time. Similarly at the group level, changes of group behaviour often are driven by external factors such as events.


         

The above framework is developed and implemented in the forthcoming chapters of this thesis, which describes the implementation of parts of the framework as well as its application on two social data sources with distinct characteristics.

Chapter four presents an implementation of the above framework for fast-paced social media platforms such as Twitter which mainly focus on the external aspects in Figure~\ref{fig:proposed_platform}. Such social media platforms are mainly characterised by its fast-paced nature where discussions are used to rapidly disseminate information across the participants of the social media platform. The participants discuss topics related to recent events and express their opinion/sentiment. However discussion topics are changed frequently. An unsupervised self-structuring and incremental learning algorithm is developed that automatically structure the social data into coherent clusters based on the semantic similarity which exists in social data, and an event detection algorithm is developed to capture temporal changes of various behavioural change indicators.


Chapter five presents an implementation of the above framework for online support groups (OSG) which are slow-paced and show contrasting characteristics to Twitter. OSG discussions focus prominently on the internal or psychological aspects such as emotions specified in Figure~\ref{fig:proposed_platform}. OSG discussions are tightly coupled to the main theme of OSG which is often an disease condition and participants engaged in longer discussions. Because of higher homophily and stronger ties people tend to engage in deeper self-disclosure such as expression of deeper emotions. Multi-granular emotional expression was structured and incrementally captured overtime to generate individual emotional trajectories. Also, a suite of machine learning and natural language processing techniques are developed to capture different levels of self-disclosure. 



Chapter six extends the algorithms developed in chapters five to gain insights from prostate cancer related OSG. Extensions include algorithms to capture the treatment decision making behaviour and decision factors.


%% file: Chapter4.tex
	
\onlyinsubfile{		
\setcounter{chapter}{3} 
}

\chapter[]{Ideation to creation: A new self-structuring artificial intelligence algorithm}\label{chap:4}
\epigraph{{\textit{No event can be judged outside of the era and the circumstances in which it took place.\\}}{\hfill Fidel Castro}}

This chapter presents the algorithmic development of the  self-structuring artificial intelligence conceptual model proposed in Section~\ref{sec:proposed_platform}. As discussed in previous chapters, social data accumulated on online social media platforms is a representation of social interactions and behaviours of individuals from diverse socio-demographic backgrounds. 

This chapter is organised as follows. Section~\ref{sec:highlevel_technique} describes the high-level design of the proposed algorithm. Section~\ref{sec:incremental_learning_ch4} focuses on incremental learning for topic separation and Section~\ref{IKASL} presents IKASL: a recent unsupervised incremental learning algorithm that forms the foundation for the proposed algorithm. Section~\ref{sec:3_extensions_IKASL} presents the first technique of the proposed algorithm, for handling complexities of  social text streams and Section~\ref{sec:event_detection_technique} presents the second component, the event detection technique. Section~\ref{sec:illustrative_example} demonstrates the full algorithms using an illustrated example and Section~\ref{sec:chap4_experiments} presents an empirical evaluation using two Twitter datasets.

\section{The design of the algorithm}\label{sec:highlevel_technique}

Given the diversity of social data generated by distinctive social media platforms, this development focuses on fast-paced online social media platforms such as Twitter which is a high volume/velocity text stream contributed by a diverse set of participants from all over the world. Also, social media messages are relatively short and highly unstructured.
	
Due to this fast-paced nature, such social media platforms are mainly employed for rapid dissemination of current and trending information through the social network to a large number of individuals~\cite{Kwak2010,Lotan2011,lovejoy2012engaging}. Due to this rapid dissemination of information, any trending topic would capture the attention of a large number of individuals quickly leading to a surge in discussions related to that topic. After a peak such trending topic often rapidly lose interest and discussions may switch to a different trending topic. As the primary intention is information dissemination rather than having continuing discussions, the ties between dyads are relatively weak and often lack homophily, reciprocity. Because of weak ties and lack of homophily, as discussed in Section~\ref{sec:self_disclosure}, self-disclosures tend to be  limited to revealing surface information  about oneself.  Also, the emotions expressed in \textit{tweets} are often superficial and intense, with a clear direction towards positive or negative sentiment. 
	
The developed algorithm consists of two techniques designed based on the above characteristics of the platforms. The first is a new unsupervised incremental machine learning technique that automatically captures discussions from a text stream of a fast-paced social media platform. This technique self-structures social text into coherent clusters which are indicative of distinct topics, and extend those clusters across time into pathways a.k.a \textit{topic pathways} to capture changes of those discussions over time due to changes in trending topics and appearance new unseen topics.

The composition of a social text stream evolves continuously over time, which includes changes in the focus of the topics, the emergence of new topics and cease of certain topics. Considering a social text stream altogether makes it difficult to capture those interesting dynamics of the discussion topics. Therefore, it is important to separate the social text stream into dynamic topics that evolve over time i.e., topic pathways. This topic pathway separation reduces noise and separates social media messages in a meaningful way so that they are categorised into topic pathways based on their topical similarity. The proposed technique is designed to achieve this task by generating a self-adaptive structure automatically separate social text stream into topic pathways.     

The second technique is a multi faceted event detection technique developed to monitor topic pathways for significant changes in social behaviours to detect significant changes over time and identify such changes as potential events of interest. Due to the above discussed characteristics changes in social behaviours are mainly on changes in self-disclosure as in changes of discussion topics and changes in emotional expression. Therefore, detection technique was developed to capture those changes. 

These new techniques were designed to be capable to effectively handle the challenges related to the unstructured nature of social data outlined in Chapter~\ref{sec:social_data_analysis_challenges}. They are unsupervised and capable of incremental learning thus overcomes issues related to unlabelled data and time sensitivity.    

This algorithm was trialled using two large datasets of tweets, one on a  public figure and the other on an international organisation. The results highlighted the capability to automatically structure tweets semantically and temporally yielding coherent and meaningful topic pathways. Also, the automatically detected events from the detection algorithm were relevant and meaningful.

\begin{figure}[!htb]
	\centering
	\includegraphics[clip=true, width=1.0\linewidth]{{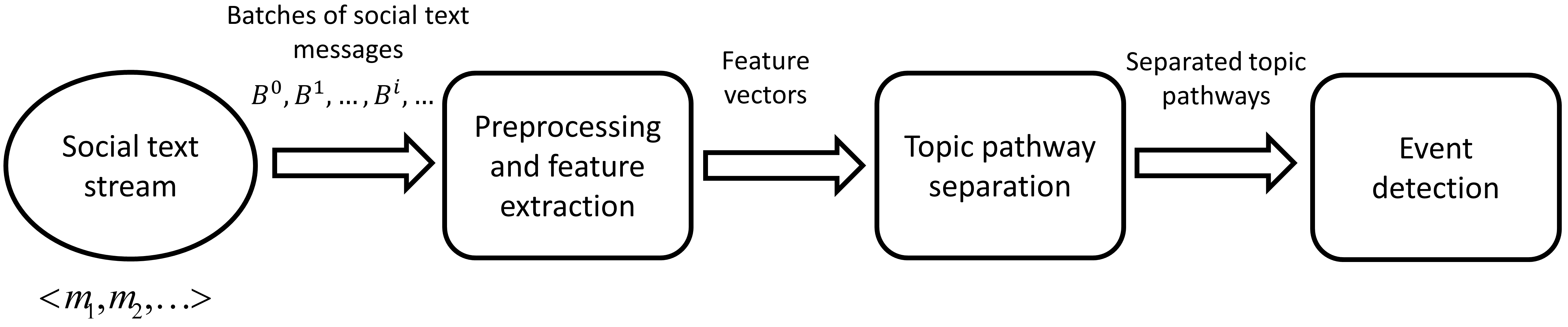}}
	\caption{High-level design of the proposed platform for topic separation and event detection from a social text stream}
	\label{fig:high_level_flow}
\end{figure}

\noindent Figure~\ref{fig:high_level_flow} presents the high-level design of the proposed platform. A social text stream is denoted as a continuous stream of messages over time $<m_1,m_2,\ldots>$ where each message $m$ consists of a \textit{short text} and a \textit{time-stamp} of its published time. 
  
Firstly, the social text stream is sampled using a fixed time-interval $\Delta t$ based on the \textit{time-stamp} of social media messages producing batches of social media messages $B^0,B^1,\ldots,B^i,\ldots$. Each batch is a set of social media messages collected for the time period $\Delta t$.
\begin{equation}
B^i = \big\{ m \big\}_{t=i\times \Delta t}^{t=(i + 1)\times \Delta t}
\end{equation}

\noindent The granularity of the time period $\Delta t$ (e.g., hourly, daily, weekly) can be adjusted to suit the velocity of the social text stream and also to match other application specific requirements. 

Each batch is preprocessed to reduce noise due to unstructured nature of social text and feature vectors $v$ are extracted from each social text message $m$. Note that technical details of preprocessing and feature extraction are discussed in Section~\ref{sec:chap4_experiments} . The feature vectors of each batch are employed by the topic pathway separator to separate the social media messages into topics and topic pathways over time. 

A \textit{topic pathway} is a series of social media messages that discuss a distinct topic. It stretches over time across several batches of social media messages. A \textit{topic segment} is a part of the topic pathway that belongs to a particular batch. Thus, each topic pathway is a chain of topic segments. A topic pathway $D$ can be depicted as follows: 

$$D= D^0 \rightarrow D^1 \rightarrow \ldots D^i\ldots\rightarrow D^n$$
\noindent where $D^i= \{m\}$ is a topic segment and $D^i \subset B^i$  

Each topic segment is a set of social media messages that are semantically similar to each other. Hence, technically it can be considered as a clustering problem within each batch. A chain of such semantically similar clusters across batches forms a topic pathway.  Therefore, this problem can be reduced to unsupervised incremental clustering algorithm that builds a set of semantically similar cluster chains across batches of social media messages.

As discussed in Section~\ref{sec:event_detection}, and event can be broadly defined as an occurrence that can be bounded by space and time~\cite{allan1998line}. Events could happen anytime and anywhere which could impact a single person, a handful of individuals, or a large number of individuals (e.g., natural disaster, election). Since the interest here is to capture any event that has caught attention in online social discussions, any specifics of a particular event is not priorly known, thus \textit{unspecified event detection} (see Section~\ref{sec:unspecified_event_detection}) technique was designed for this task. This unspecified event detection technique looks for changes in online social behaviour of the impacted individuals for the duration of the event (e.g., change of discussion topics, changes of emotions). Such changes in human social behaviours related to an event appear as sudden bursts~\cite{Barabasi2005} in social text stream in between moderate or low level of activity. However, such bursts in a particular topic often diluted inside the diverse social text stream and can go unnoticed because the dynamics of other topics act as noise. 

As a solution, we propose to detect events in the separated topic pathways because each topic pathway is a coherent stream of social media messages which continues a discussion about a certain topic. Hence, a burst of activity related to a particular topic would be easily detectable in relevant topic pathways as they are not (or minimally) affected by the noise of other topics. This proposed event detection algorithm monitors each topic pathway for topic specific events and detects such events using a multi-faceted event detector that looks for multiple event indicators.  

The following sections describe the topic separation and event detection components of the proposed platform in detail. 

\section{Unsupervised incremental learning for topic separation}\label{sec:incremental_learning_ch4}
As discussed above, topic pathway separation is technically an unsupervised incremental learning task that builds a self-adaptive structure which consists a set of cluster chains over time across the batches of data.  

As discussed in Section~\ref{sec:incremental_learning}, unsupervised incremental learning is a challenging task. It has to learn previously unseen patterns from new incoming data while preserving the knowledge acquired previously (no catastrophic forgetting). Also, it should not require access to the past data that it has already processed.

Due to this challenging nature of the unsupervised incremental learning, there are only a handful of algorithms in literature that satisfies the above three characteristics (see Section~\ref{sec:incremental_learning} for details). Out of those, the recently published Incremental Knowledge Acquisition and Self-Learning (IKASL) \cite{DeSilva2010a,DeSilva2010} is selected as the base for this work.
 
In IKASL, learning occurs as a structure of layers where each layer is learned from a new batch of data. In each layer, structure of the incoming data is self structured using the unsupervised self-learning mechanism introduced in GSOM algorithm \cite{Alahakoon2000}. GSOM employs the brain-inspired Hebbian learning rule \cite{Kohonen1998} to learn the underlying structure from a dataset.

IKASL preserves the acquired knowledge in a generalised form therefore, it does not require access to past data that it has already processed. The generalised version of the acquired knowledge from each layer (n) is used as the basis for the knowledge acquisition from the subsequent layer (n+1), thus it avoids catastrophic forgetting of the past knowledge. Moreover, while using the past acquired knowledge as the base, it incrementally acquires new knowledge that is encapsulated in the upcoming data.
The following section elaborates the functional details of the IKASL algorithm.

\section{IKASL algorithm}\label{IKASL}
This section describes the functional details of the IKASL algorithm. Table~\ref{table:IKASLNotation}, states the symbolic notation that will be used in this section to detail the IKASL algorithm.

\begin{table}[!htb]
\caption{Notations used for the functional details of IKASL algorithm}\label{table:IKASLNotation}
\begin{tabulary}{\linewidth}{|M|L|}
	\hline 
	\textbf{Symbol}	& \textbf{Description} \\ \hline 
	B^i & i\textsuperscript{th} batch of social media messages \\	\hline 
	d & number of dimensions in (i) feature vector of a social text message, and \newline (ii) weight vector of a feature map node\\	\hline 
	LE^i& i\textsuperscript{th} learning phase \\ \hline 
	GE^i& i\textsuperscript{th} generalisation phase \\ \hline 
	n^j& j\textsuperscript{th} feature map node \\ \hline  
	TE^j& total quantisation error of node j \\ \hline  
	C^i& cluster representation vector learned during GE\textsuperscript{i}\\ \hline 
	\mathcal{N}_{N_q}& Set of N\textsubscript{q} and neighbouring nodes of N\textsubscript{q} \\ \hline 
\end{tabulary} 
\end{table}

The IKASL algorithm functions in three key phases; initialisation, learning and generalisation which are delineated below.

\subsection{Initialisation}
IKASL is initialised in a similar manner to GSOM, using a map of four starting nodes $\{n\}^{4}$. Each node $n$ is initialised with a weight vector $w$ of size $d$. 
Initial values for the weight vector can be selected based on two different lines of approaches:
\begin{itemize}
	\item \textbf{random initialisations:} this type of approaches include completely random initialisations, initialise using randomly selected data points and random sampling from a statistical distribution.
	\item \textbf{data distribution based initialisation:} these type of approaches include the use of K-means clustering to initialise weights to identified cluster centres and heuristic methods such as principal component analysis to assign optimal values to weights. 
\end{itemize}
\noindent Each approach comes with pros and cons however, often such gains are marginals and only affect the rate of convergence. Interesting readers can refer to~\cite{Attik2005} for further details. 

For the sake of simplicity, IKASL weight vectors are initialised by sampling from a uniform random distribution in the range of [0, 1].

\subsection{Self structuring}\label{sec:self-learning}
This learning phase $LE^i$ happens in each layer of IKASL in which it self learns a topology preserving two-dimensional feature map form the data batch $B^i$.
This self structuring function employs the GSOM~\cite{Alahakoon2000} algorithm to learn the feature map. The GSOM algorithm is an enhanced version of Self Organising Map (SOM)~\cite{Kohonen1998} discussed in Section~\ref{sec:self_structuring}.

Similar to SOM, GSOM uses the concept of competitive learning, where a set of nodes (neurones) ins a feature map compete for the ownership of input data vectors. The winner or the \textit{best matching unit} is the node that is most similar to the input. 

Once the winner is selected, brain inspired Hebbian learning rule~\cite{Hebb1949} is used to update the weights of the winning node and its neighbourhood so that their distance to the particular input is reduced. This process iteratively learns the topology preserved feature map. One of the key limitations of SOM is that the number of nodes in the feature map need to be pre-determined, which is often done experimentally. GSOM overcomes this issue by initiating with a small number of nodes in the feature map and using a node expansion function to dynamically expand the feature map during the training phase. This self structuring phase executed in two stages (i) Self-Organisation with node expansion and (ii) Smoothing. The details are described below.

\subsubsection{Self structuring and node expansion}
This stage presents the inputs in a randomised order and learns the feature map using competitive learning and Hebbian learning rule~\cite{hebb2005organization}. Node map is dynamically expanded based on the quantisation error accumulated in individual nodes.

\begin{itemize}
\item A random input vector $v$ is selected from the dataset and presented to the network of nodes in GSOM. 
\item The network of nodes compete for the ownership of the input $v$ and the node $n_q$ that is closest to the input vector $v$ is selected as the winner: \\ $sim(v,n_q)\: \geq \: sim(v,n),\: \forall n \in \{ n \}$.\\ Note that similarity is measured using a designated distance function.
\item The feature map updates the weight vectors of the winning node and its neighbouring nodes $\mathcal{N}(n_q,i)$ to reduce the distance between nodes and the input vector $v$
The weight vector $w_j$ of each selected node $n_j$ is updated as follows:

\begin{equation}
w_j(k,i+1) = \begin{cases} w_j(k,i) + \alpha(i) \times (v(k)   -   w_j(k,i)), &	\mbox{if }  n_j \in \mathcal{N}(n_q,i) \\ w_j(k,i), & 	\mbox{otherwise} \end{cases}
\end{equation}

Note that $\alpha(i) \leq 1.0$ is the learning rate, $i$ is the iteration and $k= 1, 2,\ldots, d$. \\
The learning rate $\alpha$ is a tuning parameter which is determines the amount of weight correction in each selected node due to the input $v$. It starts with a higher values and gradually decreases as $i \rightarrow \infty$.\\
$\mathcal{N}(n_q,i)$ is the neighbourhood of $n_q$ that receives weight updates. This process simulates the neuro-biological phenomenon of \textit{lateral excitation} which happens in the neocortex of brain where the activation of a neurone excites the neurones that are connected to it~\cite{Dalva1994}. The size of the neighbourhood $\mathcal{N}(n_q,i)$ gradually decreases as $i \rightarrow \infty$ and eventually reduces to $\{n_q\}$.
  
\item The winning node updates its quantisation error based on the distance between the node and the input $v$: $QE_{n_q} = QE_{n_q} + \vert v - w_q \vert$. This accumulation of quantisation error is recorded for each node in the network. It is an indication of the level of generalisation of each node, where higher values indicate that the node is over-generalised and represents a large input feature space~\cite{Alahakoon2000}. 

\item IKASL uses the node expansion method proposed in GSOM~\cite{Alahakoon2000} which uses the accumulated quantisation error $QE_{n_q}$ to trigger expansion of the network from that node. When $QE_{n_q} > \tau_Q$, node expansions is triggered on $n_q$ and new node are added to the vacant spaces of $n_q$ in the grid formations as shown in Figure~\ref{fig:node_growing}.
\begin{figure}[!htb]
	\centering
	\includegraphics[clip=true, width=0.9\linewidth]{{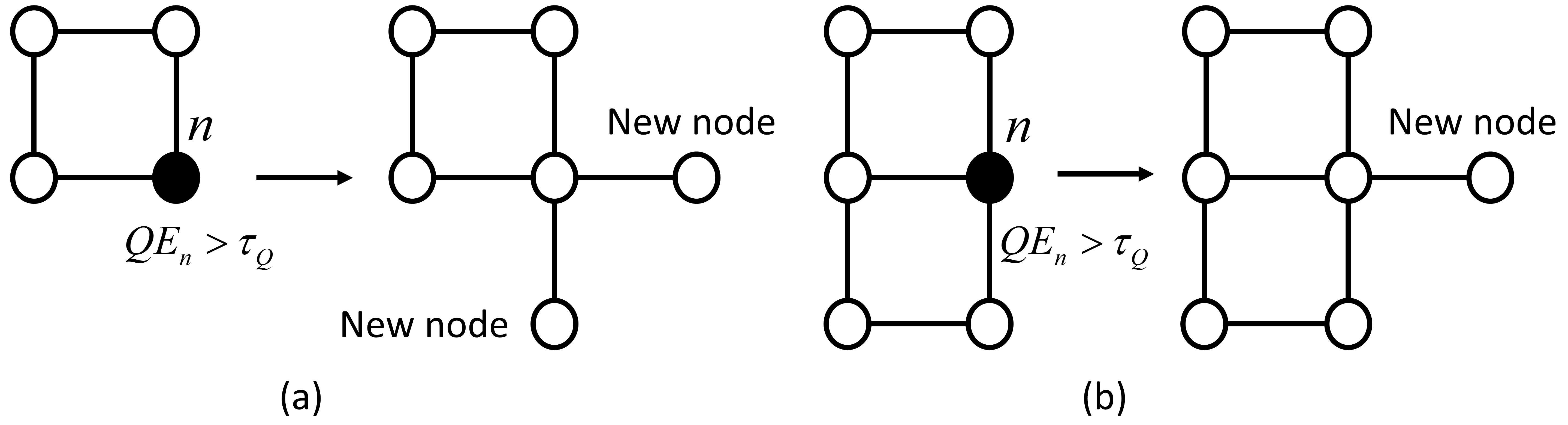}}
	\caption{Two node expansions scenarios}
	\label{fig:node_growing}
\end{figure}  
\end{itemize}

\subsubsection{Smoothing}
This stage further fine tunes the weights of the node map by re-feeding the inputs. It uses the same competitive learning and Hebbian learning rule based approach to fine tune the node weights. However, in contrast to the previous stage, this stage uses a small learning rate $\alpha$ and a small neighbourhood in order to avoid substantial weight changes due to inputs. Moreover, this stage uses a fixed node map learned from the previous stage and do not perform node expansion.  

\subsection{Generalisation}\label{sec:generalisation}
This generalisation phase $GE^i$ extracts a summarised representation as a set of cluster representation vectors $\{C\}$ from the topology preserved node map learned from the corresponding self structuring phase $LE^i$. Nodes that are able to claim a significant proportion of inputs are identified as \textit{hit nodes}. The cluster representation vectors $\{C\}$ are learned by aggregating the weights of hit nodes and weights of its neighbourhood. This phase is described below:
\begin{itemize}
	\item Input vectors are re-fed into the node network and let the nodes claim inputs based on similarity. 
	\item Hit count $h(n)$ of each node $n$ is the number of inputs claimed by each node. 	
\begin{equation}
	h(n_j) = \vert\{ \forall v \in B^i: n_j = arg\max_{n \in \{n\}}(sim(n,v)) \}\vert
\end{equation} 
	\item Nodes that claimed more than $\tau_H$ proportion of input were marked as hit nodes, where hit threshold $\tau_H$ is a tunable parameter. Note that, hit threshold $\tau_H \in [0,1]$ controls the granularity of the summarised representation where higher $\tau_H$ results few hit nodes, thus, more generalised representation. 
	\item The \textit{cluster representation vector} $C_{h1}$ is formed by aggregating the weights of an identified hit node $n_{h1}$ and its neighbourhood $\mathcal{N}_h(n_{h1})$. 
	$$C_{h1} = aggregation(n_{h1},\mathcal{N}_h(n_{h1}))$$
	Note that IKASL has employed a \textit{fuzzy integral} as the aggregation function. However, any suitable aggregation function can be used for this task.  
\end{itemize}   

\subsection{Learning from generalisation} \label{sec:lrn_from_gen}
After the first layer, learning phase $LE^i$ is based on the set of cluster representation vectors $\{C\}^{i-1}$ learned during previous generalisation phase $GE^{i-1}$. Each \textit{cluster representation vector} becomes the weights of the starting node of a separate feature map in $LE^i$. Therefore, $\vert\{C\}^{i-1}\vert$  number of feature maps are learned in $LE^i$.

Firstly, the feature vectors are assigned to their closest \textit{cluster representation vector} based on an employed similarity function. Those assigned feature vectors are then used to learn the respective feature maps initialised from that \textit{cluster representation vector.}

\begin{figure}[!htb]
	\centering
	\includegraphics[clip=true, width=0.8\linewidth]{{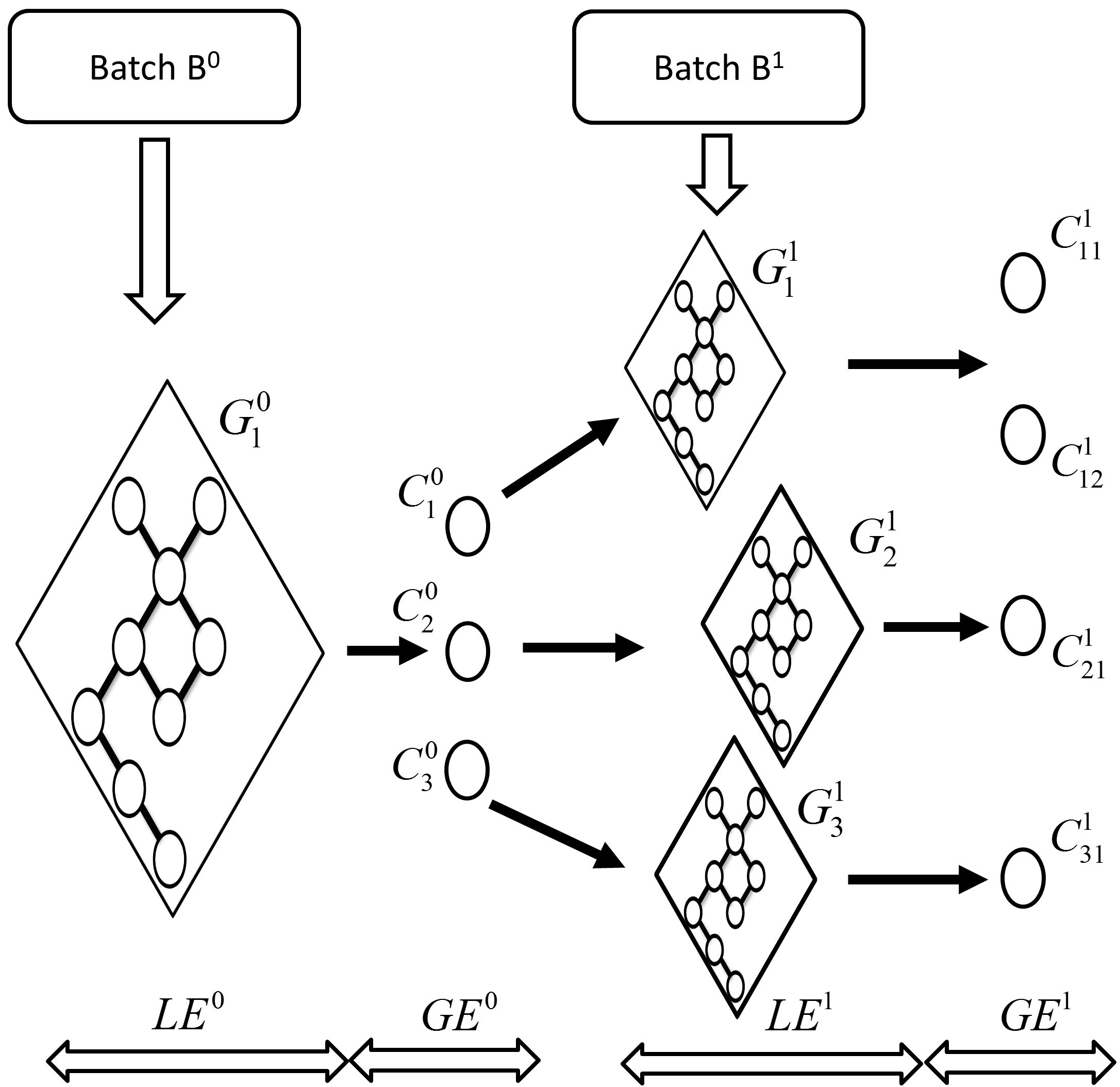}}
	\caption{Layered structure of IKASL algorithm}
	\label{fig:IKASL}
\end{figure}
Figure~\ref{fig:IKASL} highlights the self-learned layered structure of the IKASL algorithm. Learning phase $LE^1$ is initialised using the \textit{cluster representation vectors} $\{C\}^0$ of layer $0$. Hence, previously learned knowledge from $B^0$ is preserved and used as the basis for the acquisition of new knowledge. The new knowledge from batch $B^1$ of data is acquired by the learned feature maps in $LE^1$. Subsequently, generalisation layer $GE^1$ generates a summarised representation of the acquired knowledge as a set of \textit{cluster representation vectors} $\{C\}^1$. 

\subsection{Topic separation with IKASL}
The features vectors derived from batches of social media messages are used to self structure the adaptive structure that consists of chains of cluster representation vectors. Each batch of social media messages are assigned to the most similar \textit{cluster representation vector} generated from the same batch, thus forming clusters of social media messages. Chains of \textit{cluster representation vectors} form chains of social media messages clusters that become the topic pathways. 

IKASL addresses the issue of time sensitivity (see Section~\ref{sec:social_data_analysis_challenges}) in social media messages by sampling them using their published time and incrementally learning topic pathways that capture the dynamic changes of each topic. The learning of IKASL has linear complexity with number of social media messages, hence it is scalable to handle large volumes of social text data. Moreover, the acquired knowledge is stored in a summarised form as \textit{cluster representation vectors}, which makes it scalable in memory as well.

While IKASL addresses several issues related to topic pathway extraction from a social text stream, there are several identified issues that are not addressed. Hence, an extended IKASL algorithm (\ikaslext) is proposed to address those identified issues. Those issues and the resolutions in \ikaslext are presented in next section.   

\section[\ikaslext]{\ikaslext: Extended IKASL for topic separation from a social text stream}\label{sec:3_extensions_IKASL}
This section describes three key extensions included in \ikaslext, the extended IKASL algorithm for topic separation from social text stream, which is optimised to handle the challenges in social text.

Section~\ref{sec:dynamic_vocabulary} describes the developed technique to overcome the issue of dynamic vocabularies, Section~\ref{sec:brevity} describes the technique the handle the brevity in social media messages and Section~\ref{sec:new_topicpathway} presents the technique to detect new topic pathways and improve coherence in existing topic pathways.  

\subsection{Handling dynamic vocabularies}\label{sec:dynamic_vocabulary}
In online social media platforms users often coin new words that are not found in standard vocabulary~\cite{Eisenstein2013}. As discussed in Section~\ref{sec:social_data_analysis_challenges}, some of these terms are tag words that are coined to represent certain events or topics (e.g., hashtags in Twitter). Such terms are important to aggregate the social media messages of that topic/event and also to understand about the evolution of that topic/event. Often these new terms appear as bursts where its usage peaked rapidly, subsequently declines and seldom used after. 

The conventional approach would be to use a static global vocabulary of significant terms in the dataset to generate feature vectors from social media messages. However, there are two key issues associated with this conventional approach.
\begin{enumerate}
\item The tag words that are used in social data streams are not known a priori~\cite{AravindanMahendiran} thus it is not possible to create a global static vocabulary that includes all possible terms.
\item Consideration of large number of terms as features significantly increases the sparsity of the feature vectors.
\end{enumerate}

\noindent In order to resolve these issues of a global vocabulary, a new method is developed to facilitate dynamic vocabularies which can incorporate new significant terms (e.g., hashtags) to the incremental learning task as they appear in new batches of data. A vocabulary $V^i$ comprising a set of significant terms are learned from each batch $B^i$ of social media messages. This vocabulary $V^i$ is local to batch $B^i$ and used for the self structuring stage of that batch. Therefore, the learnt cluster representation vectors $\{C\}^i$ from the batch $B^i$ are based on features extracted using $V^i$. 

The candidate terms for the vocabulary $V^i$ is selected based on local term-frequency \cite{Salton1988} i.e. frequency of the term in batch $B^i$, where the terms that have a local term-frequency higher than $\tau_V$ in $B^i$ are included in $V^i$. $\tau_V$ is a tuning parameter in which setting a higher value can miss informative terms and setting a lower value can increase the sparsity of the feature representation of social media messages.

The key challenge of this approach is how to measure similarity between feature vectors based on two distinct vocabularies. This is because in IKASL, the cluster representation vectors $\{C\}^i$ learned from the batch $B^i$ are employed as the basis for the self structuring stage for subsequent batch $B^{i+1}$, in which $V^{i+1}$ might contain different terms that are not in $V^i$. 

In order to solve this problem the intersection between the two vocabularies is considered in similarity calculations. For example, let $C^i_m$ be any cluster representation vector from stage $i$ and $v^j$ be any input vector from data batch $B^j$ (where $j>i$), the Cosine similarity between $C^i$ and $v^j$ is defined as follows:

\begin{equation}
sim(C^i,v^j) = \bigg[ \dfrac{C^i \cdot v^j}{\Vert C^i\Vert \times \Vert v^j\Vert} \bigg]^{V^i \cap V^j}
\label{equ:IKASL_ext_sim}
\end{equation}

\noindent where $V^i \cap V^j \neq \emptyset$. Note that Cosine similarity is a popular similarity measure for sparse feature vectors and often used in text \textit{bag-of-word} representations.
  
\begin{figure}[!htb]
	\centering
	\includegraphics[clip=true, width=0.8\linewidth]{{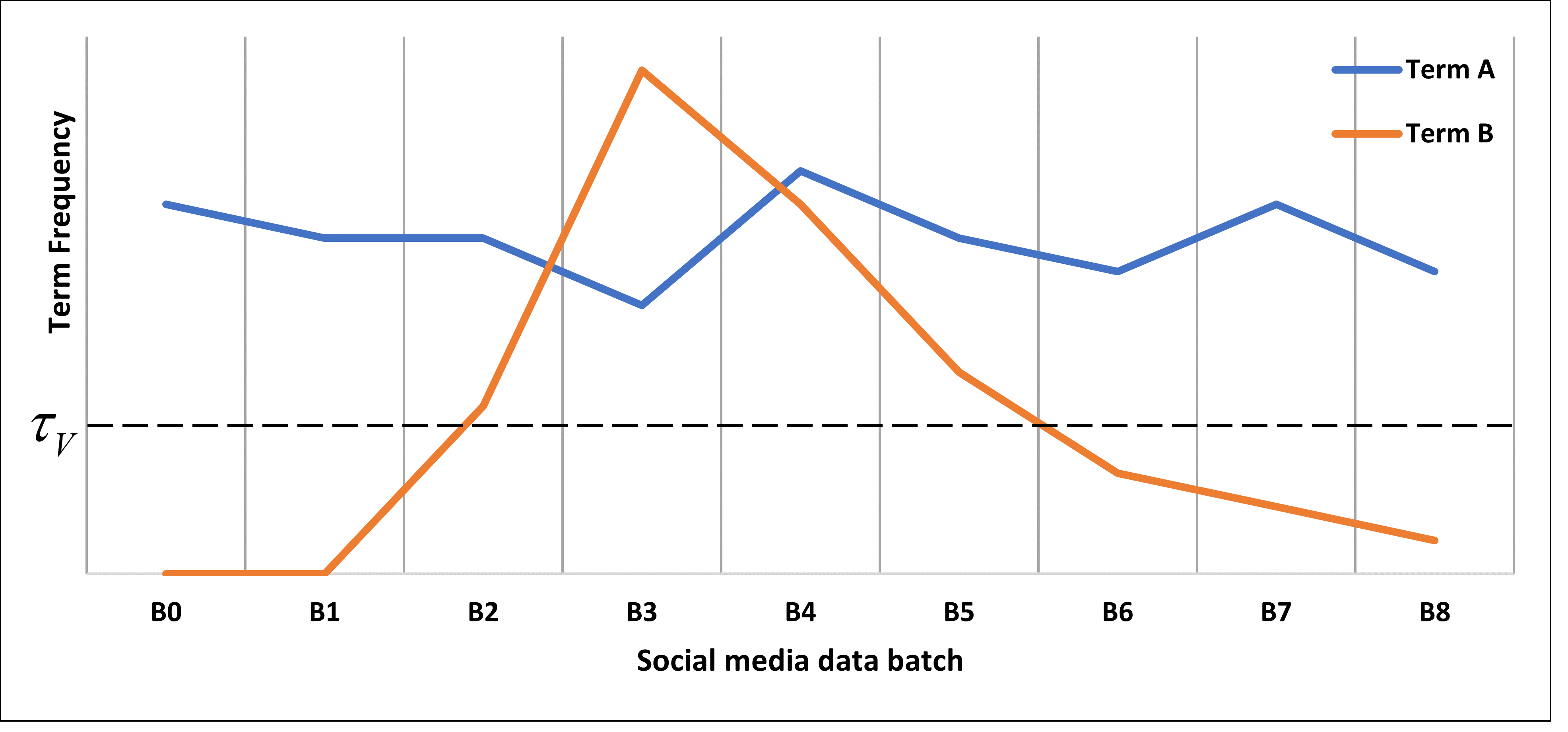}}
	\caption{Local \textit{term frequency} of terms A and B across social media message batches $B^0,\ldots,B^8$. $\tau_V$ is the local \textit{term frequency} threshold that determines inclusion of exclusion in the respective vocabulary}
\label{fig:dynamic_vocabulary}
\end{figure} 

Figure~\ref{fig:dynamic_vocabulary} shows the local \textit{term frequency} plot of two terms in the social data stream. Term A maintains a local \textit{term frequency} higher than $\tau_V$ throughout hence it is included in vocabulary of all batches shown. In contrast, term B appears as a burst in $B^3$, reached peak local \textit{term frequency} in $B^5$ and gradually declined subsequently. Its local \textit{term frequency} is greater than $\tau_V$ only in batches 3 to 7, hence included in vocabularies $V^3,\ldots,V^7$. 

This method overcomes the problem of new terms as it uses a dynamic vocabulary and incorporates new significant terms in the feature space as they appear in the new batches of social media messages. As in for \textit{term B} shown in Figure~\ref{fig:dynamic_vocabulary}, dynamic vocabularies include the new term as a feature when it is significant and relevant. Thereby, self structuring and generalisation stages on relevant stages consider the dynamics of the particular new term for topic pathway separation. Moreover, \textit{term B} is automatically excluded from the feature space when its usage drops beyond the threshold $\tau_V$ thus less relevant. This subsequent exclusion reduces the sparsity of the feature vectors and thereby improves the clustering results.  

\subsection{Addressing the brevity issue}\label{sec:brevity}
The short length of social media messages are imposed by some social media platforms (e.g., tweet is limited to 140 characters). Even in platforms without such restrictions, users prefer brevity since it allows them to express more efficiently. This results in a fewer number of words in social media messages. For example, tweets are on average 11 words long~\cite{Oconnor2010}. 

The feature vectors of the social media messages are formulated using the \textit{bag of words} technique where each feature is a word in the text corpus of that batch of social media messages. The corresponding feature of each word present in the social media messages is assigned with the feature values of that word while other features are zero (not activated). Because of the brevity, these feature vectors of the social media messages have only a handful of (non-zero) features.  

During the self structuring phase (Section~\ref{sec:self-learning}) the activated features of a feature vector activate the corresponding weights of the winning node. Throughout the self structuring phase different sets of weights of the node get activated based on the feature vectors that the node claims as the winner. However, since the winner is claimed based on similarity which is determined using the activated features of the node, the already activated features often get further strengthened during the self structuring phase. These activated weights of the node represent the acquired knowledge of that node. Because of the brevity issue, often few feature weights are activated in each node after the self structuring phase.  

When the nodes have only a few activated weights and others are close to zero, the overlapping activated features of the hit node and its neighbourhood will be less. Hence, the existing summation based aggregation function used in the generalisation phase (see Section~\ref{sec:generalisation}) would diminish the strength of the non-overlapping weights and the relevant acquired knowledge will be lost.

This issue is inherent in summation based aggregation methods which favours frequently activated features, while features activated in one or few considered nodes are discriminated~\cite{Murray2014a}. In order to minimise this issue, we adapted an aggregation method known as \textit{max-pooling} which is popular in Bag of Visual Words (BoVW) based techniques in visual recognition tasks~\cite{Peng2015}. In fact, ~\citep{Boureau2010a} have theoretically shown that \textit{max-pooling} performs better than \textit{average-pooling} (averaging based aggregation) on features with a low probability of activation.

Max-pooling takes the maximum feature value for each feature from hit node and its neighbourhood. In a $d$ dimensional feature space, a \textit{cluster representation vector} is calculated from the neighbourhood $\mathcal{N}(n_h)$ (neighbourhood includes the hit node) of a hit node $n_h$ as follows:

\begin{equation}
\forall k \in 1,\ldots,d, C(k) = \max\limits_{n \in \mathcal{N}(n_h)}(n(k))
\end{equation}

This approach does not average-out  activated features that only exist in few nodes to the near-zero values in other nodes. Intuitively, in the context of \textit{bag of words} feature extraction, \textit{max-pooling} aggregates the significant words learned by each node considered and include all of them as significant words in the respective \textit{cluster representation vector}.  

\subsection{Identifying new topic pathways}\label{sec:new_topicpathway}
The impact of dynamic and time sensitive nature of social data streams extends beyond the granularity of new terms into a much broader concept of new topic pathways. New topic pathways emerge when there is a substantial public interest on a new discussion topic i.e., \textit{social media buzz}. Such emerging topic pathways are often linked to real world events that have captured the attention of a significant portion of the general public. Hence, it is important to capture newly emerging topic pathways which are loosely related to existing pathways. 

However, the original IKASL algorithm links clusters to previous layers, thus all new clusters are formed having a cluster representation vector learned from previous layers as the base. Hence, when a new topic emerges, the relevant social media messages would map into an already existing topic pathway and only seen as incremental changes to an existing topic pathway. Such an approach would hide the new topic as well as decrease the coherence of the existing topic pathways.  

Moreover, text datasets often face the issue of feature sparsity and as discussed in Section~\ref{sec:social_data_analysis_challenges} this sparsity more so in social data. Sparse feature spaces make nearest neighbour queries unstable~\cite{Beeri1999} and less meaningful~\cite{Aggarwal2001}. Therefore, the process (described in Section~\ref{sec:lrn_from_gen}) of assigning social data messages to most relevant topic pathway i.e., the closest \textit{cluster representation vector} based on similarity would assign less similar feature vectors to respective topic pathways. Hence, the knowledge acquisition process from the new unseen messages can increase the variance within topic pathways decreasing the topic coherence of topic pathways over time. 

In order to overcome these two issues, we employed a similarity threshold based method which is often used in high-dimensional clustering algorithms~\cite{Ester1996,Ertoz2003} to filter out \textit{noise} feature vectors that are loosely related to any existing clusters. However, instead of filtering out such feature vector as \textit{noise} our proposed technique identifies any potential new topics encapsulated in them.
Therefore, in this proposed method, if the similarity with the closest \textit{cluster representation vector} is greater than the similarity threshold $\tau_{topic}$, then such features vectors are used to learn a new feature map $G_{new}$ that is randomly initialised as in $LE^0$ as below. Thus, for a feature vector $v$, the feature map $G$ is selected as bellow: 

\begin{equation}
G = \begin{cases} G_j, &	\mbox{if }  C_j= arg\max\limits_{C \in \{C\}}(sim(C,v)) \text{ and } sim(C_j,v) \geq \tau_{topic} \\ G_{new}, & 	\mbox{otherwise} \end{cases}
\label{equ:IKASL_ext_newGSOM}
\end{equation} 

$\tau_{topic}$ is the \textit{topic pathway coherence threshold} that represents the expected coherence of the identified topics. Low $\tau_{topic}$  values result in few topic pathways with less coherent content in which some might be loosely related to the particular topic. $\tau_{topic}$ positively correlates with topic coherence in topic pathways and negatively correlates with number of topic pathways. High $\tau_{topic}$ values result in more topic pathways with increased coherence while low $\tau_{topic}$ values result in less topic pathways with low coherence. 

This method improves the coherence of existing topic pathways as only the feature vectors that are similar to the respective cluster representation vector, are used in the self structuring stage to incrementally learn dynamic changes of that topic pathway. In contrast, the generation of a new feature map leads to the formulation of cluster representation vectors that are not influenced by the knowledge acquired in previous layers and thus, becomes the basis for new topic pathways. 

It should be noted that the proposed technique separates topic pathways autonomously. The coherence of these topic pathways can be evaluated using topic coherence measures~\cite{Roder2015}. The details of topic coherence are further discussed in the experiments section, alongside results from experiments. 

This extended IKASL algorithm has the same time-complexity as the vanilla IKASL algorithm. It primarily depends on the number of inputs $n$ and the number of nodes in the GSOM $k$. Each training step requires a nearest neighbour search between the input and GSOM nodes leading to a complexity of $\mathcal{O}(nk)$. However, the expansion of GSOM slows over training iterations and eventually saturates. Thus, when the input data set is sufficiently large ($n >> k$) it can be assumed that the time complexity of IKASL algorithm is linear with the input size i.e., $\mathcal{O}(n)$. It is a very useful property of IKASL which enables its use on very large datasets.      

\section{Event detection}\label{sec:event_detection_technique}
As mentioned in Section~\ref{sec:highlevel_technique}, an can be characterised as sudden burst of activity in a certain topic. In recent literature as discussed in Section~\ref{sec:unspecified_event_detection}, detecting such burst of activity in a topic is often simplified to detecting bursty key words or tag words (hashtags in Twitter)~\cite{Weng2011a} with the hypothesis that such burst key words are indicative of an event related to the topic inferred by the meaning of those key words. In contrast,~\citep{Paltoglou2015} recently introduced a sentiment based event detection technique where  significant changes of sentiment can be used as event indicators with  comparable results. 

As discussed in Section~\ref{sec:highlevel_technique}, it can be hypothesis that an \textit{event} in social data stream creates a disruption to related topic pathways so that the characteristics of social behaviours in the affected topic pathways change substantially. Considering the nature of characteristics in fast-paced social media platforms Such changes are reflected in various aspects of the topic pathway such as change of volume or change of sentiment.

Hence, in the proposed event detection approach we employed a multi-faceted event detector. It looks for different event indications such as volume changes ($I_v$), positive sentiment changes ($I_{PS}$) and negative sentiment changes ($I_{NS}$). In addition to those generic event indicators, domain specific event indicators ($I_{DS}$ ) such as (i) volume of disaster related words e.g., ‘shaking’~\cite{Sakaki2010a}, and (ii) volume of civil unrest related words e.g., ‘protest’~\cite{Ramakrishnan2014}; can also be incorporated in to the event detection module.   

This proposed technique linearly combines the event indicators to obtain the final event score $\mathcal{I}$. Let $D^i_j$ be a topic segment of topic pathway $D_j$, the final event score of $D^i_j$ is defined as follows:

\begin{equation}
\mathcal{I}(D_j^i)=r_V\times I_V(D_j^i)+r_{PS}\times I_{PS}(D_j^i)+r_{NS} \times I_{NS}(D_j^i)+\ldots+r_{DSE}\times I_{DSE}(D_j^i)
\label{equ:event}
\end{equation}

\noindent where $r_V,r_{PS},r_{NS},\ldots,r_{DSE} \in [0,1]$ and $\sum r = 1$

\noindent $r_k$ represents the sensitivity of the event indicator $I_k$ to the final event score, where high or low value of $r_k$ adapts the impact of $I_k$ to the final event score accordingly. $r$ values can be empirically set based on the importance of certain indicators to suit different applications. It is also possible to learn $r$ values by training a classifier using pre-labelled records. 

$\mathcal{I}$ can take values between $0$ to $\infty$, and its value is proportionate to the significance of the event. The potential events of interest can be identified by applying an event threshold $\tau_e$ where $\mathcal{I}(D_j^i) \geq \tau_e$ are flagged as events. 

The following subsections describes the volume based and sentiment based event indicators respectively.

\subsection{Volume based event indicator}\label{sec:volume_event}
The volume based event indicator looks for significant increase in volume of a topic pathway. Such increase is an indication of increased interest in that topic pathway, which can be due to a potential event relevant to that topic pathway.

In order to capture significant increases in volume, we set $I_V$ as the ratio between the proportion of messages in a topic segment and the moving average of proportion of messages in that topic pathway as:      

\begin{equation}
I_V(D^i_j)=\dfrac{V_P(D^i_j) \times W}{\sum\limits_{x=i-w-1}^{i-1} V_P(D^x_j)}
\end{equation}

\noindent where $W$ is the window size for the moving average.

\noindent The volume proportion of a topic segment $V_P(D^i_j)$ is determined as: 
\begin{equation}
V_P(D^i_j) = \frac{\vert D^i_j \vert}{\vert B^i \vert}
\end{equation}
The volume proportion is more robust than the absolute volume (number of social media messages) of a topic segment. This is because absolute volume of a topic segment is biased to the volume of social text message batch which varies significantly over time despite fixed sampling time $\Delta t$. These changes in volume of social text message batch are often due to seasonal changes which depends on hour of day (daytime vs night-time) and day of week (weekdays vs weekends). In contrast, volume proportion is more resilient to bias as it normalise the seasonal variations. 

\subsection{Sentiment based event indicator}\label{sec:sentiment_event}
Sentiment based event indicators capture the changes of public opinion~\cite{Thelwall2011}. The concept that events are associated with changes of positive and negative sentiment strength has been successfully used for a recent event detection~\cite{Paltoglou2015} task from a social text stream. The hypothesis here is that an event can alter the existing level of public opinion of relevant topic pathways and such changes can be detected by measuring the sentiment changes of topic pathways over time.   

As discussed in Section~\ref{sec:emotion_extraction_from_text}, sentiment analysis on social media messages is a challenging task mainly because of its inherent unstructured nature (see Section~\ref{sec:social_data_analysis_challenges}) where authors loosely follow grammatical rules of language~\cite{Baldwin2013} and heavily use emoticons to express sentiment. Moreover, word lengthening (e.g., \textit{sorrrry}) is widely used to emphasis the sentiment of statements~\cite{Brody2011}. Recent research reports that such variations have resulted in significant performance degradations in conventional sentiment analysis tools~\cite{Eisenstein2013}.     

Therefore, it is important to use sentiment tools that are designed to facilitate the above mentioned factors of social text. In Section~\ref{sec:sentiment_from_text}, we have investigated the techniques of several state-of-the art sentiment analysis tools~\cite{Esuli2006, Socher2013, Thelwall2012}, and found that SentiStrength~\cite{Thelwall2012} is well suited for our requirement; mainly because it has been specifically engineered to capture sentiment related features in social text. It employs a word list with sentiment strength and polarity to derive both positive and negative sentiment strengths. It considers emoticons, boosting words and negation words to strengthen or weaken the sentiment value. In addition, it considers repeated letters in a word (e.g., \textit{sorrrry}) as an indication of adding more emphasis to the sentiment of that word. 

SentiStrength provides positive sentiment values ranging from 1 to 4 and negative sentiment values ranging from  -1 to -4 respectively. Two sentiment based event indicators are designed utilising both positive and negative sentiment values to capture sentiment variations in both positive and negative sentiment. The sentiment value of a topic segment is determined by aggregating sentiment values of social media messages belong to that topic segment and taking the average. The variations of sentiment in a topic segment $D^i_j$ is captured by comparing the sentiment to the moving average of sentiment in that topic pathway $D^j$. The positive sentiment event indicator $I_{PS}$  and the negative sentiment event indicator $I_{NS}$  is defined as follows:

\begin{equation}
I_{PS}(D^i_j)=\dfrac{avgPositiveSentiment(D^i_j) \times W}{\sum\limits_{x=i-w-1}^{i-1} avgPositiveSentiment(D^x_j)}
\end{equation}        

\begin{equation}
I_{NS}(D^i_j)=\dfrac{avgNegativeSentiment(D^i_j) \times W}{\sum\limits_{x=i-w-1}^{i-1} avgNegativeSentiment(D^x_j)}
	\end{equation} 
\noindent where $W$ is the window size for the moving average.

\section{Illustrative Example}\label{sec:illustrative_example}
\begin{figure}[!htb]
 	\centering
 	\includegraphics[clip=true, width=1.0\linewidth]{{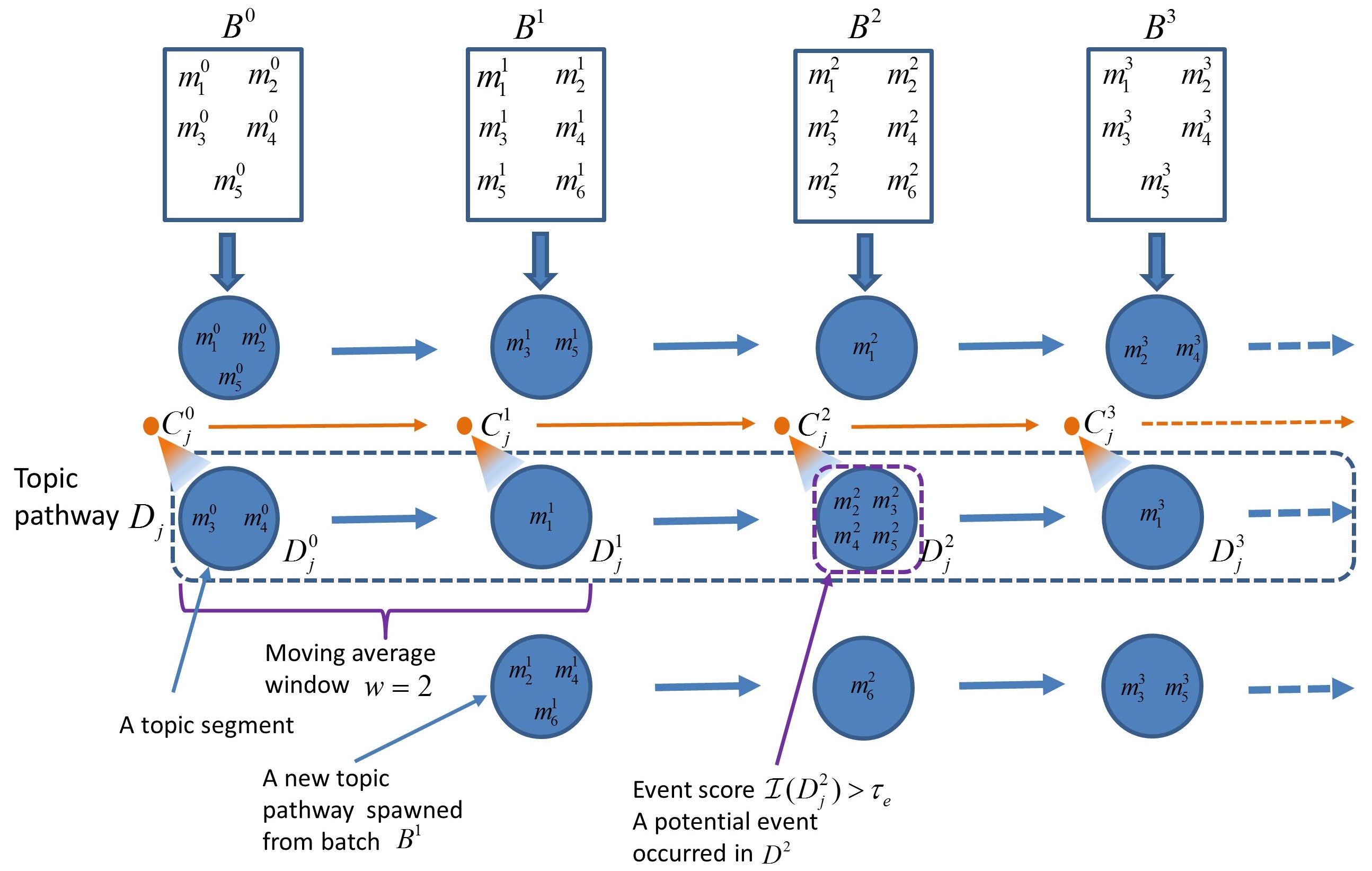}}
 	\caption{An illustrative example showing topic pathway separation and event detection}
 	\label{fig:illustrative_example}
 \end{figure}

This section demonstrates the functionality of the proposed platform for separating topic pathways and event detection using an illustrative example shown in Figure~\ref{fig:illustrative_example}.

As shown in Figure~\ref{fig:illustrative_example} social media messages are sampled into batches and extracted feature vectors are fed to the topic separation technique. Once the cluster representation vectors $\{C\}^i$ are learned, social media messages in $B^0$ are mapped to their closest cluster representation vector $C^i_j \in \{C\}^i$. The topic segment $D^i_j$ is formed by the set of social media messages that are mapped to the cluster representation vector $C^i_j$. As shown in Figure~\ref{fig:illustrative_example},  For example, $C^0_j$ is the closest for $m^0_3, m^0_4 \in B^0$, thus $D^0_j = \{m_3, m_4\}$.

$C^1_j$ is learned based on $C^0_j$ (the feature map that used to derive $C^1_j$  is initialised using $C^0_j$), and is timely evolved acquiring relevant new knowledge from $B^1$. Hence $C^0_j$ shares similarity with $C^1_j$, $C^1_j$ shares similarity with $C^2_j$, and $C^2_j$ shares similarity with $C^3_j$. Because of these relationships among the cluster representation vectors, the topic segments formed based on these also contain semantically similar social media messages. Therefore, such topic segments are linked to form a topic pathway $D_j = D^0_j \rightarrow D^1_j \rightarrow D^2_j \rightarrow D^3_j$  that contains semantically similar social media messages over time.

Moreover, the proposed \ikaslext facilitates new topic pathways hence as shown in Figure~\ref{fig:illustrative_example} such new topic pathways can be spawned from any batch.

Those identified topic pathways are continuously evaluated by the proposed event detection algorithm. It collects information about event indicators and aggregates them into an event score $\mathcal{I}$. As shown, event indicators of $D^2_j$ are determined based on changes in volume proportion, positive sentiment and negative sentiment compared to the previous topic segments $D^1_j$ and $D^0_j$ (moving average window of 2). If $\mathcal{I}(D^2_j) > \tau_e$  then $D^2_j$ contains information about a potential event that is related to the topic pathway $D_j$.

The next section presents the results of an empirical evaluation  of the proposed topic separation and event detection techniques using two Twitter datasets.

\subfile{Chapter4_Experiments}

\section{Chapter Summary}
This chapter presented an algorithmic development of the conceptual model proposed in Chapter~\ref{chap:3}. It consists of two novel algorithms designed to capture insights from fast-paced social data streams. The first is a new unsupervised incremental machine learning algorithm developed extending GSOM self-structuring algorithm and IKASL incremental learning algorithm. It automatically structures a social data stream into topics and extends those across time into topic pathways. It also captures changes in topics over time as well as distinct new topics as new topic pathways at different points of time. This algorithm was designed to overcome the challenges present in social data with respect to its brevity, unstructuredness, and diversity. The second algorithm is a multi faceted event detection algorithm developed to monitor topic pathways for significant changes in online social behaviours over time, and identify such changes as potential events of interest. The changes in social behaviours were identified using automatic event indicators such as changes in volume, positive sentiment and negative sentiment.

These techniques were trialled using two large Twitter datasets containing 6 months of tweets on two entities a politician and an organisation. As shown in the experiment results, the topic pathway separation algorithm successfully captured the key topic pathways representing ongoing discussions related to the respective entities. Also, shifts in the discussions represented by new key terms were successfully learned and associated with the relevant topic pathway. Moreover, new distinct topics were automatically captured as new topic pathways. The event detection algorithm monitors those topic pathways and automatically captured significant changes in human behaviour using changes in volume and sentiment. Those captured events were aligned with contemporary news articles that discussed those events. 

The next chapter extend the above developed algorithms into a technology platform to capture insights from online support group discussions. 

\onlyinsubfile{	
\bibliographystyle{dcu}
\bibliography{library}{}
}

%% file: Chapter4_Experiments.tex
\section{Experiments} \label{sec:chap4_experiments}
This section demonstrates the core capabilities of the above proposed platform for topic separation and event detection. Twitter was used as the experimental online social media platform due to its popularity and rich API. The first subsection describes the datasets and new two subsections describe preprocessing and feature extraction steps. Subsequently, comprehensive accounts of the core capabilities, topic pathway separation, evolution of topic pathways, emergence of new topic pathways and automatic event detection are presented in separate subsections. 

\subsection{Datasets}
This section describes the two Twitter datasets collected for the experiments.

\noindent\textbf{\#Obama Dataset}: President Obama maintains a diverse range of public relations, both locally and internationally and he is the most followed political figure on Twitter with approximately 77.8 million followers. Thereby, analysing tweets about him is a challenging task. Separating these into topic pathways would showcase the capabilities of the proposed technique and its value over existing techniques. \#Obama Dataset was collected using the keywords \textit{\#obama, @obama, obama, \#potus} for the time duration 01/12/2014 to 19/04/2015 (20 weeks). 4,230,985 tweets were collected which contain 173,903 out of which 57,998 are \textit{hashtags}.

\noindent\textbf{\#Microsoft Dataset}: Microsoft Corporation has a diverse online social presence due to a large portfolio of technology-centric products and services that are consumed by a variety of end-users. This diversity is a fitting testbed to further investigate the separation of a tweet stream into different topic pathways on products, services and competitors. \#Microsoft Dataset was collected using the keywords \textit{\#microsoft, @microsoft, microsoft} for the time duration 08/12/2014 to 28/06/2015 (30 weeks). 1,953,243 tweets were collected which contain 84,758 out of which 29,655 are \textit{hashtags}.

The characteristics of the two datasets are summarised in Table~\ref{table:twitter_datasets}. \#Obama Dataset has significantly more unique tokens compared to \#Microsoft Dataset. Interestingly a third of unique tokens are hashtags in both datasets (33\% in \#Obama Dataset and 35\% in \#Microsoft Dataset) indicating its widespread use in tweets. Out of these unique tokens only around 5\% has a frequency of 5 or more in both datasets. 

\begin{table}[!htb]
\caption{Characteristics of the twitter datasets.}
\label{table:twitter_datasets}
\centering
\begin{tabulary}{\linewidth}{|L|R|R|}\hline 
 & \#Obama Dataset & \#Microsoft Dataset\\\hline
Number of tweets & 4,230,985 & 1,953,243\\\hline
Date range & 01/12/2014 to 19/04/2015 &  08/12/2014 to 28/06/2015 \\\hline
Terms used as filters & \#obama, @obama, obama, \#potus & \#microsoft, @microsoft, microsoft\\\hline
Number of unique tokens & 173,903 & 84,758 \\\hline
Number of unique hashtags &	57,998 & 29,655 \\\hline
Tokens with term frequency >5 & 8,112 & 4,151 \\\hline
\end{tabulary}
\end{table}

\subsection{Pre-processing} 
Pre-processing is carried out on text data to reduce noise. This step is important as it helps to reduce the sparseness of text data. Note that, informal language is widely used in tweets, therefore they are often noisier than standard text documents, and hence a series of pre-processing steps are necessary.

\noindent\textbf{Duplicate removal:} tweets that are duplicated within the same batch are often spam or advertisements generated by automatic bots~\cite{Wang2010}. Duplicated tweets affect the topic pathway separation process because they tend to form separate pathways. Such pathways are not informative for understanding public opinion as they are published for advertising purposes and can confuse attempts at event understanding and detection.

\noindent\textbf{Stopword removal:} stopwords are often less informative words exists in text ( e.g., \lq a\rq, \lq the\rq). It is a common text processing practice to remove those stopwords before extracting features. In addition to those, tweets also have Twitter specific stopwords such as \lq rt\rq~(a tag word used for retweets). Web links in tweets were also removed, as they do not directly relate to the social expression. 
In addition, a set of domain specific stopwords were identified and incorporated into the standard stopword list. For example, in dataset Obama, words such as \lq obama\rq, \lq barack obama\rq, \lq president obama\rq~and \lq POTUS\rq~were filtered out.

\noindent\textbf{Twitter specific tag removal:} Tweets also contain user mentions i.e. \lq @username\rq~and hashtags \lq \#keyword\rq~which are two unique features of tweets used to mentions a specific user and tag a keyword respectively. We removed user mentions from the tweet text as they are often user names. As discussed in Section~\ref{sec:cmc}, hashtags are unique keywords coined by public for social tagging or folksonomy. Such tags are often represents the content of the tweet and therefore indicative of the the topic. Therefore, hashtags were preserved as they supports the separation of topic pathways.

\subsection{Feature extraction}

In this step, representative features were extracted from the pre-processed tweets. We employed the \textit{bag of words} model, in which terms (word or phrase) are considered as features of the tweets and features scores are determined based on the frequency of that word. We use noun phrases from tweets as the features, which are extracted using JATE toolkit~\cite{Zhang2008}. Noun phrases that exist in less than a 0.5\% of tweets in each batch (vocabulary threshold $\tau_V$ discussed in Section~\ref{sec:dynamic_vocabulary}) were omitted to reduce the sparsity of the feature space. 

We employed inverse document frequency (idf) as the representative feature value of a term. $tdf_w$ of term $w$ is given as follows:

$$idf_w = \log(\frac{N}{1 + df_w})$$

Where $N$ is the number of tweets in that batch and $df_w$ (document frequency) is the number of times the term $w$ appeared in tweets of that batch.

\subsection{Topic separation}
This subsection demonstrates the first core capability; topic pathway separation based incremental learning outcomes. 
 
Pathway separation occurs as the algorithm incrementally learns from the Twitter stream. Some pathways remain prominent, continuing to receive a significant number of tweets across each topic segment, while others lose popularity and number of tweets decreased over time. 

\subsubsection{\#Obama Dataset} 
Six prominent (high volume) topic pathways were identified in \#Obama Dataset (labelled $TP^x_{Obama}$, where $x$ is 1-6). Each topic pathway consists of a group of frequent terms that signify the topic of that pathway. These words are more frequent among the tweets of that pathway compared to the tweets of other pathways. 

Table~\ref{table:ObamaTopics} presents the top 10 frequent terms of these pathways. Column 3 of this table summarises the frequent terms into a focus area.

\begin{table}[!htb]
\caption{Frequent terms in the prominent topic pathways of \#Obama dataset}\label{table:ObamaTopics}
	\begin{tabulary}{\linewidth}{|M|L|L|}\hline 
	\text{Topic pathway} & Frequent terms & Key focus of the topic pathway\\ \hline
	 TP^1_{Obama}&putin, russia, ukraine, economy, iran, respect, west, merkel, war, head& Relations with Russia and Ukraine\\ \hline
	 TP^2_{Obama}&republican, party, question, tcot, congress, world, politics, cromnibu, gop, american people &Internal Politics relevant to Republican party \\\hline
	 TP^3_{Obama}&iran, israel, sanction, nuke, netanyahu, congress, tcot, irandeal, war, nuclear weapon, nuclear deal &Relations with Iran and Israel
	   \\\hline
	 TP^4_{Obama}&nation, liberal, racism, reason, voter, democrat, consequence, tcot, gruber, pelosi&Internal Politics relevant to Democratic party
	   \\\hline
	 TP^5_{Obama}&obama news, cbs news, obama video, obamacare, sore throat, washington dc, gop, castro, cnn obama,
	 summit &News about Obama
	  \\\hline
	 TP^6_{Obama}&war, afghanistan, isis, power, congress, protection 
	 siyra, iranian dissident, iraq&Terrorism, mainly ISIS
	 \\\hline	 	 	 	 
	\end{tabulary} 
\end{table}

It should be noted that some of the pathways focus on internal political issues ($TP^2_{Obama}$, $TP^4_{Obama}$) while others are on international political issues ($TP^1_{Obama}$, $TP^3_{Obama}$, $TP^6_{Obama}$). $TP^5_{Obama}$ is discussions about new reports on Obama.   

In order to distinguish the salient terms that define a pathway, it is necessary to investigate their presence and prominence in other topic pathways. Table~\ref{table:ObamaFrequentTermStats} tabulates this investigation. It presents the distribution of top five frequent terms of each prominent topic pathway, across the six prominent topic pathways. The rows represent the top five frequent terms of each prominent topic pathway and the columns represent the six topic pathways.

\begin{table}[!htb]
\caption{Distribution (\%) of top five frequent terms of six prominent topic pathways in \#Obama dataset}\label{table:ObamaFrequentTermStats}
\begin{tabulary}{\linewidth}{|L|L|R|R|R|R|R|R|}\hline
\multicolumn{1}{|l|}{\multirow{2}{*}{Topic pathway}}&\multicolumn{1}{|l|}{\multirow{2}{*}{Frequent terms}}&\multicolumn{6}{c|}{Distribution (\%) of terms}\\
& &\small{$TP^1_{Obama}$}&\small{$TP^2_{Obama}$}& \small{$TP^3_{Obama}$}& \small{$TP^4_{Obama}$}& \small{$TP^5_{Obama}$}& \small{$TP^6_{Obama}$}\\\hline 
\multirow{5}{*}{$TP^1_{Obama}$}	&putin&97.9&0.3&0.5&0.0&0.3&0.0\\
								&russia&85.2&0.7&9.3&0.2&0.2&1.7\\
								&ukraine&84.6&1.6&1.3&0.3&3.1&3.5\\
								&economy&54.7&3.4&1.5&1.5&5.9&0.5\\
								&iran&2.2&2.0&69.8&0.4&1.7&3.8\\\hline
\multirow{5}{*}{$TP^2_{Obama}$}	&republican&1.0&81.4&1.5&2.0&2.8&0.5\\
								&party&1.7&68.5&1.1&1.7&1.7&4.4\\
								&question&3.1&67.3&4.3&0.0&0.6&7.4\\
								&tcot&4.1&16.8&14.3&7.6&2.3&9.2\\
								&congress&1.5&11.1&14.7&1.2&9.0&9.4\\\hline
\multirow{5}{*}{$TP^3_{Obama}$}	&iran&2.2&2.0&69.8&0.4&1.7&3.8\\
								&israel&1.8&3.6&53.9&1.8&1.7&2.7\\
								&sanction&5.8&2.6&57.6&0.6&1.2&1.2\\
								&nuke&2.8&1.2&70.6&0.0&0.4&0.4\\
								&netanyahu&1.6&3.2&16.9&0.7&3.0&1.9\\\hline
\multirow{5}{*}{$TP^4_{Obama}$}	&nation&0.3&1.6&0.6&90&0.9&0.9\\
								&liberal&0.0&0.0&0.8&98.4&0.0&0.4\\
								&racism&0.0&0.5&0.0&97.5&0.0&0.0\\
								&reason&0.6&2.2&2.8&75&0.0&3.3\\
								&voter&0.0&0.9&0.0&96.4&0.0&0.0\\\hline
\multirow{5}{*}{$TP^5_{Obama}$}	&obama news&0.5&0.2&0.6&1&90.4&0.4\\
								&cbs news&0.0&0.0&0.8&0.0&90.5&0.8\\
								&obama video&2.7&6.7&0.9&1.3&49.3&1.8\\
								&obamacare&2.5&5.7&1.3&3.2&69&1.3\\
								&washington dc&0.0&0.0&0.9&11.4&86&0.9\\\hline
\multirow{5}{*}{$TP^6_{Obama}$}	&isis&1.4&1.7&1.6&0.8&2&81.4\\
								&war&7.6&2.8&10.7&2.2&1.6&56.4\\
								&iraq&2.3&1.9&9.1&1.1&0.4&74.9\\
								&siyra&7&19.9&21.6&0.0&0.6&37.4\\
								&power&5.8&7.7&7.1&8.3&1.3&43.6\\\hline		
\end{tabulary} 
\end{table}	

This table highlights that the top five terms of the six prominent pathways have a significant presence in the relevant pathway compared to other pathways. For example, in pathway  $TP^1_{Obama}$, the top three words putin, russia, and ukraine have more than 80\% within that pathway. The term iran has a low percentage in  $TP^1_{Obama}$ because it has a larger presence in  $TP^3_{Obama}$. 

\begin{figure}[!htb]
	\centering
	\includegraphics[clip=true, width=1.0\linewidth]{{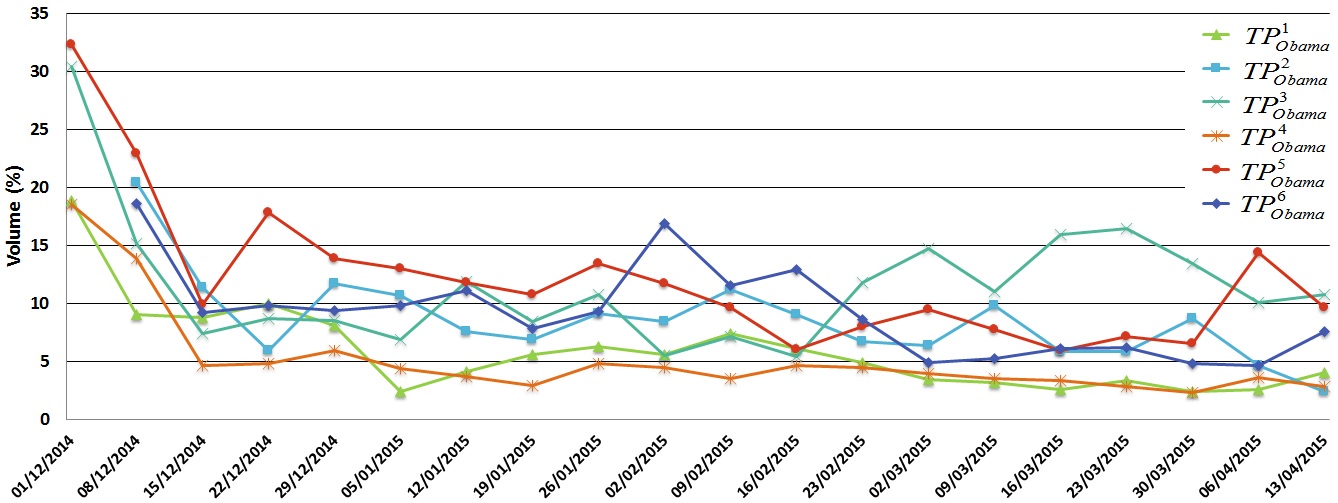}}
	\caption{Volume (\%) of tweets in six prominent topic pathways of \#Obama Dataset}
	\label{fig:tweet_volume_obama}
\end{figure} 

Figure~\ref{fig:tweet_volume_obama} shows the tweet volume of these topic pathways. It shows that different topic pathways peaked at different time periods. For example, in the latter part, the topic pathway $TP^3_{Obama}$ becomes popular to reflect the incidents related to the nuclear deal between USA and Iran.

\subsubsection{\#Microsoft Dataset}
Seven prominent topic pathways were identified in \#Microsoft Dataset (Table~\ref{table:MicrosoftTopics}), which were labelled using a similar approach to the above dataset. 

\begin{table}[!htb]
\caption{Frequent terms in the prominent topic pathways of \#Microsoft dataset}\label{table:MicrosoftTopics}
	\begin{tabulary}{\linewidth}{|M|L|L|}\hline 
		\text{Topic pathway} & Frequent terms & Key focus of the topic pathway\\ \hline
		TP^1_{Microsoft}&Google, android, amazon, apple, cyanogen, ibm, work, sony, app, window&Relations with Google\\\hline
		TP^2_{Microsoft}&Xbox, ps4, update, windows 10, xboxone, game, Cortana, sony, bitcoin, microsoft xbox &Microsoft Xbox related products and services \\\hline
		TP^3_{Microsoft}&microsoft office, ipad, android, mac, android tablet
		gmail, windowsphone, google, onenote, microsoft tech, office &Microsoft Office and its compatibility in different devices \\\hline
		TP^4_{Microsoft}&Windows, version, windows 10, os news, future, microsoft windows, upgrade, microsoft windows10
		Week, update &Microsoft Windows\\\hline
		TP^5_{Microsoft}&Facebook, bing, google, microsoft bing, zdnet
		Apple, favour, monumental deal, search result, virtual reality &Relations with Facebook\\\hline
		TP^6_{Microsoft}&windows phone, app, office, android, bitcoin, onedrive, update, bitcoin payment, tablet, ipad &Windows Phone \\\hline
		TP^7_{Microsoft}&Apple, google, cloud, amazon, Samsung, microsoft store, bigdata, fight, device, patent &Relations with Apple\\\hline	 	 	 	 
	\end{tabulary} 
\end{table}

\begin{figure}[!htb]
	\centering
	\includegraphics[clip=true, width=1.0\linewidth]{{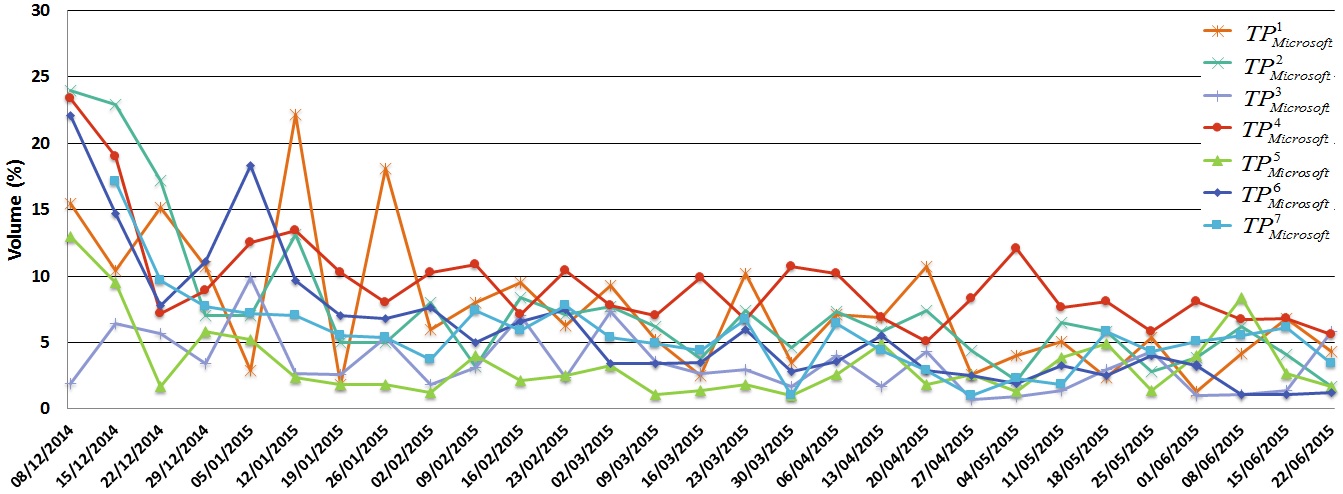}}
	\caption{Volume (\%) of tweets in six prominent topic pathways of \#Microsoft Dataset}
	\label{fig:tweet_volume_microsoft}
\end{figure} 

Out of these seven, four topic pathways focused on products and services offered by Microsoft Corporation such as Windows, Windows Phone, Microsoft Office and Xbox. Remaining three focused on relations with competitors such as with Google, Facebook and Apple. Figure~\ref{fig:tweet_volume_microsoft} shows the tweet volume of these topic pathways.

The topic pathway that focuses on Windows has the highest volume in most of the weeks as Windows operating system ($TP^4_{Microsoft}$) is the most used product of Microsoft. Similarly, the topic pathway focuses on relations with Google ($TP^1_{Microsoft}$ ) has the highest number of tweets among the pathways that focus on relations with competitors since both Microsoft and Google are direct competitors of several products including operations system, mobile device, search engine and virtual reality technology.

\subsubsection{Topic Coherence}
It is important to evaluate the quality of the automatically identified and learned topics based on their interpretability to a human analyst. There have been recent developments in quantitative metrics to evaluate the coherence of learned topics~\cite{Roder2015}.

~\citet{Mimno2011} have demonstrated such a metric based on the principle that significant terms belonging to a topic are likely to co-occur in same documents. This topic coherence metric delineates as follows:

$$C(T;V^T) = \sum_{m=2}^{M}\sum_{l=1}^{m-1}\log\dfrac{D(v_m^T,v_l^T)+1}{D(v_l^T)}$$

\noindent where $V^T = (v^T_1,\ldots,v^T_M)$ are the list of $M$ terms that are most probable (frequent) in topic $T$. $D(v_l^T)$ is the document frequency of the term $v_l^T$. $D(v_m^T,v_l^T)$ is the document frequency of the co-occurrence of terms $v_l^T$ and $v_m^T$.

This metric aggregates the term co-occurrence scores of a topic $T$ for the $M$  frequent terms of that topic. They have empirically verified that this metric correlates with the judgement of human analysts. 

\begin{figure}[!htb]
	\centering
	\includegraphics[clip=true, width=1.0\linewidth]{{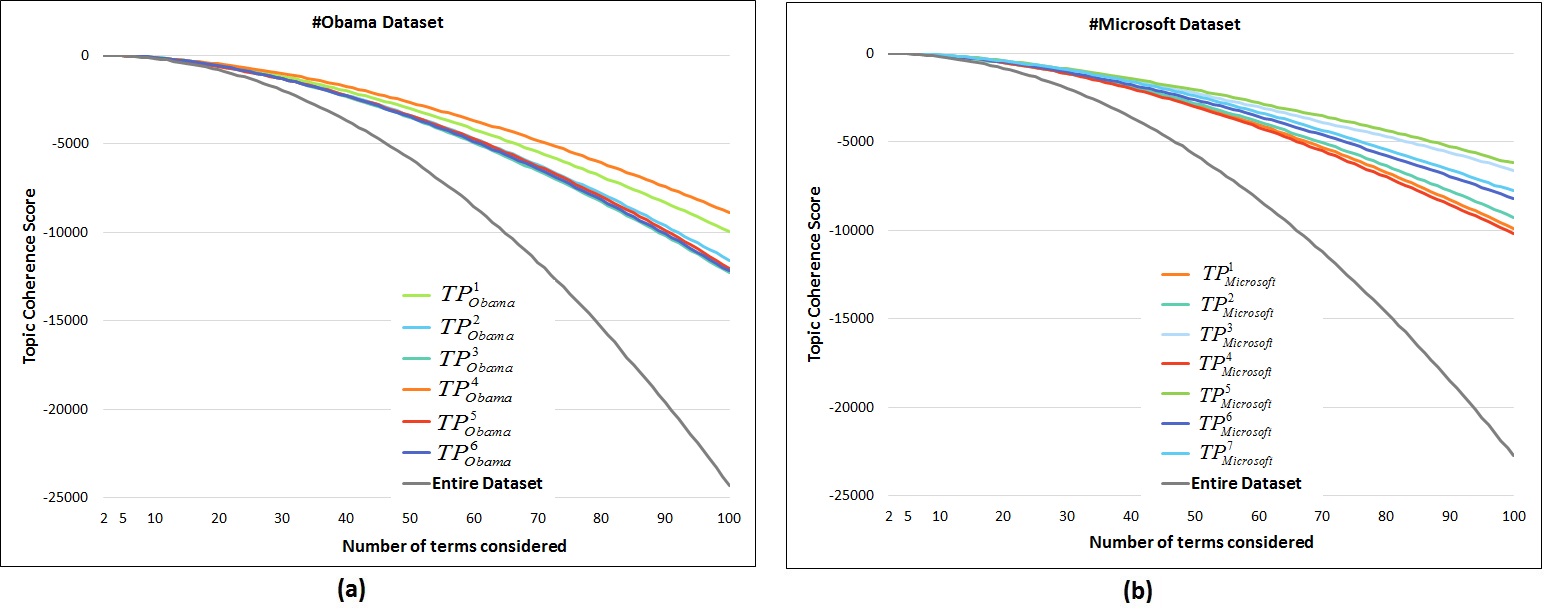}}
	\caption{Topic coherence scores for prominent topic pathways and whole dataset (for baseline) in (a) \#Obama Dataset and (b) \#Microsoft Dataset.}
	\label{fig:topic_pathways_coherence}
\end{figure}

We have adopted this metric to measure and evaluate topic coherence within individual pathways. Our experiments measure the topic coherence scores of each topic for $M \in [2,100]$ and the results are presented in Figure~\ref{fig:topic_pathways_coherence} for the two datasets; (a) \#Obama and (b) \#Microsoft. Note that, topic coherence score calculated for the entire dataset (without pathways separation) is plotted as the baseline.

The above results highlight that topic pathways have significant improvements in topic coherence compared to that of the entire datasets emphasising the semantic coherence of topic pathways learned by the proposed algorithm. 

\subsection{Evolution of topic pathways}

This subsection demonstrates the second core capability of the proposed algorithm; capturing evolution of topic pathways and temporal changes in topic segments within a topic pathway. In order to demonstrate this capability, we selected the topic pathway,  $TP^1_{Microsoft}$, which captures the relationship between Microsoft and Google. It is common knowledge that both these organisations are dynamic and innovative in the technology domain, thereby twitter activity associated with both entities is equally dynamic with new subtopics (i.e., short-term topics within the particular topic pathway) being discussed. We demonstrate that the selected topic pathway  $TP^1_{Microsoft}$, is capable of identifying each new subtopic as it emerges without a loss of focus on the overarching theme of Microsoft-Google relationship. Thereby, the entire topic pathway maintains focus on Microsoft-Google relationship, while each segment in the pathway focuses on different subtopics associated with this relationship as they emerge over time. The proposed algorithm’s autonomous capability to capture the evolution of a topic pathway in this manner provides an analyst unique insights into short-term changes within a long-term trend. In this instance, the long-term trend is the relationship with Google and the short-term changes are the subtopics associated with both entities. 

\begin{figure}[!htb]
	\centering
	\includegraphics[clip=true, width=1.0\linewidth]{{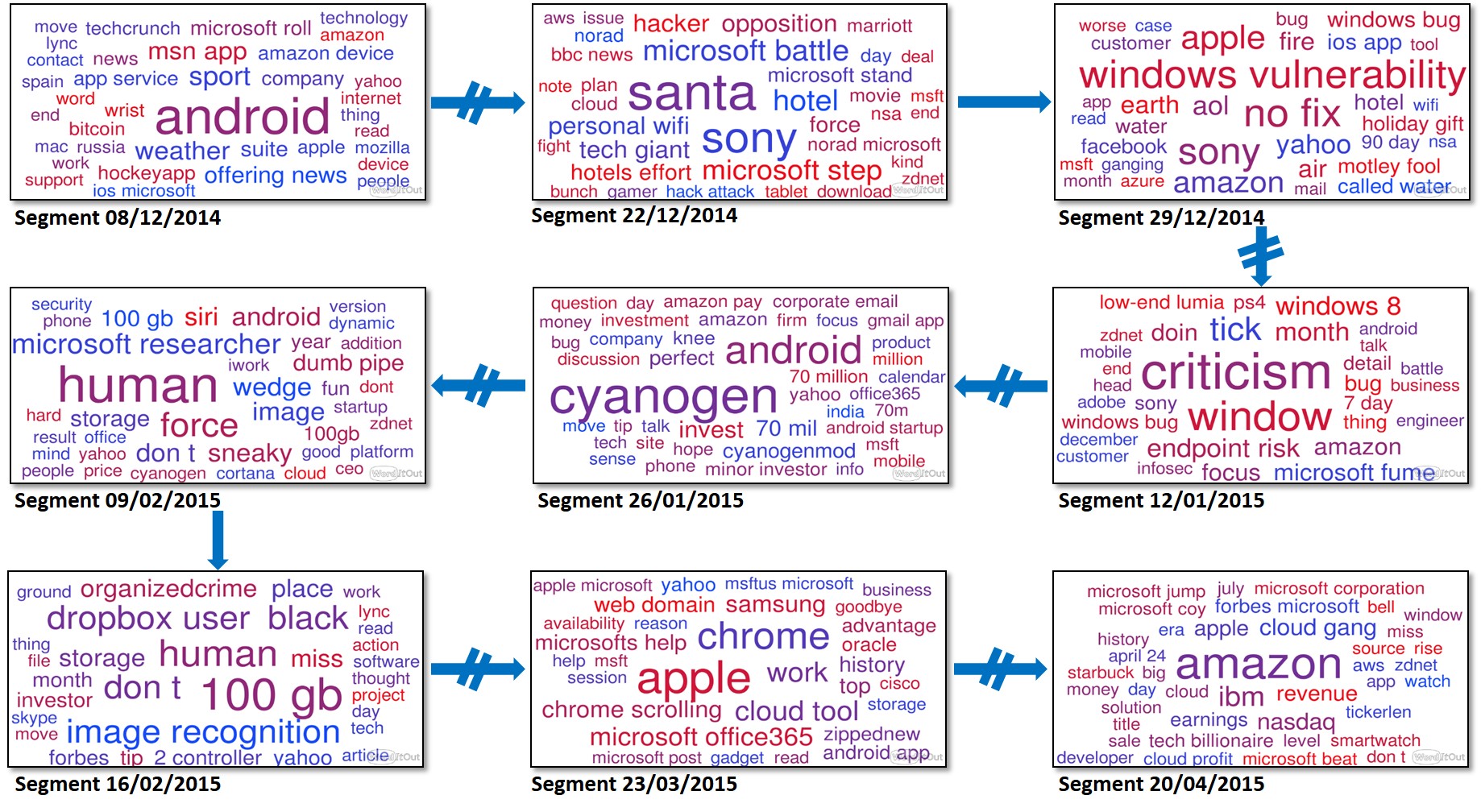}}
	\caption{Word clouds generated for nine topic segments in topic pathway $TP^1_{Microsoft}$, which represents Microsoft-Google relationship. Note that most frequent terms \emph{Microsoft} and \emph{Google} were removed from word clouds for improved clarity.}
	\label{fig:evolution_topic_pathway}
\end{figure}

We selected nine topic segments from pathway $TP^1_{Microsoft}$ for this demonstration. Figure~\ref{fig:evolution_topic_pathway} presents word clouds for these nine topic segments. Table~\ref{table:evolution_topic_pathway} further examines content of each of the nine topic segments. Word clouds were selected for the intuitive value in visualising fluctuations in the usage and frequency of terms. The primary observation here is that each word cloud has a unique set of frequent terms which are indicative of different subtopics. 

\begin{table}[!htb]
\caption{Further examination of topic segments illustrated as word clouds in Figure~\ref{fig:evolution_topic_pathway}}
\label{table:evolution_topic_pathway}
\small	
\begin{tabulary}{\linewidth}{|p{1.1cm}|p{2.5cm}|p{5cm}|p{5cm}|}\hline
Segment & Representative frequent terms & Subtopic(s) of the segment & Relevant News Article(s) \\\hline
Segment 08/12/ 2014& android, ios, msn app, sport, weather, offering news & Microsoft rolls out a set of MSN applications such as Weather, News and Sports for iOS and Android devices& MSN consumer apps arrive on Android, iOS, Amazon-CNET (11/12/2014)\\\hline
Segment 22/12/ 2014& (i) microsoft battle, personal wifi,  marriott, hotel\newline
(ii) santa & (i)Google and Microsoft step in to oppose the attempts of Marriott Hotels to block personal Wi-Fi inside the hotel\newline
(ii) Google and Microsoft provides interactive features to track Santa’s journey during Christmas & (i)Microsoft, Google join opposition to hotels' wi-fi blocking-ZNET(23/12/2014)\newline
(ii) Tracking Santa with help from Microsoft, Google-CNET (22/12/2014)\\\hline
Segment 29/12/ 2014& windows vulnerability, no fix, bug, windows bug & Google has openly published a Windows 8.1 vulnerability that allows users to get administrator privileges& Google discloses unpatched Windows vulnerability-ZDNET (31/12/2014)\\\hline
Segment 12/01/ 2015& criticism, windows, windows bug, microsoft fume, endpoint risk & Microsoft criticised Google for releasing information about security vulnerabilities to public& Microsoft slams Google for spilling the beans on Windows 8.1 security flaw-ZDNET(12/01/2015) \\\hline
Segment 26/01/ 2015& android, 70 mil, cyanogen, android startup, 70 million, invest& Microsoft is taking part in a \$70 million investment round in a startup that is developing a competitive operating system to Android & Microsoft to Invest in CyanogenMod: What Could It Mean For Google?-Tech Times (29/01/2015)\\\hline
Segment 09/02/ 2015& human, microsoft researcher, image & Microsoft and Google are working on deep learning systems that can beat human at image recognition& Microsoft researchers say their newest deep learning system beats humans and Google-VENTUREBEAT (09/02/105)\\\hline
Segment 16/02/ 2015& (i) 100 gb, dropbox user, storage, move\newline
(ii) human, image recognition & (i) Microsoft offers 100GB cloud storage free as a promotion to get more cloud users.\newline
(ii) Microsoft and Google are working on deep learning systems that can beat human at image recognition & (i) Microsoft offers free 100GB OneDrive space to Dropbox users worldwide-ZDNET  (20/02/2015)\newline
(ii) Microsoft researchers say their newest deep learning system beats humans and Google-VENTUREBEAT (09/02/2015) \\\hline
Segment 23/03/ 2015 & chrome, chrome scrolling, microsofts help& Google is going to fix a scrolling issue of the Chrome browser using a Microsoft API & Google will fix Chrome's scrolling with Microsoft's help-ENGADGET(26/03/2015) \\\hline
Segment 20/04/ 2015 & amazon, cloud gang, nasdaq, revenue, cloud profit & NASDAQ spiked to its highest in 15 years, propelled by the revenues from cloud services of Amazon, Microsoft and Google & Techs surge on earnings, boosts Nasdaq, S\&P 500 to record close-CNBC (24/04/2015) \\\hline
\end{tabulary}
\end{table}

In Table~\ref{table:evolution_topic_pathway}, each row represents a topic segment. The first column is a label to identify each segment (i.e. first row in Table~\ref{table:evolution_topic_pathway} \textit{Segment 08/12/2014} corresponds to the first word cloud in Figure~\ref{fig:evolution_topic_pathway}, also labelled \textit{Segment 08/12/2014}). The second column denotes representative frequent terms for the corresponding segment and the third column denotes the subtopic for each segment. It should be highlighted that side by side, second and third column specify the relationship between top frequent terms and the subtopic. Although the subtopics change per segment, the main theme (or focus) of the pathway remains the same. The fourth column presents a robust evaluation of the subtopic of each segment, in the form of title of corresponding news articles published in mainstream online social media platforms during the same period of time. 

For example, in \textit{Segment 08/12/2014}, the discussion is about Microsoft providing MSN apps for android and iOS platforms. The second column presents representative frequent terms from this segment, the third column presents the subtopic and the fourth column validates both terms and the subtopic with the title of the corresponding news article. Some representative frequent terms are common across consecutive segments which indicates that some subtopics are discussed over several weeks. For example, frequent terms in week 22/12/2014 such as \lq hotel\rq~and \lq wifi\rq~appear in the word cloud of week 29/12/2014, but with a relatively lower frequency. Similarly, \lq human\rq~and \lq 100 gb\rq~appears in consecutive segments \textit{Segment 09/12/2015} and \textit{Segment 16/02/2015}.

Representative frequent terms within a segment are related to each other based on the subtopic. Representative frequent terms across segments are related to each other based on the topic pathway evolution. As noted, this topic pathway focuses on Microsoft-Google relationship, and each segment represents a different subtopic (or phase) of this relationship. Top terms across segments are able to capture this evolving topic pathway, despite differences in subtopics in each segment.

\begin{figure}[!htb]
	\centering
	\includegraphics[clip=true, width=1.0\linewidth]{{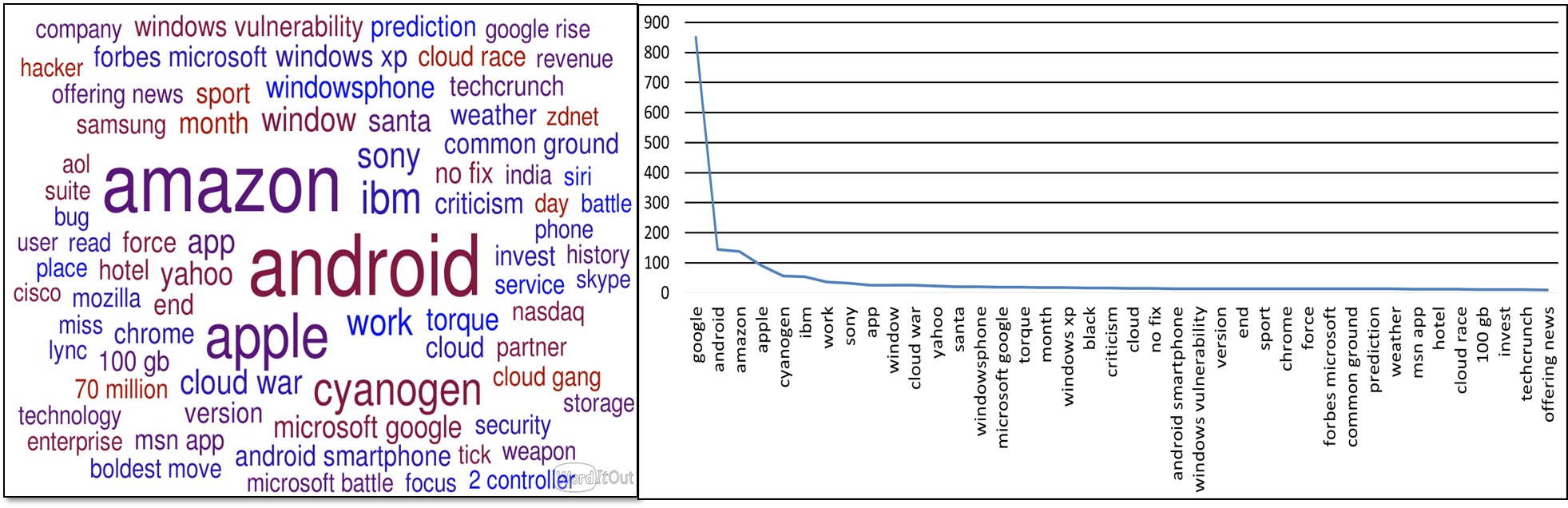}}
	\caption{A word cloud and a word frequency graph generated for the topic pathway $TP^1_{Microsoft}$. Note that most frequent terms \emph{Microsoft} and \emph{Google} were removed from word clouds for improved clarity.}
	\label{fig:global_topic}
\end{figure} 

Figure~\ref{fig:global_topic} presents a contrasting view. It illustrates outcomes from conventional analysis of a social media platform, which analyse all tweets corresponding to the topic pathway as a single dataset. It only provides a concise view of what is discussed. Unlike Figure~\ref{fig:evolution_topic_pathway} and Table~\ref{table:evolution_topic_pathway}, it is impossible to capture evolution of a topic pathway or short-term changes in topic segments. The most frequent term (apart from \lq Google\rq) is \lq android\rq~as it is one of the main competitive product for windows operating system. Other frequent terms include \lq amazon\rq, \lq ibm\rq~and \lq apple\rq~which are competitors in the same space as Microsoft and Google. Insights gained by an analyst will be severely limited if Twitter activity is explored in this manner.

Other topic pathways of both datasets show similar characteristics. Due to space restrictions, details of frequent terms for four prominent topic pathways (two each from two datasets) are provided in Table~\ref{table:topic_segments_Microsoft} and Table~\ref{table:topic_segments_Obama}. Note that, the identifying terms in each pathway (\lq xbox\rq~in  $TP^2_{Microsoft}$; \lq facebook\rq~in  $TP^5_{Microsoft}$; \lq russia\rq~and \lq ukraine\rq~in $TP^1_{Obama}$; and \lq obama news\rq~in $TP^5_{Obama}$) have been excluded to highlight subsequent frequent terms.

\begin{table}[!htb]
\caption{Two topic pathways from \#Microsoft Dataset represented using high frequent terms of the topic pathway and high frequent terms of top five most voluminous topic segments}
\label{table:topic_segments_Microsoft}
\small
\begin{tabulary}{\linewidth}{|p{1.1cm}|p{2.25cm}|p{1.75cm}|p{1.75cm}|p{1.75cm}|p{1.75cm}|p{1.75cm}|}\hline
Topic pathway&Frequent terms of topic pathway&\multicolumn{5}{l|}{Frequent terms of top five most voluminous topic segments}\\\hline
\rotatebox[origin=r]{90}{$TP^2_{Microsoft}$}&update, xboxone, game, windows 10, sony, bitcoin, plan,  lumia,  ps4, kinect, windows,  playstation & 	\textit{08/12/2014}
bitcoin, windows10, november, sony, playstation & 15/12/2014 game, pandora, apple tv, app, chromecast, xboxone&  \textit{22/12/2014} sony, christmas, ddos attack, playstation, lizardsquad & \textit{16/02/2015} plan, week, update, ps4, xboxone, lumia & \textit{11/05/2015} xboxone, ps4 , 10  april, playstation 4, sony \\\hline
\rotatebox[origin=r]{90}{$TP^5_{Microsoft}$}& bing, google, zdnet, apple, favour, monumental deal, search result, ganging, yahoo, sense, future, search & \textit{08/12/2014} bing, search tool, safe preference, monday, graph search, 240 million, built-in search & \textit{15/12/2014} bing , google, facebooks split, zdnet key event, amber alert, motley fool & \textit{29/12/2014} google, apple, ganging, bing, worse, microsoft service, hotmail, msn, yahoo search & \textit{18/05/2015} google, future, lead, bing, voice communication, sweden, eurovision, winner & \textit{08/06/2015} monumental deal, virtual reality, alibaba, sense, google, oculus rift, investor, gamescom\\\hline	
\end{tabulary}
\end{table}		

$TP^2_{Microsoft}$ globally contains frequent words that are indicative of different products related to \lq xbox\rq~like \lq xboxone\rq, \lq kinect\rq~and \lq windows 10\rq. Also, the word \lq sony\rq~denotes competitive manufacturer while \lq ps4\rq~and \lq playstation\rq~denotes competitive products. Notable subtopics include (i) \textit{Segment 08/12/2014}: Microsoft accepts \lq bitcoin\rq~as a valid payment method to pay for Xbox games; (ii) \textit{Segment 22/12/2014}: Xbox and Sony ps4 services are down due to a \lq ddos attack\rq~by a hacking group called \lq lizardsqsuad\rq~on Christmas day. 

$TP^5_{Microsoft}$ has frequent terms \lq bing\rq, \lq znet\rq~and \lq monumental deal\rq~that are frequent terms of different segments. In addition, there are other organisation names such as \lq apple\rq, \lq google\rq~and \lq yahoo\rq. Notable subtopics include (i) \textit{Segment 08/12/2014}: Facebook drops Microsoft Bing in favour of its own search tool; (ii) \textit{Segment 08/06/2015}: Facebook and Microsoft jointly working to make the Oculus Rift virtual reality headset works with Windows 10 and Xbox.

\begin{table}[!htb]
\caption{Two topic pathways from \#Obama Dataset represented using high frequent terms of the topic pathway and high frequent terms of top five most voluminous topic segments}
\label{table:topic_segments_Obama}
\small
\begin{tabulary}{\linewidth}{|p{1.1cm}|p{2.25cm}|p{1.75cm}|p{1.75cm}|p{1.75cm}|p{1.75cm}|p{1.75cm}|}\hline
Topic pathway&Frequent terms of topic pathway&\multicolumn{5}{l|}{Frequent terms of top five most voluminous topic segments}\\\hline
\rotatebox[origin=r]{90}{$TP^1_{Obama}$}&putin, economy, iran, respect, war, cuba, west, india, fight, china, weapon,   merkel, pressure, europe, sanction, leader& \textit{15/12/2014} putin,  cuba economy, sanction,  west, work,  crimea, problem, foxnews & \textit{22/12/2015} economy, dow, gas, foxnews, putin, cuba, head, golf, hawaii & \textit{02/02/2015} putin, wwiii, merkel, kiev, arms, fight, Moscow, weapon, china, Kerry, respect & \textit{09/02/2015} putin, weapon, war, merkel, poroshenko, fighting, hollande, minsksummit & \textit{16/02/2015} putin, merkel, hollande, respect, allies, weapon, sanction, shot, war, west\\\hline
\rotatebox[origin=r]{90}{$TP^5_{Obama}$}& cbs news, obama video, obamacare, sore throat, gop,  cuba, castro, acid reflux, summit, congress, prayer breakfast,  iran, hospital, seattle, cnn,  abc news &  \textit{22/12/2014} troop, 6 year, hawaii, apron, hacking row, cbs news, obamacare, afghanistan &  \textit{05/01/2015} community college, france, seattle, malia obama,  union tour, congress, housing move & 1\textit{2/01/2015} obamacare, cameron, methane emission, iran, middle class, obama video & \textit{02/02/2015} slavery, prayer breakfast,  isis, crusade,  obamas, comparison christianity & \textit{06/04/2015} castro, cuba, summit, stage, saturday, panama, abc news, historic meeting \\\hline	
\end{tabulary}
\end{table}		

$TP^1_{Obama}$ contains opinions about Obama handling Russia and Ukraine issues. Frequent terms of this pathway contains the names such as \lq putin\rq~(President of Russia), and \lq merkel\rq~(Chancellor of Germany). Also, the word \lq sanction\rq~is frequent showing that people talk about sanctions on Russia. Subtopics include (i) \textit{Segment 15/12/2014}: Obama’s support for new U.S. sanctions to Russia; (ii) \textit{Segment 09/02/2015}: ahead of Minsk summit, Obama threatened the Russian President with serious consequences for involvement in the Ukrainian conflict.

$TP^5_{Obama}$ contains news tweets about Obama often tagged as \lq \#obama \#news\rq~. Tweets posted by news agencies have tags used by them such as \lq cbs news\rq~, \lq cnn\rq~and \lq abc news\rq~. Topic segments in this pathway is often a mixture of subtopics about different news articles. Notable subtopics: (i) \textit{Segment 05/01/2015}: Obama proposes free two years at community colleges, Malia Obama supports anti-cop rap group, Obama to tout Arizona housing initiative; (ii) \textit{Segment 12/01/2015}: Obama to call for tax increase on rich to help middle class, proposes reduction in methane emissions from oil and gas, and Obama and Cameron talk counter terrorism and economy.

\subsection{Emergence of new topic pathways}
This subsection demonstrates the third core capability of the proposed algorithm; the detection of emerging topic pathways in the twitter stream. These new topic pathways are loosely related to previously identified pathways. Such new topics can be internal (e.g., new product or service launch) or external (e.g., interaction with a new organisation) to the business entity. Subsequently, some of them fade away as people lose interest, while some topic pathways prevail due to continuing interest.

\begin{figure}[!htb]
\centering
\includegraphics[clip=true, width=1.0\linewidth]{{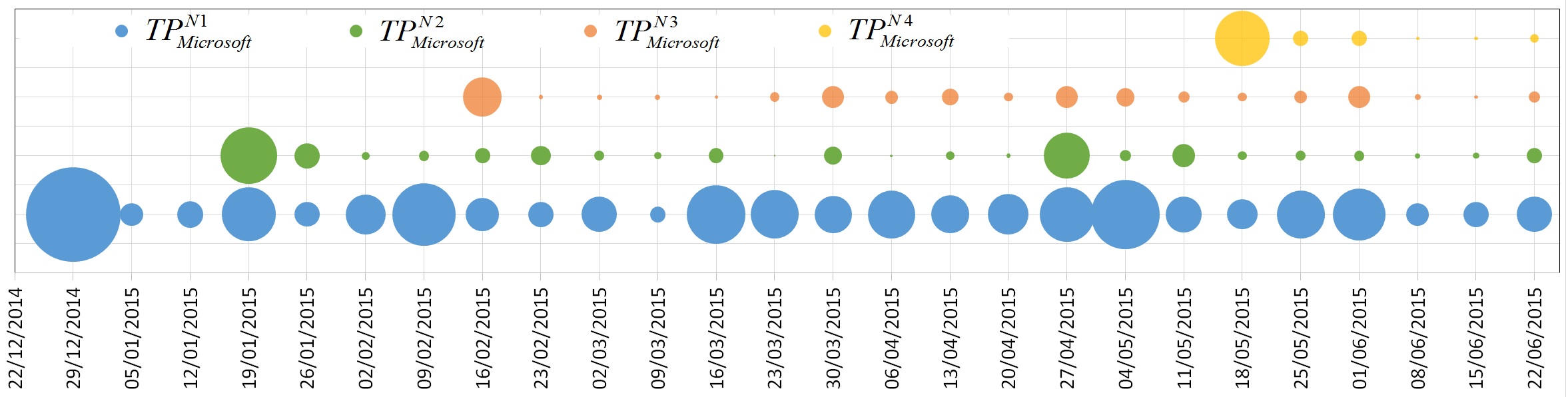}}
\caption{Four new topic pathways emerged in \#Microsoft Dataset. Each topic pathway is represented using a chain of bubbles where each bubble is a topic segment. Diameter of a bubble is set to number of tweets in the topic segment.}
\label{fig:new_topic_pathways}
\end{figure} 

\noindent $\boldsymbol{TP^{N1}_{Microsoft}}$: (frequent terms- windows 10, internet explorer, version, pirate, upgrade) 
Initiation of this new topic pathway coincides with the week where Microsoft announces that there will be a new web browser in Windows 10 operating system. Since then it has a significant amount of tweets that mainly discuss about new features in Windows 10. It is interesting to note that this pathway was separated as a new topic despite having a topic pathway focused on Windows) in general. We hypothesis that it is due to differentiating terms (e.g., windows 10, release, future, upgrade). Unlike other new topic pathways, this pathway retains its popularity throughout the time span of the dataset.

\noindent $\boldsymbol{TP^{N2}_{Microsoft}}$: (frequent terms- microsoft hololen, world, hologram, future, google glass) 
This pathway initiated when Microsoft introduced its virtual reality glasses \lq Hololens\rq~during \lq Microsoft Consumer Preview\rq~on 21/01/2015. After the initial peak it declined to few tweets a week except for the week 27/04/2015 in which the capabilities of Hololens was demonstrated at Microsoft Build Developer Conference.

\noindent $\boldsymbol{TP^{N3}_{Microsoft}}$: (frequent terms- linux, service, windows, python, bigdata) 
This pathway is initiated when Microsoft announces the incorporation of Linux operating system in its big data services provided in Microsoft Azure cloud platform. This move was widely discussed as Linux is a long term direct competitor to Windows and this is the first time that Microsoft incorporate Linux as part of its product/service suit.

\noindent $\boldsymbol{TP^{N4}_{Microsoft}}$: (frequent terms- salesforce, talk, msft, price, 55b) 
This pathway is focused on Microsoft's attempted acquisition of the company \lq Salesforce\rq~. It created hype on twitter as the news emerged and progressed for a couple of weeks. However, later declined in popularity when it failed due to the hefty price tag of \textit{55 billion USD} (\lq 55b\rq) demanded by Salesforce management.

\begin{figure}[!htb]
 	\centering
 	\includegraphics[clip=true, width=0.8\linewidth]{{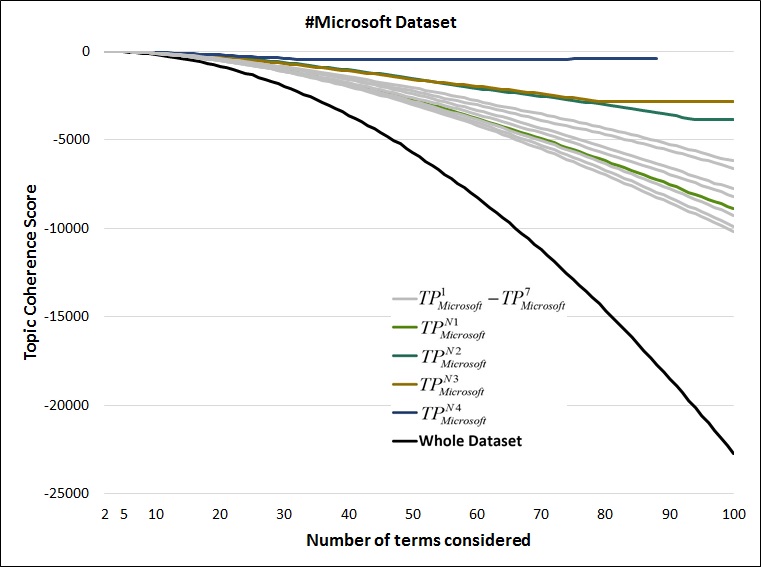}}
 	\caption{Topic coherence of newly emerged topic pathways.}
 	\label{fig:new_topic_pathways_coherence}
\end{figure}

\subsubsection{Topic Coherence}
Figure~\ref{fig:new_topic_pathways_coherence} presents the topic coherence metric for the above mentioned topic pathways plotted alongside the seven original topic pathways of the \#Microsoft dataset. It shows that new topic pathways (except $TP^{N1}_{Microsoft}$) have increased topic coherence scores than the other prominent pathways. This is because they are focused on topics which emerged as a result of particular events and thus contain more closely related terms that are uniquely used to discuss them. $TP^{N1}_{Microsoft}$ has frequent terms shared with the topic pathway focused on Windows in general ($TP^{4}_{Microsoft}$) which results in relatively low coherence scores.

Detection of these new topic pathways highlights the capability of our proposed algorithm to identify previously unseen topics at any stage of the social text stream. This capability is due to the third extension of the extended \ikaslext algorithm described in Section~\ref{sec:new_topicpathway}, which facilitates learning a new randomly initialised feature map from social media messages that are less similar to the existing topic pathways and spawning new topic pathways when there is a burst of social media messages on a new topic.

\subsection{Automatic event detection}
The aim of this section is to demonstrate the fourth capability; automatic detection of events from topic pathways using event indicators.  
 
As previously shown in Equation~\ref{equ:event}, event score is determined by combining three event indicators for each topic segment in topic pathways. 

The $r$ parameters in Equation~\ref{equ:event} decides the impact of each event indicator on the final event score. However, we have observed that the frequency and intensity changes of volume is higher compared to the changes of positive and negative sentiment. In order to compensate for this behaviour  $r_V$ is set to a lower value than others as follows  $r_V = 0.1$, $r_{PS} = 0.45 $ and $R_{NS} = 0.45$. We also employed $W=2$ for the calculations of $I_V$, $r_{PS}$ and $r_{NS}$. $\tau_e$ is set to $1.0$ for this experiment and we filtered-out events that belong to topic segments containing less than 1\% of the tweets in that batch.

\begin{table}[!htb]
\caption{Top 10 events detected based on event score $\mathcal{I}$  for \#Microsoft Dataset}
\label{table:events_Microsoft}
\small
\begin{tabulary}{\linewidth}{|p{0.4cm}|p{1.1cm}|p{0.4cm}|p{0.6cm}|p{0.63cm}|p{0.63cm}|p{3.75cm}|p{4.75cm}|}\hline
\multirow{4}{*}{\rotatebox[origin=r]{90}{Segment}} & Topic & \multirow{4}{*}{$\mathcal{I}$} & \multicolumn{3}{c|}{Event indicator} & \multicolumn{1}{c|}{Frequent terms}& \multicolumn{1}{c|}{Verification evidence}\\
& \multirow{3}{*}{pathway} & &\multicolumn{3}{c|}{scores} & \multicolumn{1}{c|}{\multirow{3}{*}{related to the event}}& \multicolumn{1}{c|}{\multirow{3}{*}{based on news articles}}\\
& & &$0.1$ & $0.45$ & $0.45$ & & \\
& & &$\times I_V$ & $\times I_{PS}$ & $\times I_{NS}$ & & \\\hline 	
\rotatebox[origin=r]{90}{27/04/2015}&\rotatebox[origin=r]{90}{$TP^{N2}_{Microsoft}$}&1.62&0.69&0.53&0.4& microsoft hololen, google glass, build2015, augmented reality, future &Microsoft demonstrated the capabilities of Hololens in Microsoft Build Developer Conference \\\hline
\rotatebox[origin=r]{90}{22/12/2014}&\rotatebox[origin=r]{90}{$TP^{2}_{Microsoft}$}&1.37&0.1&0.52&0.75& psn, ddos attack, christmas, lizardsquad, credit, lizard squad &Xbox Live online services were down on Christmas day after a DDoS attack by a group called 'LizardSquad'\\\hline
\rotatebox[origin=r]{90}{16/03/2015}&\rotatebox[origin=r]{90}{$TP^{N1}_{Microsoft}$}&1.31&0.52&0.36&0.43& windows 10, pirated version, summer, pirate, password & Microsoft announced that it will give free upgrades to Microsoft 10, even for pirate copies \\\hline
\rotatebox[origin=r]{90}{30/03/2015}&\rotatebox[origin=r]{90}{-}&1.26&0.48&0.42&0.36& Iphone, rogue window, android phone, office lens, document, powerful scanner &Microsoft launches  Office Lens scanner app to iPhone and Android for the first time\\\hline
\rotatebox[origin=r]{90}{12/01/2015}&\rotatebox[origin=r]{90}{$TP^{1}_{Microsoft}$}&1.22&0.18&0.54&0.5& criticism, window, tick, endpoint risk, windows 8 &Google released malicious code which can be used to exploit Microsoft Windows\\\hline
\rotatebox[origin=r]{90}{11/05/2015}&\rotatebox[origin=r]{90}{-}&1.16&0.2&0.4&0.56& microsoft hyperlapse , video, desktop, hyperlapse, shaky video, launch & Microsoft launches Hyperlapse, a video smoothing software for Android and Windows phones \\\hline
\rotatebox[origin=r]{90}{30/03/2015}&\rotatebox[origin=r]{90}{$TP^{4}_{Microsoft}$}&1.15&0.15&0.52&0.48& open source, windows sensor, open-source, microsoft exec, sale &Microsoft executive says that it is possible that Windows become open-source in future\\\hline
\rotatebox[origin=r]{90}{09/02/2015}&\rotatebox[origin=r]{90}{$TP^{7}_{Microsoft}$}&1.14&0.24&0.4&0.5& google app, note, iwork, windows user, microsoft office, 700 billion, watch, google threaten, work &Google just threatened to expose Apple and Microsoft security flaws.  Apple opens up iWork to Windows users for free\\\hline
\rotatebox[origin=r]{90}{09/03/2015}&\rotatebox[origin=r]{90}{-}&1.13&0.3&0.45&0.38& microsofts cortana, report, standalone app, ios device, digital assistant, siri &Microsoft announced its   intelligent personal assistant Cortana will be released for iPhone and Android\\\hline
\rotatebox[origin=r]{90}{09/02/2015}&\rotatebox[origin=r]{90}{$TP^{N1}_{Microsoft}$}&1.13&0.2&0.48&0.45& public preview, lumia , first preview, microsoft windows10forphone, lumia 635 &Microsoft delivers first public preview of Windows 10 for phones\\\hline	
\end{tabulary}
\end{table}	
		
Table~\ref{table:events_Microsoft} and Table~\ref{table:events_Obama} present the top 10 events identified based on event score $\mathcal{I}$ in the two data sets. The frequent terms that are related to an event are derived by capturing the frequent terms that appear in the topic segment containing the event but does not appear in the frequent terms of the $W$ previous topic segments of the same pathway. We analysed online news articles from major news agencies to verify the significance of events detected by the algorithm.   
 
\begin{table}[!htb]
\caption{Two topic pathways from \#Obama Dataset represented using high frequent terms of the topic pathway and high frequent terms of top five most voluminous topic segments}
\label{table:events_Obama}
\small
\begin{tabulary}{\linewidth}{|p{0.4cm}|p{1.1cm}|p{0.4cm}|p{0.6cm}|p{0.63cm}|p{0.63cm}|p{3.75cm}|p{4.75cm}|}\hline
\multirow{4}{*}{\rotatebox[origin=r]{90}{Segment}} & Topic & \multirow{4}{*}{$\mathcal{I}$} & \multicolumn{3}{c|}{Event indicator} & \multicolumn{1}{c|}{Frequent terms}& \multicolumn{1}{c|}{Verification evidence}\\
& \multirow{3}{*}{pathway} & &\multicolumn{3}{c|}{scores} & \multicolumn{1}{c|}{\multirow{3}{*}{related to the event}}& \multicolumn{1}{c|}{\multirow{3}{*}{based on news articles}}\\
& & &$0.1$ & $0.45$ & $0.45$ & & \\
& & &$\times I_V$ & $\times I_{PS}$ & $\times I_{NS}$ & & \\\hline 	
\rotatebox[origin=r]{90}{19/01/2015}&\rotatebox[origin=r]{90}{-}&1.22&0.33&0.52&0.37& tuesday, netanyahu, india visit, proof, nice, world, obamainindia &Obama visits India\\\hline
\rotatebox[origin=r]{90}{05/01/2015}&\rotatebox[origin=r]{90}{$TP^{3}_{Obama}$}&1.22&0.1&0.42&0.7& gitmo, france, freedom, paris, press, terror, war &Obama released 9 suspects held in GTMO prison. Obama condemns Charlie Hebdo attack saying  terrorists fear freedom\\\hline
\rotatebox[origin=r]{90}{26/01/2015}&\rotatebox[origin=r]{90}{$TP^{6}_{Obama}$}&1.16&0.08&0.43&0.65& white house, taliban, muslim province, whitehouse, japanese journalist, islamicstate, beheading &ISIS threatens to behead Obama and transform America into a Muslim province. ISIS killed second Japanese Hostage\\\hline
\rotatebox[origin=r]{90}{30/03/2015}&\rotatebox[origin=r]{90}{$TP^{3}_{Obama}$}&1.16&0.23&0.49&0.44& irandeal, lying , obama kerry, detail, framework, trust, irantalk, good deal, nuke agreement & \\\hline
\rotatebox[origin=r]{90}{02/02/2015}&\rotatebox[origin=r]{90}{$TP^{6}_{Obama}$}&1.15&0.12&0.41&0.62& death, january, crusade, unemployment rate, campaign, christianity, muslim leader, teaparty, immigration & Obama, at National Prayer Breakfast, compares ISIS to violence of Christian Crusades. USA unemployment rate rises to 5.7\% in January\\\hline  
\rotatebox[origin=r]{90}{19/01/2015}&\rotatebox[origin=r]{90}{-}&1.13&0.07&0.57&0.49& battle, marijuana legalization, tntweeters obama, crisis, obama cybercare &Obama on Thursday said he expects more states to experiment with marijuana legalization\\\hline
\rotatebox[origin=r]{90}{16/02/2015}&\rotatebox[origin=r]{90}{-}&1.13&0.14&0.6&0.39& fabric, founding, woven, love, obama islam, giuliani, obamas love, gov walker, mayor giuliani &Obama created controversy by saying “Islam Has Been Woven Into the Fabric of Our Country Since Its Founding”. New York Mayor Rudy Giuliani said Obama doesn't love America\\\hline
\rotatebox[origin=r]{90}{22/12/2014}&\rotatebox[origin=r]{90}{$TP^{5}_{Obama}$}&1.13&0.15&0.5&0.48& 6 year, apron, hacking row, novice, obama motorcade, afghanistan, trend, malaysian leader, afghan milestone &Obama hails end of combat operations in Afghanistan\\\hline
\rotatebox[origin=r]{90}{23/03/2015}&\rotatebox[origin=r]{90}{$TP^{5}_{Obama}$}&1.12&0.1&0.49&0.53& obamacare, price, double funding, antibiotic resistance, national plan, senate, superbug  & Obama announced a 5-year action plan to prevent lives lost from antibiotic resistance\\\hline
\rotatebox[origin=r]{90}{19/01/2015}&\rotatebox[origin=r]{90}{$TP^{1}_{Obama}$}&1.12&0.19&0.57&0.36& respect, russia, ukraine, saudi arabia, pressure, saudi king, putin , india trip, russian aggression & Obama cuts short presidential trip to India, and travelled to Saudi Arabia to pay respects to King Abdullah. Obama pledges more pressure on Russia as Ukraine clashes broaden\\\hline
\end{tabulary}
\end{table}	

The above tabulated results demonstrate and verify the authenticity of events detected by the proposed algorithm. It is interesting to note that these events were captured exclusively based on social expressions on Twitter. It is also interesting to observe that different events were triggered by high values from different indicators. Volume based indicators are high in the first event of Table~\ref{table:events_Microsoft}, which is about an event where Microsoft demonstrated the capabilities of its virtual reality product Hololens. As shown in Figure~\ref{fig:new_topic_pathways} a new topic pathway is spawned when Hololens was first introduced by Microsoft on 21/01/2015. 

Another example where $I_V$ is high is the third event in Table~\ref{table:events_Microsoft}, which is due to Microsoft’s statement that Windows 10 would be a free upgrade for all users (including those with pirated). This statement attracts attention on online social media platforms as it was a surprise move by Microsoft (quite often aggressive about software piracy). This peak of volume can also be seen in Figure~\ref{fig:new_topic_pathways} for $TP^{N1}_{Microsoft}$. 

There are several events in both tables that are significantly high in negative sentiment event indicator score. Event 2 in Table~\ref{table:events_Microsoft} is the highest triggered by the event where Xbox Live online services were offline due to a hacker attack on Christmas day. This event has caused issues for online gaming community who took it into Twitter criticising Microsoft for lack of security and sluggish response. 

Event 2 in Table~\ref{table:events_Obama} is highly negative, which was due to firstly, people condemning the release of nine suspects held in Guantanamo Bay and secondly the Charlie Hebdo Shooting in the same week. Social commentary combined both incidents stating that releasing of terror suspects can lead to more attacks. This incident is identified from topic pathway focused on Iran/Israel relations, which is loosely related to the event since most of the released suspects are from Middle East.

Events with high positive sentiment are comparatively rare. This is largely because online social media platform Twitter is more often used to express negative sentiment more than positive. It also supports the finding that popular events are often associated with negative sentiment~\cite{Thelwall2011}. 

This subsection aptly demonstrates the final capability of the proposed platform, automated event detection from social text streams. The incorporation of several event indicators allow us to capture events across diverse domains without having to develop tailored event capture mechanisms for individual applications.

%% file: Chapter5.tex
\onlyinsubfile{	
\setcounter{chapter}{4} 
}

\chapter[]{Expatiation from algorithm to technology platform}\label{chap:5}

\epigraph{{\textit{I alone cannot change the world, but I can cast a stone across the waters to create many ripples}}{ -Mother Teresa}}

This chapter presents the advancement of the  conceptual model for social behaviour understanding proposed in Chapter~\ref{chap:3}, and the new incremental machine learning algorithms, based on the principles of  self-structuring artificial intelligence, proposed in Chapter~\ref{chap:4},  into a functional technology platform, that delivers on the ambition of common good and contribution to human society. More specifically, the focus of social good is on patient-centred healthcare and the role of online support groups (OSG). 
\\\\
OSG are social forums that provide peer support, mainly for physical and mental health related issues. Unlike fast-paced mainstream social media platforms, OSG are relatively low volume social data streams where the bigger OSG generate few hundreds of social media messages daily compared to millions in Twitter. The participants of these platforms engage in longer discussions, and such discussion mainly focused around the main theme of OSG. Engaging in longer discussions lead to the formation of stronger ties between the OSG members where they support each other to cope with the hardships of the disease conditions. Because of these stronger ties participants tend to frequently do deeper self disclosures which include information about their disease conditions and deeper emotions related to  their wellbeing.   

The proposed technology platform operates in multiple stages to better facilitate the different information needs of its stakeholder groups: consumers, researchers and health professionals. It extracts information encapsulated in free-text discussions of OSG using a suite of machine learning, natural language processing and emotion analysis techniques and automatically generate individual profiles for each OSG user which enriches the information retrieval process.

The chapter is organised as follows: Section~\ref{sec:OSG_intro} describes an OSG, its purpose and functions, Section~\ref{sec:users_OSG} defines the stakeholder groups of OSG. Section~\ref{sec:limitations_OSG} delineates limitations of OSG and Section~\ref{sec:proposed_framework} presents the proposed technology platform that addresses such limitations and advances the role of OSG in understanding social behaviours. Finally, Section~\ref{sec:evaluation} presents a statistical evaluation of the technology platform, followed  by Section~\ref{sec:demo} which demonstrates its capabilities on two high-volume OSG.
  
\section{Online Support Groups (OSG)}\label{sec:OSG_intro}
Health related information seeking has become one of the major use of internet, where surveys reported that in 2010, 80\% internet users (59\% of all adults) in United States have looked online for health related information such as specific disease and treatment options~\cite{fox2011social}. Moreover, 34\% internet users have read online resources that contain consumer generated content about health issues and 18\% internet users have searched internet to find others with similar diseases or health concerns~\cite{fox2011social,fox2013after}.  

One of the growing sources of consumer generated online health information is Online Support Groups (OSG) which are also known as online health communities and health forums. OSG are virtual communities that are established to discuss or exchange knowledge about health related topics. OSG are formed in many different forms of online social media platforms such as discussion forums, blogs, wikis, groups in mainstream social media platforms (e.g., Facebook groups), and Q/A websites (e.g., Yahoo! Answers~\footnote{http://answers.yahoo.com}). While there are few professional-led OSG, most OSG are peer-to-peer and moderated by few self-appointed senior users of the OSG. It is reported that seeking such peer-to-peer support is prominent among patients living with chronic illnesses such as high blood pressure, diabetes, heart or lung problems and cancer~\cite{Fox2011a}.

OCSG discussions are organised as discussion threads where each thread starting with a question, comment or an experience about corresponding individual\lq s health concerns. Other OSG members (patients, partners of patients or other caregivers) respond to such concerns and thereby create discussion threads.

\begin{figure}[!htb]
	\centering
	\includegraphics[clip=true, width=1.0\linewidth]{{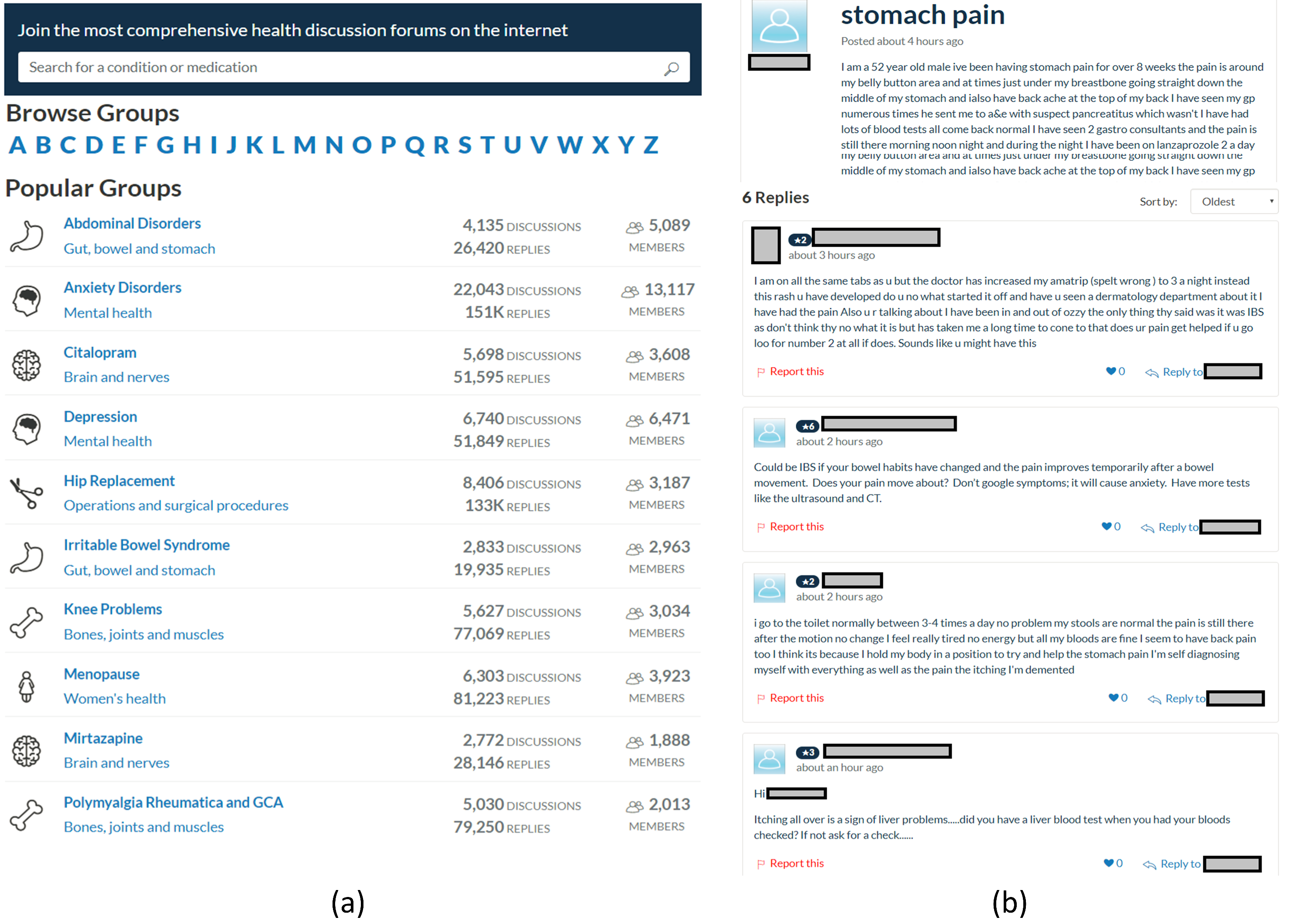}}
	\caption{a. Home page of patient.info OSG, which shows the popular discussion groups and the simple search interface, b. A sample OSG discussion thread.}
	\label{fig:osg_image}
\end{figure}

Figure~\ref{fig:osg_image}.a shows the home page of a popular OSG, \textit{patient.info}\footnote{http://patient.info/forums}. It contains a \textit{free-text} search and a list of high level topics (e.g., ‘Mental health’ and ‘Women's health’). Figure~\ref{fig:osg_image}.b shows a sample OSG thread where the initiating author starts with the title \lq stomach pain\rq~ and a description about his experience, which includes demographics of the author as well as the symptoms that he encountered. Figure~\ref{fig:osg_image}.b also shows four subsequent posts (out of the six replies in the thread), in which first, second and fourth seems to be advice and the third is further information by the initiating author.

\section{Users of Online Support Groups}\label{sec:users_OSG}
\begin{figure}[!htb]
	\centering
	\includegraphics[clip=true, width=0.8\linewidth]{{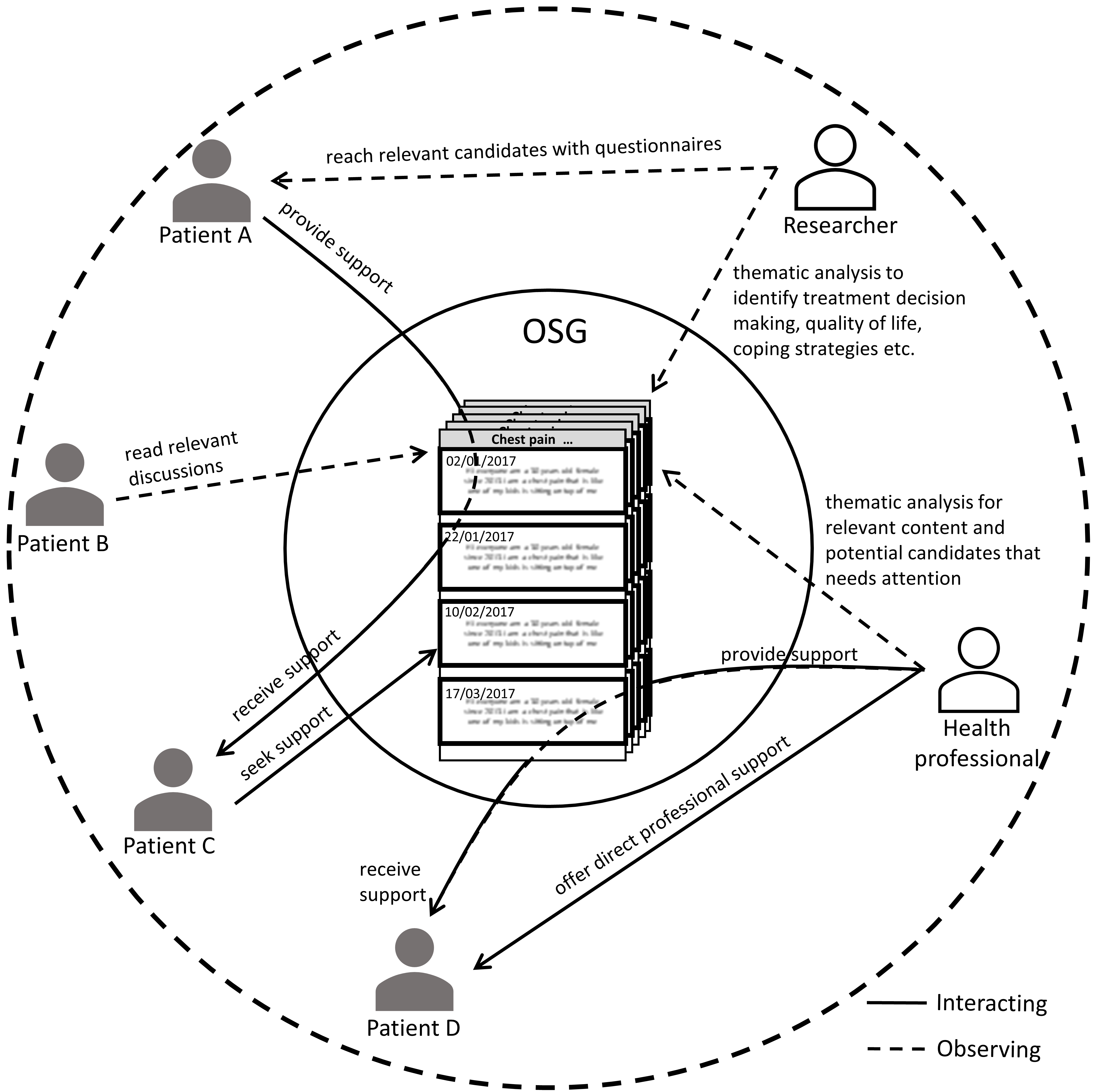}}
	\caption{Stakeholders of Online Support Groups (OSG).}
	\label{fig:osg_stakeholders}
\end{figure} 

OSG are contributed by consumers (e.g., patients or caregivers), health professionals and health researchers who are been denoted as the three main user groups of Medicine 2.0 applications~\cite{Eysenbach2008}. Figure~\ref{fig:osg_stakeholders} highlights the activities and interactions of these user groups. 
\begin{enumerate}
	\item Consumers: OSG are mostly contributed by health consumers who are often patients or caregivers of patients (e.g., partner, children, family). They seek and receive support from other OSG users (peers) in OSG as well as provide support to others.
	\item Health professionals: professional involvement in OSG varies, where some OSG have designated health professionals to monitors and mediate the health information exchanged, while in most OSG professionals involve voluntarily to provide their professional opinion to support seekers.
	\item Researchers: researchers often scrutinise the OSG discussions thematically to understands discussions topics as well as to find suitable participants for research studies.   
\end{enumerate} 

\subsection{Motivations to participate in OSG}\label{sec:motivations_OSG}
OSG are mainly used by patients and caregivers to fulfil some of their social needs discussed in Section~\ref{sec:social_needs} such as belongingness/affiliation, achievement and power over others.  These were achieved using social behaviours such as self-disclosure and social comparison.   

Functionally, motivations to participate can be categorised into three: (i) seek informational support, (ii) seek emotional support and (iii) provide support.     

\subsubsection{Seeking informational support}\label{sec:information_support}
Information support is sought for information related to a particular health condition. Individuals first self-disclose their information related to the interested health condition and either ask specific questions or similar experiences. The objective of seeking information support is two-fold. First is to gain knowledge about the health condition. The second objective is to compare (social comparison) their situation to similar situations of others. Such comparisons  enable individuals to get an understanding of their situation compared to others, and also as discussed in Section~\ref{sec:social_comparison} downward social comparisons reduce distress related to the health conditions by knowing that others have similar or even worse conditions/sufferings.  

Information support seeking happens at three different stages of the healthcare process such as initial diagnosis, treatment decision making, and post-treatment quality of life~\cite{Hu2012b,Ziebland2012}. 

Information seeking in initial diagnosis is mainly to re-affirm the diagnosis made by the clinician. It is reported that comparison of own symptoms and disease conditions against others with similar diagnosis helps to boost the confidence about the diagnosis as well as the clinician~\cite{Ziebland2012,Mazanderani2012}. 

Treatment decision making process is another key stage where people seek informational support~\cite{Huber2011, Zhang2017b,Berry2003}, especially when there are multiple treatment options with contrasting pros and cons. The accounts of people who have undergone different treatment options are scrutinised to capture the decision making criteria as well as their contemplation about those decisions afterword.      

Another most sort after information is the post treatment quality of life from the people with chronic conditions, where people look for coping strategies and lifestyle issues~\cite{Ziebland2012}. Such information seeking is more prominent when the issues are in intimate and delicate in nature~\cite{Blank2010, Huber2011}.

\subsubsection{Seeking emotional support}
Emotional support is sought for emotional issues related to disease conditions. Emotional support attempts to satisfy belonging social need by facilitating an understanding, supportive and empathetic audience where users can express their emotional issues related to the disease conditions. OSG members provide supportive, encouraging and positive response to such issues and thereby reduce the emotional burden of the disease conditions~\cite{Tanis2008,Bar-Lev2008a}.

Another key problem specifically faced by people with chronic illnesses is the disconnect or isolation from the community (work or living) where they were once part of~\cite{Williams1984,bury1982chronic}. Some individuals may even feel embarrassed and stigmatised by their illness and afraid to face their living community. For such individuals, contributing to OSG provide a sense of belonging to a community that consists of others with similar circumstances, which helps to reduce the feeling anxiety and stress resulted due to the isolation of community~\cite{Bar-Lev2008a,Harvey2007}.

Moreover, it is reported that individuals with chronic health issues reduce their anxiety by comparing their situation to others with similar circumstances~\cite{Frost2008}.~\citet{Bender2011} explains this phenomenon  using the theory of social comparison processes~\cite{festinger1954theory}, which states that during uncertain situations individuals tend to seek others with similar circumstances to compare their behaviours and abilities. 

\subsubsection{Providing support}
One of the vital element of the OSG ecosystem is the voluntary support provided by peer OSG users.~\citet{Chiu2006} argue that willingness to share knowledge to support peers is key to the fostering of virtual communities such as OSG. Providing support satisfy needs such as a sense of belongingness to the community as well as a sense of achievement for supporting others. Also, it may give a sense of power over others for influencing. The following are the ways that OSG users provide peer-support.
\begin{itemize}
\item \textbf{Answer direct questions:} Most of the direct questions are on information support discussed in Section~\ref{sec:information_support}. Users answer such direct information requests, based on their experience in similar health concerns. A study on users in a Diabetes OSG~\cite{Zhang2016}  has found that users often answer those question because they are confident about their understanding of the illness. Also, answering questions made the users felt proud or empowered as a contributor and a mentor to the OSG~\cite{Zhang2016, Kummervold2002}.

\item \textbf{Self-disclosure:} Users often self-disclose their story of coping with health concerns as support for other users facing similar situations. Such disclosures help to find important clues to cope with the situation and also it serves as emotional support for the receiving users as they tend to feel belonging to a social group with similar health concerns~\cite{Høybye2005,Ziebland2012}.  
\end{itemize}     

\subsection{Role of professionals in OSG}\label{sec:prefessionals_OSG}
Support groups (both face-to-face and online) often have different levels of professionals support which varies from professional-led to patient-led~\cite{shepherd1999continuum}. Professional-led support groups are often small closed groups (by invitation only) which facilitates support for certain patient groups of a particular healthcare organisation. Such groups are mediated and quality controlled by a board of health professionals. While such quality control reduces the risk of spreading misnomers, researchers have found that excessive professional involvement tends to reduce patients supporting each other~\cite{kurtz1990self}. In addition to mediation, professional therapist often involved in OSG to provide emotional support to patients and caregivers.  

Most of the current popular OSG are patient-led, with no designated professional body to mediate or quality control. However, it is observed that health professionals voluntarily join the OSG as members and provide their opinions as informational and emotional support to the other OSG members. In addition, some health professionals refer to relevant OSG discussions to understand the patient concerns related to disease conditions.

\subsection{Opportunities for researchers in OSG}\label{sec:researchers_OSG}
OSG contain unsolicited accounts of first person experiences, which is a rich source of information for the medical researchers to study different aspects of patient experience~\cite{Robinson2001}. There are two main avenues that OSG are used by researchers.

In the first approach, Medical researches use OSG to recruit potential candidates to administrate interviews or questionnaires for various research studies. OSG provide access to user groups aggregated with similar health concerns which are used by the researchers to send invitations to participate in various studies. One of the key advantage of OSG is it provide access to concentrated groups that are otherwise difficult reach with conventional methods, which can be due to either rareness of the inclusion criteria or social stigmatised nature of the illness~\cite{Wright2006}. Moreover, it allows the researchers to reach out to a large group of potential candidates within a short period of time. This approach uses OSG only as a recruitment platform for conventional survey based methods, without paying attention to its peer-to-peer support ecosystem or the content of the discussions.

In the second approach, medical researchers retrospectively analyse OSG discussions of the selected groups with specific selection criteria. This approach is less intervening compared to the previous, as it does not reach the users directly, but more exhaustively analyse the content of the OSG messages. The self-reported expressions of an individual are scattered across the OSG as parts of different discussion threads. However, once aggregated and ordered the OSG posts of an individual provides significant insights about the patient journey over time which can be exploited to analyse their demographics as well as the clinical outcomes and the associated emotions across the patient journey. Such analysis is more beneficial for a chronic illness as it provides the opportunity for the researchers to evaluate the self-reported quality of life of each individual against the clinical outcomes at different stages of the patient journey. 

In the current research most of these studies are conducted manually or semi automatically, often as thematic analysis of the OSG discussions of chronic illnesses such as breast cancer~\cite{Høybye2005,Huber2017,winzelberg2003evaluation}, prostate cancer~\cite{Gooden2007,Huber2011}, HIV~\cite{Bar-Lev2008a}, and Autism~\cite{Zhang2017c}. The limitation of this approach is that due to the exhaustiveness of the process, such analysis is limited to few OSG participants and often limited to analysing only the clinical outcomes without correlating them with the associated emotions expressed. This mainly due to the challenging and nature of extracting such information from noisy and unstructured free-text in patient-authored OSG posts.       
   
\section{Limitations in finding relevant information}\label{sec:limitations_OSG}

As discussed in Section~\ref{sec:motivations_OSG} OSG plays an important role as an online platform that enables users to engage with virtual peers who have (or had) similar health concerns. One of the key elements of these engagements is finding users with similar health concerns. Based on a survey of diabetes OSG users, ~\citet{Zhang2016} points out that the users look for similar or related users mainly based on similar illness conditions and comparable demographics. In the existing OSG infrastructure users achieve this task by manually scrutinising the posts in OSG threads for clues about the illness conditions and demographics of the other users. 

Moreover, as shown in Section~\ref{sec:researchers_OSG} researchers look for OSG users with particular selection criteria (e.g., high risk prostate cancer patient who has gone through robotic assisted laparoscopic radical prostatectomy for prostate removal), where such user were either contacted to administrate questionnaires or their OSG discussions were analysed thematically depending on the research methodology. Such selection process is also carried out by manually going through the posts of OSG users while evaluating such users eligibility based on the self-stated information in OSG posts.      

However, with the widespread use of internet and popularity in online health information seeking across individuals from all walks of life, use of OSG participation is growing rapidly. Table~\ref{table:osg_stats} presents several usages statistics of three large OSG. It shows that those OSG are consists of over 100,000 registered users (\textit{healthboards} has over 1,000,000 registered users). Moreover, these OSG consists of over 100,000 discussion threads and more than 1,000,000 posts. Therefore, it is impossible to manually scrutinising these colossal text croups or go though large number of users to find the information that either users or researchers are looking.

\begin{table}[!htb]
\caption{Population and volume statistics of three large OSG. Note that this information are as of 02/04/2018.}
\label{table:osg_stats}
	\centering
	\begin{tabulary}{\linewidth}{|L|R|R|R|}\hline 
		OSG & registered users & threads & posts\\\hline
		\textit{healthboards} & 1,151,152 & 909,486 & 4,975,203\\\hline
		\textit{healingwell}  & 161,571  & 323,288 & 2,946,335\\\hline
		\textit{patient.info} & 161,734 & 225,220 & 1,882,821\\\hline
	\end{tabulary}
\end{table}

Based on the existing capabilities of these large OSG, consumers (users or researchers) narrow down their information retrieval process using two main approaches: (i) by navigating through the hierarchy of topic often exist in OSG platforms to find the topics relevant to theirs health concerns, and (ii) using the free-text search function that is available in some OSG to lookup threads based on relevant keywords.  

\begin{figure}[!htb]
	\centering
	\includegraphics[clip=true, width=0.8\linewidth]{{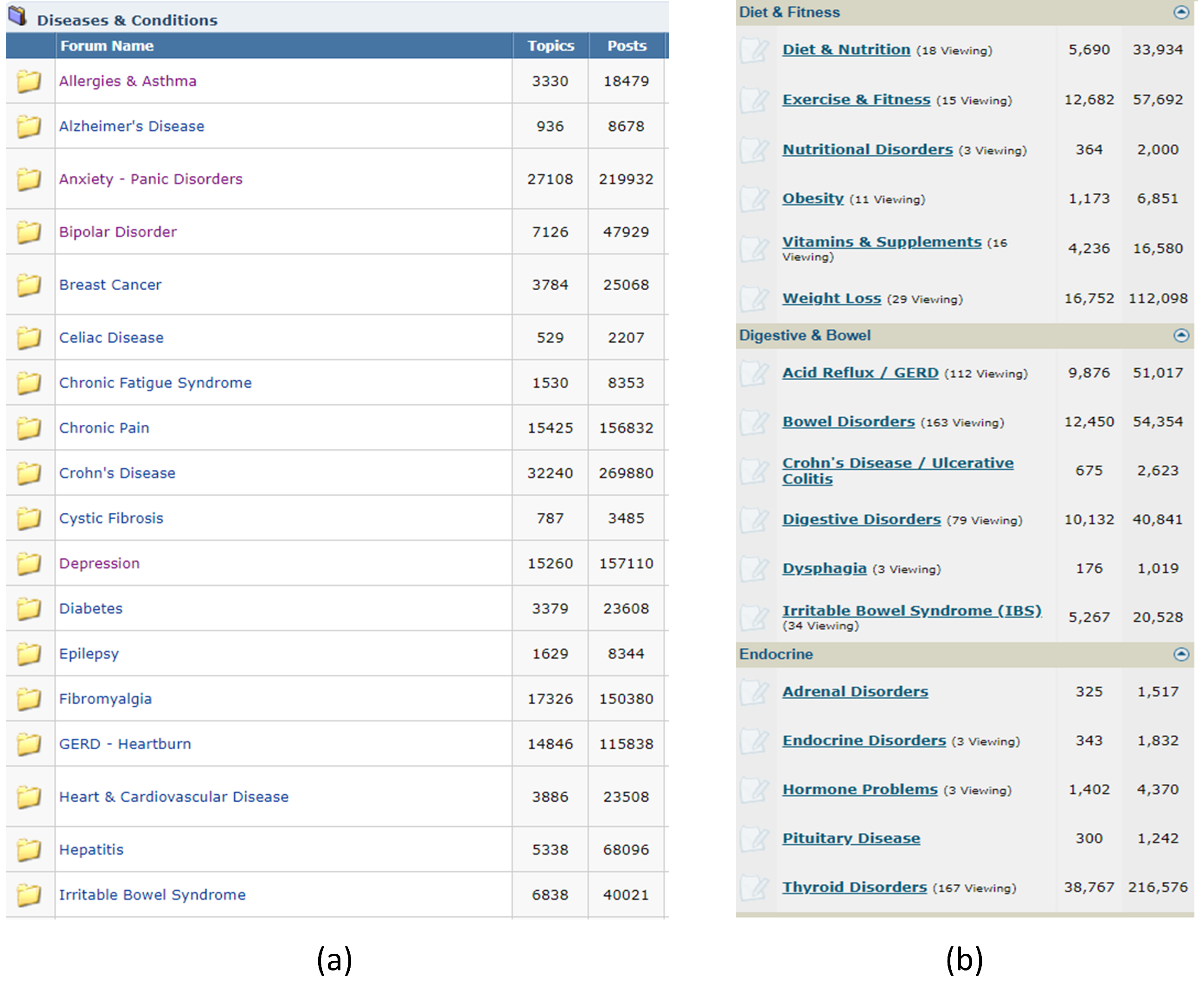}}
	\caption{Topic hierarchies in (a) healingwell and (b) healthboards OSG.}
	\label{fig:osg_topic_hierarchy}
\end{figure}
 
Figures~\ref{fig:osg_image}.a,~\ref{fig:osg_topic_hierarchy}.a, and~\ref{fig:osg_topic_hierarchy}.b show parts of the topic hierarchies in OSG \textit{patient.info, healingwell, and healthboards} respectively. As shown, the topic hierarchies in OSG are fairly shallow, often limited to two levels. Hence, the narrowing down by topic can still leave large number of posts for manual analysis. Moreover, these topic structure of the OSG are user generated, often by the moderators as per popular requests from users. Therefore, the topic hierarchy is developed incrementally, and thus can be overlapping and sometimes confusing. For example, in healthboards (Figure~\ref{fig:osg_topic_hierarchy}.b), the second level topics under \lq Digestive \& Bowel\rq can be overlapping as \lq IBS\rq can fall under both \lq Digestive Disorders\rq~ and~\lq Bowel Disorders\rq. 

Free-text search capabilities of those OSG are often basic and limited to a keyword search across the entire text corpus of the OSG, while some OSG allow keyword or phrase search across a selected topic. These search capabilities are facilitated by indexing the terms found in each post (except stop-words) so that they are searchable. Free-text search is demonstrated below using an example query \enquote{I’m a 40 year old woman taking Nexium for heartburn}.
\begin{enumerate}
\item Tokenize the query into words (I, m, a, 40, year, old, woman, taking, Nexium, for, heartburn).
\item Remove stop-words and stem-words to their root form (40, year, old, woman, take, Nexium, heartburn).
\item Search the OSG database for the posts that have any of these words.
\item Determine a relevance score for each identified post by aggregating tf-idf scores of each matching word (tf-idf is a statistical measure of how important a word is to a document in a collection of documents).
\item Present the top results with the highest relevance score. 
\end{enumerate}

The top results consist of posts that contain several matching words in the end-user query. However, it does not recognise that heartburn is a symptom, Nexium is a medication and the end-user is interested in posts that mention both of these words. In addition, it does not recognise 40 is an age mention and woman is a gender mention. 

These limitations of topic based navigation and free-text search results the precision and recall of the current information retrieval capabilities of the OSG.
 
\textbf{Low precision:} in an information retrieval task, precision is the fraction of the retrieved documents that are relevant~\cite{Manning2008IIR}. Topic based navigation ends up with a substantial amount of OSG post that is of little relevance to the health concerns of the users. Also, as demonstrated in the above example, free-text search pulls out OSG posts that have one or few relevant terms, which may not relevant. For example, there may be OSG posts with the terms \lq old\rq~ and \lq woman\rq~ but with different symptom mentions of disease conditions, thus, not relevant to health concerns in the query.    

\textbf{Low reliability in recall}: in an information retrieval task, recall is the fraction of relevant documents that are retrieved~\cite{Manning2008IIR}. Topic based navigation approach loose recall when users post relevant content under different topics, which can happen due to confusing topic structures or users misunderstanding the scope of a topic. Free-text search loose recall mainly because it is not capable of identifying the different types of information present in the query or OSG posts. For example, OSG posts from a user of age 41 with similar health concerns are a very close match for the above given query, but will not be given a higher relevance in the free-text search. Moreover, users only mention a portion of information in a single OSG post. For example, a user might have mention relevant health concerns in a one OSG post and relevant  demographics in another, but the free-text cannot automatically aggregate them and show it as a relevant result.    

The above issues are mainly due to insensitivity towards different types of information (e.g., demographic, clinical). These types of information are crucial to understand the context of OSG posts as well as the search criteria. Moreover, the information encapsulated in OSG posts needs to be aggregated by the author to get a more comprehensive picture of each author. The next section proposes such information structuring model which can overcome the  limitations of the existing information retrieval capabilities in OSG.

\section{Proposed technology platform}\label{sec:proposed_framework}
\begin{figure}[!htb]
	\centering
	\includegraphics[clip=true, width=1.0\linewidth]{{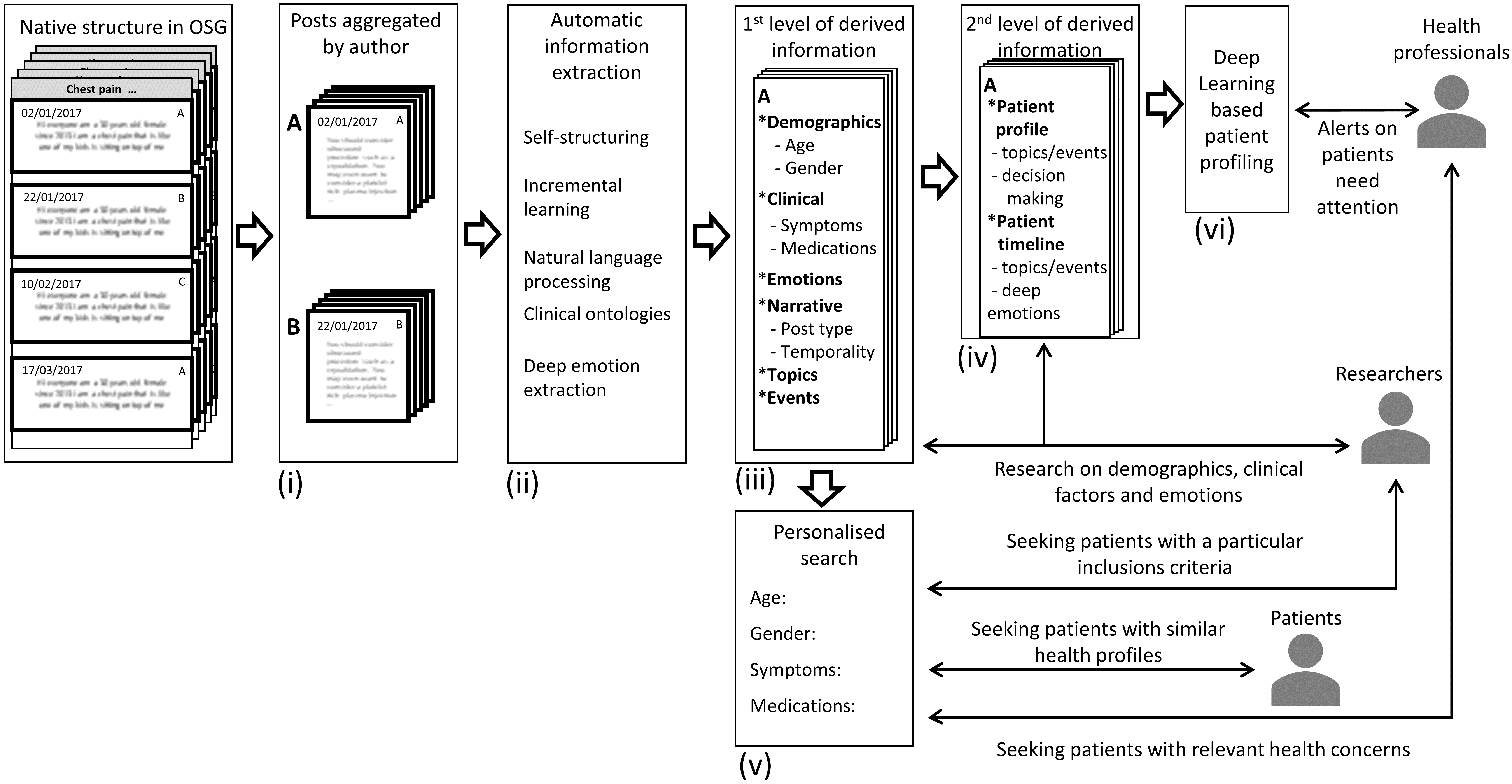}}
	\caption{The proposed multi-stage self-structuring AI platform and its benefits for the three stakeholder groups in OSG.}
	\label{fig:ir_model}
\end{figure}
Figure~\ref{fig:ir_model} presents the proposed multi-stage self-structuring AI platform which is an implementation of the proposed self-structuring artificial intelligence platform in Section~\ref{sec:proposed_platform}. It employs machine learning and natural language based information extraction methods to transform the free-text corpus of discussion threads into a structured multi-dimensional information database pivoted around each user to better facilitate the information needs of three groups of stakeholders in OSG. Stages of this framework is as follows: 
\begin{enumerate}[i]
	\item The OSG posts are aggregated by the author ID of each OSG user. As discussed, OSG users include bits and pieces of information in different posts, posted across multiple discussion threads. Therefore, aggregating the posts by each author enables more comprehensive information extraction from each user.
	\item Natural language processing and machine learning based techniques are employed; with the support of clinical ontologies and emotional thesaurus; to extract and infer information of each OSG user based on the free-text content of their aggregated OSG posts.
	\item The first level of information consists of demographic, clinical, emotional state and narrative type information about each user extracted from the previous stage. This information is stored in a multi dimensional database pivoted by the user ID of each user.
	\item The second level of information consists of complex information structures such as patient event timeline, which is constructed based on the clinical and emotional events derived in the previous stage.
	\item Structured search retrieves OSG members with a given set of clinical factors and range of  demographics.  
	\item Deep learning techniques are employed to learn the personal traits (e.g., depression) of the OSG users based on their emotions and linguistic patterns in discussion threads.
\end{enumerate}

The information structured in stages~\rom{3} and~\rom{4} opens up new opportunities for researchers to conduct cohort studies on the OSG population with particular health concerns. Emotional information can be employed in a similar manner as survey instruments to quantify the quality of life of patients. Aggregated quality of life across different dimensions can be used to compare and contrast different treatment types and side-effects. The patient profiles and timelines extracted in stage~\rom{4} can be employed for in depth research about patient decision making behaviour and decision factors as well as emotional trajectory over time of patients suffering from chronic illnesses.       

The structured search in stage~\rom{5} enables the patients to search up other patients with similar health concerns and demographics. Such functionality also benefits users who are keen to provide peer-support as they can easily look up the users with similar health concerns to provide support. Researchers can also use this structured search to apply their inclusions criteria of the research cohort to look up potential candidates for research. Health professionals can use it to lookup patients who have health concerns relevant to their field of expertise.

The stage~\rom{6} use deep learning techniques to model personal traits of the patients, based on their discussion topics, language pattern and emotions so that it can identify the behaviours of patients that needs the attention of a health professional. For example, a patient posting depressive content can be brought into the attention of a health professional on mental health. 

The implementation, evaluation and validation of stages~\rom{3} to~\rom{6} of this framework is presented in the rest of this chapter. A use case of this platform on prostate cancer related OSG are presented in Chapter~\ref{chap:6}. 

The next section presents the implementation details of stages~\rom{2} and~\rom{3} and their evaluations can be found in Section~\ref{sec:evaluation}. Section~\ref{sec:timeline_construction} presents the implementation of stage~\rom{4}, and Section~\ref{sec:ir_search_impl} provides  implementation details of stage~\rom{5} as part of the demonstration.

\subfile{Chapter5_Information_Extraction}

\subfile{Chapter5_Timeline_Extraction}

\subfile{Chapter5_Evaluation}

\subfile{Chapter5_Demonstration}

\section{Chapter Summary}
This chapter presents a multi-stage information structuring platform for online support groups which facilitate the information needs to of OSG stakeholders: consumers, researchers and health professionals. This platform process unstructured OSG messages using a suite of machine learning and natural language processing techniques to retrieve age, gender, narrative type, medical entities, emotions and patient timeline. Age, gender, and narrative type extraction modules show high precision and recall when evaluated with labelled data sets. The demonstration shows that the extracted attributes provide relevant and reliable information to consumers. Also, aggregated information is useful for research purposes. Next chapter extends this platform to structure OSG related to prostate cancer. 

\onlyinsubfile{	
\bibliographystyle{dcu}
\bibliography{library}{}
}

%% file: Chapter5_Information_Extraction.tex
\section{Information retrieval techniques}\label{sec:information_extraction}
This section presents the information retrieval techniques developed  for stages~\rom{2} and~\rom{3} of the proposed information structuring platform discussed in the previous section (see Figure~\ref{fig:ir_model}). This section is organised as follows: Section~\ref{sec:narrative_type_extraction} presents the narrative type extraction, Section~\ref{sec:age_extraction} presents age extraction, Section~\ref{sec:gender_extraction} presents gender extraction, Section~\ref{sec:medical_concept_extraction} presents medical concept extraction and Section~\ref{sec:emotion_extraction} presents the emotion extraction.  

\subsection{Narrative type extraction}\label{sec:narrative_type_extraction}
Online social platforms such as  OSG are used to seek, provide and exchange information~\cite{Zhang2015,Oh2012}. As discussed in Section~\ref{sec:motivations_OSG}, individuals share similar experiences and also provide advice to their peers in the OSG. 

Differentiating the posts as experience sharing or advice is an important task. Experience sharing posts are important as they contain self-disclosed information of patient factors~\cite{Sadovykh2015}, which includes demographic, clinical and emotional information during various stages of the patient timeline. This information can be employed for patient profiling. On the other hand, advice can be used to understand the nature of peer-advice in OSG.

OSG community is consists of not only patients but also a significant proportion of caregivers who posts on behalf of the patients that look after. Hence, the expressions of experience in OSG can be further categorised  as an expression of own experience or an expression of experience of someone else (often family member, close relative or a friend). Thereby, in this work the posts are grouped into three narrative type categories (i) experience: first-person, (ii) experience: second-person, and (iii) advice.    

Detecting experience expressing posts is being studied in various application domains. \citet{Park2010} developed a classifier based on a set of linguistic features to detect sentences that discuss human activities from Weblogs.~\citet{Liu2015} developed a similar classifier to extract sentences that express experience in online health forums (OSG) using  linguistic features such as pronouns, verb tense and modality of the verb.

\cite{Nguyen2012} has employed both bag-of-words (BoW) and linguistic features to detect experience expressing posts and identified that the most discriminative BoW features are \lq I\rq~ and \lq my~\rq and most discriminative linguistic feature is \textit{first person pronouns} (\lq I\rq~ and \lq my~\rq). This is because users frequently use \textit{first person pronouns} in expression of experience.     

In this work, the narrative type is determined based on the main subject of each OSG post. The key intuition for this approach is that different types of nouns and pronouns are used as the subject in different narrative types. For example, the first column of Table~\ref{table:narrative_type} shows three sample posts taken from a OSG. First post is a patient expressing his experience, and the main subject is a first-person pronoun. In the second post, a son is expressing about his mother, so the main subjects are ‘mother’ and third person pronoun ‘she’ (refers to ‘mother’). The last post is a piece of advice given to someone else, so the main subject is the second-person pronoun ‘you’. 

Based on the above observations, a set of rules are designed to categorise OSG posts into the three narrative types based on the main subject of each OSG post as follows:
\begin{enumerate}
	\item \textbf{experience first-person:} if the main subject is a first-person pronoun (e.g., I, my)
	\item \textbf{experience second-person:} if the main subject represents a relationship (e.g., mother, friend, son) to the narrator of the OSG post.
	\item \textbf{advice:} if the main subject is a second-person pronoun (e.g., you).
\end{enumerate}

The key challenge of this approach is that OSG posts sometimes contain several nouns and pronouns as the subject of different sentences, for example, patients expressing their experience might mention their family history using words such as ‘mother’ and third-person pronouns like ‘her’. Moreover, third-person pronouns are often used to mentions other people as well (e.g., I went to see GP and he prescribed ...).

\begin{table}[!htb]
\caption{Examples for narrative type resolution from OSG posts}\label{table:narrative_type}
\small
\begin{tabulary}{\linewidth}{|>{\raggedright}p{4.7cm}|>{\raggedright}p{4.7cm}|>{\raggedright}p{2.2cm}|>{\raggedright\arraybackslash}p{2cm}|}\hline 
Post & Pronoun resolved post & Human mention nouns& Narrative type\\\hline

My doctor ... yesterday. He ... daily. I took it ... and ... I felt ... Is this normal ...? I was ... because ... I haven't ... and I was ... with my fiancée. I got paranoid about her ... even though I knew ... she ...  & My doctor ... yesterday. \textless doctor\textgreater~ ... daily. I took it ... and ... I felt ... Is this normal ... ? I was ... because ... I haven't ... and I was ... my fiancée. I got paranoid about \textless fiancée\textgreater~ ... even though I knew ... \textless fiancée\textgreater~ ...  & I: 7 \newline doctor: 2 \newline fiancée: 2 \newline \newline prominent noun: \lq I\rq & experience: first-person\\\hline

My mother is diagnosed ... She is suffering with ... I know that ... her heart. An endocrinologist suggested her to ... Also, he recommended her to ... But now she had ... She is suffering ... & My mother is diagnosed ... \textless mother\textgreater~ is suffering with ... I know ... \textless mother\textgreater~ heart. An endocrinologist suggested \textless mother\textgreater~ to .. . Also , \textless endocrinologist\textgreater~ recommended \textless mother\textgreater~ to ... But now \textless mother\textgreater~ had ... \textless mother\textgreater~ is suffering ... & mother: 5 \newline endocrinologist: 2 \newline I: 1 \newline \newline  prominent noun: \lq mother\rq & experience: second-person\\\hline

You should consider ... such ... . You may even want to ... .  Try to ..., but if you've ... your other options you may ... & You should consider ... such ... . You may even want to ... .  Try to ..., but if you've ... your other options you may ... & you: 5 \newline \newline prominent noun: \lq you\rq  & advice \\\hline
\end{tabulary}
\end{table}

In order to overcome these ambiguities, initially, the gender specific third person pronouns in OSG posts are replaced with relevant nouns (e.g., she \textrightarrow mother). For this task, we employed a pronominal anaphora resolution~\cite{Mitkov2014} algorithm which predicts the relevant antecedents of third person pronouns. 

Pronominal anaphora resolution is initially attempted using JavaRAP tool~\cite{Qiu2004} that implements an algorithm presented by~\citet{Lappin1994}. However, this approach relies on parsing text to identify its structural elements and grammatical roles, which is found to be less effective for OSG posts because as discussed in Section~\ref{sec:social_data_analysis_challenges} they are not written conforming to the formal rules of grammar. Hence, a relatively simple rule based pronoun resolution algorithm CogNIAC~\cite{Baldwin1997} is selected. CogNIAC uses a set of rules to assign the most probable  antecedent to each pronoun. 

For this task, a partial implementation of CogNIAC is used to resolve pronouns to their antecedent. Note that the pronoun \lq it\rq~ is ignored as it is irrelevant to the given task. The second column of Table~\ref{table:narrative_type} shows the pronoun resolved output where each pronoun is replaced with the relevant noun.   

Once the pronouns are resolved it is easier to determine the key narrative type of the OSG post. For example, pronoun resolution in the second OSG post specified in Table~\ref{table:narrative_type} has aggregated \lq mother\rq and \lq she\rq, which makes the subject \lq mother\rq more evident, and that the narrative of that post is \textit{experience: second-person}. Column three of Table~\ref{table:narrative_type} shows a list of human subjects identified from the sample posts and column four shows the predicted narrative type.

Note that the pronoun resolved output of the OSG posts are further used in Section~\ref{sec:gender_extraction} for determining the gender of the patient.

\subsection{Age extraction}\label{sec:age_extraction}
Age is an important feature for patient profiling because medical information such as symptoms are interpreted differently for different age groups. Patients often mention their age within the text of the forum post. However, such mentions are highly diverse, unstructured and often expressed using shorten forms of English terms. Some sample age mentions are shown in column 1 of Table~\ref{table:age_related_phrases}. Also, they need to be differentiated from other number mentions such as duration, dose, weight etc.

\begin{table}[!htb]
\caption{Positive and negative samples of age related phrases found in forum posts}
\label{table:age_related_phrases}
\begin{tabulary}{\linewidth}{|L|L|}\hline 
Age Phrases (positive examples) &	Non Age Phrases (negative examples)\\\hline
when i was in my 20s &	My SVT started about 10 years ago\\\hline
In my 41 years of life	& I had a 48 hr monitor\\\hline
I am 42 yrs old &	I am using atenolol 50 mg\\\hline
my 25 year old son &	doc prescribes 10 tablets 1 twice a day\\\hline
operated on at 10 days old	& Currently I weigh 16 3/4 stone\\\hline
\end{tabulary}
\end{table}

Previous literature on age extraction from forums is limited mainly due to its challenging nature.~\citet{Liu2014} focused on standard age mentions such as \lq I am 35 years old\rq~and identify them as age mentions.~\citet{Kim2013} employed regular expressions such as ‘age’+number to extract age. These approaches achieve higher precision but as shown in Table~\ref{table:age_related_phrases}, age mentions are diverse thus these approaches would overlook a significant amount of age mentions resulting in a low recall.  

\citet{Zhu2012} look for clue words such as ‘years’, ‘old’, ‘aged’ etc. appear within two-word distance of each two-digit number mentions and such mentions are extracted as the age of the patient. In addition, they look for clue words such as ‘teenager’, ‘toddler’, ‘child’ to approximate the age of the patient. This approach would result in high recall however, can lead to a significant number of misclassifications (e.g., ‘about 10 years ago’ can be misclassified as an age mention).

A common issue of these approaches is that they do not resolve whether the age mentions is about the patient. For example, a patient might mention the age of relatives when talking about their family history. Also, there can be age mentions in past incidents of the patient (e.g., ‘I have this issue since 11 years old’). Moreover, there can be multiple age mentions in a forum post. Thereby, it is necessary to develop an extraction method to resolve age from multiple age mentions.

\begin{figure}[!htb]
	\centering
	\includegraphics[clip=true, width=1.0\linewidth]{{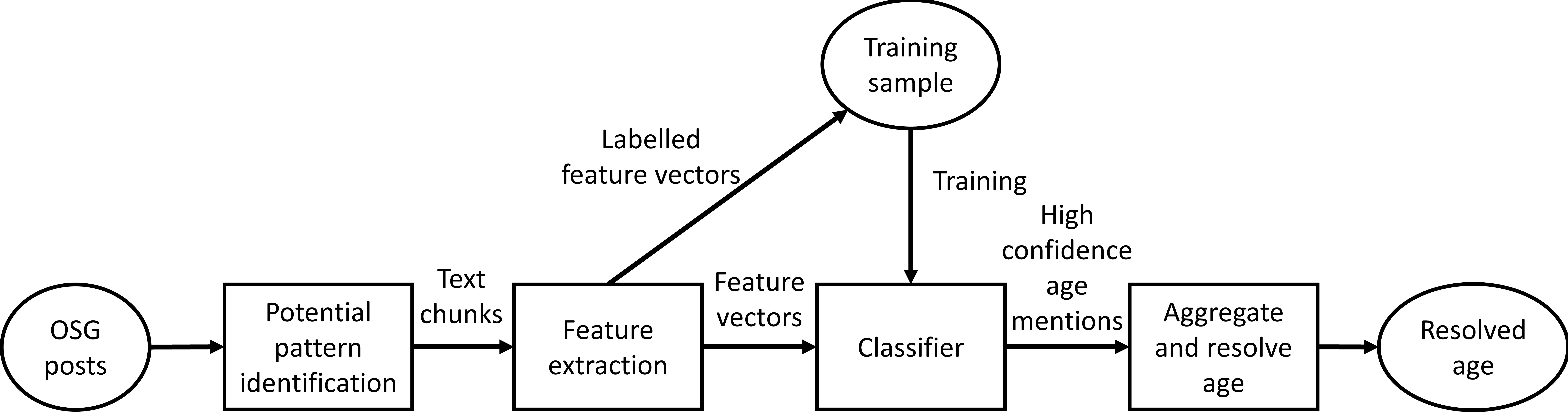}}
	\caption{Age resolution process}
	\label{fig:age_extraction_process}
\end{figure}

In order to overcome the above mentioned issues and also to improve both precision and recall we employed a multi-step process shown in Figure~\ref{fig:age_extraction_process} to resolve the age of each user in OSG. Each phase of this process is elaborated below. 

\subsubsection*{Potential pattern identification}
In this initial step, a regular expression based pattern in used to identify text chunks that are potential age mentions.
 
Firstly, each post is break into sentences using the sentence splitter in OpenNLP tool~\cite{baldridge2005opennlp}. Subsequently, the \emph{regular expression}: \verb/(\d{1,2}[a-zA-Z]{0,5})/ is used to capture two-digit number mentioned in the text. Note that, the above \emph{regular expression} is designed to accommodate numbers that co-exist with several letters such as ‘20s’ and ‘34yrs’.

A text chunk of five words before the detected two-digit number and five words after are extracted as a possible age mention. Note that start and end of sentences are padded with \verb/<S>/ symbols to accommodate age mentions in edges of the sentence. 

Let a set of OSG posts of a user $A$ be $\{p\}^A$. For each post $p$, a set of text chunks $\{t\}$ with possible age mentions were identified $\{p\}^A \leftarrow \{t\}^A$.

\subsubsection*{Feature extraction}
In this step the text chunks extracted from the previous step are transformed in to feature vectors ($\{t\}^A \leftarrow \{f\}^A$). These features are extracted to differentiate age mentioned from others (e.g., drug dose, clinical factors). 

\begin{figure}[!htb]
	\centering
	\includegraphics[clip=true, width=0.7\linewidth]{{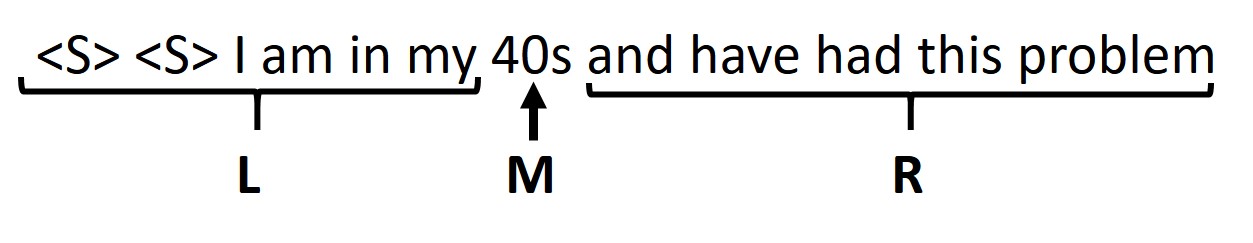}}
	\caption{A text chunk and its segments L, M, and R}
	\label{fig:text_chunk}
\end{figure}

As shown in Figure~\ref{fig:text_chunk}, the text chunk is first divided into three segments (L, M, and R) and different features were extracted separately from each segment.

We engineered 29 features that are capable of differentiating age related text chunks from others. Table~\ref{table:age_related_features} presents a sample of the extracted features. Note that, the features that are extracted from multiple segments are included in the feature vector as multiple features (one for each segment).

\begin{table}[!htb]
\caption{A Sample of features extracted from a text chunk}
\label{table:age_related_features}
	\begin{tabulary}{\linewidth}{|L|L|L|}\hline 
	Feature & Applied segments &	Representing terms\\\hline 
	First-person pronouns	& L	& I, Im, Iam\\\hline
	Possessive pronouns	& L &	my, our\\\hline
	Family relationship mentions &	L,R &	mother, mom, father, dad, brother, sister, son, daughter\\\hline
	State of being verbs &	L&	am, is was, were, will be\\\hline
	Age related prepositions&	L&	at, until, around, under, about, abt\\\hline
	Old Mentions& M, R& old, yo, y/o\\\hline
	Year Mentions&	M, R&	year, years, yo, yrs, y/o, s\\\hline
	Time mentions& 	M, R&	days, hours, hrs, minutes, min, weeks, wks\\\hline
	Dose mentions&	M, R&	mg, dose, tablets, mgs, micrograms, ug\\\hline
	\end{tabulary}
\end{table}

Most of these features are constructed to capture age mentions. For example \textit{First-person pronouns} are often used before an age mention. Some other features such as dose and time mentions are added to differentiate other key classes such as dosage (e.g., atenolol 50 mg) and duration (e.g., 10 years ago). 

\subsubsection*{Classifier}
Once the features were extracted from text chunks, we employed a classifier to identify the chunks that are most likely to be age mentions and assigned a confidence probability. To train this classifier we manually labelled 2,212 text chunks identified from the first step of this process. We labelled them as \lq age\rq~ or \lq other\rq~ based on whether it is an actual age mention or not. In the labelled dataset there are 1,186 \lq age\rq~ and 1,026 \lq other\rq~ sample.

The usual techniques for training a predictive classifier are Na\"ive Bayes, Support Vector Machines (SVM), Random Forest, and Multilayer Perceptron (MLP). We have excluded MLP as the dataset is small which could lead to overfit MLP models. Na\"ive Bayes, SVM and Random Forest classifiers are trained using the feature vectors of the above labelled dataset. The classifier implementations in WEKA~\cite{Hall2009} data mining toolkit are employed 
for this task. Each classifier is evaluated using 5-fold cross validation and Na\"ive Bayes, SVM and Random Forest result  averaged f-measures of 0.80, 0.83 and 0.85 respectively. Therefore, Random Forest is selected as the classifier for this task.

Classifier determines a confidence value ($c(t)$) for each text chunk to be an age mention and based on that text chunks that have higher confidence of being an age mention are selected for each author $A$. 
\begin{equation}
\{t_a\}^A = \{t_a\colon t \in \{t\}^A\textbf{ and }c(t) > \tau\}
\end{equation} 
\noindent Where $\tau$ is a confidence threshold which is set to be 0.7.

\subsubsection*{Aggregate and resolve age}\label{sec:resolve_age}
This step aims to resolve the age of each individual author from the selected high confidence age mention text chunks $\{t_a\}^A$ from the previous step. Note that the numerical value of age mentions might be different from the actual age value due to (i) age mentions of a past event, (ii) age mention of a different person, and (iii) misclassified text chunk. Hence, a weighted majority voting technique is employed to resolve the most probable age of each user $A$.     

First, the numerical value of each age mention ($D(t_a)$) is determined. Note that age mentions are parsed to get the age value normalised to year (e.g., $6\text{ months}  \rightarrow 0.5\text{ year}$). Also, two other parameters are derived from each age mention as follows: (i) tense of age mention (present, other), and (ii) subject type of the age mention (first-person, other).

Tense of the age mention is derived based on the tense of the verb in the text chunk. The text chunks with present tense verbs are marked as \textit{present}, and other tenses or if the tense is not available it is marked as \textit{other}). Subject type of the age mention is determined based on the subject in the chunk. The text chunks with first-person subjects are marked as \textit{first-person}, and other subject types or if the subject is not available it is marked as \textit{other}).  

The rationale of these two parameters is that if the chunk is in the present tense then it is more likely to be the current age of author. Also, if the subject is first-person then it is more likely to be about the author. Therefore, a weight boost of  0.25 is added if the age mention is present tense and a weight boost of 0.25 boost is added if it has a first-person subject. The aggregated confidence value $C(a)$ determined as follows:
\begin{equation}
C(a) = \sum^{\{t_a\}^A}I(a \leq D(t_a) \leq a + \Delta a) \times 2^{-(D(t_a) - a)} \times W(t_a)
\label{equ:age}
\end{equation}
where $I$ is the indicator function and the weight $W(t_a)$ is determined as $1 + 0.25 \times I(t_a\text{ present tense})+ 0.25 \times I(t_a\text{ first-person})$.

Equation~\ref{equ:age} is formulated in a way that an age value $D(t_a)$ contributes to the confidence of $a$ to $a+\Delta a$. This is because users often contributes to a particular OSG over a period of time. Hence, an age mention of a user can be different in older posts. For example, a user might mention age to be \lq56\rq~ two years ago and now mention that he is \lq58\rq~. In such cases, age mention of \lq56\rq~should contribute to the age value of 58. Note that $\Delta a$ is set to 2 to hypothesising that a user often contributes to OSG for a period of 2 years. 

The aggregated confidence value is determined for all the age values $\{a\}^A = \{D(t_a)\:\forall\:t_a \in \{t_a\}^A\}$ mentioned for the user $A$, and the age value of the user is resolved based on the highest aggregated confidence value:
\begin{equation}
a_{resolved} = arg\max\limits_{\{a\}^A}(C(a))
\end{equation}

\subsection{Gender extraction}\label{sec:gender_extraction}
Gender is another important demographic information that is important for personalised retrieval. Some symptoms can relate to different diagnoses depending on whether the patient is a male or a female. Gender is sometimes explicitly self-disclosed (e.g., \lq I\rq m a female\rq), and in some cases gender can be inferred based on gender mentions (e.g., \lq my mother\rq) or gender specific medical term mentions (e.g., \lq pregnant\rq). However, similar to age extraction, this task is challenging due to the unstructured and diverse nature of gender mentions in OSG posts.

\citet{Cheng2011} focused on the language style of the text to predict the gender of the author. The idea is that males and females often follow different language styles for written communication. However, the accuracy of this approach will be low as OSG attract information seekers and providers with very diverse language styles from across the world. Another approach is to predict the gender by looking at the first name of the author \cite{Herdagdelen2011}. It keeps two lists of male and female first names and resolves the gender based on that. This approach does not work in our scenario as many people do not use their real name in OSG, they often use a nickname or a part of their name. 

\citet{Zhu2012} look for gender clues: (i) gender specifying words such as \lq men\rq, \lq woman\rq, etc. and (ii) gender specific medical terms such as \lq hot flashes\rq, \lq prostate cancer\rq, etc. It was not mentioned that how they resolve the gender if terms related to both genders appear in posts of the patient. Also, another key issue is to resolve whether the gender clues are about that patient. 

\begin{figure}[!htb]
	\centering
	\includegraphics[clip=true, width=1.0\linewidth]{{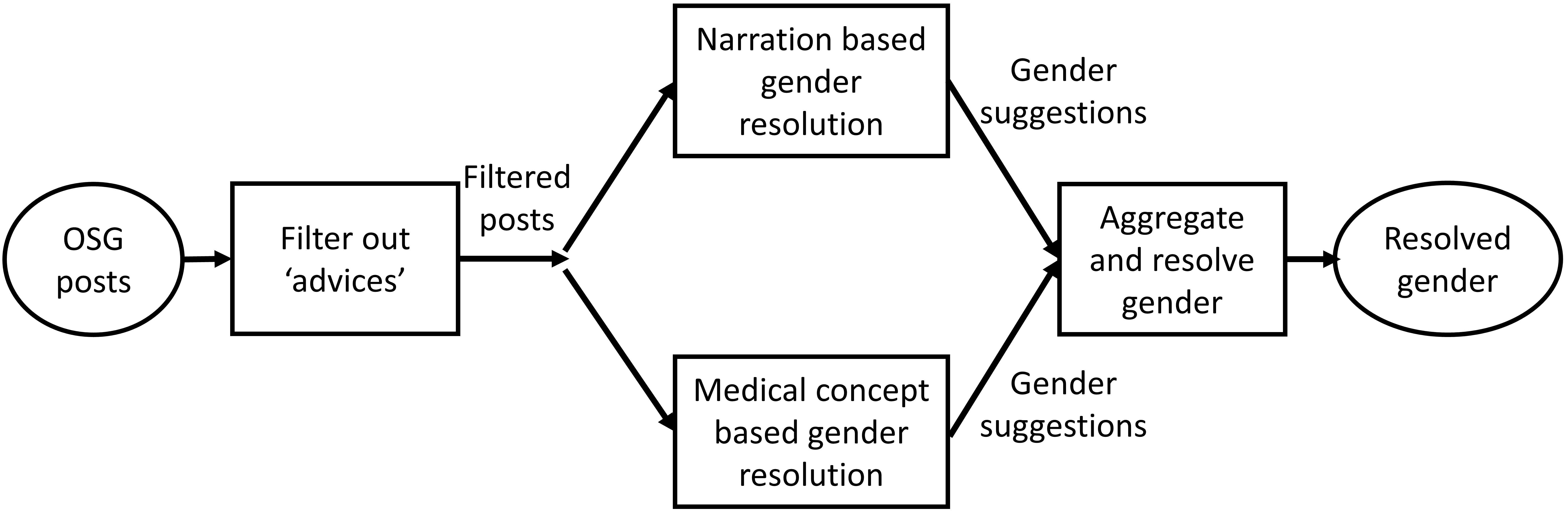}}
	\caption{Gender resolution process}
	\label{fig:gender_extraction_process}
\end{figure}

We extended the above approach of using gender clues to resolve gender, however, our technique is designed to be more robust to noisy gender cues, and also capable of fusing multiple gender cue to resolve the most probable gender of each author. Similar to~\cite{Zhu2012} this technique is two fold, (i) the first approach extracts gender specific narrations and processed them to resolve the gender, and (ii) the second approach looks for gender specific medical terms (body-parts, illnesses, symptoms). The results from both approaches are fused using a weighted majority voting method to resolve the highest confident gender of each author. This technique is presented in Figure~\ref{fig:gender_extraction_process} and described in detail below.  

As shown in Figure~\ref{fig:gender_extraction_process}, first the posts which are marked as advice by the narrative type identification are filtered out. This step is taken because advice posts often contain gender cues that are related to the receiver of the advice rather than the author of the post. Hence, gender cues in advice posts can be misleading as they often not related to the author of the post.

\subsubsection{Narration based gender resolution}
The gender is often explicitly or implicitly mentioned in the narration of OSG posts. However, gender specified in narrations can be ambiguousness since such terms may be referring to some other actor other than the patient. The proposed technique is developed to overcome such ambiguities.
 
Firstly, the narration type is identified based on the technique mentioned in Section~\ref{sec:narrative_type_extraction}. If the narration type is a \textit{first-person} narration (e.g., \lq I\rq, \lq me\rq), then the narrator is the actual patient. Therefore, the narrator gender is resolved and assigned to the patient. The gender of the narrator is resolved using two types of gender mentions found in OSG posts: 
\begin{enumerate}
	\item Direct gender mentions: directly reveals the gender of the narrator (e.g.,  \lq a male\rq, \lq a man\rq,  \lq a mother\rq,  \lq a widow\rq). Those mentions have to exist within close proximity of first-person pronoun i.e., \lq I\rq~in a sentence to verify that it is about the patient.
	\item Indirect gender mentions: narrator gender can be inferred from gender specific relationships. For example, if OSG posts talk about \lq my wife\rq~ or \lq my fiancée\rq~ that implies that the narrator is a male.  
\end{enumerate}

If the narration type is a \textit{second-person} narration, then the narrator is a caregiver (e.g., partner, parent, child, friend) who is posting on behalf of the patient. In such cases, the relationship of patient to the narrator can be used to resolve the gender most of the time. For example, patient is a male if ~\lq he\rq~ is husband, son, uncle etc. of the narrator; and female if ~\lq she\rq~ is daughter, wife, mother, etc. of the narrator. 

\begin{table}[!htb]
\caption{Few examples of narration based gender resolution process}
\label{table:gender_resolution}
\small
\begin{tabulary}{\linewidth}{|>{\raggedright}p{4.3cm}|>{\raggedright}p{5.5cm}|p{1.5cm}|>{\raggedright\arraybackslash}p{2cm}|}\hline 
Post & Pronoun resolved post & Narrative type & Inference of gender\\\hline
My doctor ... yesterday. He $\dots$ daily. I took it ... and ... I felt ... Is this normal ...? I was ... because ... I haven't ... and I was ... with my fiancée. I got paranoid about her ... even though I knew ... she ... & My doctor ... yesterday. \textless doctor\textgreater~ ... daily. I took it ... and ... I felt ... Is this normal ... ? I was ... because ... I haven't ... and I was ... my fiancée. I got paranoid about \textless fiancée\textgreater~ ... even though I knew ... \textless fiancée\textgreater~ ... & first-person & Gender cue: \textit{fiancée} \newline\newline Patient: \textbf{male} \\\hline
My mother is diagnosed ... She is suffering with ... I know that ... her heart. An endocrinologist suggested her to ... Also, he recommended her to ... But now she had ... She is suffering ... & My mother is diagnosed ... \textless mother\textgreater~ is suffering with ... I know ... \textless mother\textgreater~ heart. An endocrinologist suggested \textless mother\textgreater~ to .. . Also , \textless endocrinologist\textgreater~ recommended \textless mother\textgreater~ to ... But now \textless mother\textgreater~ had ... \textless mother\textgreater~ is suffering ... & second-person & Gender cue: \textit{mother}\newline\newline Patient: \textbf{female}\\\hline 
A friend ... when he .... He suddenly ... hurt his back. He fell down .... Despite ... him to his feet and made him walk to ... away. & A friend ... when \textless friend\textgreater~ .... \textless friend\textgreater~ suddenly ... hurt \textless friend\textgreater~ back. \textless friend\textgreater~ fell down .... Despite ... \textless friend\textgreater~ to \textless friend\textgreater~ feet and made \textless friend\textgreater~ walk to ... away. & second-person & Gender cue: he (resolved to friend)\newline\newline Patient: \textbf{male} \\\hline
\end{tabulary}
\end{table}

However, sometimes patient’s relationship to the narrator is gender-neutral (e.g., partner, friend) then the pronoun resolution process that is used to identify the narration type in Section~\ref{sec:narrative_type_extraction} is examined to check if gender specific pronouns were resolved to that noun (e.g., he $\rightarrow$ friend) and gender is assigned accordingly. Table~\ref{table:gender_resolution} shows three examples of the narration based gender resolution process.

\subsubsection{Medical concept based gender resolution}
There are medical concepts (e.g., body parts, symptoms, illnesses, procedures) that are often limited to a single gender (e.g., prostate cancer, pregnancy), which can be used to infer the gender of the patient. In this approach it is hypothesis that discussion of such a medical concept indicates that the patient is of that respective gender. For this approach it is immaterial to know the narration type, because the medical concept is about the patient, hence this approach is  relatively straightforward than the narration based approach.  

A list of seed medical terms (words or phrases) were constructed for both genders. These seeds lists were populated based on the gender related concept terms from UMLS Metathesaurus~\cite{Bodenreider2004}. However, patients sometimes use lay terms for those medical concepts, which are not included in clinical ontologies such as UMLS Metathesaurus. Therefore, a set of samples from men’s and women’s sections of OSG were examined to include such lay terms of medical concepts that are deterministic of gender. 

If the above formulated gender specific terms are present in OSG posts, relevant gender is assigned to the patient that is mentioned in that post. 

\subsubsection*{Aggregate and resolve gender}
Gender suggestions based on each post are aggregated to resolve the patient gender of each OSG user. The aggregation process is similar to the age resolution process specified in Section~\ref{sec:resolve_age}, where medical concept based gender suggestion is given a weight of 1.0 and narration based gender suggestion is given a weight of 2.0. Note that narration based gender suggestions are given a higher weight as they are relatively more accurate than medical concept based. Once aggregated, gender of the patient is resolved based on the highest weighted gender, which is then re-assigned to all the posts of that author.

\subsection{Medical concept extraction}\label{sec:medical_concept_extraction}
Medical concept extraction from text is a further formidable task. Natural language processing (NLP) tools are used to extract terms from text that can be mapped to the medical concepts found in medical thesauruses such as the UMLS Metathesaurus~\cite{Bodenreider2004}. There are several state-of-the-art tools that extracts medical concepts from text such as: MedLEE~\cite{Friedman1994}, cTAKES~\cite{Savova2010}, and MetaMap~\cite{Aronson2010overview}. 

\citet{Gupta2014}, shows that better precision and recall can be achieved by developing a tool to support special characteristics of patient-authored text. Consumer health vocabularies extracted from community generated corpora~\cite{Vydiswaran2014} are employed to identify the relevant medical terms and a different stack to NLP processes is followed to extract key phrases from the text. However, as noted by the authors, the existing system is limited to certain subcategories of the OSG (Asthma and ENT). Therefore, for our task we adhere to the well-established medical concept extraction tools.

We employed  cTAKES~\cite{Savova2010} tool for our medical concept extraction task. It identifies noun phrases in the text and then conducts a dictionary-look up in SNOMED CT  and RxNORM~\cite{Nelson2011} medical concept databases. Each identified term is then mapped into five semantic types: disorder/disease, sign/symptoms, procedures, anatomical sites, and medications. In the proposed method, we subject each OSG post to this process and the identified medical concepts are extracted. As pointed-out by~\citet{Gupta2014}, some ambiguities exist in the identified concepts. For example, ‘today’ is mapped to a drug named Today and ‘web’ is mapped to the disorder ‘congenital webbing‘. However, both these words are found frequently in OSG posts mostly referring to their usual meanings. To overcome this issue, we constructed a list of terms that are often mapped incorrectly and filtered out the identified concepts based on such terms.

\subfile{Chapter5_Emotion_Extraction}


%% file: Chapter5_Emotion_Extraction.tex
\subsection{Emotion extraction}\label{sec:emotion_extraction}

As discussed in Section~\ref{sec:emotion_extraction_from_text} emotions are an important element of social text which self-disclose the mental state of the author. In the context of OSG discussions, it can be hypothesised that the emotions expressed in OSG discussions are indicative of the mental state of OSG users (patents/caregivers), which can be used to infer their emotional well-being. Moreover, since the OSG user posts at different points of time, emotional fluctuations over time is indicative of their emotional journey during their participation in OSG. 

Analysing emotions and emotion fluctuations over time of a OSG user is important for the medical researchers to understand the emotional well-being of patients/caregivers with a particular disease conditions, as well as to compare and contrast the emotional well-being of patient cohorts with different disease conditions or other clinical factors (e.g., undergone different treatments).
 
As discussed in Section~\ref{sec:emotion_models}, multiple psychological emotional models are proposed to represent the emotional state, which ranges from the two dimensional valence-arousal model~\cite{russell1980circumplex} to multi-dimensional models such as emotion wheel~\cite{plutchik1991emotions}. While such models serve as theoretical basis for emotion representation, computational implementations are required to capture the emotional expression from discourse. For example, sentiment analysis techniques are the computational implementation of the valence-arousal model~\cite{Mohammad2016b}, which provide a signed real-value as the sentiment score, where the sign (positive/negative) represents the valence and the absolute value of score represents arousal. 

Although sentiment analysis techniques are relatively mature and commonly used for capturing emotions, we found that the two dimensional model is too coarse grained for representing complex emotional states of OSG users. Therefore, we have developed a computational technique based on the Emotion Wheel~\cite{plutchik1980emotion,plutchik1991emotions} to capture a multi-dimensional representation of emotions from OSG discourse. 

As discussed in Section~\ref{sec:emotion_models} Emotion Wheel has eight primary emotions (joy, trust, surprise, sadness, disgust, anger, anticipation and fear) and further eight secondary emotions which are derived using combinations of primary emotions (e.g., love: joy+ trust). These 16 emotions (primary and secondary) specified in the Emotion Wheel is incorporated as the emotional dimensions in the proposed computational model. Note that several modifications are applied to secondary emotions to suit the healthcare domain. The emotional intensity of each emotion is determined based on the proportion of relevant emotional terms present in each OSG post resulting a 16-dimensional real-valued emotion vector for each OSG post.   

\begin{figure}[!htb]
	\centering
	\includegraphics[clip=true, width=1.0\linewidth]{{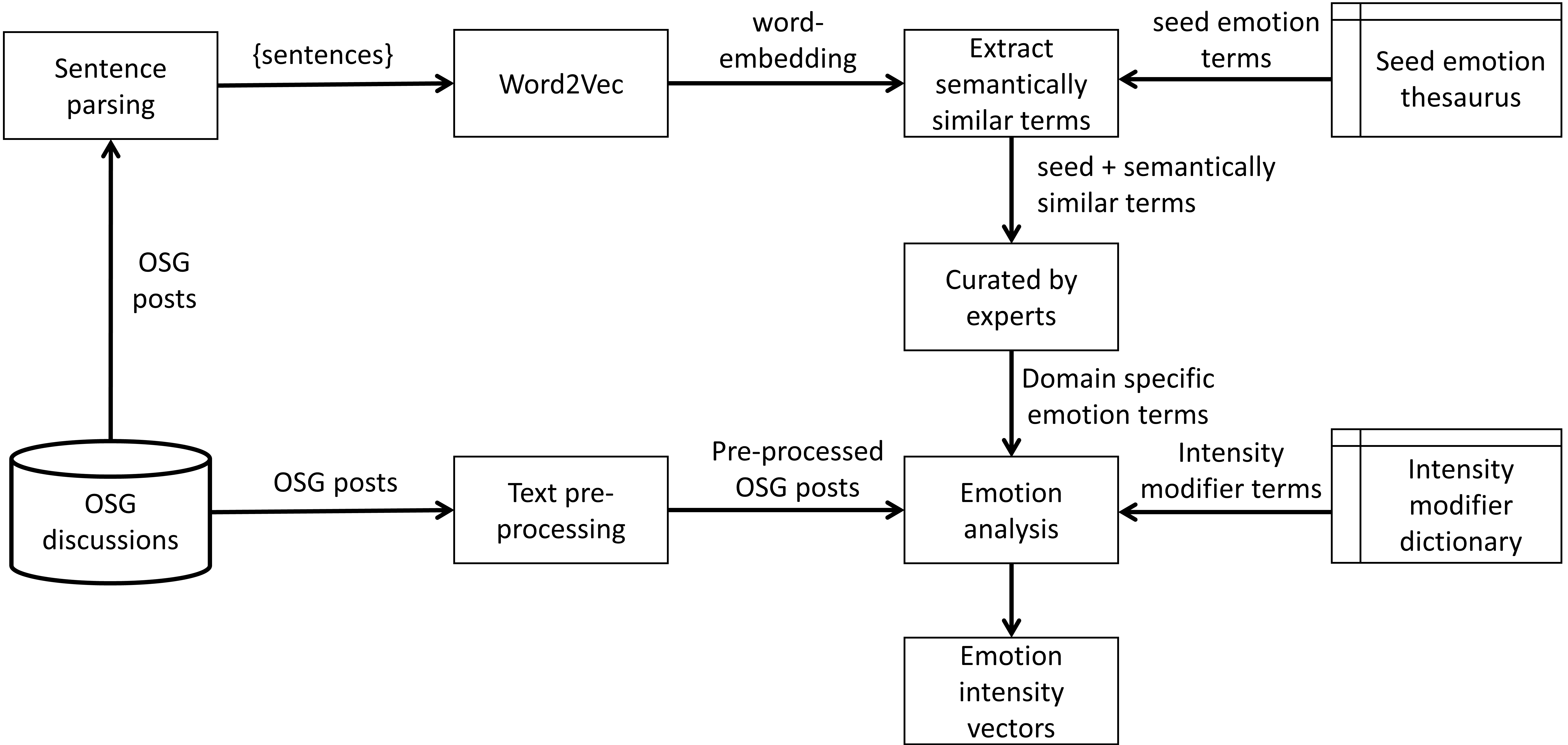}}
	\caption{The proposed technique for emotion extraction}
	\label{fig:emotion_extraction}
\end{figure}

Figure~\ref{fig:emotion_extraction} presents the proposed technique for emotion extraction. The relevant terms for each emotion is obtained using a two step process. 

First, a seed emotion term thesaurus is constructed for each emotion. We constructed this thesaurus based on a list of feeling words used for mental status exams~\cite{RichardNiolon}, which contains  emotional terms for each of the 16 emotions. In Table~\ref{table:emotion_terms} the second column provides the 16 emotions and the third column provides a sample of seed emotions terms employed for each category.

\begin{table}[!htb]
\caption{Emotion categories and a sample of representative terms used for each emotion.}
\label{table:emotion_terms}
	\begin{tabulary}{\linewidth}{|>{\raggedright}p{0.15cm}|>{\raggedright}p{1.6cm}|>{\raggedright}p{4.5cm}|>{\raggedright\arraybackslash}p{7.5cm}|}\hline
		& \textbf{Emotion} & \textbf{Emotional terms from thesaurus}& \textbf{Emotional terms from word-embedding}\\\hline
		\multirow{8}{*}{\rotatebox{90}{Positive}}
		& Happy & happy, great, joyous, glad, delighted & fab, chuffed, terrific, great news, looking forward, heart warming, uplifting, upbeat\\\cline{2-4}
		& Good & good, pleased, comfortable, relaxed, content& comfy, nice,  chill, chipper, ok, okay, clear headed, cool \\\cline{2-4}
		& Alive & alive, playful, energetic, spirited, animated& chatty, perky, sociable, vibrant, vivacious, witty, easy going, peppy\\\cline{2-4}
		& Love & love, attracted, warm, passionate, affectionate & romantic, cuddly, compassionate,intimate, adore, supportive, caring\\\cline{2-4}
		& Positive & positive, eager, keen, bold, brave& smart, ambitious, proactive, cynical, insistent, willing, upbeat\\\cline{2-4}
		& Open & open, understanding, accepting, satisfied, receptive & open minded, empathetic, cooperative, accommodating, approachable, forgiving, attuned, rational\\\cline{2-4}
		& Interested  & interested, fascinated, inquisitive, curious, intrigued&keen, impressed, cautious, leery, eager, intuitive, savvy, thoughtful\\\cline{2-4}
		& Strong  & strong, certain, dynamic, sure,  tenacious&resilient, independent, adamant, fierce, self reliant, decisive, fighter,  pragmatic\\\hline
		\multirow{8}{*}{\rotatebox{90}{Negative}} 
		& Sad & sad, tearful, grief, sorrowful, grief&heart break, teary, lonely, weepy, crying, despairing, hurtful\\\cline{2-4}
		& Afraid & afraid, fearful, terrified, panic, worry&petrified, freaking out, apprehensive, dread, obsess, fret, nervous wreck\\\cline{2-4}
		& Hurt & hurt, deprived, pained, dejected, agonised &traumatised, bruised, shattered, ached, exhausted, cramped, numb, fatigued, strained\\\cline{2-4}
		& Angry & angry, annoy, provoke, aggressive, enraged &agitated, hostile, pissed off, argumentative,  aggressive, rude, paranoid, ticked off, lashing out\\\cline{2-4}
		& Depressed & depressed, disappointed, miserable, despair, powerless &despondent, distraught, suicidal, unloved, worthless, emotionally drained, snappy\\\cline{2-4}
		& Helpless & helpless, incapable, alone, vulnerable, fatigued &insecure, tired, hopeless, powerless, defeated, overwhelmed, listless, incapacitated\\\cline{2-4}
		& Confused  & confused, upset, doubtful, uncertain, hesitant&unsure, perplexed, wary, leery freaked out, iffy, bummed, taken aback \\\cline{2-4}
		& Indifferent  &indifferent, insensitive, dull, reserved, lifeless &grumpy, apathetic, blunt, ignorant, emotionless, callous, crass, standoffish\\\hline
	\end{tabulary}
\end{table}

Expanding a seed list of lexicons is a labours task which is often achieved using crowdsourcing techniques such as Amazon Mechanical Turk~\cite{buhrmester2011amazon}. However, recent research report techniques~\cite{Hamilton2016,Fast2016} to expand seed term list using a semi-supervised approach based on the word-embedding~\cite{mikolov_word2vec} technique. Word-embeddings learn dense vector representations of words and phrases while automatically preserving the semantic relationships that exist in the text corpus by incorporating such relations into the vector space of the word-embedding. This enables the use of linear algebra to capture different semantic relationships within word-vectors in the word-embedding. The famous example by~\citet{Mikolov2013a} shows that the vector arithmetic of word vectors \lq King -Man + Queen\rq~ results a word vector similar to the word vector of \lq Woman\rq.

Developing such word-embedding using an OSG post corpus enables to capture terms used by the OSG users that are semantically similar to the seed emotional terms. We have developed a word-embedding from a large text corpus collected from two large OSG \textit{patient.info} and \textit{healingwell} mentioned in Table~\ref{table:osg_stats}, which contained a total of 4,795,428 OSG posts. This corpus is preprocessed to remove any URLs, converted  to lower case and then separated into sentences using the python NLTK sentence tokenizer~\footnote{http://www.nltk.org/api/nltk.tokenize.html}, which has resulted a text corpus of 36,222,536 sentences. This text corpus is employed to train a 200 dimensional word-embedding using Word2Vec technique with skip-gram model~\cite{mikolov_word2vec} and negative-sampling~\cite{Mnih2012}.  Note that common phrases are tagged up to three terms and used them as phrases during the training of the word-embedding. We employed the python Gensim~\cite{gensim} library for this implementation. The resulted word-embedding contains 312,196 unique terms (words or phrases).

Once the word-embedding is trained, most similar terms for each seed terms in the emotion thesaurus is identified using a nearest neighbour search in the embedding space using Cosine similarity. These identified terms are semantically similar terms to the seed emotion terms, in which some of the terms have the same emotional sense of the seed term while some others may not. For example, the top five nearest neighbours of \textit{sorrowful} are \textit{sadness, sincerity, joyful}, and \textit{deeply saddened}, in which \textit{joyful} is semantically similar but has the opposite emotional sense. Therefore, we manually looked at the selected terms and filtered out the terms that do not align with the emotional sense of the respective seed term. The third column of Table~\ref{table:emotion_terms} presents a sample of emotional terms captured using the above technique. Note that, this emotions term identification is a one time task and the final emotion terms list is added to the emotion term thesaurus.

Intensity modifier terms are a set of terms that increase or decrease the intensity of the emotional term. For example, the term \lq very\rq~ increases the intensity of the emotion \lq good\rq~ when used together, whereas, the term \lq kind of\rq~ decreases the intensity of the emotion \lq okay\rq~ when used together. Moreover, some terms completely negate the emotions e.g., \lq not okay\rq~ negates the emotion expressed by \lq okay\rq.  A thesaurus of such terms is often used in rule based sentiment analysis tools such as SentiStrength~\cite{Thelwall2012} and VADER~\cite{Hutto2014} to improve the accuracy of the sentiment score. In this work, we have employed the intensity modifier term thesaurus used in VADER~\cite{Hutto2014}.

When assessing the emotional expression of an OSG post, first it is sent through a preprocessing step which converts text lower case and removes URLs, alphanumeric characters, and punctuations. Stop words are not removed as stop words may contribute to the emotional expression. It is then tokenised into words. The pre-processed text is then used for the emotion analysis, which identifies the relevant emotion terms for each emotion and any intensity modifier terms that are associated with the emotion terms. Algorithm 1 presents the pseudo-code for this emotions vector $E_P$ calculation of a particular OSG post $P$.

\begin{algorithm}[!htb]
\DontPrintSemicolon 
\KwIn{ $P$- OSG post, $\mathcal{T}_E$- Emotion thesaurus, $\mathcal{T}_m$- Intensity modifier thesaurus}
\KwOut{$E(P)$ - Emotion vector of OSG post $P$}
$E(P) \gets \emptyset$\; 
$P \gets Preprocess(P)$  \tcp*{remove URLs, numbers, punctuations} 
$\{w\}_P \gets Tokenize(P)$ \tcp*{tokenise the post into words}
\;
\For{$i \gets 1$ \textbf{to} $16$} {
	$\mathcal{T}_{Ei} \gets \mathcal{T}_E$ \;
	$e_i \gets 0$ \;
	$w_{prev} \gets null$ \;
	\ForEach{$w$ \textbf{in} $\{w\}_P$}{
		\tcp{lookup $w$ in emotion $i$ term list}
		\If{$w \in \mathcal{T}_{Ei}$}{
			$e_i \gets e_i + 1$ \;
			\;
			\tcp{lookup $w_{prev}$ in intensity modifier thesaurus}
			\If{$w_{prev} \neq null$ \textbf{and} $w_{prev} \in \mathcal{T}_m$}{
				$e_i \gets e_i + \mathcal{T}_m(w_{prev})$ 
			}
		}
		$w_{prev} \gets w$ 
	}
	$e_i \gets e_i \div \vert \{w\}_P \vert $ \;
	$E(P) \gets E(P) \cup e_i$\; 
}
\Return{$E(P)$}\;
\caption{Determine emotion intensity vector}
\label{algo:emotion_calc}
\end{algorithm}

%% file: Chapter5_Timeline_Extraction.tex
\section{OSG user timeline construction}\label{sec:timeline_construction}
This section describes the automatic construction of user timelines from their self-disclosed narratives in OSG posts. A timeline consists of time sensitive events organised in chronological order.  Apart from demographics which are not time sensitive, all other information extracted in the previous section can be employed to the construction of the user timeline. However, this work is limited to the timeline construction based on clinical events and emotional events (emotional trends) only.

A timeline of events helps to examine the user behaviour over time as well as temporal associations between multiple events. For example, such a timeline is an important resource for a health professional to identify the temporal ordering of clinical events over the disease progression of a patient. Such insights can be employed during prognosis as well as therapy planning~\cite{Augusto2005}. Also, the medical researchers can use such information to investigate disease progression of certain illnesses across different cohorts of the patients in the population. Having emotions alongside clinical events enables the medical researchers to analyse the associations between clinical events and emotions, and find insights on the emotional burden of a disease condition over time.

\subsection{Clinical event timeline}~\label{sec:clinical_event_timeline}
Clinical event is defined in literature as a clinically relevant symptom, state, perception, procedure or occurrence~\cite{Tao2010,Tao2010a}. Clinical event extraction is a specified event extraction task as the event related information such as relevant key terms is priorly known. As discussed in Section~\ref{sec:specified-event}, specified event detection uses an engineered feature dictionary or labelled datasets to train a classifier to identify documents/texts that contain relevant event information, and then capture the relevant named entities (such as event time) that are related to the identified event. 

Existing work on clinical event extraction is mostly developed for the clinical narratives in Electronic Health Records (EHR) and discharge summaries. There is a recent interest on such techniques due to the availability of several human annotated datasets such as the dataset released for i2ib challenge~\cite{Sun2013} and SemEval-2015 Task-6~\cite{Bethard2015}.~\citet{Tao2010a} developed CNTRO (Clinical Narrative Temporal Relation Ontology) which uses an ontology based approach to model time sensitive information in clinical narratives. For each identified event it extracts temporal information such as time and duration by parsing the time indicative terms/phrases (e.g., \lq2 week~\rq for duration of two weeks) using time ontologies~\cite{hobbs2006time}. 

These existing techniques are specifically designed for the narratives generated by health professionals, where such narratives often adhere to certain guidelines and professional practices. Therefore, vocabulary, reporting of clinical events and expression of temporarily is fairly uniform across those narratives. On the other hand, OSG content is user generated (patient/caregiver). As discussed in Section~\ref{sec:social_data_analysis_challenges} such user generated content is so diverse and contain different language patterns prevalent among different socio-geographic groups~\cite{Eisenstein2014}. Therefore, the existing techniques need to be extended to facilitate timeline extraction from OSG.

\citet{Wen2012} have developed such an extended technique to construct clinical timeline i.e., cancer trajectories from the OSG discussions in breast cancer support groups. They have engineered a consumer health vocabulary of breast cancer related keywords which can be phrases or certain abbreviations (e.g., diagnosed: dx, dx'd). From the OSG posts, the sentences with the keywords in the above developed vocabulary are extracted and further parsed to extract the temporal expression related to the time of that event. Time is inferred using both specific time expressions (e.g., 4th May 2018) and relative time expressions (e.g., a week ago). The relative time expressions are converted to a specific time with the support of the post date of the OSG post (e.g., \lq a week ago\rq \textrightarrow post date - 7 days). This approach is further extended by~\citet{naik2017extracting} with a semi-automated approach for engineering relevant keyword dictionaries and also have employed HEidleTime~\cite{Strotgen2010} for time resolution. 

We employed~\citet{Wen2012} approach for clinical event extraction from OSG. The medical concept extraction discussed in Section~\ref{sec:medical_concept_extraction} already extracts clinical concepts such as symptoms and procedures based on the terms found in medical thesauruses. However, as discussed in Section~\ref{sec:medical_concept_extraction} such thesauruses are optimised for the vocabulary in clinical discourse  used by health professionals, therefore often lack the terms used by general health consumers i.e., patient or caregiver. This issue highlights the need for a consumer health vocabulary to more effectively extract clinical concepts from consumer generated discourse such as OSG. 

Developing a generic consumer health vocabulary is a laborious task which is beyond the scope of this work. However, we have  developed a consumer health vocabulary for several key cancer related clinical concepts, in order to demonstrate its utility as well as for the further application of this work in cancer domain (presented in Chapter~\ref{chap:6}). We have selected five key clinical events of a cancer patient as follows:
\begin{enumerate}
	\item diagnosis
	\item medical tests (e.g., biopsies, other pathology tests)
	\item surgery
	\item radiotherapy
	\item recurrence
\end{enumerate}  
A consumer health vocabulary was engineered based on the terms (words/phrases) used by the users in OSG that are indicative of the above designated event types. In order to capture such terms from a very large text corpus, we employed the same approach used in Section~\ref{sec:emotion_extraction} for expanding emotion thesaurus, which is a semi-supervised approach based on the word-embedding~\cite{mikolov_word2vec} technique. Similar to the emotion term capturing, such word-embedding based on OSG post corpus enables to capture terms used by the OSG users that are semantically similar to the clinical terms that are indicative of the above designated event types.  

We employed the same word-embedding developed in Section~\ref{sec:emotion_extraction} and looked up semantically similar terms for a given term in the embedding space using Cosine similarity. For example, the ten most similar terms of \lq radiotherapy\rq~ ordered based on Cosine similarity score are \lq radiation\rq, \lq chemotherapy\rq, \lq hormone\_therapy\rq, \lq radio\_therapy\rq, \lq salvage\_radiation\rq, \lq external\_beam\_radiation\rq, \lq chemo\rq, \lq brachytherapy\rq, \lq external\_radiation\rq, \lq brachy\rq~ and \lq ebrt\rq. radiation and  radio therapy are synonyms while salvage radiation,  external beam radiation and brachytherapy are different variants of radiotherapy. EBRT is a abbreviation for external beam radiation therapy, while chemo and brachy are shortened forms for chemotherapy and brachytherapy respectively. 

The inclusion of the phrase \lq hormone therapy\rq~ as a similar term for ~\lq radiotherapy\rq~ is interesting as it is neither a synonym nor a variant, but just used in a semantically similar manner in OSG discussions. This inclusion highlights the fact that the candidate terms selected using word-embeddings need to further validated for their meaning by an expert. Therefore, we employed the support of a cancer surgeon to further curate the automatically generated term sets of each event category. Table~\ref{table:event_terms} shows ten term samples of this curated consumer health vocabulary for each event category delineated above. 

\begin{table}[!htb]
\caption{A sample of representative terms of each clinical event type. Note that some of the terms are common misspelled words.}
\label{table:event_terms}
\centering
\begin{tabulary}{\linewidth}{|>{\raggedright}p{2.5cm}|>{\raggedright\arraybackslash}p{11.6cm}|}\hline 
Clinical event & Representative terms\\\hline
Diagnosis & diagnosed, dxed, diagonsed, diagnosedwith, diag, dx, clinically diagnosed, officially diagnosed\\\hline
Medical tests & biopsy, biopsies, colonscopy, biop, fna, endoscopy, scope, mammogram, mpmri, cystoscopy\\\hline
Surgery & surgery, operation, op, sugery, surgury, opp, corrective surgery, ops, surgical procedure, opperation, keyhole surgery, bunion surgery\\\hline
Radiotherapy & radiotherapy, radiation, chemotherapy, radio therapy, salvage radiation, external beam radiation, chemo, brachytherapy, external radiation, brachy\\\hline
Recurrence & recurrence, reoccurrence, recurrance, bcr, biochemical recurrence, biochemical failure, local recurrence, biochemical relapse,  secondary cancers\\\hline
\end{tabulary}
\end{table}

The identified event sentences are then further processed to identify any temporal expression mentioned in the free-text form. We employed SUTime~\cite{Chang2012} to detect such temporal expressions. SUTime is a rule based classifier which contains a set of hand-crafted rules and dictionary of time sensitive terms.
It first identifies the time related terms and then extend them to chunks and apply the rules to resolve the time. Note that we extended the time sensitive term list of SUTime by including some of the relevant terms identified based on the above created word-embedding. For example, \lq weeks\rq~ is often referred to using terms such as \lq wks\rq, \lq wk\rq, and \lq months\rq~ as \lq mnths\rq. Such additions improve the recall of temporal expression. SUTime resolves both specific time expressions and relative time expressions (e.g., \lq 2 wks ago\rq~ \textrightarrow \lq OFFSET P-2W\rq). The relative time expression is then resolved based on the post date of the OSG post.  

OSG users, often mention the same event multiple times in different OSG posts. Such duplication can be resolved with slightly different specific times mainly due to the differences in the granularity of the temporal expression. In such instances, mentions of same clinical event are aggregated and its representative time is resolved using the median time of the respective cluster.

\subsection{Emotion timeline}~\label{sec:emotion_timeline}

The emotions timeline is an emotional representation of the state of mind of a user based on his discourse over time. Unlike clinical events, emotions are better represented as a continues flow over time i.e., an emotional trajectory of a user. 

For the construction of the emotion timeline, initially, the OSG posts of a user are sampled into fixed-time bins of time $\Delta t$ (e.g., week, month etc.). The emotions of the OSG posts in a single time-bin is aggregated to determine the emotional expression over that time period. We hypothesis that the approach of aggregating emotional expressions over a time-period would provide a more unbiased account of the emotional state of that user over that time, as the averaging smoothed-out frequent fluctuations in the emotions but retain macro emotional trends that correlate with the clinical events.     

The emotions are captured as a 16-dimensional vector for each OSG post based on the technique introduced in Section~\ref{sec:emotion_extraction}. The emotional representation of user $U$ for the time period $t - t + \Delta t$ is obtained by averaging the emotional vectors of the OSG posts that has the time stamp within that time period.

%% file: Chapter5_Evaluation.tex
\section{Evaluation of the information extraction techniques}\label{sec:evaluation}
This section evaluates the three knowledge extraction modules described in Section~\ref{sec:information_extraction}: (i) narrative type classification, (ii) age extraction, and (iii) gender extraction. We obtained the services of qualified domain experts for manual classification of test datasets. Narrative type classification is evaluated using a labelled set of posts as advice or experience. Age and gender resolution is evaluated using a labelled set of OSG post authors using their published posts.

\subsection{Narrative type classification performance} 
Narrative type classification is evaluated using 500 posts labelled by domain experts as experience or advice. Note that we ignored the sub-classification of experience (first person or second person) for this evaluation, because second person experiences are relatively rare in the dataset. It is evaluated as a classification problem where the two classes are Experience and Advice. Table~\ref{table:narrative_type_evaluation} presents the evaluation results.

\begin{table}[!htb]
\caption{Evaluation results of the narrative type extraction technique}
\label{table:narrative_type_evaluation}
	\centering
	\begin{tabulary}{\linewidth}{|L|R|R|R|}\hline 
		Label & Number of posts & Precision & Recall\\\hline
		Experience & 329 & 0.92 & 0.96\\\hline
		Advice  & 171 & 0.91 & 0.81\\\hline
		Combined & 500 & 0.92 & 0.89\\\hline
	\end{tabulary}
\end{table}

The results show that both \textit{experience} and \textit{advice} are identified with a precision above 0.9. Recall of \textit{advice} is relatively low mainly because some advising posts are mixed with the authors experience and therefore hard to identify them as advice. 

\subsection{Age and gender resolution performance}
Age and gender resolution was evaluated using a set of 300 labelled author profiles. Posts of each author were examined to identify the age or gender of the author if such information is present. Each author profile is annotated based on the identified age and gender. 

The labelled data is then compared to the output of age and gender resolution modules. We employed precision and recall statistics for this evaluation. Note that age is often mentioned in incremental values for some authors as a result of prolonged contribution to the OSG over several years. Therefore, age resolution is considered correct if it falls within two integer values of the labelled age. 

Similar to the previous evaluation, performances of the gender and resolution modules are evaluated as classification problems. For gender, the classes are \textit{Female}, \textit{Male}, and \textit{Unknown} and for age, classes are \textit{Age mentioned} and \textit{Age not mentioned}. Note that, in \textit{Age mentioned} class the classifier has to correctly resolve the age value (within two integer values to the labelled age value) in order to be a true positive.

\noindent Table~\ref{table:age_evaluation} and Table~\ref{table:gender_evaluation} present the age and gender classification results respectively.

\begin{table}[!htb]
\caption{Evaluation results of the age extraction technique}
\label{table:age_evaluation}
\centering
	\begin{tabulary}{\linewidth}{|L|R|R|R|}\hline 
		Label & Number of profiles & Precision & Recall\\\hline
		Age mentioned & 131 & 0.78 & 0.89\\\hline
		Age not mentioned  & 169 & 0.95 & 0.84\\\hline
		Combined & 300 & 0.86 & 0.87\\\hline
	\end{tabulary}
\end{table}

\begin{table}[!htb]
\caption{Evaluation results of the gender extraction technique}
\label{table:gender_evaluation}
	\centering
	\begin{tabulary}{\linewidth}{|L|R|R|R|}\hline 
		Label & Number of profiles & Precision & Recall\\\hline
		Female & 109 & 0.91 & 0.87\\\hline
		Male & 35 & 0.90 & 0.77\\\hline
		Unknown & 	156 & 0.87 & 0.94\\\hline
		Combined & 300 & 0.90 & 0.86\\\hline
	\end{tabulary}
\end{table}
Both age and gender resolution have average precision and recall over 0.85. Precision in \textit{Age mentioned} class is relatively low because when a profile does not have actual age mentions, the classifier tends to pick up low confidence age mentions that are often age mentions in past incidents or age mentions about other people. The same issue resulted a relatively low recall in \textit{Age not mentioned} class as well. 

Recall for \textit{Male} is relatively low. Most of those misses are classified as \textit{Unknown} as the classifier misses the gender specific clues. This is mainly because males tend to expose very few clues about their gender compared to females.

%% file: Chapter5_Demonstration.tex
\section{Demonstration}\label{sec:demo}
This section demonstrates the capabilities of the proposed platform in addressing the information needs of patients (end-users) and researchers. Patient information needs are based on finding similar cases that are more relevant to them (e.g., patients with similar conditions). Researchers aim to extract aggregated insights on medical conditions based on different dimensions such as age, gender and time. Two use cases are presented in the following subsections, one each for a patient and a researcher, to demonstrate how their information needs can be addressed using knowledge extraction layer generated by the proposed method.

The rest of this section is organised as follows. First subsection describes the test dataset collected from an active OSG and then present the implementation of the proposed knowledge extraction layer. Next subsection delineates the implementation of personalised search with a patient use case. Finally, the last subsection demonstrates OSG analytic capabilities for the researcher use case.  

\subsection{OSG data collection}
The popular OSG \textit{patient.info}\footnote{http://patient.info/forums} and \textit{healingwell}\footnote{https://www.healingwell.com/community/default.aspx} mentioned in Table~\ref{table:osg_stats} were selected for data collection because of their high volume of posts and large number of participants. 
 
A web scraping tool is developed to automatically traverse through each topic and collect all the threads. From each thread, title, first post and subsequent reply posts are collected. Each OSG post is collected with its time-stamp and author id. As shown in Figure~\ref{fig:ir_model}, the posts are then aggregated by the author id. We have filtered-out the users who have less than five posts in the OSG, because the information exposed by such users are often not sufficient to understand their demographic and clinical factors. The final dataset contains 4,469,107 posts from 79,829 OSG users, where 1,662,312 posts (47,712 users) are from \textit{patient.info} and 2,806,795 posts (32,117 users) from \textit{healingwell}.

The collected OSG posts are stored as JSON documents using a setup of the open source search platform Elasticsearch\footnote{http://www.elastic.co/products/elasticsearch}. Elasticsearch search is employed both as the document store and search platform  since it handles both full-text and structured search. Elasticsearch is a distributable full-text search engine, designed to be scalable to handle very large datasets~\cite{Kononenko2014}. 

\subsection{Information extraction}

The information extraction techniques developed in Section~\ref{sec:information_extraction} are employed to enrich the extracted OSG posts.  
 
The narrative type of each post is identified individually for each post, where each OSG post is categorised as \textit{experience: first person}, \textit{experience: second person}, and \textit{advice}.

The age and gender are resolved for each author. All the posts of an author are aggregated using the associated author-id of each post. These aggregated posts are then processed to resolve age and gender of each author. 

\begin{table}[!htb]
\caption{Counts and percentages of extracted age, gender and narrative type}
\label{table:OSG_stats_authors}
	\centering
	\begin{tabulary}{\linewidth}{|L|R|R|}\hline 
		& Users (percentage) & OSG posts (percentage)\\\hline
		Total & 79,829 (100) & 4,469,107 (100)\\\hline
		Age resolved & 48,126 (60.3) & 4,045,781 (90.6)\\
		$<$ 20 & 10,380 (13.0) & 824,405 (18.5)\\
		21-30 & 10,303 (12.9) & 664,950 (14.9)\\
		31-40 & 6,849 (8.6) & 534,157 (12.0)\\
		41-50 & 6,939 (8.7) & 569,000 (12.7)\\
		51-60 & 6,581 (8.2) & 732,864 (16.4)\\
		61-70 & 4,536 (5.7) & 519,317 (11.6)\\
		$>$ 70 & 2,538 (3.2) & 201088 (4.5)\\\hline
		Gender resolved	& 44,700 (56.0) & 3,963,523 (88.7)\\
		female & 32,268 (40.4) & 2,943,776 (65.9)\\
		male & 12,432 (15.6) & 1019747 (22.8)\\\hline
		Age and gender resolved & 33501 (42.0) & 3770618 (84.4)\\\hline
	\end{tabulary}
\end{table}

Table~\ref{table:OSG_stats_authors} provides a summary of the demographics identified. The proposed demographic extraction modules managed to resolve age and gender of 42\% users which comprises of 84.4\% of the posts in the collected dataset. It is a 12\% increase compared to the reported 30\% success by~\citet{Cho2007}. 

Among the gender resolved users, 72.2\% are females whereas only 27.8\% are males. The gender resolved OSG posts are further skewed, where among the gender resolved posts 74.3\% from females and 25.7\% from males.  We have further investigated this in the research literature.~\citet{Kummervold2002} report that women are more participative in OSG than men.~\citet{Li2015} state that women tend to self-disclose more information than men. Therefore, we assume there is less male participation in OSG and also even the participating males tend to expose less information to identify their gender. 

\subsection{Personalised search to extract relevant information}\label{sec:ir_search_impl}
One of the key driver to participate in OSG is the ability to engage with individuals who has similar conditions~\cite{Tanis2008}. In fact,~\citet{Fox2011} reports that in 2011, 20\% of adult internet users have used internet to find individuals with similar health concerns. 
 
As discussed in Section~\ref{sec:proposed_framework}, stage~\rom{5} of the proposed framework provides personalised (relevant and reliable) information in response to patient search queries based on their medical and demographic information.

We first employed the medical information extracted from the query to identify the posts that are experiences and contain matching medical information. Such posts are then ranked based on the demographic similarity of the author to the demographics of the patient. For this ranking, we use a custom relevance measure based on age and gender.

Let age and gender of an information seeking patient  $P^i$ be $P^i_a$  and $P^i_g$  respectively. Using the same notation, let age and gender of an existing patient $P^e$  be $P^e_a$  and $P^e_g$. The relevance measure $r_E$  is defined as follows:

\begin{equation}
r_E = W_a \times \exp(\frac{-\vert P^i_a -P^e_a \vert}{\sigma^2_a}) + W_g \times \text{I}(P^i_g = P^e_g) 
\end{equation}

\noindent where $W_a$ and $W_g$ are the weights for age and gender respectively. I is the indicator function which is 1 if $P^i_g$ equals $P^e_g$ and 0 otherwise. Weight of age is associated with a Gaussian decay function which is 1 if $P^i_a$ equals $P^e_a$. $\sigma^2_a$ is used to control the granularity of age matching where smaller values make the decay function to decrease rapidly with the age difference and vice versa. $W_a$ and $W_g$ are set to 0 if the patient does not provide their demographic details. The OSG posts are ranked based on this relevance measure and presented to the patient.

In this patient use case, we demonstrate how personalised retrieval can provide relevant and reliable information to the patient compared to the existing full-text search. We use the same query: \enquote{I’m a 40 year old woman taking Nexium for heartburn}. Contextual information is initially extracted from the query and used to identify medical information (symptom: heartburn, medication: Nexium) and demographic information (age: 40, gender: female). Most relevant experiences are retrieved from the database using the abovementioned method.
 
For comparison of the results we employed two approaches of full-text search: (i) default search in the OSG, and (ii) search key terms with the Boolean aggregation ‘AND’ (retrieves the posts that contain all the search terms). 
Excerpts from the top five results from our method and the two approaches of the full-text search are provided in Table~\ref{table:patient_search}. The key terms that are relevant to the query is highlighted in each excerpt. Note that, some experiences retrieved by the proposed method does not have age or gender mentioned in that post. This is because age and gender is resolved for each author using all posts by that author, so age or gender of that author is inferred from other posts and not the retrieved post.    

\begin{table}[!htb]
\caption{Excerpts from the top five results obtained using the three search approaches (including the proposed method) to retrieve similar experiences for the query \enquote{I'm a 40 year old woman taking Nexium for heartburn}}\label{table:patient_search}
\centering
\small
\begin{tabulary}{\linewidth}{|>{\raggedright}p{3cm}|>{\raggedright\arraybackslash}p{12cm}|}\hline 
Querying method & Excerpts from top five results \\\hline
The proposed structured search\newline 
Breaks the query into the following  structure:\newline
{symptom:heartburn medication: Nexium\newline age: 40 gender: female} & (\textit{female,40}): How I cured my gastritis ... side effect of fish oil is \textbf{heartburn}... my doctor said I could try \textbf{Nexium} as well I decided not to... \textbf{My husband} made gluten-free banana bread... \textbf{I am 40}...\newline
(\textit{female,40}): I had a very severe attack during a 24 hour ph probe test. I used to have \textbf{heartburn}... I am on nexium, ranitadine and donperidone... \textbf{I too am 40 years old} but I feel 80 ...\newline
(female,43): My ...it's been in a long time and the heartburn seems to be easing off... inhibitors are medication for reducing acid in your tummy like nexium and zoton... I'm only 43 and feel my life is...\newline
(female,50): I'm a 50yo Aussie female with Barrett's... I still have heartburn if I eat/drink the wrong things or forget to take my medication for a few hours... I was on Nexium 40 forever until I saw a different GP...\newline
(female,30): i am bloated and get heartburn all the time...iam 30yrs old wiv 4kids...take them today along with nexium. Nexium in my opinion is of no use at all...\\\hline
Full-text search using \lq heartburn Nexium woman 40\rq~\newline (searched in the actual forum search of patient.info website) & Nexium and side effect, anyone here while taking Nexium suffer diarrhea...\newline
I was taking another PPI tablet for heartburn for about 2 years (Nexium)\newline
Constant burping ...no pain or heartburn. I have been taking nexium for the past two weeks to relieve constant burping...\newline
just wondering if ... if you have been on Nexium for a long time? Women would be the ones who might find their iron levels low...\newline
I am a 40 year old woman with no notable health problems aside from the nephrotic syndrome. At 10 years old I was diagnosed with primary focal segmental glomerulosclerosis\\\hline
Full-text search using \lq heartburn AND Nexium AND woman AND 40\rq \newline
(searched within the collected posts using Elasticsearch) & the meds slowly over two years improved symtoms, during the symptom phase l started with different pain, chronic heartburn, leading to gall bladder removal...bought nexium as l’d read theyre same as omp, they are expensive... they use 40-80gm.... but tomorrow will try cabbage juice, just told woman on cfs site who has chronic nausea to try it \\\hline
\end{tabulary}
\end{table}

Above results show that the proposed method retrieves more relevant posts for the given query. Instead of taking key terms of the query as-is, the proposed method identifies the patient is a female and her age is 40. Therefore, it retrieves similar experiences from females who are aged close to 40. Also, the posts do not necessarily need to have age and gender mentions in the post itself as they were resolved for each author. 

In comparison, the second query (full-text search in the OSG) attempts a direct string matching with the search terms and retrieves partially matched results that contain any (unknown) combination of search terms. It is apparent that the last match of this query is irrelevant, because it is only matching ‘woman’ and ‘40’ but does not have the symptom ‘heartburn’ or the medication \lq Nexium\rq. 
The third query has a very low recall with only one retrieved post, as it is rare to have all four terms in a single post. Also, it is clearly noticeable that the match term \lq 40\rq~is not an age mention.

\subsection{OSG analytics for researchers}
Medical research is often conducted using small samples of patients due to the associated cost (both time and money) of such research. On the other hand, such information is accumulated in OSG, crowd sourced by real patients. These untapped resources are inaccessible to researchers due to inherent noise, unstructured nature and diversity of information representation. Researchers have to attempt the formidable task of executing full-text queries and manually extract information from the resulting posts.

As discussed in Section~\ref{sec:proposed_framework}, stages~\rom{3} and~\rom{4} of the proposed framework builds a structured layer on top of the unstructured text of OSG posts which can be utilised for OSG analytics. As shown in stage~\rom{3} of Figure~\ref{fig:ir_model}, each OSG user can be represented using demographic, clinical and emotional dimensions that enables researchers to conduct OSG analytics and gain insights. It provides unprecedented access to OSG data from different viewpoints.

In order to demonstrate the OSG analytics capability, we performed several analyses on patients who report the symptom heartburn. Note that this attempt is solely to showcase the analytical capabilities and not a comprehensive medical research on heartburn.

\begin{figure}[!htb]
	\centering
	\includegraphics[clip=true, width=0.7\linewidth]{{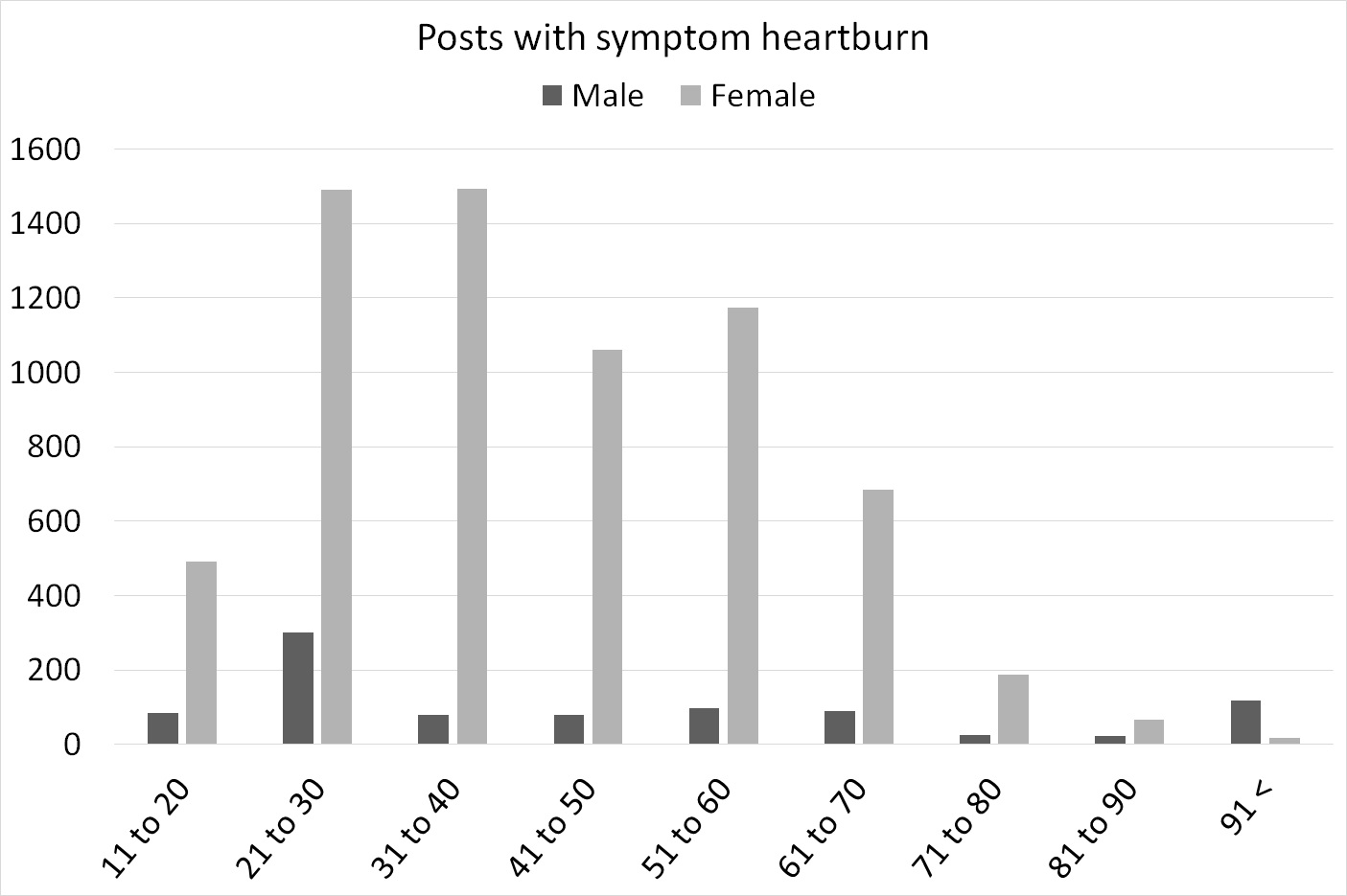}}
	\caption{Demographic distribution of the posts that mention symptom \lq heartburn\rq}
	\label{fig:researcher_use_case_1}
\end{figure}

\textbf{Dimensional analysis:} In this analysis, we combine age and gender dimensions and present demographic distribution of patients who report the symptom heartburn. Figure~\ref{fig:researcher_use_case_1} shows the demographic distribution of posts that mention the symptom heartburn. This type of analysis is useful to identify the age groups that are more affected by a particular symptom and also observe potential demographic biases.

\textbf{Association mining:} The OSG analytics layer is also useful for association mining. It can be used to analyse associations between different symptoms in order to identify co-existing symptoms. Table~\ref{table:researcher_use_case_2} presents the top five other symptoms that co-exists with the symptom heartburn in different age groups.

\begin{table}[!htb]
\caption{Top five symptoms co-exists with the symptom 'heartburn' in different age-groups}\label{table:researcher_use_case_2}
\centering
\begin{tabulary}{\linewidth}{|>{\raggedright}p{3cm}|>{\raggedright\arraybackslash}p{12cm}|}\hline 
Age group  & Top five symptoms co-exist with \lq heartburn\rq \\\hline
$<$ 20 &	less sleep, depressed, tiredness, anxiety, stress, living alone\\\hline
21 to 40 & reflux, anxiety, less sleep, nausea, stress\\\hline
41 to 60 & anxiety, stress, depression, reflux, indigestion\\\hline
61 to 80 & reflux, indigestion, anxiety, constipation, less sleep\\\hline
\end{tabulary}
\end{table}

\begin{figure}[!htb]
\centering
\includegraphics[clip=true, width=1.0\linewidth]{{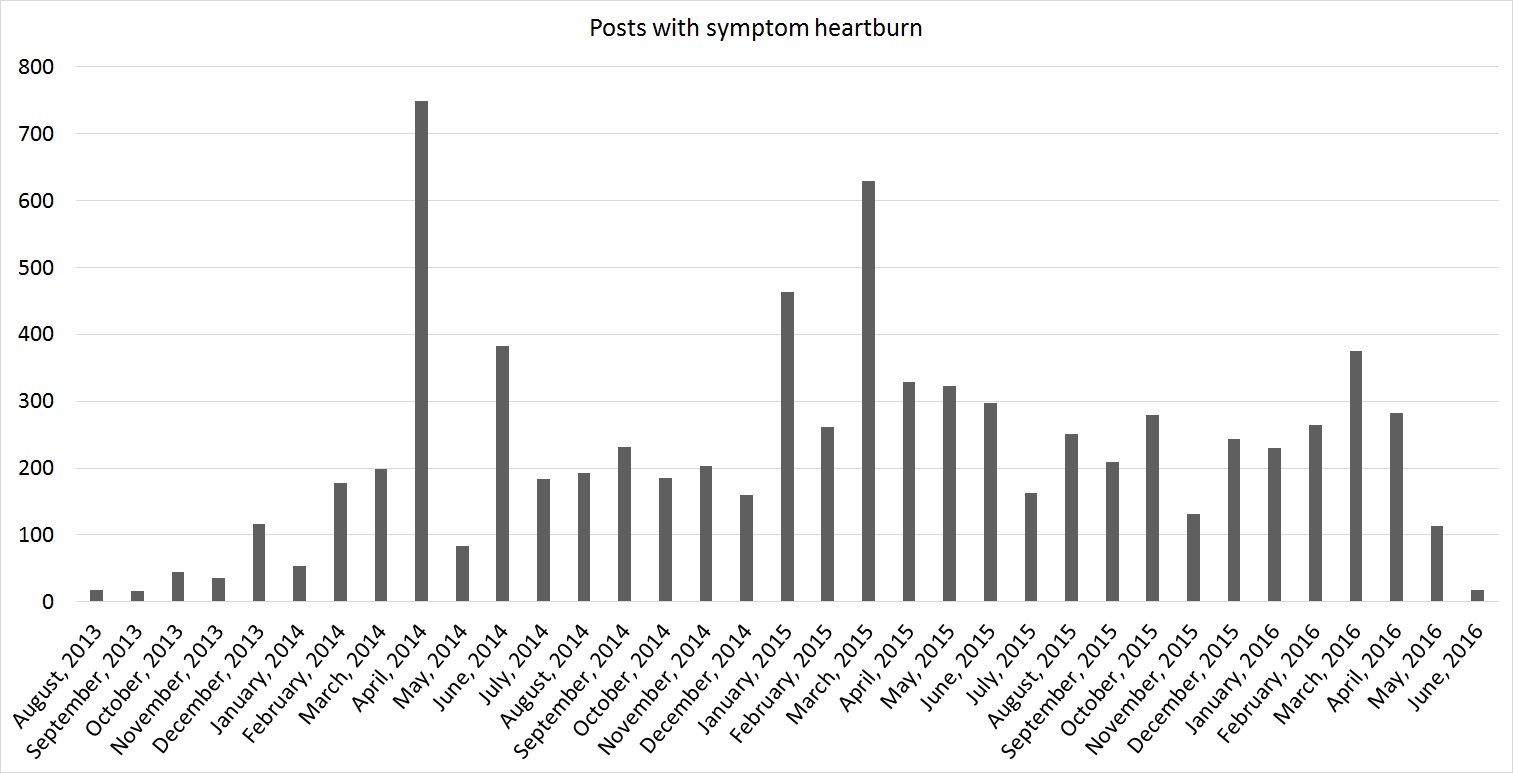}}
\caption{Temporal distribution of the posts that mention symptom \lq heartburn\rq~ over a three-year period }
\label{fig:researcher_use_case_3}
\end{figure}

\textbf{Temporal analysis:} The Date dimension can be used to perform temporal analysis to identify seasonal patterns in the OSG. Figure~\ref{fig:researcher_use_case_3} shows the temporal distribution of the post counts that report the symptom heartburn drawn for each month for a period of three years. It shows that over the three-year period reported heartburns are relatively high during March and April. 

%% file: Chapter6.tex
\onlyinsubfile{	
\setcounter{chapter}{5} 
}

\chapter[]{A digital health platform}\label{chap:6}

\epigraph{{\textit{When you come to the end of your rope, tie a knot and hang on.\\}}{\hfill Franklin D. Roosevelt}}

This chapter employs the information extraction and structuring platform presented in the previous chapter to analyse the free-text discussions in Online Support Groups (OSG) related to prostate cancer and conduct a multi-dimensional exploration on the emotional expression of various groups and the underlying causalities for such expressions. This analysis informs the key groups that are in need of support as well as the aspects upon which such support needs to be delivered. As the number of cancer survivors is increasing globally within finite healthcare systems, this understanding is pivotal to optimise the care schemes and also to optimally provide care to vulnerable groups.  

The platform presented in Chapter~\ref{chap:5} is a platform designed for OSG on any health topic and it extracts demographic, clinical and emotional information. In this chapter, this platform is further extended to capture prostate cancer specific clinical and decision making related information. This extended platform is applied on a large corpus of prostate cancer related OSG discussions collected from ten active OSG and the collected insights are presented as two case studies where one explores the groups that are in need of more psychological support and the other investigates different decision making behaviours and decision factors. 

The rest of the chapter is organised as follows. Section~\ref{sec:cancer_and_cancer_care} briefs on cancer incidence, unmet needs of cancer survivorship and OSG as a potential resource to understand such needs. Section~\ref{sec:why_prostate_cancer} justifies the selection of prostate cancer for the analysis due to its higher survivorship and unique characteristics related to treatment choices and side-effects. Section~\ref{sec:prime} presents the extended platform to analyse prostate cancer OSG discussions.  Section~\ref{sec:prostate_data} discusses the prostate cancer OSG data collection. Sections~\ref{sec:prostate_cancer_casestdy1} and~\ref{sec:prostate_cancer_casestdy2} presents the two case studies on the emotional expression of different groups and treatment decision making. Section~\ref{sec:chap6_discussion} concludes the chapter with a discussion.   

\section{Cancer, cancer burden and cancer care}\label{sec:cancer_and_cancer_care}
Cancer is a diverse group of diseases in which cells in some part of the body begin to multiply abnormally. In contrast to normal cells, cancer cells do not die when they should and new cells are produced when they are not needed, thereby developing a lump of cancer tissues i.e., tumour. This abnormal and out of control growth  of cancer cells damage nearby tissue and also reduce the supply of nutrients to nearby normal cells. Such damage hinders the functionality of the body part with the abnormal growth and may fail to function eventually causing death. Also, cancer cells are \textit{malignant} which means that they can spread into other body parts using the blood or  lymph systems~\footnote{https://www.cancer.gov/about-cancer/understanding/what-is-cancer}.

According to Global Cancer Observatory, in 2018 there are an estimated 18.1 million new cancer cases diagnosed and 9.6 million cancer related deaths worldwide~\cite{bray2018global}. In US, American Cancer Society estimated that there are 1.7 million new cancer cases diagnosed and 0.6 million cancer related deaths in 2018~\cite{american2018cancer,Siegel2018}. Cancer incident rates are rising as the growth of population and increase of life expectancy, as well as  changes of lifestyles (e.g., obesity, lack of exercises) in developing countries due to socioeconomic development~\cite{bray2018global}.  
 
Although cancer incident rates increases, cancer related deaths are decreasing due to the advances of treatment options as well as the advances in screening and early detection programs~\cite{bray2018global}. Therefore, \textit{cancer} which was once uniformly fatal in a short period of time has been transformed into a diverse set of diseases with different survival rates. Long-term survival is possible in the majority of major cancers types when diagnosed earlier. In US alone by 2018 it is estimated that there are 15.5 million cancer survivors, which is expected to rise to 20.3 million (10 million males and 10.3 million females) in 2026~\cite{desantis2016cancer}.

Besides for the cancer patient/survivor; partners, family and friends often provide the necessary care   
during the primary or adjuvant treatments as well as during  the longterm management of cancer as a chronic disease. It is reported that in 2015, at least 2.8 million Americans provided care to a cancer patient~\cite{national2016cancer}.  

\subsection{Unmet needs of cancer}\label{sec:unmet_needs}
The transformation of most cancers from a fatal to chronic disease and the growing number of cancer survivors have become a key challenge to the modern health care systems as it struggles to understand and allocate resources to support the key needs of cancer patients and survivors. 

Cancer patients are increasingly involved in the treatment related decision making and are in need of better support on selecting treatment options. The cancer survivors, although being relived of the primary treatment, are unprepared to live with cancer and its treatment related chronic outcomes (e.g., changes to the body image, infertility, stroke, intimacy issues), and also often worry about cancer recurrence~\cite{national2005cancer,alfano2006recovery}. Moreover, cancer caregivers report that long-term caring of the loved one is physically, emotionally, and financially demanding~\cite{northouse2010interventions,adelman2014caregiver,kim2008family} often leading to high levels of stress and anxiety~\cite{alfano2006recovery} and in need of help to manage their physical and emotional issues~\cite{national2016cancer}.

The healthcare providers and institutions are progressively limited in their scope of reach and service, due to increased demand, financial constraints and resource limitations~\cite{aiken2002hospital,luxford2011promoting}. Therefore, it is important to understand the core needs properly, prioritise the allocation of resources based on the importance, and identify the groups that essentially in need of support so that the support can be optimally delivered with the available resources.     

\subsection{Understanding the psychological burden of cancer}
Although there have been so many studies that led to advances in cancer preventing, screening and treatment which treats cancer clinically, the studies on the social and psychological burden of cancer have been limited. This lack of  studies is one of the key reason for lack of resource allocation in the current health care system to provide the required support to overcome the social and psychological challenges of the cancer patients, survivors and caregivers. 

The studies on social and psychological burden of cancer are often carried out as randomised control trials and cohort studies where the social and psychological well-being is assessed using survey instruments either designed for general quality of life or specialised for certain moods such as depression and anxiety~\cite{groth2009handbook}. For example, some generic instruments are Rotterdam Symptom Check List (RSCL)~\cite{de1990measuring}, Functional Assessment of Cancer Therapy-General (FACT-G)~\cite{cella1993functional} and European Organization for Research Treatment in Cancer: Quality of Life Questionnaire (EORTC-QLQ-C30)~\cite{aaronson1993european}. Some depression and anxiety specific instruments are  The Distress Thermometer (DT)~\cite{roth1998rapid}, The Hospital Anxiety and Depression Scale (HADS)~\cite{zigmond1983hospital}, and The Beck Depression Inventory (BDI)~\cite{beck1996beck}.
 
These studies require the individuals (patients, survivors, caregivers) to self-report their social and psychological well-being time-to-time and correlate to their clinical outcomes from cancer and cancer treatments, thereby developing an understanding of the social and psychological issues at different time points of the cancer journey for different cohorts individuals. However, such long-term analysis is cumbersome, expensive and often suffered from high drop-out rates as the individuals are less keen to support such studies that stretch for a long period of time. Moreover, as discussed in Section~\ref{sec:why_study_social_data}, such studies are often subjected to issues in generalisability due to small sample sizes, and various recall biases. 

\subsection{OSG as a complementary approach to understand the psychological burden of cancer}
As discussed in Section~\ref{sec:motivations_OSG}, OSG are a form of a social media platform that provides anonymous comfortable virtual spaces for patients survivors and carers to share experiences, seek advice, express emotions and provide emotional support. Participants with similar experiences provide each other with informational and emotional support. 

As pointed-out in Section~\ref{sec:researchers_OSG}, OSG contain unsolicited accounts of self-reported first person experiences at different time points of the illness journey. Those self-reported expressions are scattered across the OSG as parts of different discussions. However, once aggregated they provide significant insights about the patient/caregiver and their illness journey over time. This resource is seen to be instrumental in addressing the limitations and challenges of the current approach to understanding the unmet issues of cancer. The free-text corpus in OSG discussions can be mined to uncover the needs and challenges of the cancer patients, survivors and caregivers based on the issues that they discuss at different stages of their cancer journey. The emotions expressed when discussing the issues can be employed as a proxy to the quality of life to understand the severity of the issues. The individuals can be categorised based on their demographics (age, gender), role (patient, caregiver), and other clinical factors (e.g., treatment, cancer aggressiveness). The issues expressed by different categories of individuals can be prioritised based on their emotional expression and employed in the decision making process to identify the pressing issues and the groups that essentially in need of support so that the support can be optimally delivered with the available resources. 

This approach requires the relevant information about the individuals to be extracted from the free-text OSG discussions, which is challenging due to the noisy and unstructured nature of OSG content. 

In order to validate this premise that OSG can be better utilised to gain insights about the diverse needs of cancer care among the patients, survivors and caregivers, a use case of prostate cancer has been selected. The next section provides an overview of prostate cancer and justifies its use as a case study to analyse the needs and challenges of prostate cancer patients, survivors and caregivers, discussed in prostate cancer OSG.

\section{Prostate cancer and related OSG usage}~\label{sec:why_prostate_cancer}
Prostate cancer is the development of cancer in the prostate gland, a part of the male reproductive system. It is often slow growing and relatively less fatal compared to other types of cancers, but it may spread into other parts of the body (e.g., bones, lymph nodes). The key incidence factors of prostate cancer are age (mainly on males $>$ 50) and family history.  

\subsection{Incidence}
It is the second most frequent cancer among men (after lung cancer), and fifth leading cancer for mortality among men~\cite{bray2018global}. In 2018 there has been an estimated 1.3 million new incidents  and 359,000 mortalities of prostate cancer across the globe~\cite{bray2018global}. It is the most frequently diagnosed cancer among men in 105 countries (out of 185 countries), including most of the high income countries such as US, UK, Canada, Australia, New Zealand and most of the northern and western Europe~\cite{bray2018global}.  

Despite being one of the most frequent cancers among men, its 5-year survival rate is highest in the developed countries, which is 99\% (2017) in US~\cite{american2018cancer}, 84\% (2010) in UK~\footnote{https://www.cancerresearchuk.org/health-professional/cancer-statistics/statistics-by-cancer-type/prostate-cancer/survival}, and 95\% (2014) Australia\footnote{https://prostate-cancer.canceraustralia.gov.au/statistics}. This increased survival is mainly due to the introduction of early detection  schemes such as abnormal prostate-specific antigen (PSA) testing~\cite{kvaale2007interpreting} so that cancer can be removed or contained during early stages of cancer. In US it is estimated that there are over 3.3 million men with a history of prostate cancer~\cite{desantis2016cancer}. 

\subsection{Diagnose}
PSA and Gleason score are the key determinants of the prostate cancer stage. PSA which stands for \textit{prostate specific antigen} is a protein made in the prostate gland. Increase of PSA level than normal in the blood is an indication of having prostate cancer, although such an effect can be due to various reasons, which may often lead to misdiagnose (false negatives). Due to the simplicity of the procedure, PSA test is often used as a routine test in older adults where abnormalities were directed to further tests. 

Gleason grading system measures the development of cancer by analysing the cell morphology of cancer cells and grading them 1-5 where 5 being most aggressive~\cite{gleason1974prediction}. The final score is determined by combining the score of the most common cell pattern and non-dominant cell pattern with the highest grade. The final score ranges between 2-10 which is indicative of the growth of cancer where 10 being the most aggressive. 

\subsection{Treatment and side-effects}\label{sec:prostate_cancer_treatment}
There are three major treatment options prostate cancer:
\begin{enumerate}
	\item \textbf{Surgery:}  a surgical procedure known as radical prostatectomy that completely removes the prostate gland. It is the most widely used treatment for prostate cancer and often administrated as open surgery, laparoscopically or robotic-assisted.   
	\item \textbf{Radiation therapy:}  using ionizing radiation destroy the malignant cells of cancer, it uses as a primary treatment method or as a secondary treatment following a surgery to remove any remaining cancer cells. 
	\item \textbf{Active surveillance:} for men with a less aggressive cancer (low-risk) or older individuals who have other serious conditions that prevent them from a curative treatment. It involves continues monitoring of cancer for its progression with the intent of switching to a curative treatment if cancer becomes aggressive.
\end{enumerate} 

As these treatments lead to different outcomes in terms of curing (or containing cancer) as well as the subsequent treatment related side-effects~\cite{Albertsen2016,cooperberg2015trends}. Therefore, the patients are eager to chose the best treatment for them, which is often selected based on multiple factors such as the stage of the cancer, age, other medical conditions, financial conditions as well as the personal preference of the individual~\cite{Berry2003,xu2011men}. In fact, there have been reports of decision regrets among the post-treatment survivors~\cite{clark2001living,Gwede2005} mainly attributed to treatment related side-effects. This complexity makes patients to seek multiple opinions on the treatment options from clinicians, family, friends and other prostate cancer survivors~\cite{Denberg2006,Berry2003}. Also, patients acquire more information by reading relevant books, magazines and web pages~\cite{Berry2003} as well as interacting online with other individuals in OSG~\cite{Huber2011,ihrig2011treatment}. 

The treatments of prostate cancer come with chronic side effects which are mainly urinary, sexual and bowel impairments. Among them urinary side effects such as urinary incontinence and sexual side-effect erectile dysfunction are intimate in nature, thus patients are reluctant or rather embarrassed to self-disclose and talks about these issues with their physicians, family and friends~\cite{weber2000exploring}. This issue is more prevalent among men as they are less eager to seek support for social and psychological issues~\cite{wang2013men,Lintz2003} and discuss intimate health issues with physician or peers~\cite{Lintz2003,weber2000exploring}. In contrast, online social media platforms such as OSG provide patients with the ability to communicate under a pseudo name without revealing their true identity. As discussed in Section~\ref{sec:self_disclosure} this anonymous nature leads to \textit{online disinhibition effect}~\cite{suler2004online} where individuals are comfortable in self-disclose more frequently and intensely in online communication in contrast to face-to-face communication. Therefore, prostate cancer patients (or survivors) are more comfortable and likely to use OSG to discuss their intimate side-effect related problems~\cite{nanton2018men}.  

\subsection{Prostate cancer OSG}
The discussions related to prostate cancer happens in OSG either dedicated for prostate cancer discussions (e.g, Prostatecanceruk), or in generic OSG with a specific section for prostate cancer  (e.g, Healingwell sub-forum on Prostate Cancer). 

The above discussed high incidence rates and the high survival rates globally result a significantly large population of prostate cancer patients, survivors as well as caregivers of prostate cancer patients participating in prostate cancer OSG. Especially given that prostate cancer is prominent in the developed countries where most of the patient are frequent users of internet and online social media platforms. 
Therefore, a large volume of prostate cancer related discussions are accumulated in OSG platforms. 

The participants use prostate cancer OSG at four key stages of their cancer journey which are diagnosis, pre-treatment, post-treatment side-effects and recurrence. Those stages of cancer are being identified as the most in need of psychological support~\cite{Weis2003} due to significant transitions of the quality of life as well as the significant decisions to be made during those stages. The participants use OSG discussions to express their concerns and receive information and emotional support on how to cope with the situation. This process result accumulation of the information as well as the emotional state of the individuals during those stages.  

\begin{figure}[!htb]
	\centering
	\includegraphics[clip=true, width=1.0\linewidth]{{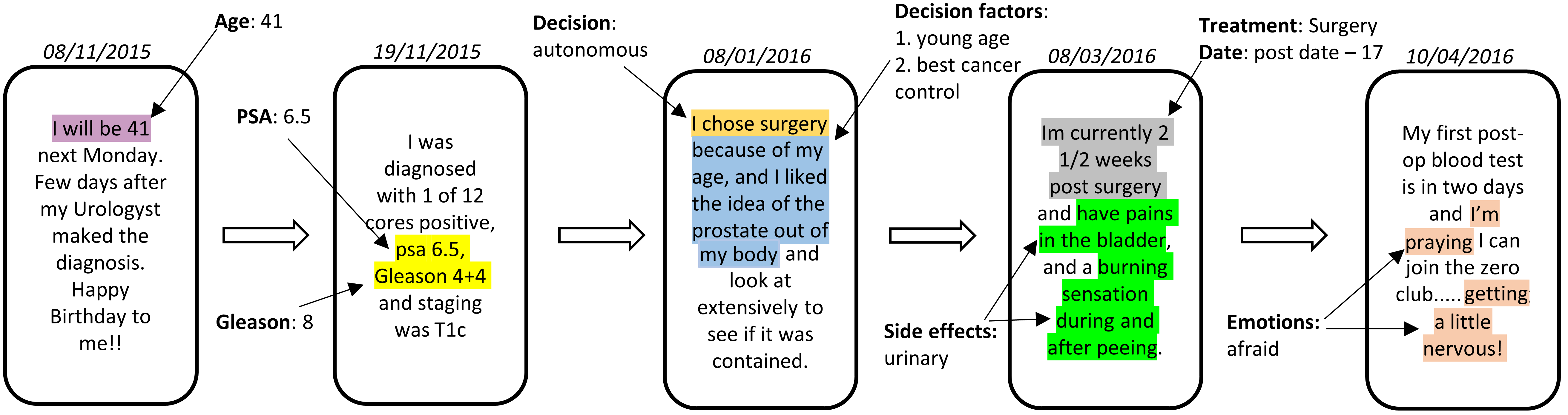}}
	\caption{An anonymised sample (parts are omitted and rephrased to preserve privacy) of five posts by a prostate cancer patient. The highlighted excerpts are demographic, clinical, emotion expressions and decision making process related information, stated in the form of free-text.}
	\label{fig:prostate_osg_user_posts}
\end{figure}

Figure~\ref{fig:prostate_osg_user_posts} presents an anonymised sample of five OSG posts by a patient, from diagnosis of cancer to four months post-surgery. 

Figure~\ref{fig:prostate_osg_user_posts} demonstrates the wealth of implicit information contained within OSG posts. Patients begin by self-disclosing demographic and clinical information, followed by their decision-making process, relevant decision factors and emotions, in order to seek validation from other patients~\cite{Jayles2017,Mishra2013}. The timeline of clinical and emotion information is implicit in the time-stamp of the post and in addition often explicitly mentioned in the post content. However, this entire body of information is encapsulated within large volumes of unstructured text data which lacks a domain-specific structure required for investigation or intervention and support by primary care providers~\cite{Murdoch2013}. 

In nutshell, the prostate cancer OSG are voluminous due to its scale of occurrence and high survival rate, as well as information rich due to the unique set of characteristics of the prostate cancer patient trajectory. The next section presents the Patient reported information multidimensional exploration platform (PRIME) which is an extended version of the proposed  information structuring framework in Section~\ref{sec:proposed_framework} to analyse prostate cancer specific OSG discussions.

\section[PRIME]{PRIME: Patient reported information multidimensional exploration platform}~\label{sec:prime}

This section presents the proposed Patient Reported Information Multidimensional Exploration (PRIME) platform which is developed to automatically analyse a large collection of cancer related OSG posts and extract demographic, clinical and emotional factors with their associated temporality. These factors were employed for automated investigation of patient behaviours, clinical factors and patient emotions, across the temporalities of diagnosis, treatment and recovery. More specifically, we focus on the automated multi-granular extraction, analysis, classification and aggregation of decision-making behaviours, decision factors, the temporality of patient interactions, the temporality of clinical information and side effects, and trajectory of positive and negative emotions, in the context of decision groups, demographics and treatment type. 

\begin{figure}[!htb]
	\centering
	\includegraphics[clip=true, width=1.0\linewidth]{{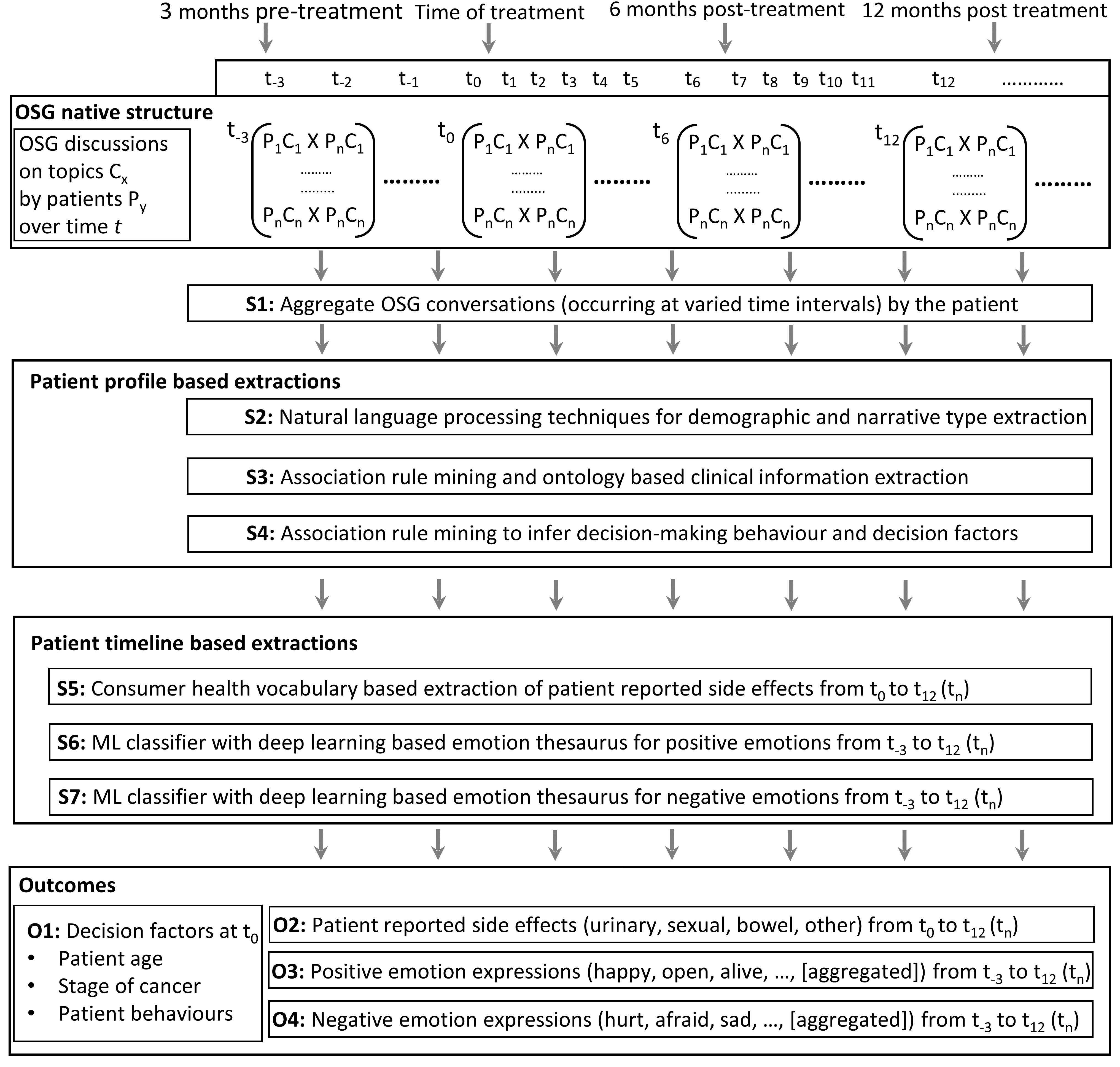}}
	\caption{Patient Reported Information Multidimensional Exploration (PRIME) platform, which contains a suite of machine learning and natural language processing based information extraction modules to process cancer related OSG data.}
	\label{fig:PEAP}
\end{figure} 

As shown in Figure~\ref{fig:PEAP} the PRIME framework functions in seven \textit{Stages~S1-S7} to capture multiple layers of information from free-text OSG posts. As discussed in Section~\ref{sec:OSG_intro}, OSG are organised as discussion threads where each thread is initiated by a user (often a patient, partner of a patient or a caregiver) with a particular question or health concern. Other participants respond to that with answers, suggestions, expression of their own experience or a follow-up question for further information. This process creates a discussion thread in OSG.

An OSG comprises a large number of discussions where patients contribute their decisions, experiences and opinions at different stages of their cancer journey from diagnosis to post-treatment and potential recurrence of cancer. Hence OSG contain a multitude of granular and aggregate information on patient behaviours, side effects and emotion expressions over time. 

\subsection{Aggregating OSG posts by the patient} 

The patient behaviour in an OSG over time can be represented as $t_x\:(P_1C_1\:\times\: P_yC_z)$, where $t_x$ denotes time, $P_y$ denotes a contributing patient, $C_z$ denotes an OSG thread and $P_yC_z$ denotes an OSG post by patient $P_y$ as a contribution to the discussion in thread $C_z$. In this naturally occurring order of OSG discussions, the posts by a single patient $P_y$ are scattered over multiple discussions ($C_1,\ldots,C_z$). Hence, as the initial step in the \textit{Stage~1} the OSG conversations by a single patient are collocated and chronologically ordered based on the timestamp.

The timestamp of OSG post depicts its published time. However, as different cancer patients start their cancer journey at different times, published time cannot be used to compare and contrast patients. Therefore, the time of each OSG post is normalised based on a key cancer related event (e.g., diagnosis, treatment) of each patient, which is experimentally set to treatment event as it was found that many patients join OSG close to their treatment. Note that the cancer events and their temporalities are automatically extracted during the construction of patient timeline which will be discussed later in \textit{Stage3}. 

\subsection{Patient profile based extractions}
\textit{Stages~2-4} extracts various aspects of the patient profile such as demographics, clinical information and treatment decision making. 
 
\subsubsection{Capturing demographics}
\textit{Stage~2} is based on natural language processing and machine learning based information retrieval techniques presented in Section~\ref{sec:information_extraction}, which extracts age, gender and narrative type. Based on the narrative type being \textit{first-person} or \textit{second-person} the role of the participant has being identified as the patient or caregiver, where female caregivers are predominantly the partners of patients. If the gender cannot be inferred with sufficient confidence, such individuals were marked as male patients as prostate cancer is a male only cancer. Patients are grouped based on age into five age groups ($<$40, 40-50, 50-60, 60-70 and $>$70) to be in line with the standard prostate cancer cohort studies.  

\subsubsection{Capturing clinical information}  
\textit{Stage~3} enriches patient profiles with clinical information, which are important to categorise patients based on their treatment type and the stage of cancer. 

The treatment type that the patient has undergone is often disclosed during the discussions about treatment choice and their respective consequences. As mentioned in Section~\ref{sec:why_prostate_cancer}, multi-modality of treatment options makes the selection of treatment a complex process. Hence, it is important to compare and contrast the outcomes of a patient with different modalities. 

The treatment information is extracted using the clinical event extraction technique (as part of the clinical timeline extraction) presented in Section~\ref{sec:timeline_construction} which is a specified event detection technique (See Section~\ref{sec:specified-event} on specified event detection). It extracts the OSG posts that mention any of the treatment modalities and then determines the temporality of that event based on the explicitly mentioned time (e.g., \lq 2 weeks ago\rq) and the implicit time based on the timestamp of the OSG post. Note that if the patient had undergone multiple treatments during their cancer journey (e.g., surgery and then radiotherapy) the first treatment is considered to represent his modality as it is the primary treatment in his cancer journey. 

In OSG, participants disclose their PSA and Gleason scores as a means of informing their stage of cancer to the OSG community. However, such mentions employ diverse narrative styles. For example, Gleason score is mentioned as \lq gleason\rq, \lq gleson\rq, \lq gleeson\rq, \lq g\rq~ and \lq gs \rq followed by the value in either as a total score (e.g., \lq 7\rq) or as a combination of the two components (e.g., \lq 4+3\rq). In \textit{Stage~3}, association rules and extracts from clinical ontologies are employed to develop regular expressions to identify these mentions of Gleason and PSA and subsequently, capture the numerical details of Gleason and PSA scores.

\subsubsection{Capturing treatment decision making behaviour and decision factors}~\label{sec:patient-decison-making}
\textit{Stage~4} infers the decision making behaviour of each patient and the corresponding decision factors considered for the decision making. As discussed in Section~\ref{sec:why_prostate_cancer}, multiple modalities in prostate cancer treatment process make the treatment decision making a complicated process. Due to this complexity, patients are often eager to get involved in the decision making process to select what is best for them. Information of this decision making process is often posted in OSG with the intention of receiving peer-validation from those who had similar circumstances. 

These decision making behaviours are captured based on the established patient decision making models~\cite{Charles1999,emanuel1992four,veatch1972models} which highlights different types of decision making behaviours depending on the degree of involvement by clinician and patient. The three prominent categorises of the decision making behaviours are as follows:
\begin{enumerate}
	\item \textbf{Paternalistic}: also known as \textit{priestly} which is the traditional behaviour in clinical decision making where clinician assumes a dominant role in making the decision based on his expertise and experience. In this approach patient delegates the decision making authority on the treatment to the clinician and provide consent to the recommendations of the clinician, assuming that the clinician would use his expertise and experience to consider all potential possibilities and select the treatment that is the best option for the patient. 
	\item \textbf{Autonomous}: also known as \textit{informative} or \textit{consumer model} is the decision making behaviour where the patient assumes the dominant role. In this behaviour, the clinicians provide the relevant information, let the patient make the decision, and subsequently executes the suggested treatment option. In this approach, it is assumed that the patient with sufficient knowledge is capable of selecting the treatment option that best suits his condition.
	\item \textbf{Shared}: also known as \textit{deliberative} is the decision making behaviour where both patient and the clinician jointly work to find the best treatment options by weighing all options using multiple aspects such as clinical, social and financial.     
\end{enumerate}  

\noindent These decision making behaviours are reflected in self-disclosed statements that announce the treatment decision of the individual (see third OSG post in Figure~\ref{fig:prostate_osg_user_posts}). In order to capture such statements, a set of template patterns was engineered to capture sentences that describe that either individual has taken the decision (Autonomous) or the treatment option was recommended by a clinician (Paternalistic). The template patterns are as follows:
\begin{itemize}
	\item Autonomous template ($T_A$): \verb|<I/We> <words>* <DECIDE> <words>* <TREATMENT>|
	\item Paternalistic template ($T_P$): \verb|<DOCTOR> <words>* <RECOMMEND> <words>* <TREATMENT>|
\end{itemize}

Note that \verb|<words>*| denotes zero or multiple words in-between, and upper case terms are \textit{template terms} which consider a set of synonym terms (word or phrases). Table~\ref{table:decision_making_terms} shows a selected sample of terms employed as candidates for each template term.   

\begin{table}[!htb]
\caption{A sample of candidate terms for each template term used in the template sentences of decision making behaviour. Note that some of the terms are abbreviations or shorten versions of the actual terms or phrases.}
\label{table:decision_making_terms}
	\centering
	\begin{tabulary}{\linewidth}{|>{\raggedright}p{2.3cm}|>{\raggedright\arraybackslash}p{11.8cm}|}\hline 
		Template term & Sample candidate terms\\\hline
		\verb|DECIDE| & decided, chosen, wind up going, made the call, settled, opted, went for, took the option, end up\\\hline
		\verb|RECOMMEND| & recommend, recommended, prescribe, prescribed, advised, advise, endorse, endorsed, advocate\\\hline
		\verb|DOCTOR| & doctor, doc, surgeon, urologist, uro, specialist, consultant, radiologist, oncologist, radiotherapist\\\hline
		\multirow{3}{*}{\verb|TREATMENT|} & surgery, davinci, da vinci, robotic, prostatectomy, ralp, rrp, lrp, rpp, key hole, open op\\\cline{2-2}
		 & radiation, imrt, brachytherapy, radiotherapy, seed therapy, brachy, seed implant, ebrt\\\cline{2-2}
		 & surveillance, AS, watch and wait\\\hline
	\end{tabulary}
\end{table}

These template sentences from both templates $T_A$ and $T_P$ were used to capture the matching sentences in the OSG posts of each individual. The proportion  $P(T_X) = T_X/(T_A + T_P)$ (where $X$ is $A$ or $P$) appeared in the OSG posts of a user is used to determine the decision making behaviour of that user as follows:
\begin{enumerate}
	\item \textbf{Paternalistic}: if $P(T_P) >= 0.25$ and $P(T_A) < 0.25$
	\item \textbf{Autonomous}: if $P(T_P) < 0.25$ and $P(T_A) >= 0.25$
	\item \textbf{Shared}: if $P(T_P) >= 0.25$ and $P(T_A) >= 0.25$
\end{enumerate}
For example, if a user only has $T_P$ or $T_A$, his decision making behaviour is categorised as Paternalistic or Autonomous respectively. However, if he has both $T_P$ and $T_A$ significantly ($>=0.25$), then his decision making behaviour is Shared.

Multiple decision factors, both clinical/non-clinical affect the treatment decision-making process. As shown in the third OSG post of Figure~\ref{fig:prostate_osg_user_posts}, patients often mention underlying decision factors alongside with the mentions of the treatment decision often as a way of justifying their decision to the OSG community and seeking peer-validation. 

The decision factors were searched in the OSG posts which are identified from the above technique as containing sentences related to decision making and captured using a thesaurus of consumer health terms. Initially, a list of common decision-making factors related to prostate cancer was created based on existing literature~\cite{Berry2003,Gwede2005,Huber2011,Huber2017,Ihrig2011} and further validated by clinicians. This list includes medical concerns such as side effects, doctor skills, and best cancer control as well as socio-demographic reasons such as age, fast recovery and financial concerns. Once the decision factors were identified, a  thesaurus of consumer health terms was developed by analysing a set of sample OSG posts identified from the above technique as containing information related to treatment decision making. 

\subsection{Patient timeline based extractions}
\textit{Stage~5} onwards, PRIME framework incorporates the time dimension of OSG discussions and patient interactions. The patient timeline extraction technique presented in Section~\ref{sec:timeline_construction} is adapted with prostate cancer specifics to automatically generated a clinical event and emotion timeline of each individual based on the self-disclosed side effects captured in \textit{Stage~5} and positive/negative emotions captured in \textit{Stages~6-7}. 

\textit{Stage~5} captures the self-disclosure of side effects and grouped them into four key categories: \textit{urinary}, \textit{sexual}, \textit{bowel} and \textit{other} in which \textit{urinary}, \textit{sexual}, and \textit{bowel} are the key side effect categories of prostate cancer treatments~\cite{Hamdy2016,Donovan2016}, and \textit{other} represents the miscellaneous side effects arises due to prostate cancer and its treatments.

The relevant side-effects for each category were selected with the support of clinicians and a thesaurus of relevant terms (words/phrases) for each side-effect were engineered based on the medical concepts found in the UMLS Metathesaurus~\cite{Bodenreider2004}. However as discussed in Section~\ref{sec:medical_concept_extraction}, health consumers (patients and caregivers) discuss side effects using everyday layman language (e.g., urinary incontinence described as leakage, leak, drip). Therefore, the above thesaurus is extended by adding consumer health terms identified using a semi-supervised approach where a candidate set of terms were identified using a word-embedding~\cite{mikolov_word2vec} created from prostate cancer OSG and further validated by a team of clinical experts. This developed thesaurus is employed to capture any mentions of an occurrence of side effects and map such mentions to the timeline based on the associated timestamp.  

In \textit{Stages 5-6} the positive and negative emotions were captured using the emotion timeline generation technique presented in Section~\ref{sec:emotion_timeline}. It provides a 16-dimensional real-valued emotion vector for each OSG post which included 8 positive and 8 negative emotions extracted based on the psychological emotional model Emotion Wheel~\cite{plutchik1980emotion,plutchik1991emotions}. The value of each emotion is a representation of the strength of that emotion expressed in the respective OSG post.

Each patient timeline is time-normalised based on the treatment month captured in \textit{Stage 4} as $t_0$. The events (side effects and emotions) are aggregated monthly based on the reported timestamp, and the timeline is generated from three months pre-treatment ($t_{-3}$) to 12 months post-treatment ($t_{12}$) based on the available information.

The PRIME platform has been implemented using JAVA language and available at \url{https://github.com/tharindurb/PRIME}. 

The following sections empirically evaluate the PRIME platform using a large corpus of prostate cancer related OSG discussions collected from 10 active OSG. The evaluation is carried as two case studies where the first case-study evaluates the emotional expression of those who have undergone treatment for low-intermediate risk prostate cancer and second case study evaluates decision making behaviours and their associated decision factors and emotional expression.

\subfile{Chapter6_Results}

\section{Chapter Summary}
This chapter extends the multi-stage information structuring platform (presented in previous chapter) to capture insights specific to prostate cancer related online support groups (OSG). It structures and transforms prostate cancer related online support group discussions into a multidimensional representation based on demographics, emotions, clinical factors. This representation is employed to investigate the self disclosed quality of life against time, demographics and clinical factors. Moreover, it was used to assess different decision making behaviours and decision factors related to treatment decision making and their associated emotions over time pre- and post-treatment. These investigations have provided insights on the emotional expression of different groups and highlight several highly emotionally expressive groups as more in need of support. Also, differences in decision making behaviours and decision factors across distinct decision making groups. These insights help to shape-up optimum delivery of necessary care to the vulnerable prostate cancer patients/survivors and caregivers.

\onlyinsubfile{	
\bibliographystyle{dcu}
\bibliography{library}{}
}

%% file: Chapter6_Results.tex
\section{Prostate related OSG data collection}~\label{sec:prostate_data}
The social data related to online prostate cancer discussions were collected from high volume active OSG on prostate cancer discussions. An OSG is considered active if it has at least 100 new conversations per week. From these active OSG, conversations were selected using the specific topic \lq prostate cancer\rq.  Table~\ref{table:prostate_osg_stats} presents the ten OSG selected for the data collection and their URLs. Note that, some of these OSG are dedicated for prostate cancer discussions (e.g, Prostatecanceruk), while others are generic OSG with a specific section for prostate cancer discussions (e.g, Healingwell sub-forum on Prostate Cancer). 

The identified discussions were extracted using a set of web scrapers developed for each OSG platform. The collected dataset contains 609,960 conversations from 22,233 OSG users, comprising a text corpus of 93,606,581 word tokens. Table~\ref{table:prostate_osg_stats} shows the number of users and number of posts collected from each OSG platform.  

\begin{table}[!htb]
\caption{Population and volume statistics of the ten selected OSG on prostate cancer discussions.}
\label{table:prostate_osg_stats}
	\centering
	\begin{tabulary}{\linewidth}{|p{4.4cm}|L|R|R|}\hline 
		OSG & URL & Users & Posts \\\hline
		Healingwell & healingwell.com/community & 6,829 & 401,325\\
		Cancer Survivors Network & csn.cancer.org/forum &   2,775 & 33,166\\
		Cancerforums & cancerforums.net & 2,761 & 52,840\\
		Cancercompass & cancercompass.com & 2,524 & 12,141\\
		Healthboards & healthboards.com/boards & 1,938 & 17,144\\
		Patientinfo & patient.info/forums & 1,727 & 33,304 \\
		Macmillanuk & community.macmillan.org.uk & 1,178 & 9,210\\
		Prostatecanceruk & community.prostatecanceruk.org & 1,119 & 28,041\\
		Prostatecancerinfolink & prostatecancerinfolink.ning.com/forum & 862 & 7,187\\
		Ustoo & inspire.com/groups/us-too-prostate-cancer & 519 & 15,602\\\hline
	\end{tabulary}
\end{table}

\subsection{Ethical considerations}
This study has been conducted under the ethics approval from La Trobe University Research Ethics Committee. All patient-reported data used in this study are non-identifying and publicly available from the corresponding OSG. The OSG do not provide access to identifying information of patients, and PRIME does not process any identifying information. This work only publishes aggregates of the analysed data, which cannot be reverse engineered using any means for any form of re-identification. 

\section{Analysis of the quality of life of prostate cancer patients}~\label{sec:prostate_cancer_casestdy1}
As discussed in Section~\ref{sec:why_prostate_cancer}, the treatment of low-intermediate risk clinically localised PCa is becoming increasingly complex due to comparable cure rates of the available treatment options; radical prostatectomy (RP), external beam radiation therapy (EBRT), and active surveillance (AS). Therefore, in addition to the tumour characteristics, significant emphasis is  placed on customising the side effect profiles of each treatment option to the patient, based on their preference, anxiety levels and experience of the medical professional~\cite{Zeliadt2006}.

Recent studies~\cite{Barocas2017,Chen2017,Donovan2016} compare patient quality of life (QoL) in men randomised to AS, RP and EBRT demonstrating that QoL post-treatment mirrored reported changes in function, yet no significant differences were observed among the groups in measures of anxiety, depression, and general health-related or cancer-related QoL. While study settings are controlled for many variables, they are dependent on questionnaires completed in a \textit{trial setting} and may not accurately capture \textit{real-life} issues experienced by patients in diverse circumstances undergoing different treatment options for localised prostate cancer~\cite{Bowling1995,Carr2001}. 

In contrast, the complex phenomenon as QoL is effectively communicated as free flowing text containing expressions of emotions instead of predetermined responses in a questionnaire~\cite{Carr2001,Seale2010}. A complementary approach to conventional studies  is provided in this case study which employs the emotional expression extracted from PRIME as a proxy to patient reported QoL, and compare/contrast the expression of emotions across the three modalities and other demographic factors. Note that the \textit{time} factor is not considered in this study, thus, side-effects and emotions captured in the patient timeline are aggregated. 

\subsection{Inclusion criteria}
Following the completion of the automated intelligent extraction of demographic, clinical information and expressions of emotion, inclusion criteria set in order to focus on the analysis of a specific cohort. Thereby, this study focuses on those who have undergone treatment (RP, EBRT, AS) for low-intermediate risk PCa. Low risk PCa (Gleason $\leq$6) and intermediate risk PCa (Gleason 7) according to D\rq Amico risk classification~\cite{DAmico1998}. This inclusion criteria is applied to all the individuals with known modality and Gleason/PSA scores, which resulted in a participant cohort of 6,084 individuals (27.3\% of the total profiles extracted). 

\subsection{Participant characteristics}
The selected participant cohort has RP as the highest modality with 4,241 (69.7\%) participants, followed by 1,528 (25.1\%) EBRT and 315 (5.2\%) AS. Table~\ref{table:patient_characteristics} presents demographic and clinical information of participants (patient or partners of patients) in total and across each modality. The percentages in each column are determined relative to the total participants in each cohort (total or each modality). The \textit{p-value} presented is from chi-square tests that evaluated the statistical significance across the three modalities.

\begin{table}[!htb]
\caption{Demographics and clinical characteristics of the identified low-intermediate risk prostate cancer patients and partners of patients from the ten selected OCSGs. Note that, the percentages in each column is determined relative to the total participants in each cohort (total or each modality). The \textit{p-value} presented in the last column is from chi-square tests that evaluated the statistical significance across the three modalities.}
\label{table:patient_characteristics}
\centering
\begin{tabulary}{\linewidth}{|L|R|R|R|R|R|}\hline 
& RP  &  EBRT  &  AS  &  Total &	\\\hline
&n (\% in RP)&n (\% in EBRT)&n (\% in AS)&n (\% in Total)&p value\\\hline
Total participants  &  4241   &  1528  &  315   &  6084   &  \\\hline
\textit{Gleason score}  &    &    &    &    &  \\\hline
$<=$6  &  2543 (60)  &  831 (54)  &  269 (85)  &  3643 (60)  &  $<$0.001 \\\hline
7  &  1698 (40)  &  697 (46)  &  46 (15)  &  2441 (40)  &  $<$0.001 \\\hline
&    &    &    &    &  \\\hline
Median PSA  &  6  &  7.8  &  6.5  &    &  \\\hline
&    &    &    &    &   \\\hline
\textit{Age}   &    &    &    &    &   \\\hline
$<$40  &  110 (3)  &  37 (2)  &  8 (2)  &  155 (3)  &  0.937 \\\hline
41-50  &  688 (16)  &  144 (9)  &  32 (10)  &  864 (14)  &  $<$0.001 \\\hline
51-60  &  1638 (39)  &  411 (27)  &  89 (28)  &  2138 (35)  &  $<$0.001 \\\hline
61-70  &  933 (22)  &  456 (30)  &  91 (29)  &  1480 (24)  &  $<$0.001 \\\hline
$>$70  &  197 (5)  &  192 (13)  &  39 (12)  &  428 (7)  &  $<$0.001 \\\hline
Unknown  &    &    &    &  1019 (17)  &  \\\hline
&    &    &    &    &  \\\hline
\textit{Participant role} &    &    &    &    &  \\\hline
Patient  &  3702 (87)  &  1319 (86)  &  288 (91)  &  5309 (87)  &  0.676 \\\hline
Partner  &  539 (13)  &  209 (14)  &  27 (9)  &  775 (13)  &  0.069 \\\hline
&    &    &    &    & \\\hline 
\textit{Side effects} &    &    &    &    &  \\\hline
Urinary symptoms  &  2229 (53)  &  625 (41)  &  75 (24)  &  2929 (48)  &  $<$0.01 \\\hline
Bladder Irritation  &  804 (19)  &  407 (27)  &  32 (10)  &  1243 (20)  &  $<$0.01 \\\hline
Bladder neck contracture  &  586 (14)  &  196 (13)  &  17 (5)  &  799 (13)  &  $<$0.01 \\\hline
Urinary incontinence  &  1953 (46)  &  390 (25)  &  41 (13)  &  2384 (39)  &  $<$0.01 \\\hline
Urethral strictures  &  312 (7)  &  108 (7)  &  11 (3)  &  431 (7)  &  0.045 \\\hline
Bowel symptoms  &  84 (2)  &  98 (6)  &  1 (0)  &  183 (3)  &  $<$0.01 \\\hline
Rectal irritation  &  33 (1)  &  42 (3)  &  0 (0)  &  75 (1)  &  $<$0.01 \\\hline
Rectal bleeding  &  66 (2)  &  72 (5)  &  1 (0)  &  139 (2.28)  &  $<$0.01 \\\hline
Sexual symptoms  &  1679 (40)  &  424 (28)  &  40 (13)  &  2143 (35)  &  $<$0.01 \\\hline
Erectile dysfunction  &  1679 (40)  &  424 (28)  &  40 (13)  &  2143 (35)  &  $<$0.01 \\\hline
Other  &  579 (14)  &  254 (17)  &  14 (4)  &  847 (14)  &  $<$0.01 \\\hline
Bleeding  &  453 (11)  &  202 (13)  &  14 (4)  &  669 (11)  &  $<$0.01 \\\hline
Clot  &  221 (5)  &  66 (4)  &  3 (1)  &  290 (5)  &  0.002 \\\hline
Hernia  &  497 (12)  &  115 (7)  &  14 (4)  &  626 (10)  &  $<$0.01 \\\hline
Infection  &  796 (19)  &  297 (19)  &  40 (13)  &  1133 (19)  &  0.038 \\\hline
Stroke  &  119 (3)  &  54 (3)  &  3 (1)  &  176 (3)  &  0.041 \\\hline
Tiredness  &  411 (10)  &  333 (22)  &  5 (2)  &  749 (12)  &  $<$0.01 \\\hline
\end{tabulary}
\end{table}

Table~\ref{table:patient_characteristics} presents a comparison of different participant characteristics across the three modalities. Among the participants who selected AS, 85\% have a Gleason score of less than 6, while only 15\% have 7. This is because AS is preferred for patients who are \textit{low-risk} (have a slow growing cancer). The age distribution among EBRT and AS are skewed towards older participants as some are not suitable for surgical procedures in RP due to their old age. The partner cohort across the modalities are around 10\% while AS has lowest of 9\%.  

Side effect results show that those undergoing RP had comparatively high urinary symptoms and sexual side effects compared to the other groups, while those who had EBRT had comparatively high bowel symptoms. These findings on side effects are corroborated by key outcomes from a large randomised controlled study and two large population based prospective cohort studies~\cite{Barocas2017, Chen2017, Donovan2016}. 

\subsection{Emotions by age}
\begin{figure}[!htb]
	\centering
	\includegraphics[clip=true, width=1.0\linewidth]{{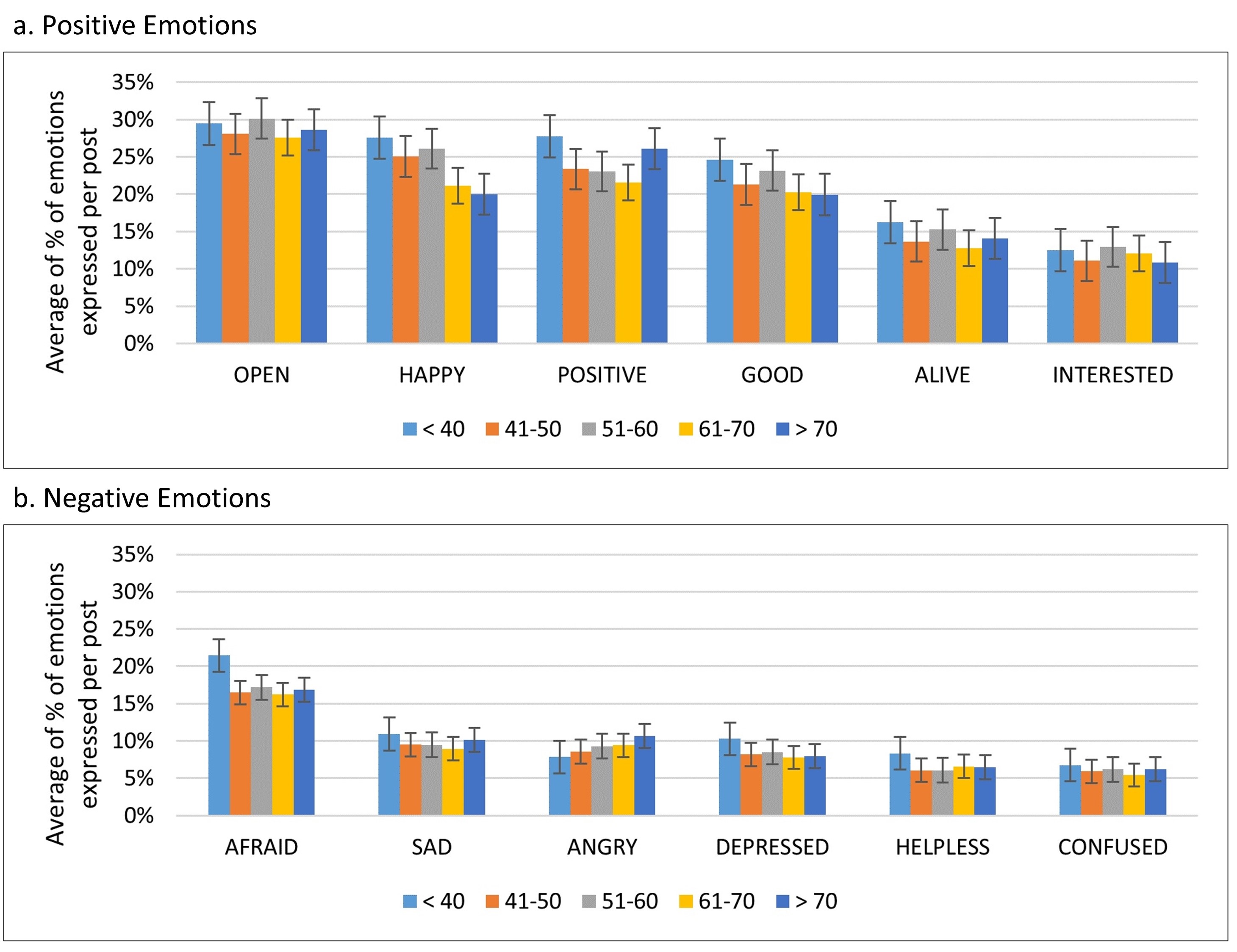}}
	\caption{Positive and negative emotions by age group.}
	\label{fig:Figure_emotion_age}
\end{figure}

Figure~\ref{fig:Figure_emotion_age}a depicts the positive emotion categories of five key age groups. The categories; \textit{open}, \textit{happy}, \textit{positive} and \textit{good} are consistently close to 20\% whereas \textit{alive} is close to 15\% and \textit{interested} is close to 10\%. The cohort aged $<$40 has on or above average emotion levels across all positive emotions.

Figure~\ref{fig:Figure_emotion_age}b presents the negative emotions of the same age groups. In general, negative emotions are less expressed compared to positive emotions. The expression of emotion \textit{afraid} is significantly high for individuals aged $<$40 which is 10\% higher than other age groups. Also, this group is close to 5\% higher than others in \textit{depressed} and \textit{helpless}. However, the emotion \textit{angry} is less expressed by $<$40 group in comparison to all other groups. The remaining two negative emotions \textit{sad} and \textit{confused} are expressed consistently across all age groups. 

\subsection{Emotions by modality}
\begin{figure}[!htb]
	\centering
	\includegraphics[clip=true, width=1.0\linewidth]{{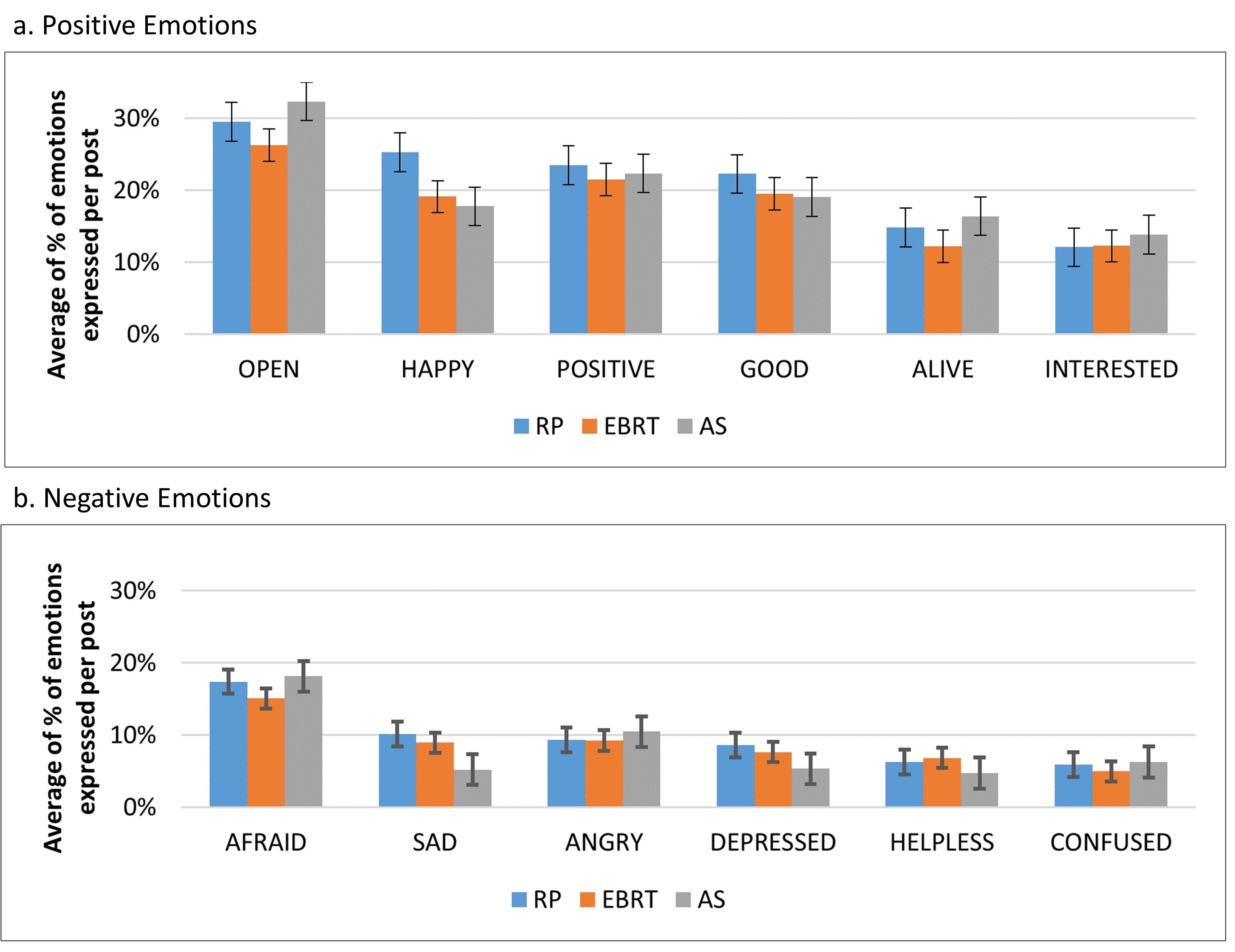}}
	\caption{Positive and negative emotions by the treatment modality (RP, EBRT and AS).}
	\label{fig:Figure_emotion_modality}
\end{figure}

Figure~\ref{fig:Figure_emotion_modality} presents positive and negative emotions by modality. The positive emotion \textit{open} is the most significant, while \textit{alive} and \textit{interested} are the least significant. 

Figure~\ref{fig:Figure_emotion_modality}b depicts negative emotions by modality and \textit{afraid} is the most significant with an on average of 10\% higher across all modalities. Noteworthy fluctuations are observed for AS; where \textit{sad}, \textit{depressed} and \textit{helpless} are significantly low compared to the other two modalities.  


\subsection{Emotions by patient and partner}
\begin{figure}[!htb]
	\centering
	\includegraphics[clip=true, width=1.0\linewidth]{{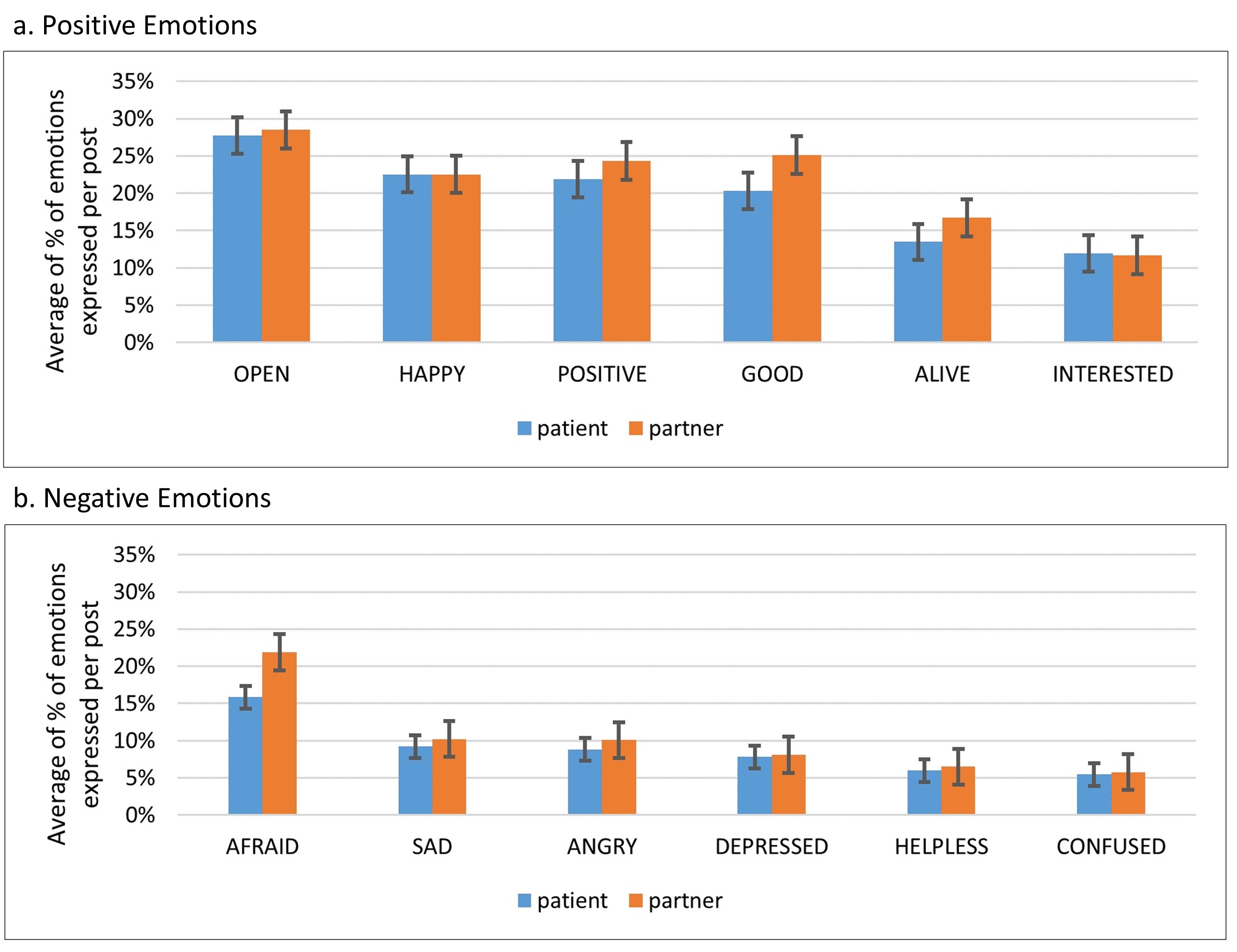}}
	\caption{Positive and negative emotions by patient and partner.}
	\label{fig:Figure_emotion_gender}
\end{figure}

Figure~\ref{fig:Figure_emotion_gender} presents positive and negative emotions by patients and partners. Both groups were more \textit{open}, and less \textit{alive} and \textit{interested}. Partners were slightly more \textit{positive}, \textit{good} and \textit{alive}. On average, partners have expressed more negative emotions than patients, where \textit{afraid} is expressed 7\% more than patients. 

Figure~\ref{fig:Figure_emotion_gender}b presents partner negative emotions by treatment modality. Partners of participants under active surveillance are more \textit{afraid}, more \textit{angry}, and less \textit{hurt} than other two modalities. 

\subsection{Significant emotion terms} 
Table~\ref{table:frequent_emotion_terms} shows the frequently used emotion terms by the three selected cohorts ($<$40, $>$70, partners), which were determined based on the percentage increase of frequency of each emotion term in the selected cohort compared to all the participants. 

\begin{table}[!htb]
\caption{Frequently-used emotion terms for three selected cohorts: (i) patients aged $<$40, (ii) patients aged $>$70 and (iii) partners of patients.}\label{table:frequent_emotion_terms}
\centering
\begin{tabulary}{\linewidth}{|>{\raggedright}p{3.2cm}|p{5.2cm}|p{5.2cm}|}\hline 
Group & Terms expressing positive emotions  &  Terms expressing negative emotions\\\hline
Patients aged $<$40 (30\% above baseline)  &  merry, lucky, intent, challenged, content, affected, loved, peaceful, brave, determined, pleased, tender, secure, reassured, wonderful  &  sore, fatigued, nervous, suspicious, frustrated, alarmed, bad, worried, frightened, scared\\\hline
Patients aged $>$70 (10\% above baseline)  &  reliable, accepting, touched, brave, courageous, satisfied, optimistic  &  unpleasant, threatened, frightened, fatigued\\\hline
Partners of patients (20\% above baseline)  &  quiet, spirited, reassured, loved, calm, admiration, brave, amazed, kind  &  tearful, upset, frightened, scared, guilty, worried, miserable \\\hline
\end{tabulary}
\end{table}


Patients aged $<$40 have expressed significantly high positive and negative emotions compared to other age groups. Aligning with studies that report high levels of psychological distress in young cancer patients~\cite{Compas1999,Mosher2005}, this group demonstrates significantly high negative emotions and emotion terms (Table~\ref{table:frequent_emotion_terms}).  Positive categories (\textit{happy}, \textit{positive}) and positive emotion terms (Table~\ref{table:frequent_emotion_terms}) are also high in this group. This is noted as post-traumatic growth in trials on other types of cancer patients~\cite{Blank2008,Manne,Pudrovska2010}. As such, this cohort appears to be the ideal beneficiaries of OSG and could gain from additional health care resources and discussions focused on treatment decision making and cultivating positivity. 

Emotions of patients and partners differ significantly in several categories. Partners express higher levels of \textit{afraid}, \textit{positive}, these differences can be explained by (1) female partners being emotionally expressive than males~\cite{Kring1998}, thereby actively use OSG to share emotional experiences~\cite{Seale2006} and (2) caregiving partners report more emotional distress and anxiety than the patient~\cite{Given2001,Northouse2007}. Negative terms significantly used by partners e.g., \lq tearful\rq, \lq frightened\rq, \lq miserable\rq~ further confirms this. While the cancer care process is directed towards the patient with regular interactions with healthcare professionals and support groups; partner\rq s burden of caregiving and anxiety of potentially losing their loved one, are often overlooked, highlighting the need for better support for partners of patients. 

This study reports that significantly higher and more negative emotions were expressed in younger age cohorts and partners of patients who may benefit from increased psychological support from healthcare providers. 

\section{Analysis of decision making behaviour and decision factors}~\label{sec:prostate_cancer_casestdy2}
As discussed in Section~\ref{sec:prostate_cancer_treatment}, treatment decision making in prostate cancer is a complex process due to the availability of multiple treatment options and comparable outcomes. Therefore, the selection of treatment is not only based on clinical factors such as the stage of cancer, age, other medical conditions, but also other factors such as financial conditions and personal preference of the individual~\cite{Berry2003,xu2011men}. Moreover, as discussed in Section~\ref{sec:patient-decison-making}, over the last few decades patients have assumed a more dominant role in decision making deviating from clinician dominated \textit{paternalistic} behaviour to a more patient involved \textit{autonomous} and \textit{shared} decision making behaviours. 

The PRIME framework has captured the patient decision making behaviours and the associated decision factors mentioned in the OSG by the OSG participants, which enables a large scale analysis of the decision making behaviours and key decision factors across different cohorts based on demographic and clinical factors.  

\subsection{Inclusion criteria}
In this case study, the inclusion criteria is set to included patients those who self-disclosed their primary PCa treatment and discussed the decision-making process that led to the particular treatment. A total of 6,457 participants matched this criteria which is 29.0\% of the total profiles extracted. Among them 420 (6.5\%) have shown \textit{paternalistic}, 3883 (60.1\%) \textit{autonomous} and 2154 (33.4\%) \textit{shared} decision making behaviours. This distribution shows that OSG participants predominantly get involved in the treatment decision making process (\textit{autonomous} or \textit{shared}). Note that there may be a representative bias on \textit{paternalistic} individuals, as they are more reliant on clinical sources and comparatively less likely to participate in OSG discussions. 

\subsection{Decision making behaviour groups}
\begin{figure}[!htb]
	\centering
	\includegraphics[clip=true, width=1.0\linewidth]{{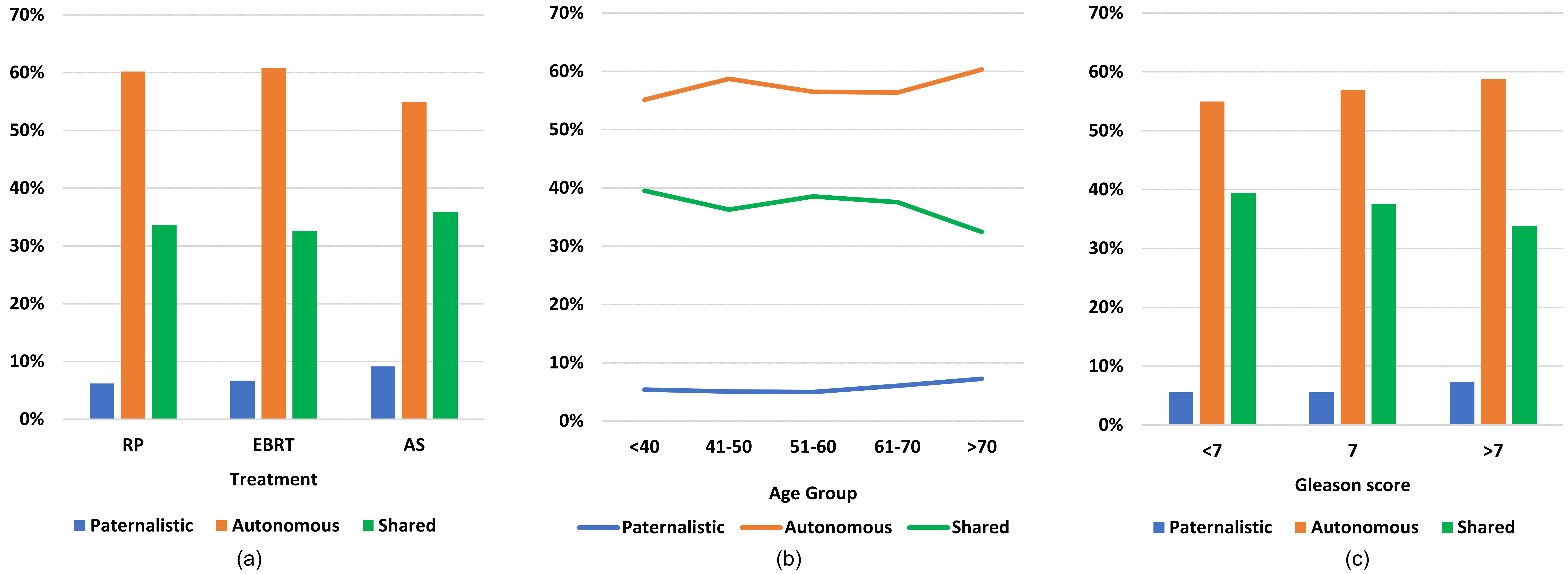}}
	\caption{The distribution of decision making behaviour groups across: (a) Treatment modality, (b) Age group, and (c) Gleason score. Note that the percentage value for a behaviour group in a particular cohort is determined using the total (across the three behaviour groups) in each cohort as the denominator.}
	\label{fig:Figure_decision_behaviour_1}
\end{figure}

Figure~\ref{fig:Figure_decision_behaviour_1} presents a comparison of the distribution of decision behaviour groups across treatment modality, age group and Gleason score. The three Gleason score groups ($<7$,7,$>7$) are based on D\rq Amico risk classification~\cite{DAmico1998} as low-risk, intermediate-risk and high-risk respectively. 

The results highlight that the \textit{autonomous} group dominates each cohort, followed by the \textit{shared} group leaving less than 10\% for the \textit{paternalistic} group. This shows that patient involvement in decision making is significant across all the cohorts. Active Surveillance (AS) has the lowest percentage on the \textit{autonomous} group, but highest in  \textit{paternalistic} and \textit{shared}. The \textit{paternalistic} group has a fairly uniform distribution across the age groups. The percentage of the \textit{shared} group is highest among $<$40 and the \textit{autonomous} group is highest among $>70$, which shows that young individuals are more keen to jointly work with the clinician to make the treatment decision, while older individuals are more keen to take the decision themselves. The \textit{shared} group is highest among Gleason $<$7 (low-risk) group. They may be more keen for shared decision making as they often have all three treatment options. 

\begin{figure}[!htb]
	\centering
	\includegraphics[clip=true, width=0.7\linewidth]{{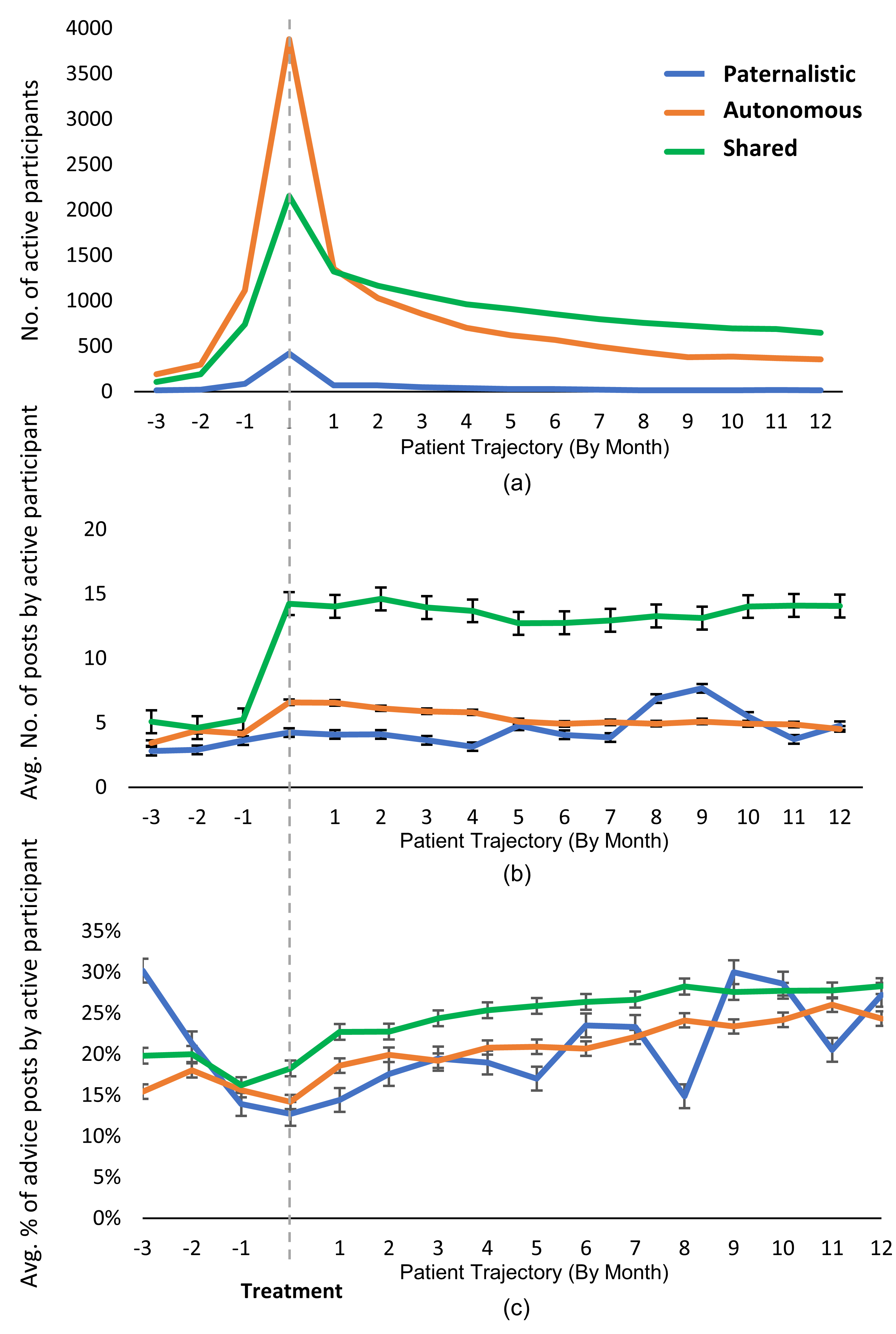}}
	\caption{The OSG participation activities by the three decision making behaviour groups over the patient trajectory from 3 months pre-treatment to 12 months post-treatment. Note that an active participant during a particular month is an individual who has participated in OSG discussions during that month. The OSG posts are classified as an expression of experience or providing advice by the classifier presented in Section~\ref{sec:narrative_type_extraction}.}
	\label{fig:Figure_decision_behaviour_2}
\end{figure}

The decision making behaviour groups exhibit different behaviours in OSG activities over time in their patient trajectory such as actively participating to OSG discussions and type of activity (e.g., seeking support or providing support). The Figure~\ref{fig:Figure_decision_behaviour_2} presents insights on these participant behaviours over time for the decision making behaviour groups.         

Figure~\ref{fig:Figure_decision_behaviour_2}a shows the number of active participants over time which highlights that the participation peaked for all three groups during the month of treatment decision-making which is an indication of active information seeking by all groups. \textit{Paternalistic} and \textit{autonomous} groups reduce activity soon afterwards, but the \textit{shared} group more consistently participated in OSG discussions throughout the next 12 months. This trend is reflected in Figure~\ref{fig:Figure_decision_behaviour_2}b as well, where the average number of posts by an active participant is significantly high for the \textit{shared} group. Figure~\ref{fig:Figure_decision_behaviour_2}c shows that in all groups the proportion of \textit{advice} posts increases over time which shows that the participants gradually become support providers to others who need. However, the shared group with a consistently high proportion of \textit{advice} is shown to be more collaborative in OSG discussions. This transformation of the initial information/support seeker gradually becoming an information/support provider happens as the participants gain more relevant knowledge and wisdom (often by receiving support from peers), which is very important for the ecosystem in any support group as it provides the continues supply of information and support for those who need~\cite{Ziebland2012}.

\subsection{Decision factors}
It is important to understand the decision factors that influenced the treatment decisions of the individuals. ~\textit{Stage 4} of PRIME delineated in Section~\ref{sec:patient-decison-making} extracted these self-disclosed decision factors alongside the selected modality and the decision making behaviour. This section analyse such decision factors against the selected modality and decision making behaviour shown by the participant. 

Table~\ref{table:decision_factors_vs_modality} present the discussed decision factors against the selected modality, and Table~\ref{table:decision_factors_vs_decision_behaviour} present the discussed decision factors against decision making behaviour. The \textit{p-value} in the last column of both tables denotes the statistical significance of the mentioned decision factors among the cohorts.    

\begin{table}[!htb]
\caption{The decision factors employed for the  treatment decision against the primary treatment modality of the patient.}\label{table:decision_factors_vs_modality}
\centering
\begin{tabulary}{\linewidth}{|L|R|R|R|R|R|}\hline 
&  RP  &  EBRT  &  AS  &  Total  &  \multirow{2}{*}{p value}\\
&n (\% in RP)&n (\% in EBRT)&n (\% in AS)&n (\% in Total)&\\\hline
Doctor skill  discussed &  2617 (66.76)  &  1392 (61.78)  &  190 (66.9)  &  4199 (65.03)  &  0.06 \\\hline
Surgeon mentioned &  1975 (50.38)  &  880 (39.06)  &  141 (49.65)  &  2996 (46.4)  &  $<$0.01 \\\hline
Radiologist mentioned &  371 (9.46)  &  488 (21.66)  &  27 (9.51)  &  886 (13.72)  &  $<$0.01 \\\hline
GP mentioned  &  142 (3.62)  &  75 (3.33)  &  11 (3.87)  &  228 (3.53)  &  0.80 \\\hline
Medical Indication  &  377 (9.62)  &  162 (7.19)  &  16 (5.63)  &  555 (8.6)  &  $<$0.01 \\\hline
Nerve  mentioned&  675 (17.22)  &  124 (5.5)  &  10 (3.52)  &  809 (12.53)  &  $<$0.01 \\\hline
Young age &  272 (6.94)  &  93 (4.13)  &  16 (5.63)  &  381 (5.9)  &  $<$0.01 \\\hline
Bowel symptoms  &  13 (0.33)  &  46 (2.04)  &  1 (0.35)  &  60 (0.93)  &  $<$0.01 \\\hline
Urinary incontinence  &  624 (15.92)  &  196 (8.7)  &  8 (2.82)  &  828 (12.82)  &  $<$0.01 \\\hline
Erectile dysfunction  &  535 (13.65)  &  151 (6.7)  &  14 (4.93)  &  700 (10.84)  &  $<$0.01\\\hline
Recovery time  &  214 (5.46)  &  50 (2.22)  &  0 (0)  &  264 (4.09)  &  $<$0.01 \\\hline
Best cancer Control  &  611 (15.59)  &  295 (13.09)  &  34 (11.97)  &  940 (14.56)  &  0.02 \\\hline
Financial concerns &  336 (8.57)  &  179 (7.94)  &  17 (5.99)  &  532 (8.24)  &  0.29 \\\hline
Less invasive  &  117 (2.98)  &  31 (1.38)  &  1 (0.35)  &  149 (2.31)  &  $<$0.01 \\\hline
\end{tabulary}
\end{table}

As shown in Table~\ref{table:decision_factors_vs_modality} \lq doctors skill\rq~ has been the most discussed decision factor and equally important across all the modalities. Similarly, best cancer control and financial concerns are equally discussed across all modalities reflecting they are equally important to all modalities. RP and AS groups talk about the surgeon more often while EBRT group is more interested on the radiologist. In terms of side effects preserving nerves, urinary incontinence and erectile dysfunction are mostly discussed by the RP group as they are the prominent side-effects of surgical procedures. EBRT group discuss bowel symptoms more often as they are the prominent side effects for radiotherapy. Young age has been given as a reason by mostly RP group followed by AS, since both are the most preferred by young patients.     

\begin{table}[!htb]
\caption{The decision factors employed for the  treatment decision against the decision making behaviour group.}\label{table:decision_factors_vs_decision_behaviour}
	\centering
	\begin{tabulary}{\linewidth}{|L|R|R|R|R|R|}\hline 
 &  Paternalistic  &  Autonomous  &  Shared  &  Total  &  p value \\\hline
 &n (\% in Paternalistic)&n (\% in Autonomous)&n (\% in Shared)&n (\% in Total)&\\\hline
Doctor skill discussed&  200 (47.62)  &  2162 (55.68)  &  1837 (85.28)  &  4199 (65.03)  &  $<$0.01 \\\hline
Surgeon mentioned &  219 (52.14)  &  1047 (26.96)  &  1730 (80.32)  &  2996 (46.4)  & $<$0.01 \\\hline
Radiologist mentioned &  51 (12.14)  &  186 (4.79)  &  649 (30.13)  &  886 (13.72)  & $<$0.01 \\\hline
GP mentioned  &  12 (2.86)  &  51 (1.31)  &  165 (7.66)  &  228 (3.53)  & $<$0.01 \\\hline
Medical indication  &  8 (1.9)  &  187 (4.82)  &  360 (16.71)  &  555 (8.6)  & $<$0.01 \\\hline
Nerve mentioned &  20 (4.76)  &  299 (7.7)  &  490 (22.75)  &  809 (12.53)  & $<$0.01 \\\hline
Young age  &  6 (1.43)  &  110 (2.83)  &  265 (12.3)  &  381 (5.9)  & $<$0.01 \\\hline
Bowel symptoms  &  0 (0)  &  18 (0.46)  &  42 (1.95)  &  60 (0.93)  & $<$0.01 \\\hline
Urinary incontinence  &  20 (4.76)  &  301 (7.75)  &  507 (23.54)  &  828 (12.82)  & $<$0.01 \\\hline
Erectile dysfunction  &  13 (3.1)  &  205 (5.28)  &  482 (22.38)  &  700 (10.84)  &$<$0.01\\\hline
Recovery time  &  3 (0.71)  &  77 (1.98)  &  184 (8.54)  &  264 (4.09)  & $<$0.01\\\hline
Best Cancer Control  &  22 (5.24)  &  331 (8.52)  &  587 (27.25)  &  940 (14.56)  & $<$0.01\\\hline
Financial  &  10 (2.38)  &  136 (3.5)  &  386 (17.92)  &  532 (8.24)  & $<$0.01\\\hline
Less Invasive  &  2 (0.48)  &  42 (1.08)  &  105 (4.87)  &  149 (2.31)  & $<$0.01 \\\hline
\end{tabulary}
\end{table}

The results in Table~\ref{table:decision_factors_vs_decision_behaviour} shows that the shared groups overwhelmingly discussed all the decision factors than the other two groups (p $<$ 0.001). This is because the \textit{shared} group actively utilised the OSG discussions in treatment decision making process to weight the available options with their outcomes. Thus, they tend to talk more about the decision factors. In contrast, paternalistic and autonomous groups either less comprehensive on decision factors or may not tend to engage on OSG discussions about the considered decision factors.

When comparing the autonomous and paternalistic groups, it can be seen that the paternalistic group has high mention rate of clinicians (surgeon, radiologist, and GP) than the autonomous group, as they tend to discuss more about clinicians that took the decision on behalf of them. In contrast, the autonomous group discuss relatively more on side effects aspects (e.g., nerve mentions, urinary, bowel and sexual symptoms) than the paternalistic groups as they tend to discuss about their decision factors more than the paternalistic group. 

The results generated the PRIME framework strongly correlate with the existing studies on decision making behaviour~\cite{Charles1999} and decision factors~\cite{Berry2003,Gwede2005}. All groups actively sought information on OSG: The Shared group provided consistent, prolonged interactions and contributed more as support providers. The Autonomous group only sought advice mainly around the treatment and subsequently contributed minimally to conversations on OSG. 

\subsection{Patient timeline analysis}
It is important to analyse the patient timeline to understand the temporalities of the clinical events and emotions of the different behaviour groups. Such analysis enables to compare and contrast the characteristics of the behaviour groups.
 
\begin{figure}[!htb]
	\centering
	\includegraphics[clip=true, width=0.7\linewidth]{{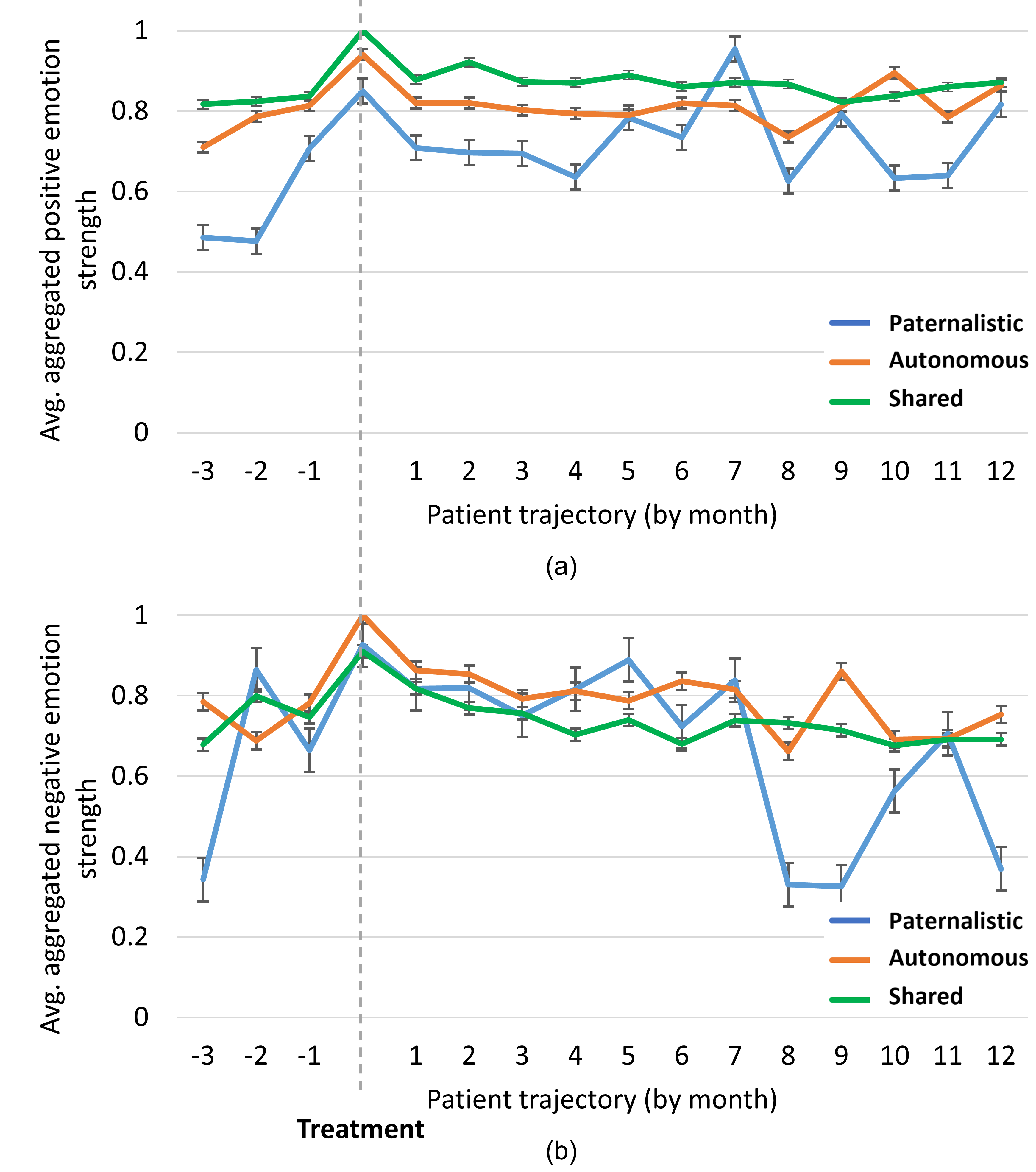}}
	\caption{The positive and negative emotion trajectories of the three decision making behaviour groups over the patient trajectory from 3 months pre-treatment to 12 months post-treatment.}
	\label{fig:Figure_decision_behaviour_emotion_trajectory}
\end{figure}

Figure~\ref{fig:Figure_decision_behaviour_emotion_trajectory} presents the aggregated positive and negative emotions timelines of the three decision behaviour groups. Note that the aggregated emotions are obtained by accumulating all positive emotions (e.g., \textit{open}, \textit{alive}) into an aggregated positive emotion and accumulating all negative emotions (e.g., \textit{afraid}, \textit{hurt}) into an aggregated negative emotion.   

As shown in Figure~\ref{fig:Figure_decision_behaviour_emotion_trajectory} both positive and negative emotions peaked at the treatment month, which due to the elevated emotional state associated with a traumatic event like surgery. The \textit{shared} group is the most positive and \textit{autonomous} group is the most negative during the treatment month. After the treatment, there is a rapid drop in positive and negative emotions across all groups which stabilised in subsequent months. However, the \textit{shared} group consistently expressive in positive emotions than the other two groups. The paternalistic group have shown high variance in the latter part of the timeline in both positive and negative emotions. It seems to be due to the drop of active \textit{paternalistic} participants as shown in Figure~\ref{fig:Figure_decision_behaviour_2}a resulting in the averaged emotions values impacted by a few emotional individuals.   

\begin{figure}[!htb]
	\centering
	\includegraphics[clip=true, width=0.8\linewidth]{{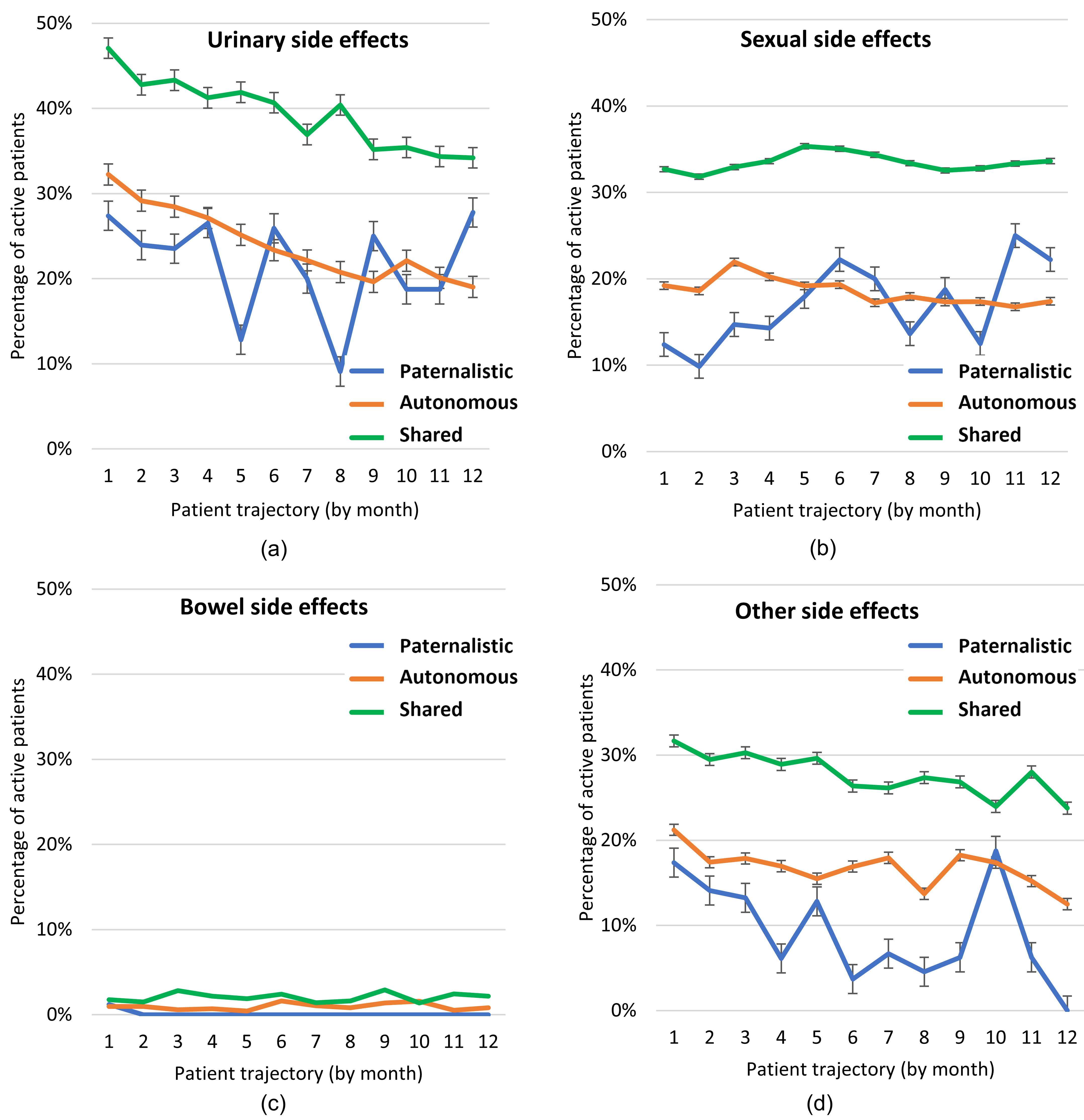}}
	\caption{The self-disclosed side effect timeline aggregated monthly across the three decision making behaviour groups from 1st month post-treatment to 12 months post-treatment.}
	\label{fig:Figure_decision_behaviour_side-effect_trajectory}
\end{figure}

Although treatment decision making does not impact the occurrence of treatment related side-effects, it impacts the awareness and acceptance of the treatment related side-effects. Figure~\ref{fig:Figure_decision_behaviour_side-effect_trajectory} presents the self-disclosed side effect timeline aggregated monthly across the three decision making behaviour groups. It highlights that \textit{shared} group reports more on side effects than the other two groups across all side effect categories. \textit{Shared} and \textit{autonomous} groups are initially reported more urinary side effects (Figure~\ref{fig:Figure_decision_behaviour_side-effect_trajectory}a) which gradually declined over time, in contrast to the \textit{paternalistic} group which has fluctuations over time. \textit{shared} and \textit{autonomous} groups are consistently reported sexual side effects (Figure~\ref{fig:Figure_decision_behaviour_side-effect_trajectory}b) while \textit{paternalistic} group shows an increased reporting over time, reaching a level equivalent to that of the \textit{autonomous} group by month 12. The bowel side effects (Figure~\ref{fig:Figure_decision_behaviour_side-effect_trajectory}c) are the least mentioned across all groups as they are relatively less prominent. This significantly more reporting of side-effects by the \textit{shared} group is mainly due to their openness to discuss such issues and high collaborativeness in OSG. Similarly, the \textit{paternalistic} group is less open and collaborative to discuss side-effects in OSG. However, it is interesting to observe the increase of the report of sexual side effects by the \textit{paternalistic} group over time. It may indicate that over time they become more open about the chronic sexual side effects and use OSG discussion to seek emotional support from OSG as a coping mechanism.         

These results highlight the characteristics of different decision making behaviours, the decision factors considered, reported side effects and the emotions by the individuals. The multiplicity of the decision factors including both clinical, psychological and financial aspects highlights the complex nature of the treatment decision making which has deviated from strictly clinical outcome based to a more holistic approach. The \textit{shared} group has shown to discuss significantly more decision factors indicating that they are more aware of decision factors and effectively use OSG discussions in the decision making process. The \textit{shared} group reported more side effects. However, they have demonstrated a better quality of life (increased positive and decreased negative emotions) over time more than the other two groups. These characteristics of the shared group align with two coping strategies of traumatic events such as cancer. First is being more informed about the expected outcomes leads to an informed decision that develops acceptance to the outcomes of the treatment. The second is, discussing about a traumatic event among individuals with similar or worse outcomes leads to positive social comparisons (see Section~\ref{sec:social_comparison}) highlighting that the individual is not alone reduces the emotional distress~\cite{taylor1989social}. 

These results also highlight the importance of the shared decision-making. It is argued that patients should be provided necessary tools to gather information, know their decision options, scenarios and consequences for shared decision-making to be effective~\cite{elwyn2012shared}. The significance of emotional support that allows patients to freely express values and preferences and ask questions without clinician obstruction is also highlighted~\cite{Stacey2011}. The proliferation of OSG is a clear indication that patients and carers are bridging this gap by seeking (and providing) this service extraneous to healthcare providers and institutions. Further, OSG provide information, decision options and emotional support with the added advantage of a geographically dispersed community of individuals who are undergoing/have undergone similar circumstances. 

\section{Discussion}~\label{sec:chap6_discussion}
The above two case studies have shown the wealth of information encapsulated in OSG discussions. The first case-study investigated the positive and negative emotional expression of different groups and highlights several highly emotionally expressive groups as more in need of support. The second case-study explored different decision making behaviours and evaluated their decision factors, emotional expression and report of side-effects; and identified that \textit{shared} group reports better emotional expression compared which could be attributed to their extensive consideration of decision factors and active role in decision making. Also, \textit{shared} groups is pivotal for the continuation of OSG as they are more likely to stay longer with the OSG and become information/support providers over time. These insights help to shape-up optimum delivery of necessary care to the vulnerable prostate cancer patients/survivors and caregivers.       

With the increasing presence of social media empowered patients, OSG make a paramount contribution as a medium for discussions on information exchange and emotional support. This is seen to be instrumental in addressing the \lq out of sight out of mind\rq~dilemma that arises due to periodic or occasional clinician consultations during the cancer journey. The participants who have undergone similar circumstances (e.g., similar treatment) are willing to share their experience, offer advice and emotional support to those who seek in OSG. Hence, the medical support network for care in any type of cancer  should look into relevant OSG platforms in order to understand the unmet needs and provide optimal and individualised care that is clinically appropriate for patients with cancer.

As explicated in this study, the PRIME platform provides this required functionality to automatically analyse OSG discussions and transform into a multi-dimensional information source which can be explored to understand the unmet needs of prostate cancer. PRIME can be succinctly extended into any form of cancer by adapting the clinical information extraction layers of the platform to incorporate the clinical information of the respective cancer.


%% file: Chapter7.tex
\onlyinsubfile{	
	\setcounter{chapter}{6} 
}
	
\chapter[]{Conclusion}\label{chap:7}

\epigraph{{\textit{Begin thus from the first act, and proceed; and, in conclusion, at the ill which thou hast done, be troubled, and rejoice for the good.\\}}{\hfill Pythagoras}}

This chapter concludes the thesis by presenting a summary of the research contributions, as well as directions for future work. Section~\ref{sec:summary_contributions} provides a summary of conceptual, technical and application contributions, Section~\ref{sec:address_research_questions} describes how those contributions have addressed the key research questions, and Section~\ref{sec:future_directions} concludes this chapter by providing future research directions arising from this thesis.
 
\section{Summary of contributions}\label{sec:summary_contributions}
This thesis is an exploration to develop techniques that can be used to gain deeper insights from social data, which has yield theoretical, technical and application contributions.

As the conceptual contribution, a comprehensive literature study has been conducted on current state-of-the-art machine learning and natural language processing techniques employed to generate insights from social data in Chapter~\ref{chap:2}. This study is focused around three types of insights often captured from social data which are topics, events and emotions. Section~\ref{sec:topic_extraction} investigates current techniques used for topic extraction which include topics modelling and text clustering. Section~\ref{sec:event_detection} assessed two types of event detection techniques: specified event detection techniques to capture known event types using event related taxonomies and unspecified event detection techniques to capture unknown/unforeseen event types using various change indicators. Section~\ref{sec:emotion_extraction_from_text} first reviews theories emotions and emotional models developed in social sciences and subsequently review how those models are incorporated into computational  techniques to extract emotions  from text. Finally, Section~\ref{sec:social_data_analysis_challenges} provides an exhaustive analysis of the challenges and limitations of existing techniques in relation to gaining insights from social data.

As the second theoretical contribution, this thesis developed a layered conceptual framework for interpreting data generated in online social media platforms a.k.a. social data. This framework is presented in Chapter~\ref{chap:3}. It is built upon existing theories of cognition, social needs and social behaviour, which depicts that social data is generated by social interactions happening in online social media platforms, and such interactions reflect different social behaviours driven by different social needs. This conceptual framework provides a deeper meaning to social data as archived traces of online social behaviours rather than just a corpus of unstructured text. In addition, four key social behaviours were conceptualised based on existing literature and a comprehensive comparison is provided comparing and contrasting exhibition of each behaviour in the physical world and online social media platforms.   

On the technical front, two suites of algorithms were developed to harness insights from different types of online social media platforms. First is two unsupervised incremental learning algorithms presented in Chapter~\ref{chap:4}. These algorithms were designed to capture insights from fast-paced social data streams. The first is a new unsupervised incremental machine learning algorithm that automatically structures a social data stream into topics and extends those across time into topic pathways. This algorithm is developed based on GSOM self-structuring algorithm and IKASL incremental learning algorithm. It captures changes in topics over time as well as distinct new topics as new topic pathways at different points of time. Also, it was designed to overcome the challenges present in social data with respect to its brevity, unstructuredness, and diversity. The second algorithm is a multi faceted event detection algorithm developed to monitor topic pathways for significant changes in online social behaviours over time, and identify such changes as potential events of interest. The changes in social behaviours were identified using automatic event indicators such as changes in volume, positive sentiment and negative sentiment. These algorithms are presented in the journal publication entitled \textit{Automatic event detection in microblogs using incremental machine learning}~\cite{Bandaragoda2017}.

The second technical contribution is another a set of algorithms presented in Chapter~\ref{chap:5}, built with a suite of machine learning and natural language processing algorithms developed to structure slow-paced social data streams. These algorithms are specifically developed for online support groups which are social discussions related to health issues by patients and caregivers. These algorithms extract demographics, deeper emotions and clinical events from users self-disclosed information encapsulated in unstructured text. Subsequently, the extracted information is further structured by formulating patient event timelines based on clinical events and emotions with their associated time. This extracted information is used to build a multi-dimensional database that can be used to query relevant and reliable information for different use cases. Moreover, multiple aggregates of information can be obtained for research purposes. These algorithms are presented in journal publications \textit{Text mining for personalized knowledge extraction from online support groups}~\cite{Bandaragoda2018} and \textit{Machine learning to support social media empowered patients in cancer care and cancer treatment decisions}~\cite{de2018machine}.      

As application contributions, the above developed algorithms have been successfully trialled with social data streams from two online social media platforms. The suites of algorithms developed for fast-paced text streams has been tested with two Twitter datasets containing 6 months of tweets on two entities-- a politician and an organisation. As shown in the experiment results in Section~\ref{sec:chap4_experiments}, the topic pathway separation algorithm successfully captured the key topic pathways representing ongoing discussions related to the respective entities. Also, shifts in the discussions represented by new key terms were successfully learned and associated with the relevant topic pathway. Moreover, new distinct topics were automatically captured as new topics pathways. The event detection algorithm monitors those topic pathways and automatically captured significant changes in human behaviour using changes in volume and sentiment. Those captured events were aligned with contemporary news articles that discussed those events. These results are published in the journal publication entitled \textit{Automatic event detection in microblogs using incremental machine learning}~\cite{Bandaragoda2017}.

The second application contribution is based on the second suite of algorithms developed to structure slow-paced social data streams. It was applied to a large text corpus collected from two large online support groups (OSG) which consist of over 4 million posts. It successfully structures those posts pivoted by user id, automatically building user profiles and user timelines from the extracted information. The use of these algorithms is demonstrated using two uses cases where first shows how it can be used by patient/caregivers to look for relevant and reliable information based on different criteria, better than the standard methods available. Also, the researcher use case demonstrated that researchers can use this structure to capture aggregated information on different study cohorts which can be used to generate population level insights. These results are published in the journal publication entitled \textit{Text mining for personalized knowledge extraction from online support groups}~\cite{Bandaragoda2018}.

The third application contribution is presented in Chapter~\ref{chap:6}, which is an extension to the researcher use case discussed above. In this application, the platform developed to structure slow-paced social data stream is extended to capture insights specific to prostate cancer related online support groups (OSG). It structures and transforms prostate cancer related online support group discussions into a multidimensional representation based on demographics, emotions,  clinical factors and investigated the self disclosed quality of life against time, demographics and clinical factors. Moreover, it assessed different decision making behaviours and decision factors related to treatment decision making and their associated emotions over time pre- and post-treatment. These investigations have provided insights on the emotional expression of different groups and highlight several highly emotionally expressive groups as more in need of support. Also, differences in decision making behaviours and decision factors across distinct decision making groups. These insights help to shape-up optimum delivery of necessary care to the vulnerable prostate cancer patients/survivors and caregivers. These investigations were carried out in collaboration with researchers from a key urology institute and the results were published in three journals papers: (i) \textit{The patient-reported information multidimensional exploration (PRIME) framework for investigating emotions and other factors of prostate cancer patients with low intermediate risk based on online cancer support group discussions}~\cite{Bandaragoda2018a}, (ii) \textit{Machine learning to support social media empowered patients in cancer care and cancer treatment decisions}~\cite{de2018machine}, and (iii) \textit{Robotic-assisted vs. open radical prostatectomy: A machine learning framework for intelligent analysis of patient-reported outcomes from online cancer support groups}~\cite{Ranasinghe2018}.           

\section{Addressing the research questions}\label{sec:address_research_questions}
This section describes how the above contributions have addressed the research questions delineates in Chapter~\ref{chap:1}. 

\begin{enumerate}
\item \textit{What are the limitations of existing artificial intelligence algorithms and natural language processing techniques in the study of social interactions and social behaviours using representative online social data in digital environments?} 

This research question investigates applications of existing techniques on social data and consolidates key challenges and limitations faced in such applications. Existing machine learning and natural language processing techniques applied to social data is assessed in Chapter~\ref{chap:2}. This assessment captures key applications such as topic extraction, event detection, emotion extraction, self-structuring and incremental learning. It discussed different supervised and unsupervised state-of-the-art techniques and what are their limitations when applied to social data. Subsequently section~\ref{sec:social_data_analysis_challenges} consolidates the key challenges of social data leading to the limitations in existing techniques. 

\item \textit{How can theories of social behaviour from social sciences contribute towards a conceptual model of enhanced understanding of social interactions in digital environments, as well as the representative online social data?} 

This research question is formulated to investigate how existing social theories on human social behaviour, social needs and cognition can be applied to social data. In order to address this research question, a conceptual framework is developed in Section~\ref{sec:conceptual_framework_social_data} based on existing theories in social sciences to consider social data as representing the surface layer of a hierarchy of human social behaviour, social needs and cognition. This model assumes social data is generated as a result of social interactions that happens in online social media platforms, social behaviours are abstractions of social interactions, and cognition drives social interactions to achieve social needs. The Section~\ref{sec:conceptual_framework_social_data} further explains four key behaviours that present in most of the social interactions and compared those behaviours in natural and digital environments. Based on this conceptual framework a platform is proposed in Section~\ref{sec:proposed_platform} which provides structure to unstructured social data for machine processing and also as a means for coupling established social theories with the new forms of digital social behaviour and interactions. 

\item \textit{How can new incremental machine learning algorithms, founded on the principles of self-structuring artificial intelligence, address the challenges of using social data to understand social behaviours?}

This research question is formulated to investigate how existing conventional machine learning and natural language processing techniques can be extended using the concepts of ‘self-structuring AI’ and ‘incremental machine learning’ developed to address question 1  can be employed to address the limitations identified from question 2.  Two suites of algorithms specified in the previous section as technical contributions were developed to address this question. Both were developed by extending existing machine learning and natural language processing algorithms using the above concepts to overcome the identified challenges in social data and capture deeper insights. 

The set of algorithms presented in Chapter~\ref{chap:4} is a new platform developed by extending existing unsupervised techniques to self-structure and incrementally learn from fast-paced social data streams. As discussed in previous section, the first algorithm is capable of structuring a social data stream into coherent groups i.e., topics and extend them across time as topic pathways. It is specifically designed for the highly unstructured nature of the text present in fast-paced social data streams and is capable of handling sparse features due to brevity as well as dynamically changing vocabulary due to changes in social interactions. Also, a new multi-faceted event detection algorithm is developed to monitor changes in human behaviours inside topic pathways using multiple changes indicators.

The second suite of algorithms presented in Chapter~\ref{chap:5} is designed to captured deeper insights from slow-paced social data streams. It consists of machine learning and natural language processing algorithms specifically designed to capture self-disclosed information from unstructured text corpus in online support groups. Those algorithms capture self-disclosed demographics, author role (patient, caregiver), clinical events and deeper emotions. The emotion capturing algorithm capture higher granular emotional expression using 16 emotions derived from Plutchik's emotions wheel. Further extensions added to these algorithms in Chapter~\ref{chap:6} where algorithms were developed to capture treatment choice related decision making behaviours and decision factors.

\item \textit{How can the algorithms developed in 3. advanced into technology platforms that deliver actionable insights for societal advancement?} 

This research question is formulated to investigate how the developed algorithms can be employed to develop technology platforms that are capable of generating actionable insights which can be used for societal advancements. These technology platforms were specifically designed to facilitate distinct and contrasting characteristics/challenges of different online social media platforms. As detailed in the previous section as application contributions, this thesis presents three applications where the developed platforms are employed to generate valuable insights. In these applications, large text corpora collected from different online social media platforms were processed using the developed platforms. In the first application, two datasets collected from Twitter are used to extract topic pathways and significant events using the platform developed in Chapter~\ref{chap:4}. The second application is structuring information present as self-disclosed unstructured text in online support groups. This structured information provides relevant and reliable information to other users and aggregates pivoted by different variables to health researchers. Chapter~\ref{chap:6} expands the researcher application into  gain insights from prostate cancer related online support groups.     

\end{enumerate}

\section{Future directions}\label{sec:future_directions}
The platforms and techniques developed in this thesis have addressed numerous problems of gaining insights from social data accumulated in online social media platforms, and have successfully applied to different applications. However, considering the scale of data and the amount of insights encapsulated, gaining insights from social data on underlying social behaviours is still in its infancy. As discussed, swaths of social data accumulated in social media platforms every day and people from all walks of life are rapidly embracing online social media platforms to perform online social interactions. Hence, there exists substantial challenges to overcome and massive potential to gain.    

One of the key avenue for future research is the predictive aspect, where predictive techniques can be integrated into the developed self-learning platforms to predict future dynamics of the human behaviours. These predictions can be individualistic where predictions can be generated for an individual based on different layers of the conceptual model presented in this thesis. For instance, predictions can be generated related to the cognitive process of an individual to predict the state of the mind and transition of emotions overtime based on the past indicators of such transitions captured from social data. Such predictions are invaluable to manage chronic mental disorders such as depression and anxiety which spreads among the society at epidemic levels. A predictive model trained to capture warning signs of mental disorders from social data/social interactions can be deployed to social media platforms to identify individuals who are in need of support. 

Also, predictions can be at group or population level to predict different group dynamics of behaviour. For instance, the event detection technique developed in Chapter~\ref{chap:4} can be extended for event prediction where predictive models can be trained to learn transitions of event indicators to learn whether or when a discussion gains popularity and generate predictions on future popularity. Such predictions are useful in many applications such as in law-enforcement to predict potentially disruptive movements as well as in marketing to predict gain of a marketing campaign.

Another direction for future work is on how to extend this work to a fully automated system that can self-adapt and scale to generate insights from social media relevant to any domain. The self-structuring AI framework presented in Chapter 3 is generalisable to any domain. Also, the unsupervised self-structuring algorithms presented in Chapter 4 and 5 are scalable and applicable to any domain. However, some of the natural language processing and machine learning algorithms developed to gather insights are specifically developed and fine-tuned for the applications presented in the thesis. As future work, this can be extended with a further layer of self-structuring and self-adaptive AI to make the insight generation fully automated. 

\onlyinsubfile{	
\bibliographystyle{dcu}
\bibliography{library}{}
}